\pdfoutput=1
\documentclass[rmp,aps,amssymb,twocolumn,nofootinbib,floatfix]{revtex4}

\usepackage{amsmath,amssymb,bm}
\usepackage{graphics}
\usepackage{graphicx}
\usepackage[colorlinks,linkcolor=blue,citecolor=blue]{hyperref}
\usepackage{dcolumn}
\usepackage{rotating}
\usepackage{multirow}
\usepackage{epsfig}
\usepackage{color}
\usepackage{float}
\usepackage{setspace}
\usepackage{natbib}

\bibpunct{[}{]}{,}{n}{}{;}

\begin{document} 
\bibliographystyle{plainnat}

\title{Spatial Networks}
\author{Marc Barth\'elemy}
\email{marc.barthelemy@cea.fr}
\affiliation{Institut de Physique Th\'eorique, CEA, IPhT CNRS, URA 2306 F-91191 Gif-sur-Yvette
France}

\affiliation{Centre d'Analyse et de Math\'ematique Sociales
(CAMS, UMR 8557 CNRS-EHESS)\\ 
Ecole des Hautes Etudes en Sciences
 Sociales, 54 bd. Raspail, F-75270 Paris Cedex 06, France.}

\begin{abstract}  

  Complex systems are very often organized under the form of networks
  where nodes and edges are embedded in space. Transportation and
  mobility networks, Internet, mobile phone networks, power grids,
  social and contact networks, neural networks, are all examples where
  space is relevant and where topology alone does not contain all the
  information. Characterizing and understanding the structure and the
  evolution of spatial networks is thus crucial for many different
  fields ranging from urbanism to epidemiology. An important
  consequence of space on networks is that there is a cost associated
  to the length of edges which in turn has dramatic effects on the
  topological structure of these networks. We will expose thoroughly
  the current state of our understanding of how the spatial
  constraints affect the structure and properties of these
  networks. We will review the most recent empirical observations and
  the most important models of spatial networks. We will also discuss
  various processes which take place on these spatial networks, such
  as phase transitions, random walks, synchronization, navigation,
  resilience, and disease spread.

\end{abstract}

\pacs{89.75.-k, 89.20.-a, 05.10.-a}
\keywords{Networks, Graphs, spatial properties, statistical physics,
  geography, urban systems}

\maketitle
\tableofcontents

\section{Networks and space}
\label{sec1}

\subsection{Introduction}
\label{sec1A}

Networks (or graphs) were for a long time the subject of many studies
in mathematics, mathematical sociology, computer science and in
quantitative geography. In the case of random networks, the first and
most important model was proposed by Erdos and Renyi
\cite{Erdos:1959,Erdos:1960} at the end of the $1950$s and was at the
basis of most studies until recently. The interest in networks was
however renewed in $1998$ by Watts and Strogatz \cite{Watts:1998} who
extracted stylized facts from real-world networks and proposed a
simple, new model of random networks. This renewal interest was
reinforced after the publication, a year later, of an article by
Albert and Barabasi \cite{Barabasi:1999a} on the existence of strong
degree heterogeneities.  Strong heterogeneities are in sharp contrast
with the random graphs considered so far and the existence of strong
fluctuations in real-world networks triggered a wealth of studies. A
decade later, we can now find many books
\cite{Pastor:2003,Dorogovtsev:2003a,BBVBook:2008,Caldarelli:2008,Newman:2010,Havlin:2010}
and reviews on this subject
\cite{Albert:2002,Dorogovtsev:2002a,Newman:2003b,Boccaletti:2006}. These
books and reviews discuss usually very quickly spatial aspects of
networks. However, for many critical infrastructures, communication or
biological networks, space is relevant: most of the people have their
friends and relatives in their neighborhood, power grids and
transportation networks depend obviously on distance, many
communication network devices have short radio range, the length of
axons in a brain has a cost, and the spread of contagious diseases is
not uniform across territories. In particular, in the important case
of the brain, regions that are spatially closer have a larger
probability of being connected than remote regions as longer axons are
more costly in terms of material and energy
\cite{Bullmore:2009}. Wiring costs depending on distance is thus
certainly an important aspect of brain networks and we can probably
expect spatial networks to be very relevant in this rapidly evolving
topic. Another particularly important example of such a spatial
network is the Internet which is defined as the set of routers linked
by physical cables with different lengths and latency times. More
generally, the distance could be another parameter such as a social
distance measured by salary, socio-professional category differences,
or any quantity which measures the cost associated with the formation
of a link.

All these examples show that these networks have nodes and edges which
are constrained by some geometry and are usually embedded in a two- or
three-dimensional space and this has important effects on their
topological properties and consequently on processes which take place
on them. If there is a cost associated to the edge length, longer
links must be compensated by some advantage such as for example being
connected to a well-connected node--- that is, a hub. The topological
aspects of the network are then correlated to spatial aspects such as
the location of the nodes and the length of edges.

\subsection{Quantitative geography and networks}
\label{sec1B}

Spatial networks were actually the subject a long time ago of many
studies in quantitative geography. Objects of studies in geography are
locations, activities, flows of individuals and goods, and already in
the $1970$s the scientists working in quantitative geography focused
on networks evolving in time and space. One can consult for example
books such as the remarkably modern `Networks analysis in Geography'
by Haggett and Chorley (published in $1969$ \cite{Haggett:1969}) or
`Models in Geography' \cite{Chorley:1967} to realize that many modern
questions in the complex system field are actually at least $40$ years
old. In these books, the authors discuss the importance of space in
the formation and evolution of networks. They develop tools to
characterize spatial networks and discuss possible models. Maybe what
was lacking at that time, were datasets of large networks and larger
computer capabilities, but a lot of interesting thoughts can be found
in these early studies. Most of the important problems such as the
location of nodes of a network, the evolution of transportation
networks and their interaction with population and activity density
are addressed in these earlier studies, but many important points
still remain unclear and will certainly benefit from the current
knowledge on networks and complex systems. Advances in complex
networks already helped researchers to gain new insights on these
difficult problems (see for example the recent Handbook on Theoretical
and Quantitative Geography \cite{Handbook:2009}) and the present
review is an attempt to collect modern results on networks and to help
researchers in various fields to reach quantitative answers and
realistic modeling. In geography and urban studies, it would be about
understanding the evolution of transportation networks, the human
mobility, the spatial structure of urban areas, etc. and how these
different factors are entangled with each other, in order to propose
an integrated approach of scale, mobility, and spatial distribution of
activities at various scales.

\subsection{What this review is (not) about}
\label{sec1C}

The importance of spatial networks in current problems, together with
the lack of a review specific on this topic brought us to propose the
present review article. We will first review the tools to characterize
these networks and the empirical properties of some important spatial
networks. We then review the most important models of spatial network
formation which allows to understand the main effects of the spatial
constraints on the network properties. We will also discuss how space
affect various processes taking place on these networks such as
walking and searching, resilience, or disease spread.

As mentioned above, spatial networks appear in many different fields
and we will try to cover in some detail the studies in these various
areas.  However, it is obviously impossible to review all the existing
results related to spatial networks. This implies to make some choices
and we basically restricted ourselves to (i) recent topics, (ii) with
a sufficient body of literature, (iii) of fundamental research, and
(iv) with preferably applications to real-world systems. The goal is
to summarize what we understand at this point about the effect of
space in networks. In particular, we decided not to discuss here the
following subjects (which is not an exhaustive list).

{\it River networks.} Rivers form spatial networks (which in most
cases are essentially trees), and result from the interplay between
gravity and the elevation distribution. These networks were the
subject of many studies and we refer the interested reader to the
books \cite{Iturbe:1997,Caldarelli:2008} and references therein for a
recent account on this subject.

{\it Sensor, ad hoc, and wireless networks.} There is of course a huge
literature on these subjects and which is mostly found in computer and
engineering science. In this review, we will discuss some theoretical
aspects of these problems and for more applied problems we refer the
interested reader to more specialized reviews and books such as
\cite{Akyildiz:2002}.

{\it Spatial games.} Game theory was recently applied to situations
where agents are located on the nodes of a networks. We refer the
interested reader in this particular field to the articles
\cite{Nowak:1994,Hauert:2005} for example.

{\it Mathematical studies of planar maps.} Planar graphs and maps
\cite{Tutte:1963} are combinatorial objects which are the object of
many studies in mathematics and also in physics where they appear as a
natural discretization of random surfaces used in two-dimensional
quantum gravity. In particular these mathematical methods allow to
understand the critical behavior of the Ising model on a random planar
lattice \cite{Boulatov:1987}.  We refer the interested reader to the
classical book of Ambjorn, Durhuus and Jonsson \cite{Ambjorn:1997}.

\medskip

The detailed outline of the review is the following. In the chapter
\ref{sec2}, we introduce the main tools to characterize spatial
networks. Many spatial networks are planar and we recall the main
results on planar networks and how to characterize them in section
\ref{sec2A}. In addition, since spatial networks mix space and
topology, we need specific tools to characterize them that we describe
in section \ref{sec2B}.

In section \ref{sec3}, we review the properties of real-world spatial
networks such as transportation networks (\ref{sec3A}), infrastructure
networks (\ref{sec3B}), mobility networks (\ref{sec3C}), and neural
networks (\ref{sec3D}). In these sections, we insist on salient
stylized facts which allow us to draw some general features of
real-world networks in section \ref{sec3E}.
 
In section \ref{sec4}, we review the most important models of spatial
networks. We divide these models in five large categories: we start
with random geometric graphs (\ref{sec4A}), followed by spatial
generalizations of the Erdos-Renyi graph (\ref{sec4B}) and the
Watts-Strogatz model (\ref{sec4C}). We then review models of growing
spatial networks in section \ref{sec4D} and we end this chapter with
optimal networks \ref{sec4E}.

Finally, we discuss various processes which take place on spatial
networks in section \ref{sec5}. In particular, we will review the
effects of space on transitions in the Ising model (\ref{sec5A}),
random walks (\ref{sec5B}), navigation and searching (\ref{sec5C}),
robustness (\ref{sec5D}), and disease spread (\ref{sec5E}).

In the last section (\ref{sec6}), we propose a summary of the review
about the effect of space on networks and we end the discussion with a
list of open problems that we believe are interesting directions for
future research.

\section{Characterizing spatial networks}
\label{sec2}

\subsection{Generalities on planar networks}
\label{sec2A}

\subsubsection{Spatial and planar networks}
\label{sec2A1}

Loosely speaking, spatial networks are networks for which the nodes
are located in a space equipped with a metric. For most practical
applications, the space is the two-dimensional space and the metric is
the usual euclidean distance. This definition implies in general that
the probability of finding a link between two nodes will decrease with
the distance. However, it does not imply that a spatial network is
planar. Indeed, the airline passenger networks is a network connecting
through direct flights the airports in the world, but is not a planar
network. 

With this definition of a spatial network links are not necessarily
embedded in space: social networks for example connect individuals
through friendship relations. In this case, space intervenes in the
fact that the connection probability between two individuals usually
decreases with the distance between them. 

For many infrastructure networks however, planarity is
unavoidable. Roads, rail, and other transportation networks are
spatial and \- to a good accuracy \- planar networks. For many
applications, planar spatial networks are the most important and most
studies have focused on these examples. In this section, we will thus
recall some basic results about planar networks and we then recall the
standard tools for characterizing networks. In particular, we will
focus on the not-so-standard tools which can help in characterizing
spatial networks, mixing topological and metric features.

\subsubsection{Classical results for planar networks}
\label{sec2A2}

Basic results on planar networks can be found in any graph theory 
textbook (see for example \cite{Clark:1991}) and we will very briefly
recall the most important ones.

Basically, a planar graph is a graph that can be drawn in the plane in
such a way that its edges do not intersect. Not all drawings of planar
graphs are without intersection and a drawing without intersection is
sometimes called a plane graph or a planar embedding of the graph. In
real-world cases, these considerations actually do not apply since the
nodes and the edges represent in general physical objects.

\begin{figure}[!h]
\centering
\begin{tabular}{c}
\includegraphics[angle=0,scale=.60]{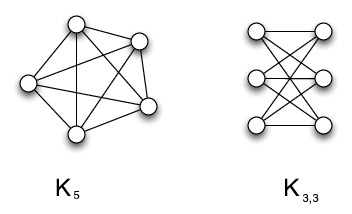}
\end{tabular}
\caption{Complete graphs $K_5$ and $K_{3,3}$. }
\label{fig:planarity}
\end{figure}

In general it is not a trivial thing to check if a graph is planar and
the Kuratowski theorem (see for example the textbook
\cite{Clark:1991}) states that a finite graph is planar if and only if
it does not contain a subgraph that is homeomorphic to the graphs
$K_5$ or $K_{3,3}$ shown in Fig.~\ref{fig:planarity}.
Many algorithms were developed to test the planarity of a given
network (see for example \cite{Jungnickel:1999}) and most of these 
methods operate in ${\cal O}(N)$ time.

There are some very general facts that can be demonstrated
for planar graphs, and among them Euler's formula is probably the best
known. Euler showed that a finite connected planar graph satisfies the
following formula
\begin{equation}
N-E+F=2
\label{eq:euler}
\end{equation}
where $N$ is the number of nodes, $E$ the number of edges, and $F$ is
the number of faces.  This formula can be easily proved by induction
by noting that removing an edge decreases $F$ and $E$ by one, leaving
$N-E+F$ invariant.  We can repeat this operation until we get a tree
for which $F=1$ and $N=E+1$ leading to $N-E+F=E+1-E+1=2$.

Moreover for any finite connected planar graph we can obtain a
bound. Indeed, any face is bounded by at least three edges and every
edge separates two faces at most which implies that $E\geq 3
F/2$. From Euler's formula we then obtain
\begin{equation}
E\leq 3N-6
\end{equation}
In other words, planar graphs are always sparse with a bounded average
degree $\langle k\rangle\leq 6$.

We note here that if we start from a planar point set, we can
construct various planar graph by connecting these points (see the
next section on tessellations for example) and by definition the
maximal planar graph is the triangulation of a planar point set such
that the addition of any edge results in a non-planar graph. Such a
network is useful for example to assess the efficiency of a real-world
planar network and provides an interesting null model. For such a
maximal planar graph the number of edges and faces are maximal and are
equal to the bounds $E=3N-6$ and $F=2N-5$, respectively.

Planar sets thus form faces or cells which have a certain shape. In
certain conditions, it can be interesting to characterize
statistically these shapes and various indicators were developed in
this perspective (see \cite{Haggett:1969} for a list of these
indicators). In particular, if $A$ is the area of a cell, and $L$ the
major axis, the form ratio is defined as $A/L^2$ or equivalently the
elongation ratio given by $\sqrt{A}/L$. In the paper
\cite{Lammer:2006} on the road network structure, L\"ammer et al. use
another definition of the form factor of a cell and define it as
\begin{equation}
\phi=\frac{4A}{\pi D^2}
\end{equation}
where $\pi D^2$ is the area of the circumscribed circle.

\subsubsection{Voronoi tessellation}
\label{sec2A3}

\begin{figure}[h!]
\begin{tabular}{c}
\epsfig{file=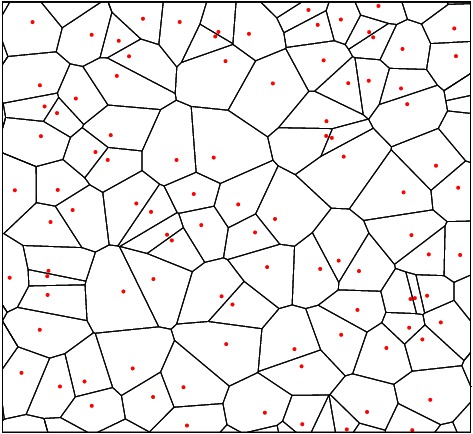,width=0.6\linewidth,clip=}
\end{tabular}
\caption{ Example of the Voronoi tessellation on a set of points. From \cite{Hilhorst:2008}.}
\label{fig:voronoi}
\end{figure}

A {\it tessellation} or tiling of the plane is a collection of plane
figures that fills the plane with no overlaps and no gaps. One may
also speak of tessellations of parts of the plane or of other surfaces
(generalizations to higher dimensions are also possible). A Voronoi
diagram is determined by distances to a specified discrete set of
points (see for example the figure \ref{fig:voronoi}). Each site has a
Voronoi cell $V(s)$ consisting of all points closer to $s$ than to any
other site. The segments of the Voronoi diagram are all the points in
the plane that are equidistant to the two nearest sites. The Voronoi
nodes are the points equidistant to three (or more) sites and unless
some degeneracies, all nodes of the Voronoi tessellation have a degree
equal to three. The dual graph for a Voronoi diagram corresponds to
the Delaunay triangulation for the same set of points.

Voronoi tessellations are interesting for spatial networks in the sense
that they provide a natural null model to which one can compare a real
world network.

In particular, an important case is obtained when the points are distributed
uniformly and independently in the plane. In this case, one can speak
of a Poisson-Voronoi tessellation for which some statistical properties
are known (see \cite{Hilhorst:2005a,Hilhorst:2005,Hilhorst:2008} for a short review
on this subject). In particular, the probability that a point has a
n-sided Voronoi cell is given for large $n$ by \cite{Hilhorst:2005a,Hilhorst:2005}
\begin{equation}
p_n=\frac{C}{4\pi^2}\frac{(8\pi^2)^n}{(2n)!}\left[1+{\cal O} \left(\frac{1}{\sqrt{n}}\right)\right]
\end{equation}
which at the dominant order behaves as $p_n\approx n^{-2n}$.

\subsection{Mixing space  and topology}
\label{sec2B}

\subsubsection{Basic measures}
\label{sec2B1}

\paragraph{Adjacency matrix, dual graph.}
\label{sec2B1a}

A graph with $N$ nodes and $E$ edges can be described by its $N\times
N$ adjacency
matrix $A$ which is defined as
\begin{equation}
A_{ij}=
\begin{cases}
=1\;\;\;{\rm if}\;\;\;i\;\;{\rm and}\;\;j\;\;{\rm are}\;\;{\rm connected}\\
=0\;\;{\rm otherwise}
\end{cases}
\end{equation} 
If the graph is undirected then the matrix $A$ is symmetric. 
The degree of a node is the  number of its neighbors and is then given by
\begin{equation}
k_i=\sum_jA_{ij}
\end{equation}
The first simple indicator of a graph is the average degree
\begin{equation}
\langle k\rangle=1/N\sum_ik_i=2E/N
\end{equation}
where here and in the following the brackets $\langle\cdot\rangle$ denote
the average over the nodes of the network. In particular the scaling
of $\langle k\rangle$ with $N$ indicates if the network is sparse
(which is the case when $\langle k\rangle\to const.$ for $n\to\infty$). 

The distribution of degree $P(k)$ is usually a quantity of interest
and can display large heterogeneities such as it is observed in scale-free
networks (see for example \cite{Albert:2002}). We indeed observe that
for spatial networks such as airline networks or the Internet the
degrees are very heterogeneous (see \ref{sec3}). However, when
physical constraints are strong or when the cost associated with the
creation of new links is large, a cut-off appears in the degree
distribution \cite{Amaral:2000} and in some case the distribution can
be very peaked. This is the case for the road network for example and more
generally in the case of planar networks for which the degree
distribution $P(k)$ is of little interest.

In some real-world cases such as the road network for example, it is
natural to study the usual (or `primal') representation where the
nodes are the intersections and the links represent the road segment
between the intersection. However another representation, the dual
graph can be of interest (see \cite{Porta:2006}. In this road network
example, the dual graph is constructed in the following way: the nodes
are the roads and two nodes are connected if there exist an
intersection between the two corresponding roads. One can then measure
the degree of a node which represents the number of roads which
intersect a given road. Also, the shortest path length in this network
represents the number of different roads one has to take to go from
one point to another. Finally, we note that even if the road network
has a peaked degree distribution its dual representation can display
broad distributions \cite{Kalapala:2006}.

\paragraph{Clustering, assortativity, and average shortest path.}
\label{sec2B1b}

Complex networks are essentially characterized by a small set of
parameters which are not all relevant for spatial networks.  For
example, the degree distribution which has been the main subject of
interest in complex network studies is usually peaked for planar
networks, due to the spatial constraints. However, the clustering
coefficient as we will see is important in the characterization of
spatial networks. For a node $i$ of degree $k_i$ it is defined as
\begin{equation}
C(i)=\frac{E_i}{k_i(k_i-1)/2}
\end{equation}
where $E_i$ is the number of edges among the neighbors of $i$.  This
quantity gives some information about local clustering and is the
object of many studies in complex networks. It is also a quantity of
interest for spatial networks. Indeed for Erdos-Renyi (ER) random
graphs (see section \ref{sec4B}) with finite average degree, the
average clustering coefficient is simply given by
\begin{equation}
\langle C\rangle\sim\frac{1}{N}
\end{equation}
where the brackets $\langle\cdot\rangle$ denote the average over the
network.  In contrast, for spatial networks, loser nodes have a larger
probability to be connected, leading to a large clustering
coefficient. The variation of this clustering coefficient in space
can thus bring valuable information about the spatial structure of
the network under consideration.

The clustering coefficient depends on the number of triangles or
cycles of length $3$ and can also be computed by using the adjacency
matrix $A$. Powers of the adjacency matrix give the number of paths of
variable length. For instance the quantity $\frac{1}{6}\rm{Tr}(A^3)$
is the number $C_3$ of cycles of length tree and is related to the
clustering coefficient. Analogously we can define and count cycles of
various lengths (see for example \cite{Bianconi:2005b,Rozenfeld:2005}
and references therein) and compare this number to the ones obtained
on null models (lattices, triangulations, etc).

Finally, many studies define the clustering coefficient per degree
classes which is given by
\begin{equation}
C(k)=\frac{1}{N(k)}\sum_{i/k_i=k}C(i)
\end{equation}
The behavior of $C(k)$ versus $k$ thus gives an indication how the
clustering is organized when we explore different classes of
degrees. However, in order to be useful, this quantity needs to be
applied to networks with a large range of degree variations which is
usually not the case in spatial networks.

In general the degrees at the two end nodes of a link are correlated
and to describe these degree correlations one needs the two-point
correlation function $P(k'|k)$. This quantity represents the
probability that any edge starting at a vertex of degree $k$ ends at a
vertex of degree $k'$. Higher order correlation functions can be
defined and we refer the interested reader to \cite{BBVBook:2008} for
example. The function $P(k'|k)$ is however not easy to handle and one can
define the assortativity \cite{Pastor:2001c,Vazquez:2002b}
\begin{equation}
k_{nn}(k)=\sum_{k'}P(k'|k)k'
\end{equation}
A similar quantity can be defined for each node as the average degree
of the neighbor
\begin{equation}
k_{nn}(i)=\frac{1}{k_i}\sum_{j\in\Gamma(i)}k_j
\end{equation}
There are essentially two classes of behaviors for the
assortativity. If $k_{nn}(k)$ is an increasing function of $k$,
vertices with large degrees have a larger probability to connect to
similar nodes with a large degree. In this case, we speak of an {\it
  assortative} network and in the opposite case of a {\it
  disassortative} network. It is expected in general that social
networks are mostly assortative, while technological networks are
disassortative. However for spatial networks we will see (see section
\ref{sec3}) that spatial constraints usually implies a flat function
$k_{nn}(k)$.

Usually, there are many paths between two nodes in a connected
networks and if we keep the shortest one it defines a distance on the network
\begin{equation}
\ell(i,j)=\min_{paths( i\to j)}|path|
\end{equation}
where the length $|path|$ of the path is defined as its number of
edges. The {\it diameter} of the graph can be defined as the maximum
value of all $\ell(i,j)$ or can also be estimated by the average this
distance over all pairs of nodes in order to characterize the `size'
of the network. Indeed for a $d$-dimensional regular lattice with $N$
nodes, this average shortest path $\langle\ell\rangle$ scales as
\begin{equation}
\langle\ell\rangle\sim N^{1/d}
\end{equation}
In a small-world network (see \cite{Watts:1998} and section
\ref{sec4C}) constructed over a $d-$dimensional lattice
$\langle\ell\rangle$ has a very different behavior
\begin{equation}
\langle\ell\rangle\sim \log N
\end{equation}
The crossover from a large-world behavior $N^{1/d}$ to a small-world 
one with $\log N$ can be achieved for a density $p$ of long
links (or `shortcuts') \cite{Barthelemy:1999} such that
\begin{equation}
pN\sim 1
\end{equation}

The effect of space could thus in principle be detected in the
behavior of $\langle\ell\rangle(N)$. It should however be noted that
if the number of nodes is too small this can be a tricky task. In the
case of brain networks for example, a behavior typical of a
three-dimensional network in $N^{1/3}$ could easily be confused with a
logarithmic behavior if $N$ is not large enough.

\paragraph{Spectral graph theory.}
\label{sec2B1c}

Spectral graph theory is an important branch of mathematics and allows
to get insights about the structure of a graph with quantities
computed from the eigenvalues of the graph Laplacian
\cite{Chung:1997}. For example, the so-called algebraic connectivity
is the smallest non-zero eigenvalue, the distribution of cycle lengths
can be computed using a moment expansion of the eigenvalues, the
stationary state of a random walk and synchronization properties are
governed by the largest eigenvalue of the adjacency matrix. In a
nutshell, spectral graph theory thus studies the adjacency matrices of
graphs and connect their eigenvalues to other properties. There are
different conventions but essentially one is interested in the
discrete Laplacian on the network defined by
\begin{equation}
L=D-A
\end{equation}
where $D_{ij}=k_i\delta_{ij}$ is the identity matrix times the degree
$k_i$ of node $i$ and $A$ is the adjacency matrix of the
graph. If the graph is undirected the Laplacian is
symmetric and if the graph is an infinite square lattice grid of
spacing $a$, this definition of the Laplacian coincides with the
continuous Laplacian $\nabla^2$ when $a\to 0$. We list here some basic
properties of the Laplacian and we refer the interested reader to more
advanced material such as the book \cite{Chung:1997}.
\begin{itemize}
\item{} In the undirected case, $L$ is symmetric and has real positive
  eigenvalues $\lambda_0=0\leq\lambda_1\leq\dots\leq\lambda_N$. The
  multiplicity of $\lambda_0=0$ (whose eigenvector has all components
  equal) is equal to the number of connected components of the graph.
\item{} The eigenvalue $\lambda_1$ is called the algebraic
  connectivity \cite{Fiedler:1973} and plays an important role in the characterization of
  the graph and, roughly speaking, the larger $\lambda_1$, the more connected the
  graph. 
\item{} The distribution of eigenvalues is important in many
  applications. For instance for random walks, it is related to the
  return probability and it can also be used (together with the
  eigenvectors) for visual representations of graphs known as spectral
  embeddings (see for example \cite{Luo:2003}). Another example
  concerns the synchronization of linearly coupled identical
  oscillators where a master stability function relates the stability
  of synchronized solutions to the eigenvalues of the Laplacian (more
  precisely to the ratio $\lambda_N/\lambda_1$, see \cite{Pecora:1998}
  and \cite{Baharona:2002} for the small-world case).
\end{itemize}

\paragraph{Betweenness centrality.}
\label{sec2B1d}

The importance of a node is characterized by its so-called
centrality. There are however many different centrality indicators
such as the degree, the closeness, etc., but we will focus here on the
{\it betweenness centrality} $g(i$) which is defined as
\cite{Freeman:1977,Newman:2001b,Goh:2001,Barthelemy:2003b,Barthelemy:2003c}.
\begin{equation}
g(i)=\sum_{s\neq t}\frac{\sigma_{st}(i)}{\sigma_{st}}
\label{eq:bc}
\end{equation}
where $\sigma_{st}$ is the number of shortest paths going from $s$ to
$t$ and $\sigma_{st}(i)$ is the number of shortest paths going from
$s$ to $t$ through the node $i$. This quantity $g(i$) thus
characterizes the importance of the node $i$ in the organization of
flows in the network. Note that with this definition, the betweenness centrality of terminal nodes is
zero.  The betweenness centrality can similarly be defined for edges
\begin{equation}
g(e)=\sum_{e\in E}\frac{\sigma_{st}(e)}{\sigma_{st}}
\end{equation}
where $\sigma_{st}(e)$ is the number of shortest paths going from $s$ to
$t$ and going through the edge $e$. 

\begin{figure}[h!]
\begin{tabular}{cc}
\epsfig{file=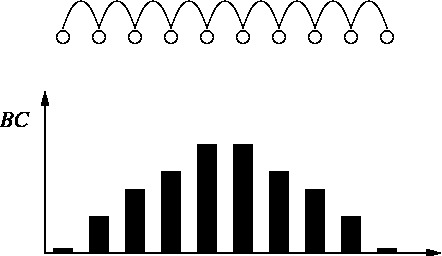,width=0.5\linewidth,clip=} &
\epsfig{file=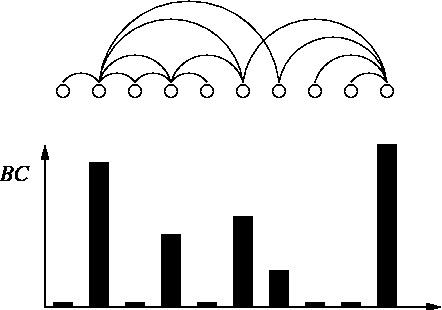,width=0.5\linewidth,clip=}
\end{tabular}
\caption{ (a) Betweenness centrality for the (one-dimensional) lattice
case. The central nodes are close to the barycenter. (b) For a
general graph, the central nodes are usually the ones with large
degree.}
\label{fig:spacebc}
\end{figure}

The betweenness centrality of a vertex is determined by
its ability to provide a path between separated regions of the
network.  Hubs are natural crossroads for paths and it is natural to
observe a marked correlation between the average
$g(k)=\sum_{i/k_i=k}g(i)/N(k)$ and $k$ as expressed in the
following relation \cite{Barthelemy:2003c}
\begin{equation}
g(k)\sim k^\eta
\end{equation}
where $\eta$ depends on the characteristics of the network. We expect
this relation to be altered when spatial constraints become important
and in order to understand this effect we consider a one-dimensional
lattice which is the simplest case of a spatially ordered network. For
this lattice the shortest path between two nodes is simply the
euclidean geodesic and for two points lying far from each other, the
probability that the shortest path passes near the barycenter of the
network is very large. In other words, the barycenter (and its
neighbors) will have a large centrality as illustrated in
Fig.~\ref{fig:spacebc}a.
In contrast, in a purely topological network with no underlying
geography, this consideration does not apply anymore and if we rewire
more and more links (as illustrated in Fig.~\ref{fig:spacebc}b) we
observe a progressive decorrelation of centrality and space while the
correlation with degree increases. In a lattice, the betweenness
centrality depends essentially on space and is maximum at the
barycenter, while in a network the betweenness centrality of a node
depends on its degree. When the network is constituted of long links
superimposed on a lattice, we expect the appearance of `anomalies'
characterized by large deviations around the behavior $g\sim
k^\eta$. In order to characterize quantitatively these anomalies
\cite{Barrat:2005}, one can compute the fluctuations of the
betweenness centrality $\Delta_{MR}(k)$ for a randomized network with
the same degree distribution than the original network and constructed
with the Molloy-Reed algorithm \cite{Molloy:1995}. We can then
consider a node $i$ to be anomalous if its betweenness centrality
$g(i)$ lies outside the interval $[\langle
g(k)\rangle-\nu\Delta_{MR}(k),\langle g(k)\rangle+\nu\Delta_{MR}(k)]$
where the value of $\nu$ determines the confidence interval. For
$\nu\approx 1.952$, the considered interval represents $95\%$ of the
nodes in the case of Gaussian distributed centralities around the
average.

\begin{figure}[h!]
\begin{tabular}{c}
\epsfig{file=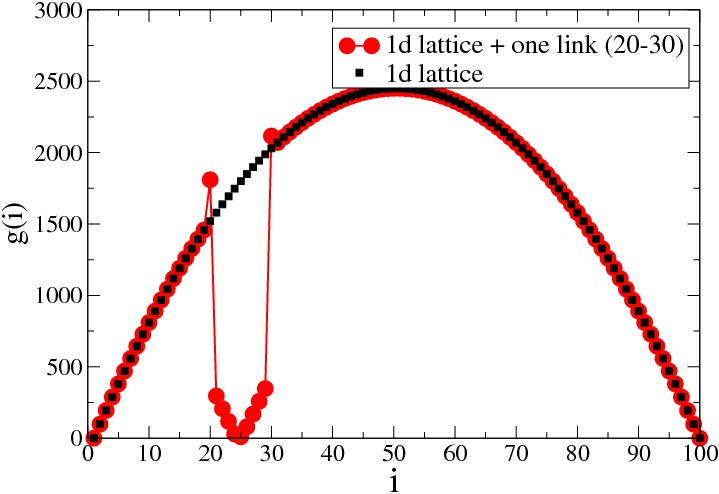,width=0.8\linewidth,clip=}
\end{tabular}
\caption{ Example of how the addition of a link perturbs the
  centrality. In black, the betweenness centrality for the 1d lattice (of size ($N=100$) has a maximum at the
  barycenter $N/2=50$. The addition of a link between $a=20$ and
  $b=30$ decreases the betweenness centrality between $a$ and $b$ and increases the betweenness centrality of
  nodes $a$ and $b$.}
\label{fig:bconelink}
\end{figure}

The effect of shortcuts and the appearance of anomalies can already be
observed in the simple case of a one-dimensional lattice
$i=1,\dots,N$. The betweenness centrality in this case is given by $g_0(i)=(i-1)(N-i)$
with a maximum at $i=N/2$ (see Fig.~\ref{fig:bconelink}).
When one link is added between nodes $a$ and $b$, the betweenness
centrality of the nodes in the interval $]a,b[$ decreases. For example,
if $a<N/2<b$ it is not difficult to see that the variation of the
centrality of the node $N/2$ is bounded by
\begin{equation}
\delta g(N/2)<-a(N-b)
\end{equation}
which basically means that the shortest paths from a node $i\in[1,a]$
to a node $j\in[b,N]$ follow the shortcut and avoid the node $N/2$. In
contrast, the betweenness centrality of the contact points increases
and the betweenness centrality of node $a$ is now $g(a)=g_0(a)+\delta
g(a)$ where $\delta g(a)$ is positive and essentially counts the new
pairs of nodes which are connected by shortest going through the new
link $(a,b)$.

\subsubsection{Mixing topology and space}
\label{sec2B2}

\paragraph{Strengths.}
\label{sec2B2a}

In many cases, the adjacency matrix is not enough to fully
characterize a network. This is particularly clear for airline
networks for example for which the links are also characterized by the
average number of passengers $w_{ij}$ flying on the connection
$(ij)$ (for other networks, such as the Internet it can be the traffic
on a cable, etc.). For these weighted networks, the distribution of
weights is a first indication of the existence of possible strong
heterogeneities \cite{Barrat:2004b}. Another important information is
the existence of correlation between the weights and the degrees. Such
a correlation can be easily displayed by the study of the {\it weight
  strength} of a node $i$ as defined by~\cite{Yook:2001,Barrat:2004b}
\begin{equation}
s_i^w=\sum_{j \in \Gamma (i)} w_{ij} \ .
\end{equation}
where the sum runs over the set $\Gamma(i)$ of neighbors of $i$. The
strength generalizes the degree to weighted networks and in the case
of the air transportation network quantifies the traffic of passengers
handled by any given airport. This quantity obviously depends on the
degree $k$ and increases (linearly) with $k$ in the case of random
uncorrelated weights of average $\langle w\rangle$ as
\begin{equation}
s^w\sim k\langle w\rangle
\end{equation}
A relation between the strength $s^w(k)$ averaged over the nodes of degree
$k$ of the form
\begin{equation}
s^w = A k^{\beta_w} \ ,
\label{eq:sw}
\end{equation}
with an exponent $\beta_w >1$ (or with $\beta_w=1$ but with $A \ne \langle w
\rangle$) is then the signature of non-trivial statistical correlations
between weights and topology.  In particular, a value $\beta_w>1$
signals that the typical number of passengers per connection is not
constant and increases with $k$.

The notion of strength can obviously be extended to many different
types of weights. In particular, for spatial networks, one can define \cite{Barrat:2005}
the  {\em distance strength} of node $i$ by
\begin{equation}
s_i^d=\sum_{j \in \Gamma(i)} d_E(i,j)
\end{equation}
where $d_E(i,j)$ is the euclidean distance between nodes $i$ and $j$.
This quantity $s^d$ represents the cumulated distances of all the
connections from (or to) the considered airport.  Similarly to the
usual weight strength, uncorrelated random connections would lead to a
linear behavior of the form $s^d(k)\propto k$ while otherwise the
presence of correlations would be signaled by a behavior of the form
\begin{equation}
s^{d}(k) \sim k^{\beta_d}
\label{eq:sd}
\end{equation}
with $\beta_d>1$. In such a case, there are correlations
between the topology and geography which implies that the typical length
of the connection is not constant as it would be in the case for
$\beta_d=1$ but increases with the number of connections.

\paragraph{Indices $\alpha$, $\gamma$ and variants. Ringness.}
\label{sec2B2b}

Different indices were defined a long time ago mainly by scientists
working in quantitative geography since the $1960$s and can be found
in \cite{Haggett:1969,Taaffe:1973,Rodrigue:2006} (see also the more
recent paper by Xie and Levinson \cite{Xie:2007}).  Most of these
indices are relatively simple but give valuable information about the
structure of the network, in particular if we are interested in planar
networks. These indices were used to characterize the topology of
transportation networks and for example Garrison \cite{Garrison:1960}
measured some properties of the Interstate highway system and Kansky
\cite{Kansky:1969} proposed up to $14$ indices to characterize
these networks. We will here recall the most important
indices which are called the `alpha' and the `gamma' indices.  The
simplest index is called the gamma index and is simply defined by
\begin{equation}
\gamma=\frac{E}{E_{max}}
\end{equation}
where $E$ is the number of edges and $E_{max}$ is the maximal number
of edges (for a given number of nodes $N$). For non-planar networks,
$E_{max}$ is given by $N(N-1)/2$ for non-directed graphs and for
planar graphs we saw in the section \ref{sec2A2} that $E_{max}=3N-6$
leading to
\begin{equation}
\gamma_P=\frac{E}{3N-6}
\end{equation}

The gamma index is a simple measure of the density of the network but
one can define a similar quantity by counting the number of elementary
cycles instead of the edges. The number of elementary cycle for a
network is known as the cyclomatic number (see for example
\cite{Clark:1991}) and is equal to
\begin{equation}
\Gamma=E-N+1
\end{equation}
For a planar graph this number is always less or equal to $2N-5$ which
leads naturally to the definition of the alpha index (also coined
`meshedness' in \cite{Buhl:2006})
\begin{equation}
\alpha=\frac{E-N+1}{2N-5}
\end{equation}
This index lies in the interval $[0,1]$ and is equal to $0$ for a tree and equal
to $1$ for a maximal planar graph. Using the definition of the average
degree $\langle k\rangle=2E/N$ the quantity $\alpha$ reads in the
large $N$ limit as
\begin{equation}
\alpha\simeq \frac{\langle k\rangle -2}{4}
\label{eq:alphalimit}
\end{equation}
which shows that in fact for a large network this index $\alpha$ does
not contain much more information than the average degree.

We note that more recently, other interesting indices were proposed in order to
characterize specifically road networks \cite{Xie:2007,Courtat:2010}. For example, 
in \cite{Courtat:2010}, Courtat, Gloaguen, Douady noticed that in some
cities the degree distribution is very peaked around $3-4$ and they
then define the ratio
\begin{equation}
r_N=\frac{N(1)+N(3)}{\sum_{k\neq 2}N(k)}
\end{equation}
where $N(k)$ is the number of nodes of degree $k$. If this ratio is
small the number of dead ends and of `unfinished' crossing ($k=3$) is
small compared to the number of regular crossings with $k=4$. In the
opposite case of large $r_N$, there is a dominance of $k=4$ nodes
which signals a more organized city.

The authors of \cite{Courtat:2010} also define the `compactness' of a
city which measures how much a city is `filled' with roads. If we
denote by $A$ the area of a city and by
$\ell_T$ the total length of roads, the compactness $\Psi\in [0,1]$ can be defined
in terms of the hull and city areas
\begin{equation}
\Psi=1-\frac{4A}{(\ell_T-2\sqrt{A})^2}
\end{equation}
In the extreme case of one square city of linear size $L=\sqrt{A}$
with only one road encircling it, the total length is
$\ell_T=4\sqrt{A}$ and the compactness is then $\Psi=0$. At the other
extreme, if the city roads constitute a square grid of spacing $a$,
the total length is $\ell_T=2L^2/a$ and in the limit of $a/L\to 0$ one
has a very large compactness $\Psi\approx 1-a^2/L^2$.

We end this section by mentioning the ringness. Arterial roads
(including freeways, major highways) provide a high level of mobility
and serve as the backbone of the road system
\cite{Xie:2007}. Different measures (with many references) are
discussed and defined in \cite{Xie:2007} and in particular the
{\it ringness} is defined as
\begin{equation}
\phi_{ring}=\frac{\ell_{ring}}{\ell_{tot}}
\end{equation}
where $\ell_{ring}$ is the total length of arterials on rings and
where the denominator $\ell_{tot}$ is the total length of all
arterials. This quantity ranging from $0$ to $1$ is thus an indication
of the importance of a ring and to what extent arterials are organized
as trees.

\paragraph{Route factor, detour index.}
\label{sec2B2c}

\begin{figure}[h!]
\begin{tabular}{c}
\epsfig{file=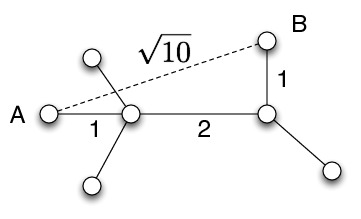,width=0.6\linewidth,clip=}
\end{tabular}
\caption{ Example of a detour index calculation. The `as crow flies'
  distance between the nodes A and B is $d_E(A,B)=\sqrt{10}$ while the
  route distance over the network is $d_R(A,B)=4$ leading to a detour
  index equal to $Q(A,B)=4/\sqrt{10}\simeq 1.265$.}
\label{fig:detour}
\end{figure}

When the network is embedded in a two-dimensional space, we can define
at least two distances between the pairs of nodes. There is of course
the natural euclidean distance $d_E(i,j)$ which can also be seen as
the `as crow flies' distance. There is also the total `route' distance
$d_R(i,j)$ from $i$ to $j$ by computing the sum of length of segments
which belong to the shortest path between $i$ and $j$. The route factor
(also called the detour index or the circuity or directness
\cite{Levinson:2007})  for this pair of nodes $(i,j)$ is then
given by (see Fig.~\ref{fig:detour} for an example)
\begin{equation}
Q(i,j)=\frac{d_R(i,j)}{d_E(i,j)}
\end{equation}
This ratio is always larger than one and the closer to one, the more
efficient the network.
From this quantity, we can derive another one for a single node
defined by
\begin{equation}
\langle Q(i)\rangle=\frac{1}{N-1}\sum_jQ(i,j)
\end{equation}
which measures the `accessibility' for this specific node $i$. Indeed
the smaller it is and the easier it is to reach the node
$i$. Accessibility is a subject in itself (see for example
\cite{Batty:2009}) and there are many other measures for this concept
and we refer the interested reader to the article
\cite{Handy:1997,Levinson:2008,Levinson:2009}. This quantity $\langle
Q(i)\rangle$ is related to the quantity called `straightness
centrality' \cite{Crucitti:2006b}
\begin{equation}
C^S(i)=\frac{1}{N-1}\sum_{j\neq i}\frac{d_E(i,j)}{d_R(i,j)}
\end{equation}
And if one is interested in assessing the global efficiency of the
network, one can compute the average over all pairs of nodes (also
used in \cite{Vragovic:2005})
\begin{equation}
\langle Q\rangle=\frac{1}{N(N-1)}\sum_{i\neq j}Q(i,j)
\end{equation}
The average $\langle Q\rangle$ or the maximum $Q_{max}$, and more
generally the statistics of $Q(i,j)$ is important and contains a lot
of information about the spatial network under consideration (see
\cite{Aldous:2010} for a discussion on this quantity for various
networks). For example, one can define the interesting quantity
\cite{Aldous:2010}
\begin{equation}
\phi(d)=\frac{1}{N_d}\sum_{ij/d_E(i,j)=d}Q(i,j)
\end{equation}
(where $N_d$ is the number of nodes such that $d_E(i,j)=d$) whose
shape can help characterizing combined spatial and topological
properties.

\paragraph{Cost and Efficiency.}
\label{sec2B2d}

The minimum number of links to connect $N$ nodes is $E=N-1$ and the
corresponding network is a tree. We can also look for the tree
which minimizes the total length given by the sum of the lengths of
all links
\begin{equation}
\ell_T=\sum_{e\in E}d_E(e)
\end{equation}
where $d_E(e)$ denotes the length of the link $e$. This procedure
leads to the minimum spanning tree (MST) which has a total length
$\ell_T^{MST}$ (see also section \ref{sec4E}). Obviously the tree is
not a very efficient network (from the point of view of transportation
for example) and usually more edges are added to the network, leading
to an increase of accessibility but also to of $\ell_T$. A natural
measure of the `cost' of the network is then given by
\begin{equation}
C=\frac{\ell_T}{\ell_T^{MST}}
\end{equation}
Adding links thus increases the cost but improves accessibility or the
{\it transport performance} $P$ of the network which can be measured as the
minimum distance between all pairs of nodes, normalized to the same
quantity but computed for the minimum spanning tree
\begin{equation}
P=\frac{\langle\ell\rangle}{\langle\ell_{MST}\rangle}
\end{equation}
Another measure of efficiency was also proposed in \cite{Latora:2001b,Latora:2003}
and is defined as 
\begin{equation}
E=\frac{1}{N(N-1)}\sum_{i\neq j}\frac{1}{\ell(i,j)}
\end{equation}
where $\ell(i,j)$ is the shortest path distance from $i$ to $j$. This
quantity is zero when there are no paths between the nodes and is
equal to one for the complete graph (for which
$\ell(i,j)=1$). Combination of these different indicators and
comparisons with the MST or the maximal planar network can be
constructed in order to characterize various aspects of the networks
under consideration (see for example \cite{Buhl:2006}).

Finally, adding links improves the resilience of the network to
attacks or dysfunctions. A way to quantify this is by using the {\it fault
  tolerance (FT)} (see for example \cite{Tero:2010}) measured as the
probability of disconnecting part of the network with the failure of a
single link. The benefit/cost ratio could then be estimated by the
quantity $FT/\ell_T^{MST}$ which is a quantitative characterization of
the trade-off between cost and efficiency \cite{Tero:2010}.

\subsubsection{Community detection. Motifs}
\label{sec2B3}

\paragraph{Community detection.}
\label{sec2B3a}

Community detection in graphs is an important topic in complex network
studies (see the review \cite{Fortunato:2010}), but after almost a
decade of efforts, there is no definitive method of identification of
communities, but instead many different methods with their respective
advantage and drawbacks.

Loosely speaking, a community (or a `module') is a set of nodes which
have more connections among themselves than with the rest of
nodes. One of the first and simplest method to detect these modules is
the modularity optimization and consists in maximizing the quantity
called modularity defined as \cite{Newman:2004}
\begin{equation}
Q=\sum_{s=1}^{n_M}\frac{\ell_s}{E}-\left(\frac{d_s}{2E}\right)^2
\end{equation}
where the sum is over the $n_M$ modules of the partition, $\ell_s$ is
the number of links inside module $s$, $E$ is the total number of
links in the network, and $d_s$ is the total degree of the nodes in
module $s$. The first term of the summand in this equation is the
fraction of links inside module $s$ and the second term represents the
expected fraction of links in that module if links were located at
random in the network (and by keeping the same degree
distribution). The number of modules $n_M$ is also a variable whose
value is obtained by the maximization. If for a subgraph S of a
network the first term is much larger than the second, it means that
there are many more links inside S than one would expect by random
chance, so S is indeed a module. The comparison with the null model
represented by the randomized network is however misleading and small
modules connected by at least a link might be seen as one single
module. This resolution limit was demonstrated in
\cite{Fortunato:2007} where it is shown that modules of size
$\sqrt{E}$ or smaller might not be detected by this method. Modularity
detection was however applied in many different domains and is still
used. In the case of spatial networks, it is the only method which was
used so far but it is clear that community detection in spatial
networks is a very interesting problem which might receive a specific
answer. In particular, it would be interesting to see how the
classification of nodes proposed in \cite{Guimera:2005b} applies to
spatial networks.

\paragraph{Motifs.} 
\label{sec2B3b}

The {\it motifs} are particular subgraphs that are over-represented in
the network with respect to an uncorrelated random network with the same degree
distribution. The idea of motifs is particularly important for
biological networks \cite{Milo:2002} but can be applied to any type of
networks, including spatial networks. In many networks found in
biology, or ecology for example, a surprisingly small number of motifs
exists and each type of network has its own characteristic set of
motifs. The occurrence of a particular motif is characterized by its
normalized Z-score, a measure for its abundance in the network. The
similarity of the motif distributions between different networks can
be a signature of the ubiquity of simple processes occuring in human
interactions, good exchanges, etc.

\section{Empirical observations}
\label{sec3}

In this chapter we will review the most salient properties of real-world
spatial networks. We will insist on features connected to the spatial
aspects of these networks and we will give references for other
aspects.

We will begin with transportation networks which are emblematic of
spatial networks. Generally speaking, transportation networks are
structures that convey energy, matter or information from one point to
another. They appear in a variety of different fields, including city
streets~\cite{Cardillo:2006,Buhl:2006}, plant
leaves~\cite{Rolland:2005}, river networks~\cite{Iturbe:1997},
mammalian circulatory systems~\cite{West:2003}, networks for
commodities delivery~\cite{Gastner:2006}, technological
networks~\cite{Schwartz:1986}, and communication networks
\cite{Pastor:2003}. The recent availability of massive data sets has
opened the possibility for a quantitative analysis and modeling of
these different patterns. In a second part, we review the recent
results obtained for infrastructure networks such as roads, power
grids or the Internet. In particular, we will also discuss the
geographical aspects of social networks. We then review the recent
results on mobility networks which describe the statistics of human
movements. We end this chapter with studies on neural networks,
including the brain which is an important example of a complex network
with a spatial component. Finally, we propose at the end of this
chapter a recap under the form of a table (see table \ref{table2}).

\subsection{Transportation networks}
\label{sec3A}

The paper of Watts and Strogatz \cite{Watts:1998} triggered many
studies on networks and in particular motivated empirical analysis of
various networks such as transportation networks. These networks
control many aspect in our societies and govern many modern problems such
as disease spread, congestion, urban sprawl, structure of cities. In
this section we review the main results concerning their structure.

\subsubsection{Representations}
\label{sec3A1}

\begin{figure}[h!]
\centering
\begin{tabular}{cc}
\epsfig{file=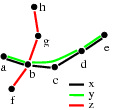,width=0.4\linewidth,clip=} &
\epsfig{file=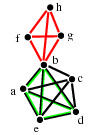,width=0.4\linewidth,clip=}\\
\epsfig{file=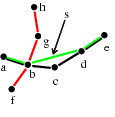,width=0.4\linewidth,clip=} &
\epsfig{file=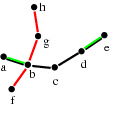,width=0.4\linewidth,clip=}
\end{tabular}
\caption{(a) Direct representation of the routes (here for three
  different routes). (b) Space-of-changes
  (sometimes called $P$ space
  \cite{Sen:2003b,Latora:2001a,Sienkiewicz:2005})). A link connects two
  nodes if there is at least one vehicle that stops at both nodes.  (c)
  Space-of-stops. Two nodes are connected if they are consecutive
  stops of at least one vehicle. (d) Space-of-stations. Here two
  stations are connected only if they are physically connected
  (without any station in between) and this network reflects the real
  physical network. From \cite{Kurant:2006b}.}
\label{fig:kurant}
\end{figure}

In \cite{Kurant:2006b}, Kurant and Thiran discuss very clearly the different
possibilities to represent a transportation system (Fig.~\ref{fig:kurant}).
Indeed, the simplest representation is obtained when the nodes
represent the stations and links the physical connections. One could
however construct other networks such as the space-of-stops or the
space-of-changes (see Fig.~\ref{fig:kurant}). One also finds in the
literature on transportation systems, the notions of L and P-spaces
\cite{Sen:2003b,Hu:2009b} where the L-space connects nodes if they are
consecutive stops in a given route. The degree in L-space is then the
number of different nodes one can reach within one segment and the
shortest path length represents the number of stops. In the P-space,
two nodes are connected if there is at least one route between them so
that the degree of a node is the number of nodes that can be reached
directly. In this P-space the shortest path represents the number of
connections needed to go from one node to another.

\subsubsection{Airline networks}
\label{sec3A2}

The airline transportation infrastructure is an important example of a
spatial network. The nodes are identified to the airports which are
located in a two-dimensional space. The location of the nodes are not
uniformly distributed and are determined by exogenous factors. The
links represent the existence of a direct flight among them and
obviously this network even if clearly spatial is not planar.

\paragraph{Weight and spatial heterogeneity.}
\label{sec3A2a}

The usual characteristics of the world-wide air-transportation network
using the International Air Transportation Association (IATA) database
IATA \footnote{\url{http://www.iata.org.}} have been presented
in~\cite{Barrat:2004b}. The network is made of $N=3880$ vertices and
$E=18810$ edges (for the year $2002$) and displays both small-world
and scale-free properties as also confirmed in different datasets and
analysis \cite{Li:2003,Guimera:2004,Guimera:2005}. In particular, the
average shortest path length, measured as the average number of edges
separating any two nodes in the network shows the value
$\langle\ell\rangle=4.37$, very small compared to the network size $N$
but largely overestimated due to the presence of long paths between
far remote areas (we can expect it to decrease if we weight the paths
by their number of passengers).

The degree distribution, takes the form 
\begin{equation}
P(k)=k^{-\gamma}f(k/k_*)
\end{equation}
where $\gamma\simeq2.0$ and $f(k/k_*)$ is an exponential cut-off
function. The degree distribution is therefore heavy-tailed with a
cut-off that finds its origin in the physical constraints on the
maximum number of connections that a single airport can
handle~\cite{Amaral:2000,Guimera:2004,Guimera:2005}. The airport
connection graph is therefore a clear example of a spatial
(non-planar) small-world network displaying a heavy-tailed degree
distribution and heterogeneous topological properties.

\begin{figure}[h!]
\begin{tabular}{c}
\epsfig{file=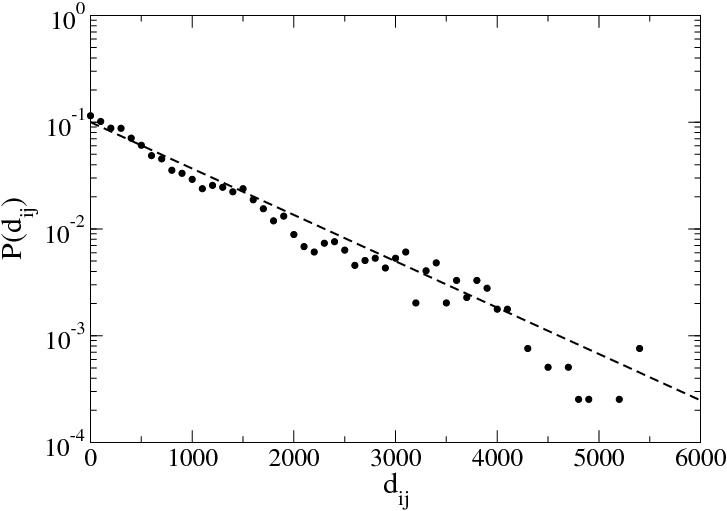,width=0.9\linewidth,clip=}
\end{tabular}
\caption{ Distribution of distances (in kms) between airports
linked by a direct connection for the North American network. The
straight line indicates an exponential decay with scale of order
$1,000$ kms. From \cite{Barrat:2005}.}
\label{fig:Pd}
\end{figure}

The world-wide airline network necessarily mixes different effects. In
particular there are clearly two different scales, global
(intercontinental) and domestic.  The intercontinental scale defines
two different groups of travel distances and for the statistical
consistency the study \cite{Barrat:2005} focused on a single
continental case with the North-American network constituted of
$N=935$ vertices with an average degree $\langle k\rangle \approx 8.4$
and an average shortest path $\langle\ell\rangle\approx 4$. The
statistical topological properties of the North American network are
consistent with the world-wide one: the degree is distributed
according to a power law with exponent $\approx 1.9$ on almost two
orders of magnitude followed by a cut-off indicating the maximum
number of connections compatible with the limited airport capacity and
the size of the network considered. The spatial attributes of the
North American airport network are embodied in the physical spatial
distance, measured in kilometers, characterizing each
connection. Fig.~\ref{fig:Pd} displays the distribution of the distances
of the direct flights. These distances correspond to Euclidean
measures of the links and clearly show a fast decaying behavior
reasonably fitted by an exponential. The exponential fit gives a value
for a typical scale of the order $1000$ kms. The origin of the finite
scale can be traced back to the existence of physical and economical
restrictions on airline planning in a continental setting.
In a recent paper \cite{Bianconi:2009} however Bianconi, Pin, Marsili
study the connection probability $p_{ij}$ for two airports and found
that for the US airline network (with $N=675$ airports and $E=3253$
connections) it behaves as a power law dependence of the form
$p_{ij}\propto d_E(i,j)^{-\alpha}$ with $\alpha\approx 3$ for $d>100$
kms. Although the two quantities $p_{ij}$ and $P(d)$ are different
($p_{ij}$ is the probability used to construct the network and $P(d)$
is the resulting distribution obtained on the specific set of airports
distributed across the US territory) it is unclear at this stage if
these results are consistent.

\begin{figure}[h!]
\begin{tabular}{c}
\epsfig{file=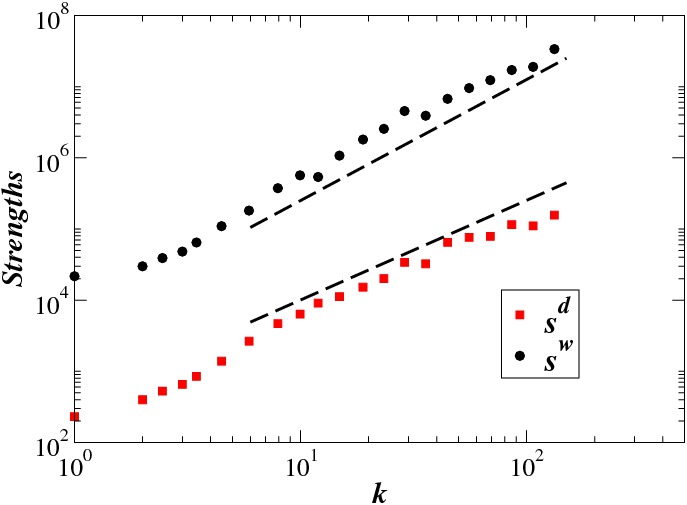,width=0.9\linewidth,clip=}
\end{tabular}
\caption{ Weight and distance strengths versus degree for the
  North American network. The dashed lines correspond to the
  power-laws with exponents $\beta_d\simeq 1.4$ and $\beta_w\simeq 1.7$. From
  \cite{Barrat:2005}.}
\label{sw-sd}
\end{figure}

In addition these networks are weighted and the weight on each link
represents the number of maximum passengers on the connection. The
traffic at a node can then be studied by the weight strength (see
section \ref{sec2B2a}) and behaves here as a power law of the degree
\begin{equation}
s^w \sim k^{\beta_w} \ ,
\label{eq:sw2p}
\end{equation}
with an exponent $\beta_w \simeq 1.7$ (Fig.~\ref{sw-sd})
demonstrating the existence of strong correlations between the degree
and the traffic. The distance strength (see the section \ref{sec2B2a}) also behaves
as a power law of the degree
\begin{equation}
s^{d}(k) \sim k^{\beta_d}
\label{eq:sd2p}
\end{equation}
with $\beta_d\simeq 1.4$ (Fig.~\ref{sw-sd}). This result shows the
presence of important correlations between topology and geography.
Indeed, the fact that the exponents appearing in the relations
Eq.~(\ref{eq:sw2p}) and Eq.~(\ref{eq:sd2p}) are larger than one have different
meanings. While Eq.~(\ref{eq:sw2p}) means that larger airports have
connections with larger traffic, Eq.~(\ref{eq:sd2p}) implies that they have
also farther-reaching connections. In other terms, the traffic and the
distance per connection is not constant but increases with $k$.  As
intuitively expected, the airline network is an example of a very
heterogeneous network where the hubs have at the same time large
connectivities, large weight (traffic) and long-distance
connections~\cite{Barrat:2004b}, related by super-linear scaling
relations.

\paragraph{Assortativity and Clustering.}
\label{sec3A2b}

\begin{figure}[h!]
\begin{tabular}{c}
\epsfig{file=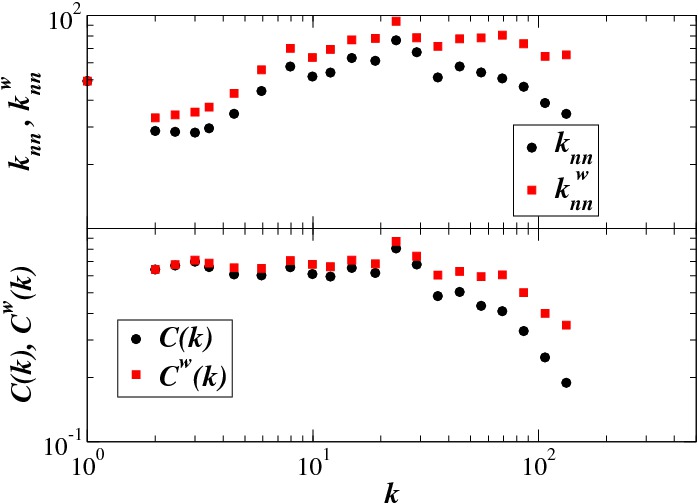,width=1.0\linewidth,clip=}
\end{tabular}
\caption{Assortativity and clustering for the North American network.
Circles correspond to topological quantities while the squares
represent the weighted assortativity and clustering. From \cite{Barrat:2005}.}
\label{fig:Assor}
\end{figure}

A complete characterization of the network structure must take into
account the level of degree correlations and clustering present in the
network. Fig.~\ref{fig:Assor} displays for the North-American airport
network the behavior of these various quantities as a function of the
degree. An essentially flat $k_{nn}(k)$ is obtained and a slight
disassortative trend is observed at large $k$, due to the fact that
large airports have in fact many intercontinental connections to other
hubs which are located outside of North America and are not considered
in this `regional' network. The clustering is very large and is
slightly decreasing at large $k$. This behavior is often observed in
complex networks and is here a direct consequence of the role of large
airports that provide non-stop connections to different regions which
are not interconnected among them. Moreover, weighted correlations are
systematically larger than the topological ones, signaling that large
weights are concentrated on links between large airports which form
well inter-connected cliques (see also~\cite{Barrat:2004b} for more
details).

\paragraph{Betweenness Centrality.}
\label{sec3A2c}

As discussed in the section \ref{sec2B1d}, the betweenness centrality
is a very interesting quantity which characterizes the importance of a node
in the network. The more central the node and the larger the number of
shortest paths going through this node. it is thus natural to observe
strong correlations between the betweenness centrality and the
degree. However for a lattice, the betweenness centrality is maximal
at the barycenter of all nodes and we thus expect for real spatial
networks an interesting interplay between the degree of the node and
its distance to the barycenter.

\begin{figure}[h!]
\begin{tabular}{c}
\epsfig{file=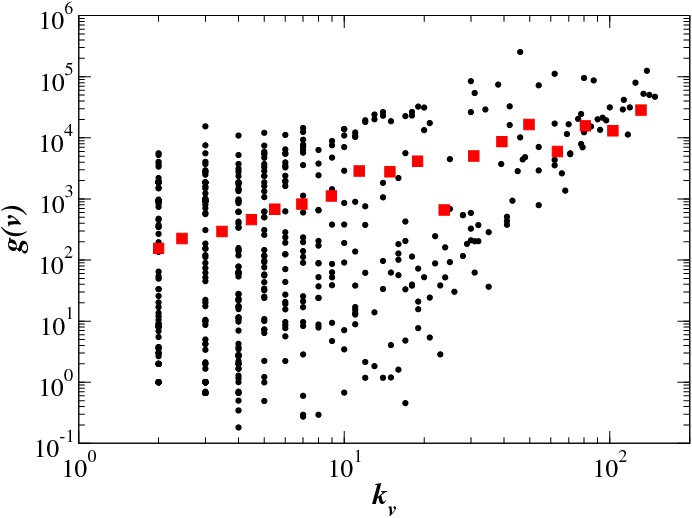,width=0.9\linewidth,clip=}
\end{tabular}
\caption{Scatter-plot of the betweenness centrality versus degree
for nodes of the North American air-transportation network. The (red) squares
correspond to the average betweenness centrality versus degree. From \cite{Barrat:2005}.}
\label{fig:bc_na}
\end{figure}

More precisely, it is generally useful to represent the average
betweenness centrality for vertices of the same degree
\begin{equation}
g(k)=\frac{1}{N(k)}\sum_{v/k_v=k}g(v) \ .
\end{equation}
where $N(k)$ is the number of nodes of degree $k$. For most networks,
$g(k)$ is strongly correlated with the degree $k$. In general, the
larger the degree, the larger the centrality. For scale-free
networks it has been shown that the centrality scales with $k$ as
\begin{equation}
g(k) \sim k^\eta
\label{eq:gk}
\end{equation}
where $\eta$ depends on the network \cite{Goh:2001,Barthelemy:2003c}.
For some networks however, the betweenness centrality fluctuations around the behavior
given by Eq.~(\ref{eq:gk}) can be very large and `anomalies' can
occur, in the sense that the variation of the centrality versus degree
is not a monotonous function.  Guimer\`a and
Amaral~\cite{Guimera:2005} have shown that this is indeed the case for
the worldwide air-transportation network. This is a very relevant
observation in that very central cities may have a relatively low
degree and vice versa. Fig.~\ref{fig:bc_na} displays the average
behavior along with the scattered plot of the betweenness versus
degree of all airports of the North American network.
In this case we find also very large fluctuations with a behavior similar to
that observed in~\cite{Guimera:2004}. 

These different observations call for the need of a network model
embedded in real space which can reproduce these betweenness
centrality features observed in real networks. More generally these
different results point out the importance of space as a relevant
ingredient in the structure of networks. In the chapter about models
(section \ref{sec4}) we focus on the interplay between spatial embedding,
topology and weights in a simple general model for weighted networks
in order to provide a modeling framework considering these three
aspects at once.

\paragraph{Communities.}
\label{sec3A2d}

In \cite{Guimera:2005}, Guimera et al. used modularity optimization
with simulated annealing (see section \ref{sec2B3}) in order to
identify communities defined by the worldwide air network.
\begin{figure}[h!]
\begin{tabular}{c}
\epsfig{file=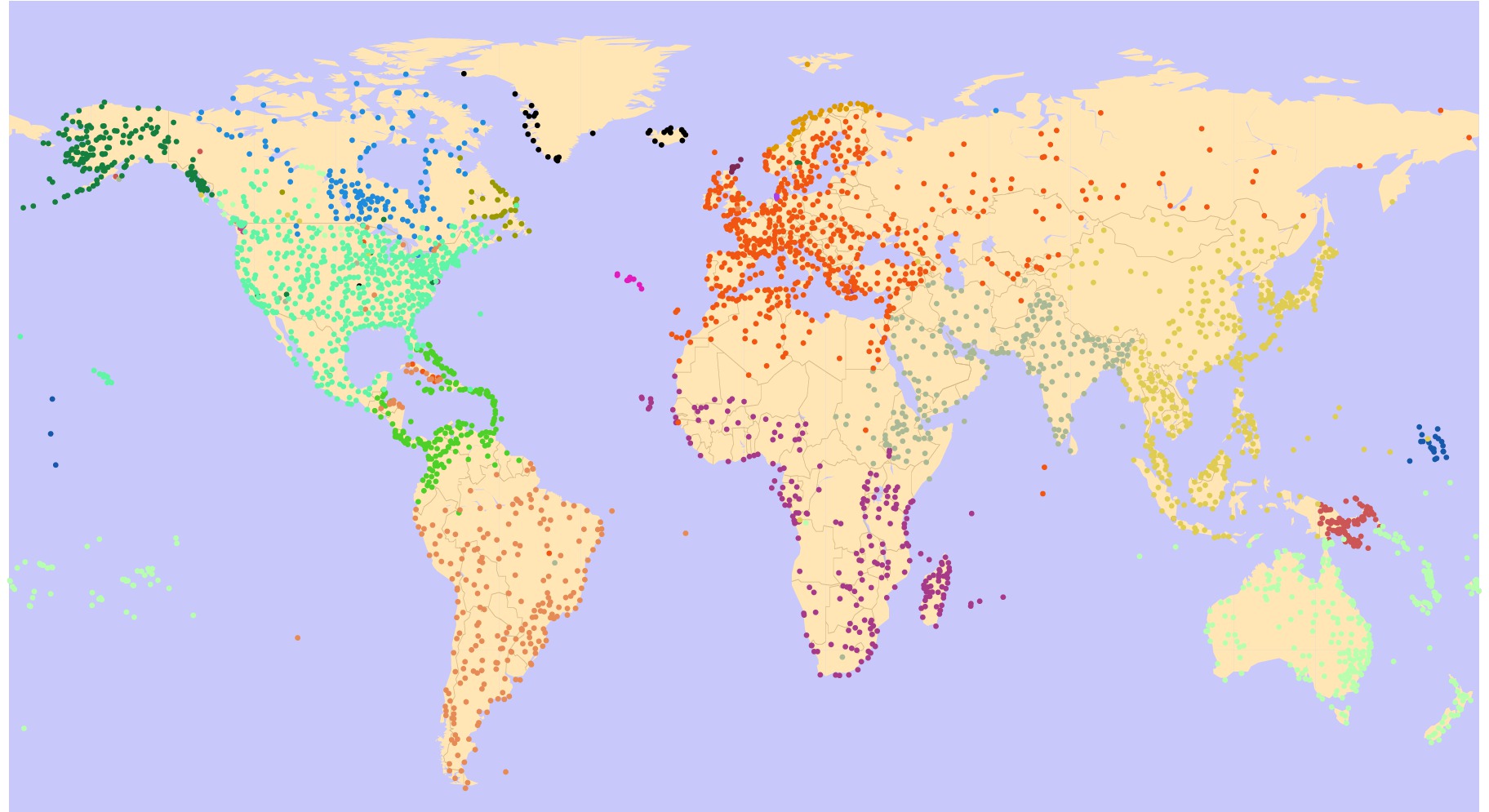,width=0.9\linewidth,clip=}
\end{tabular}
\caption{Communities of the worldwide air transportation network
  obtained with modularity optimization. Each node represents an
  airport and each color corresponds to a community.  From \cite{Guimera:2005}.}
\label{fig:commu_guimera}
\end{figure}
The obtained result is shown in Fig.~\ref{fig:commu_guimera} where
each color corresponds to a community. We immediately can see that
most of these communities are actually determined by geographical
factors and therefore is not very informative: the most important
flows are among nodes in the same geographical regions. More
interesting are the spatial anomalies which belong to a community from
the modularity point of view but which are in another geographical
region. For example, the `red' community of western Europe also
contains airports from Asian Russia. More generally, it is clear that
in the case of spatial networks, community detection offers a visual
representation of large exchange zones. It also allows to identify the
inter-communities links which probably play a very important role in
many processes such as disease spread for example.

\subsubsection{Bus, subway, railway, and commuters}
\label{sec3A3}

\paragraph{Subways.}
\label{sec3A3a}

One of the first studies (after the Watts-Strogatz paper) on the
topology of a transportation network was proposed by Latora and
Marchiori \cite{Latora:2001a} who considered the Boston subway network. It is a
relatively small network with $N=124$ stations. The average shortest path
is $\langle \ell\rangle\sim 16$ a value which is large compared to
$\ln 124\approx 5$ and closer to the two-dimensional result
$\sqrt{124}\approx 11$.

In \cite{Sienkiewicz:2005} Sienkiewicz and Holyst study a larger set
made of public transportation networks of buses and tramways for $22$
Polish cities and in \cite{vonFerber:2009} von Ferber et al. study the
public transportation networks for $15$ world cities. The number of
nodes of these networks varies from $N=152$ to $2811$ in
\cite{Sienkiewicz:2005} and in the range $[1494,44629]$ in
\cite{vonFerber:2009}. Interestingly enough the authors of
\cite{Sienkiewicz:2005} observe a strong correlation between the
number of stations and the population which is not the case for the
world cities studied in \cite{vonFerber:2009} where the number of
stations seems to be independent from the population\footnote{Incidentally,
  connecting socio-economical indicators to properties of these
  networks is certainly a very interesting and difficult question
  which probably will be addressed in the future.}. For the polish
cities the degree has an average in the range $[2.48,3.08]$ and in a
similar range $[2.18,3.73]$ for \cite{vonFerber:2009}. In both cases,
the degree distribution is relatively peaked (the range of variation
is usually of the order of one decade) consistently with the existence
of physical constraints \cite{Amaral:2000}.

Due to the relatively small range of variation of $N$ in these various
studies \cite{Latora:2001a,Sienkiewicz:2005,vonFerber:2009}, the
behavior of the average shortest path is not clear and could be fitted
by a logarithm or a power law as well. We can however note that the
average shortest path is usually large (of order $10$ in
\cite{Sienkiewicz:2005} and in the range $[6.4,52.0]$ in
\cite{vonFerber:2009}) compared to $\ln N$, suggesting that the
behavior of $\langle\ell\rangle$ might not be logarithmic with $N$ but
more likely scales as $N^{1/2}$, a behavior typical of a
two-dimensional lattice.

The average clustering coefficient $\langle C\rangle$ in \cite{Sienkiewicz:2005} varies
in the range $[0.055,0.161]$ and is larger than a value of the order
$C_{ER}\sim 1/N\sim 10^{-3}-10^{-2}$ corresponding to a random ER graph. The ratio
$\langle C\rangle/C_{ER}$ is explicitly considered in \cite{vonFerber:2009} and is
usually much larger than one (in the range $[41,625]$).  The
degree-dependent clustering coefficient $C(k)$ seems to present a
power law dependence, but the fit is obtained over one decade only.

In another study \cite{Seaton:2004}, the authors study two urban train
networks (Boston and Vienna which are both small $N=124$ and $N=76$,
respectively) and their results are consistent with the previous ones.

Due to the limited availability of data, there are less studies on the passenger
flows and the data so far seem to indicate that the flow is broadly
distributed \cite{Lee:2008,Roth:2010} a feature which seems to be
common to many transportation systems (see Table II).

\paragraph{Rail.}
\label{3A3b}

One of the first studies of the structure of railway network
\cite{Sen:2003b} concerns a subset of the most important stations and
lines of the Indian railway network and has $N=587$ stations. In the
P-space representation (see section \ref{sec3A1}), there is a link
between two stations if there is a train connecting them and in this
representation, the average shortest path is of order
$\langle\ell\rangle\approx 5$ which indicates that one needs $4$
connections in the worst case to go from one node to another one. In
order to obtain variations with the number of nodes, the authors
considered different subgraphs with different sizes $N$. The
clustering coefficient varies slowly with $N$ is always larger than
$\approx 0.7$ which is much larger than a random graph value of order
$1/N$. Finally in this study \cite{Sen:2003b}, it is shown that the
degree distribution is behaving as an exponential and that the
assortativity $\langle k_{nn}\rangle$(k) is flat showing an absence of
correlations between the degree of a node and those of its neighbors.

\begin{figure}[h!]
\begin{tabular}{c}
\epsfig{file=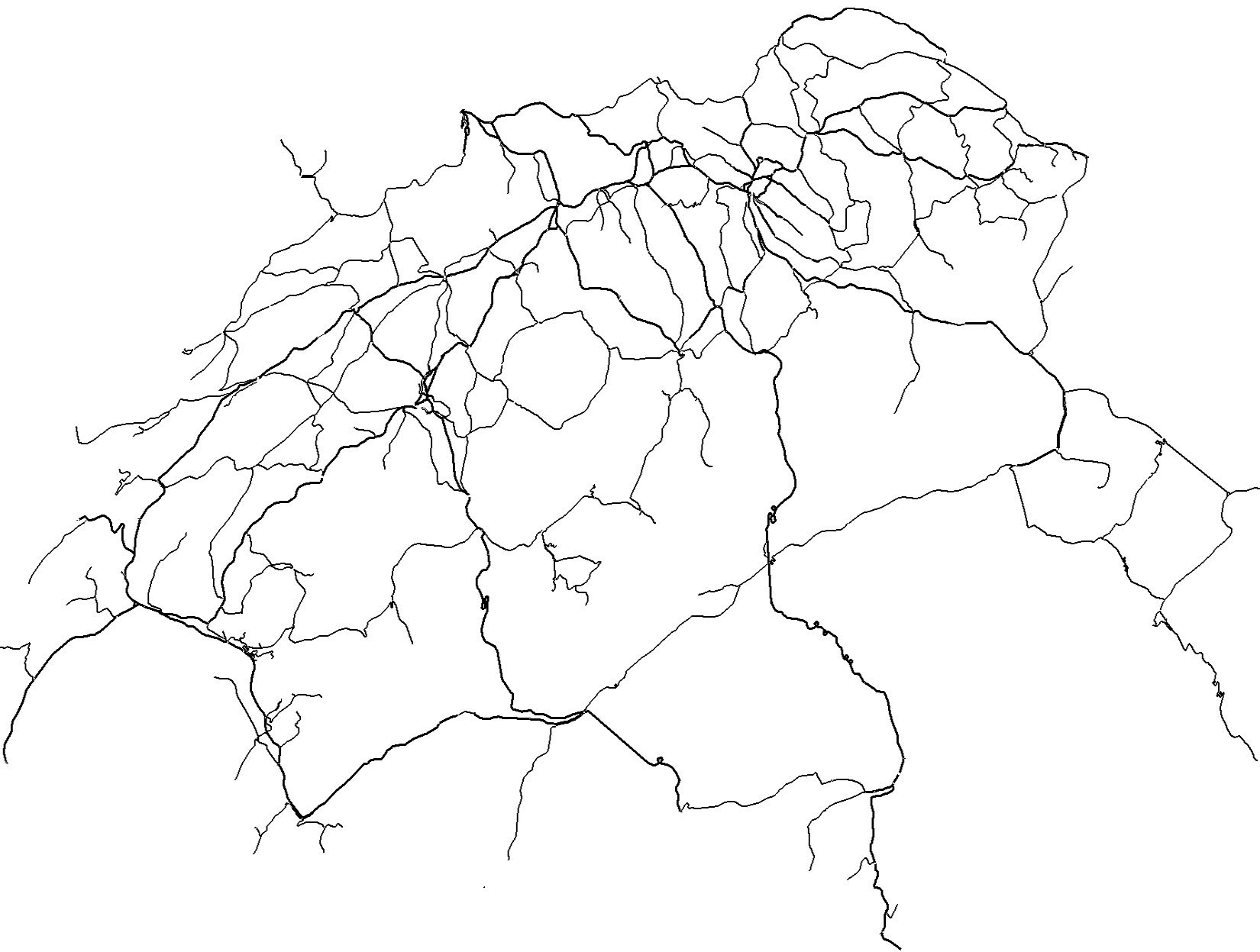,width=0.8\linewidth,clip=}
\end{tabular}
\caption{Physical map of the Swiss railway networks. From
  \cite{Kurant:2006b}.}
\label{fig:kurant2}
\end{figure}

In \cite{Kurant:2006b}, Kurant and Thiran study the railway
system of Switzerland and major trains and stations in Europe (and
also the public transportation system of Warsaw, Poland).  The Swiss
railway
network contains $N=1613$ nodes and $E=1680$ edges (Fig.~\ref{fig:kurant2}).
All conclusions drawn here are consistent with the various cases presented
in this chapter. In particular, the average degree is $\langle
k\rangle\approx 2.1$, the average shortest path is $\approx 47$
(consistent with the $\sqrt{N}$ result for a two-dimensional lattice),
the clustering coefficient is much larger than its random counterpart,
and the degree distribution is peaked (exponentially decreasing).

\begin{figure}[h!]
\begin{tabular}{c}
\epsfig{file=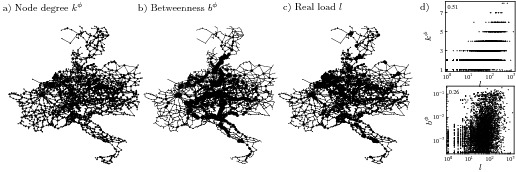,width=1.0\linewidth,clip=}
\end{tabular}
\caption{EU rail dataset. (a) Physical layout of the network showing node
  of size proportional to (a) the degree, (b) the node betweenness
  centrality , (c) the  real load. In (d, top) the degree versus the
  centrality is shown and in (d, bottom) the betweenness versus the
  real load. \cite{Kurant:2006}.}
\label{fig:kurant3}
\end{figure}

Finally, in \cite{Kurant:2006} Kurant and Thiran studied networks by
decomposing it into layers (see Fig.~\ref{fig:kurant3}).
In particular, they compare the spatial distribution of the degree,
the node betweenness centrality and the real load. As shown in
Fig.~\ref{fig:kurant3}d the correlation between these different
indicators is poor (with a Pearson correlation coefficient equal to
$0.5$ for betweenness centrality$-k$ and $0.26$ for the correlation betweenness centrality-real load). As
expected from general considerations developed in the section
\ref{sec2B1d}, we see large fluctuations in the relation $k$ and the
betweenness centrality, but surprisingly here, the degree seems to be a better indicator
of the real load than the betweenness centrality \cite{Kurant:2006}.

\paragraph{Urban commuters network.}
\label{sec3A3C}

\begin{figure}[!h]
\centering
\begin{tabular}{c}
\includegraphics[angle=0,scale=.50]{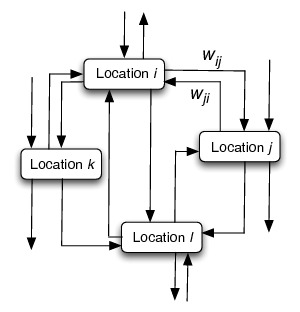}
\end{tabular}
\caption{ Construction of the network between the locations in
  Portland as reported in \cite{Chowell:2003}. The weight $w_{ij}$ on
  the link $i-j$ represents the daily traffic from $i$ to $j$. After \cite{Chowell:2003}.}
\label{fig:chowell1}
\end{figure}

\begin{figure}[!h]
\centering
\begin{tabular}{c}
\includegraphics[angle=0,scale=.40]{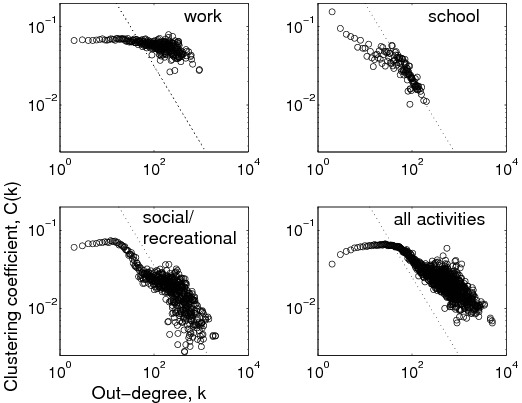}
\end{tabular}
\caption{ Loglog plots of the clustering coefficient $C(k)$ as a
  function of the out-degree for subnetworks constructed from work
  activities, school activities, social activities, and all the
  activities grouped together (the dotted line has slope $−1$). For social and
  recreational activities $C(k)\sim 1/k$ but for the networks
  constructed from work activities, the clustering coefficient is
  almost constant. From \cite{Chowell:2003}.}
\label{fig:chowell2}
\end{figure}

As discussed above, it is important to know how
the flows of individuals are structured in a dense urban area
(Theoretical discussions about this problem can be found in the
section \ref{sec3C3} about the so-called gravity law). In
\cite{Chowell:2003}, Chowell et al. study the simulated movements of
$1.6$ million individuals during a typical day in the city of Portland
(Oregon, USA) and construct a network where the nodes represent the
locations (buildings, homes, etc) and where the edges represent the
flow of individuals between these locations (see
Fig.~\ref{fig:chowell1}).
The distribution of the outdegree (which is the number of edges
leaving a node), and of the corresponding out-traffic, are well-fitted
by power laws. The clustering coefficient for different types of
locations are all large and when all activities are aggregated, the
corresponding clustering coefficient behaves as $1/k$
(Fig.~\ref{fig:chowell2}). It thus seems that at this relatively small
scale, space is not constraining enough (or in other words the cost
variations are too small) and the important features of spatial
networks do not seem to appear here.

At a larger scale, de Montis et al. \cite{deMontis:2005} studied the
network representing the interurban commuting traffic of the Sardinia
island (Italy), which amounts to 375 municipalities and $1.6$ millions
inhabitants. The nodes correspond to towns and the edges to the actual
commuting flows among those. This network has a peaked distribution of
degrees and similarly to the urban case studied in
\cite{Chowell:2003}, the flows are very heterogeneous and can be
fitted by a power law with exponent $\approx 1.8$. In contrast with
\cite{Chowell:2003} where the relation between the strength and the
degree is linear, it is strongly non-linear with $\beta\approx 1.9$ in
this inter-urban case.

Still in the case of commuters, a recent study \cite{Caschili:2009}
used modularity optimization to determine communities from this
inter-urban commuters weighted network and showed that there is a
consistency between the obtained communities and the administrative
regions. A similar result was obtained using bank notes dispersal in
\cite{Thiemann:2010}. Community detection was also used on the Belgian
mobile phone network to identify language communities
\cite{Blondel:2008}. Community detection thus appears as an
interesting tool for defining scientifically administrative boundaries
and more generally for policy making and planning.

\subsubsection{Cargo-ship}
\label{3A4}

The last example of transportation network that we will discuss in
this chapter is the cargo ship network \cite{Xu:2007,Hu:2009b,Kaluza:2010}. This
network is particularly important as more than $90\%$ of the world
trade is carried by the sea.

\begin{figure}[h!]
\begin{tabular}{c}
\epsfig{file=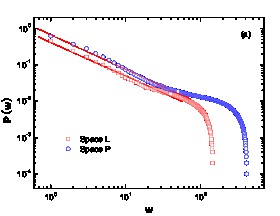,width=0.8\linewidth,clip=}\\
\epsfig{file=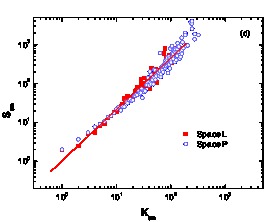,width=0.8\linewidth,clip=}
\end{tabular}
\caption{ (a) Weight distribution for the worldwide maritime network
  (the power law fit gives an exponent $\approx 0.95$). (b) Strength
  versus degree showing a nonlinear behavior (the fit gives an
  exponent $\approx 1.3$). From \cite{Hu:2009b}.}
\label{fig:hu}
\end{figure}

Cargo ships are not all the same and can be either bulk dry carriers,
oil tankers or can be designed to transport containerized cargo. These
different cargo types define different networks which can display
different properties \cite{Kaluza:2010}. In the case of containerized
cargo, many important statistical parameters were measured in
\cite{Hu:2009b} and we summarize here their results. The dataset
consists in $N=878$ ports and $1802$ different lines giving $7955$
edges. In the L-space representation (see section \ref{sec3A1}), the
nodes are the ports and there is a link between two ports when they
are connected by a direct route. The shortest path between two ports
in this representation thus represents the number of stops needed to
travel between these two nodes. The average degree of the undirected
version of this network is $\langle k\rangle \approx 9$, the average
clustering coefficient is $\langle C\rangle\approx 0.40$ which is much
larger than $1/878\approx 10^{-3}$. The average shortest path is
$\langle\ell\rangle\approx 3.6$ and the out-, in- and undirected
degree distributions seem to be relatively broad with maximum values
as large as $k\sim 10^2$.  More clearly the weight (defined as the
number of container lines for a given time period) and the strength
(see section \ref{2B2}) are broadly distributed. The power law fit
proposed in \cite{Hu:2009b} gives a behavior for the weight $P(w)\sim
1/w$ (for $w<100$, see Fig.~\ref{fig:hu}).
The strength (see \ref{sec2B2}) versus the degree display a nonlinear
behavior (Fig.~\ref{fig:hu}) typical of these networks. The exponent
is of order $\beta\simeq 1.3$ not far from the values $\beta\simeq
1.4-1.5$ measured for the worldwide network.

For this weighted network, weighted clustering and weighted
assortativity were also measured \cite{Hu:2009b} showing that
$C^w(k)>C(k)$ and $k_{nn}^w(k)>k_{nn}(k)$ which is the typical sign
that most of the traffic takes place on link connecting various
hubs. Also this network is weakly assortative with an almost flat
$k_{nn}(k)$.

\begin{figure}[h!]
\begin{tabular}{c}
\epsfig{file=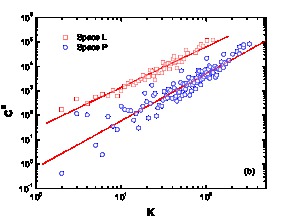,width=0.8\linewidth,clip=}
\end{tabular}
\caption{ Betweenness centrality versus the degree $k$. The power law
  fit for the L-space gives $1.66$ and $1.93$ in the P-space. From \cite{Hu:2009b}.}
\label{fig:hu2}
\end{figure}

The behavior of the betweenness centrality with the degree is interesting and is shown in
Fig.~\ref{fig:hu2} where we 
see that anomalies are very rare: the larger the degree and the larger
the betweenness centrality. This particular feature (which makes this
network very different from the worldwide air travel network) might
come from the fact that geographical constraints are less severe for
cargo ships. We saw that the existence of anomalies could be linked to
the fact that long links are difficult to create and must be
compensated by a large degree. Longer links here are maybe less costly
and constraints therefore smaller leading to less anomalies and to a
behavior closer to a power law $\langle g(k)\rangle\sim k^\eta$
typical of standard non-spatial networks.

Recently, in \cite{Kaluza:2010} Kaluza et al. studied a larger dataset
(obtained for the year $2007$) with $490,517$ non-stop journeys
between $36,351$ distinct pairs of departure and arrival ports. They
also found a small average shortest path $\langle \ell\rangle\approx
2.5$ with more than $50\%$ of possible origin-destination pairs that can be
connected by two steps or less. This study confirms the facts that the
average clustering is large, and that the degree, weight, and strength
distributions are large. The exponent of the weight (defined here as
the available space on all ships) distribution is equal to $\approx
1.7$ and is different from the study \cite{Hu:2009b}.

The weight is also very asymmetric (ie. $w_{ij}\neq w_{ji}$) with
$59\%$ of all linked pairs which exist in one direction only. This
feature can also be observed in other transportation networks (such as
mail or freight \cite{Colizza:2008}) and is probably the sign of
distribution networks as opposed to travel networks where basically
every individual performs a roundtrip implying symmetrical weights
$w_{ij}\approx w_{ji}$.

\begin{figure}[h!]
\begin{tabular}{c}
\epsfig{file=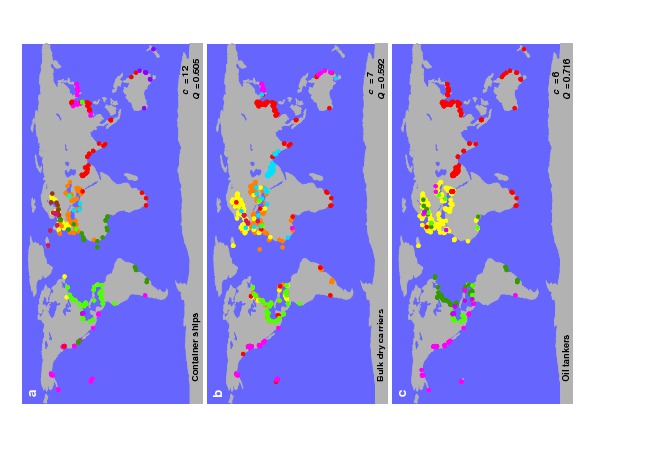,width=1.0\linewidth,angle=-90,clip=}
\end{tabular}
\caption{ Communities of ports obtained for the cargo networks
  obtained for different types of ships: (a) container ships, (b) bulk
  dry carriers, and (c) oil tankers. The optimal values of the number
  of communities $c$ and the modularity $Q$ are shown in the
  bottom right corner of each panel. From \cite{Kaluza:2010}.}
\label{fig:cargo2}
\end{figure}

The strength in \cite{Kaluza:2010} also scales as a power law of the
degree with an exponent $\simeq 1.46$ not far from the values of the
exponents found in other networks. Finally the betweenness centrality
scales well with the degree and displays very few anomalies in
agreement with \cite{Hu:2009b}.
%
%
After these rather standard measurements, the authors of
\cite{Kaluza:2010} proposed other measurements. They first
characterized the community structure (Figure \ref{fig:cargo2}) for
three subnetworks of this cargo network obtained for different ship
types (containers, bulk dry carriers, and oil tankers).
These communities were obtained by optimizing the modularity $Q$ (see
section \ref{sec2B3}). This procedure might miss some small structures
\cite{Fortunato:2007} but at large scale can be trusted. Indeed in
this case, groups are observed which are geographically consistent
demonstrating the relevance of this method in this case. One could
even think that community detection might be an important tool in
geography and in the determination of new administrative or economical
boundaries.

Finally, the authors of \cite{Kaluza:2010} studied the
statistical occurrence of motifs, a method developed in a biological
context \cite{Milo:2002}. 
\begin{figure}[h!]
\begin{tabular}{c}
\epsfig{file=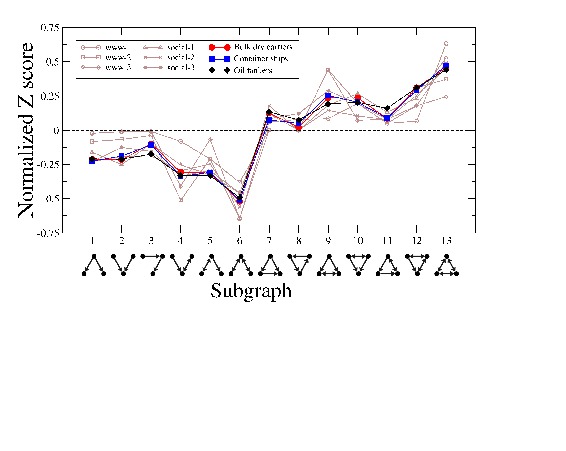,width=1.0\linewidth,angle=0,clip=}
\end{tabular}
\caption{ Z-score for the different motifs shown at the bottom of the
  figure. Results are shown for the three main cargo fleets and for
  comparison with the WWW and social networks. From
  \cite{Kaluza:2010}.}
\label{fig:cargo3}
\end{figure}
In Fig.~\ref{fig:cargo3}, we reproduce the motif distribution for
the three main cargo fleets compared with the results for the Web and
social networks displaying surprising similarities pointing to the
existence of `superuniversal' features in complex networks.

\subsection{Infrastructure networks}
\label{sec3B}

Our modern societies depend crucially on infrastructure
networks. Movement of people on roads and streets, water and gas
distribution, power grids, and communication networks are all networks
vital to the good functioning of the societies. In this chapter, we
will review the most important and recent results on their structure in
relation with spatial aspects.

\subsubsection{Road and street networks}
\label{sec3B1}

Despite the peculiar geographical, historical, social-economical
mechanisms that have shaped distinct urban areas in different ways
(see for example \cite{Levinson:2006} and references therein), recent
empirical studies
\cite{Batty:2005,Makse:1998,Clark:1951,Crucitti:2006,Jiang:2004,Cardillo:2006,Lammer:2006,Jiang:2007}
have shown that, at least at a coarse grained level, unexpected
quantitative similarities exist between road networks of very
different cities. The simplest description of the street network
consists of a graph whose links represent roads, and vertices
represent roads' intersections and end points. For these graphs, links
intersect essentially only at vertices and are thus
planar.\footnote{For roads, highways, etc. planarity can be violated
  due to bridges but can be considered as a good approximation
  \cite{Lammer:2006}.}  Measuring spatial properties of cities through
the analysis of the street network is not new and was popularized by
Hillier and Hanson \cite{Hillier:1984} under the term `space
syntax'. In this chapter, we will discuss different recent measures of
these networks in the light of our current understanding of the
structure of networks.

\paragraph{Degrees, lengths, and cell areas.}
\label{sec3B1a}

\begin{figure}[h!]
\begin{tabular}{c}
\epsfig{file=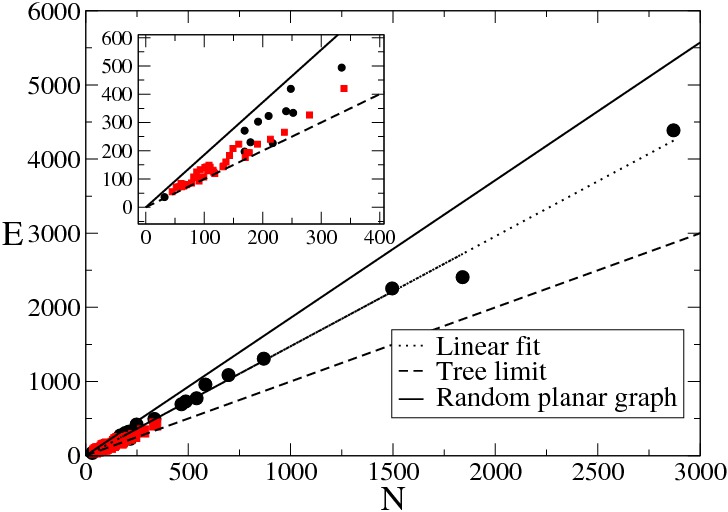,width=0.8\linewidth,clip=}\\
\epsfig{file=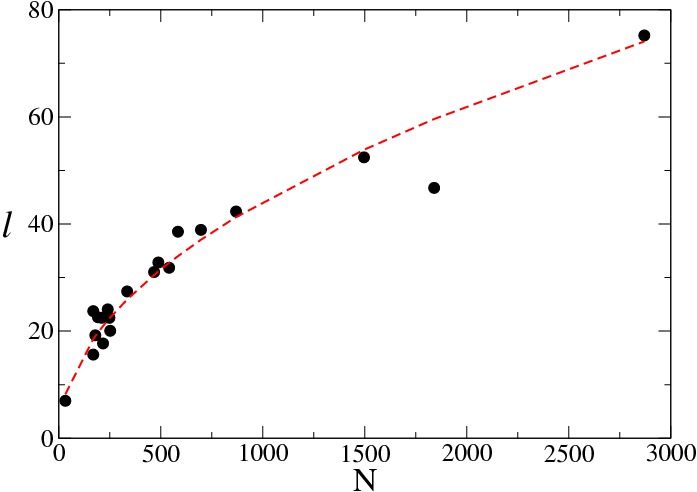,width=0.8\linewidth,clip=}
\end{tabular}
\caption{(a) Numbers of roads versus the number of nodes
  (ie. intersections and centers) for data from
  \protect\cite{Cardillo:2006} (circles) and from \cite{Buhl:2006} (squares). In
  the inset, we show a zoom for a small number of nodes. (b) Total
  length versus the number of nodes. The line is a fit which predicts
  a growth as $\sqrt{N}$ (data from \cite{Cardillo:2006} and figures from \cite{Barthelemy:2008}).}
\label{fig:k_cost}
\end{figure}

\begin{figure}[h!]
\begin{tabular}{c}
\epsfig{file=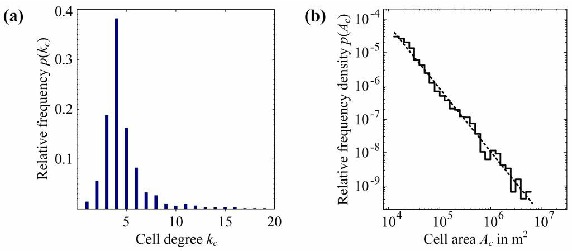,width=1.0\linewidth,clip=}
\end{tabular}
\caption{ (a) Degree distribution of degrees for the road network of
  Dresden. (b) The frequency distribution of the cell’s surface areas
  $A_c$ obeys a power law with exponent $\alpha\approx 1.9$ (for the
  road network of Dresden).  From \cite{Lammer:2006}.}
\label{fig:lammer}
\end{figure}

In \cite{Cardillo:2006,Buhl:2006} measurements for different cities in
the world are reported. Based on the data from these sources, the
authors of \cite{Barthelemy:2008} plotted (Fig. \ref{fig:k_cost}a)
the number of roads $E$ (edges) versus the number of intersections
$N$.  The plot is consistent with a linear fit with slope $\approx
1.44$ (which is consistent with the value $\langle k\rangle\approx
2.5$ measured in \cite{Buhl:2006}). The quantity $e= E/N= \langle k
\rangle / 2$ displays values in the range $1.05 < e < 1.69$, in
between the values $e=1$ and $e=2$ that characterize tree-like
structures and $2d$ regular lattices, respectively. In a study of $20$
German cities, L\"ammer et al.~\cite{Lammer:2006} showed that most
nodes have four neighbors (the full degree distribution is shown in
Fig.~\ref{fig:lammer}) and that for various world cities the degree
rarely exceeds $5$ \cite{Cardillo:2006}. These values are however not
very indicative: planarity imposes severe constraints on the degree of
a node and on its distribution which is generally peaked around its
average value. Few exact values and bounds are available for the
average degree of classical models of planar graphs. In general it is
known that $e\le 3$, while it has been recently
shown~\cite{Gerke:2004} that $e > 13/7$ for planar Erd\"os-Renyi
graphs (\cite{Gerke:2004} and section \ref{sec4B}).

\begin{figure}[h!]
\begin{tabular}{c}
\epsfig{file=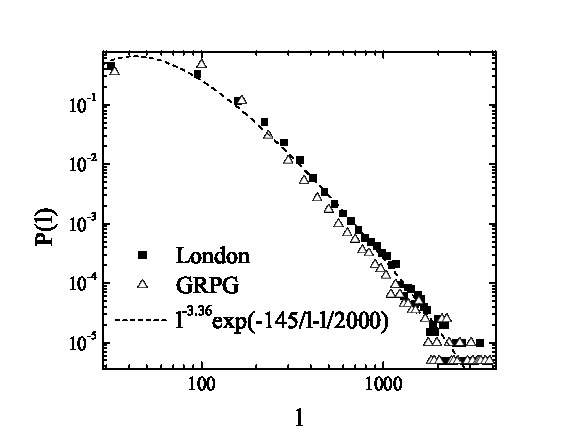,width=1.0\linewidth,clip=}
\end{tabular}
\caption{ Length distribution $P(\ell_1)$ for the street network of London
  (and for the model GRPG proposed in this study). From \cite{Masucci:2009}).}
\label{fig:pllondon}
\end{figure}

In Fig.~\ref{fig:k_cost}b, we can see that the total length $\ell_T$ of the
network versus N for the towns considered in~\cite{Cardillo:2006}. Data are
well fitted by a power function of the form
\begin{equation}
\ell_T=\mu N^{\beta}
\end{equation}
with $\mu\approx 1.51$ and $\beta\approx 0.49$. In order to
understand this result, one has to focus on the street segment length
distribution $P(\ell_1)$. This quantity  has been measured for London in \cite{Masucci:2009} and
is shown in Fig.~\ref{fig:pllondon}.
This figure shows that the distribution decreases rapidly and the fit
proposed by the authors of \cite{Masucci:2009} suggests that
\begin{equation}
P(\ell_1)\sim \ell_1^{-\gamma}
\end{equation}
with $\gamma\simeq 3.36$ which implies that
both the average and the dispersion are well-defined and finite. If we
assume that this result extends to other cities, it means that we have
a typical distance $\ell_1$ between nodes which is meaningful. This
typical distance between connected nodes then naturally scales as
\begin{equation}
\ell_1\sim \frac{1}{\sqrt{\rho}}
\end{equation}
where $\rho=N/L^2$ is the density of vertices and $L$ the linear dimension
of the ambient space. This implies that the total length scales as
\begin{equation}
\ell_T\sim E\ell_1\sim \frac{\langle k\rangle}{2}L\sqrt{N}
\end{equation}
This simple argument reproduces well the $\sqrt{N}$ behavior observed
in Fig.~\ref{fig:k_cost} and also the value (given the error bars) of the prefactor
$\mu\approx \langle k\rangle k/2$.

The simplest hypothesis consistent with all the data presented so far,
at this stage, is that the road network is an homogeneous and
translational invariant structure. However, this network naturally
produces a set of non overlapping cells, encircled by the roads
themselves and covering the embedding plane, and surprisingly, the
distribution of the area $A$ of such cells measured for the city of
Dresden in Germany (Fig.~\ref{fig:lammer}) has the form
\begin{equation}
P(A)\sim A^{-\alpha}
\end{equation}
with $\alpha\simeq 1.9$. This is in sharp contrast with the simple
picture of an almost regular lattice which would predict a
distribution $P(A)$ peaked around $\ell_1^2$.
It is interesting to note that if we assume that $A\sim 1/\ell_1^2\sim
1/\rho$ and that $\rho$ is distributed according to a law $f(\rho)$
(with a finite $f(0)$), a simple calculation gives
\begin{equation}
P(A)\sim \frac{1}{A^2} f(1/A)
\end{equation}
which behaves as $P(A)\sim 1/A^2$ for large $A$. This simple argument
thus suggests that the observed value $\approx 2.0$ of the exponent is
universal and reflects the random variation of the density. More
measurements are however needed at this point in order to test the
validity of this hypothesis.

The authors of~\cite{Lammer:2006} also measured the distribution of
the form factor $\phi=4A/(\pi D^2)$, (the ratio of the area of the
cell to the area of the circumscribed circle) and found that most
cells have a form factor between $0.3$ and $0.6$, suggesting a large
variety of cell shapes, in contradiction with the assumption of an
almost regular lattice. These facts thus call for a model radically
different from simple models of regular or perturbed lattices.

Finally, we note that the degree distribution in the dual
representation has been studied \cite{Kalapala:2006} for the road
network in the US, England, and Denmark and displays broad
fluctuations with a power law distribution with exponent
$2.0<\gamma<2.5$.

\paragraph{Betweenness centrality.}
\label{sec3B1b}

The importance of a road can be characterized by its traffic which can
be measured by sensors (see for example
\cite{Schreckenberg:2004})\footnote{The traffic between two zones
  given by the origin-destination matrix (see section \ref{sec3C}) is
  however much more difficult to obtain.}. If we assume that the
traffic between all pairs of nodes is the same a natural proxy for the
traffic is the betweenness centrality. Even if the underlying
assumptions are not correct the spatial distribution of the
betweenness centrality gives important information about the coupling
between space and the structure of the road network.

L\"ammer et al \cite{Lammer:2006} studied the German road network and
obtained very broad distributions of betweenness centrality with a power law exponent in
the range $[1.279,1.486]$ (for Dresden $\approx 1.36$).
\begin{figure}[h!]
\begin{tabular}{c}
\epsfig{file=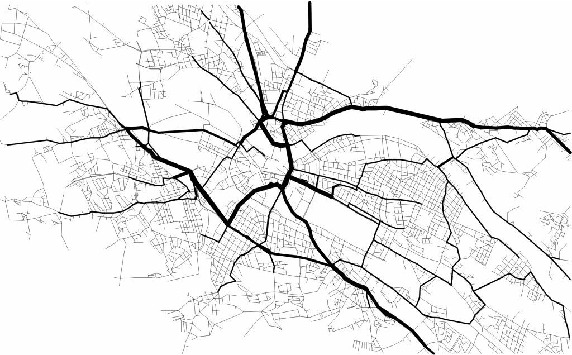,width=1.0\linewidth,clip=}
\end{tabular}
\caption{ Betweenness centrality for the city of Dresden. The width of the links
corresponds to the betweenness centrality. From \cite{Lammer:2006}.}
\label{fig:lammer2}
\end{figure}
These broad distributions of the betweenness centrality signals the
strong heterogeneity of the network in terms of traffic, with the
existence of a few very central roads which very probably points to
some congestion traffic problems. Also the absence of a scale in a
power law distribution suggests that the importance of roads is
organized in a hierarchical way, a property expected for many
transportation networks \cite{Yerra:2005}. The broadness of the
betweenness centrality distribution does not seem however to be
universal. Indeed, in \cite{Crucitti:2006,Scellato:2006}, the
betweenness centrality distribution is peaked (depending on the city
either exponentially or according to a Gaussian) which signals the
existence of a scale and therefore of a finite number of congested
points in the city.
\begin{figure}[h!]
\begin{tabular}{c}
\epsfig{file=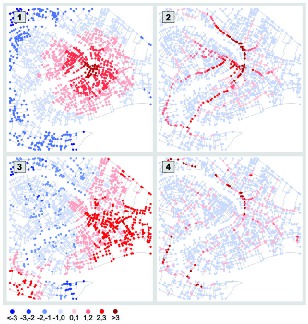,width=1.0\linewidth,clip=}
\end{tabular}
\caption{ Colour-coded maps for different centralities in Venice,
  Italy (see \cite{Crucitti:2006}). (1) Closeness, (2) Betweenness centrality, (3) Straightness,
  (4) Information (from blue to red the centrality increases).  From \cite{Crucitti:2006}.}
\label{fig:cruci}
\end{figure}
The betweenness centrality is in itself interesting since it points to
the important zone which potentially are congested. The
Figs.~\ref{fig:lammer2},\ref{fig:cruci} display the spatial
distribution of the betweenness centrality for various cities. As
expected zones which are central from a geographical point of view
also have a large betweenness centrality. We however see that other
roads or zones can have a large betweenness centrality pointing to a
complex pattern of flow distribution in cities.

In addition to have a relation with the traffic and possibly the
congestion, a recent paper \cite{Strano:2009} proposes an interesting
direction which is in the general context of connecting topological
measures of the networks and socio-economical indices. In particular,
Strano et al. show that there is a clear correlation between the
betweenness centrality and the presence of commercial activities.

\paragraph{Other measures.}
\label{sec3B1c}

Buhl et al. \cite{Buhl:2006} measured different indices for $300$ maps
corresponding mostly to settlements located in Europe, Africa, Central
America, India. They found that many networks depart from the grid
structure with an alpha index (or Meshedness) usually low. For various
world cities, Cardillo et al. \cite{Cardillo:2006} found that the alpha
index varies from $0.084$ (Walnut Creek) to $0.348$ (New York City)
which reflects in fact the variation of the average degree. Indeed for
both these extreme cases, using Eq.~(\ref{eq:alphalimit}) leads to
$\alpha_{NYC}\simeq (3.38-2)/4\simeq 0.345$ and for Walnut Creek
$\alpha_{WC}\simeq (2.33-2)/4\simeq 0.083$. This same study seems to
show that triangles are less abundant than squares (except for cities
such as Brasilia or Irvine).

\begin{figure}[h!]
\begin{tabular}{c}
\epsfig{file=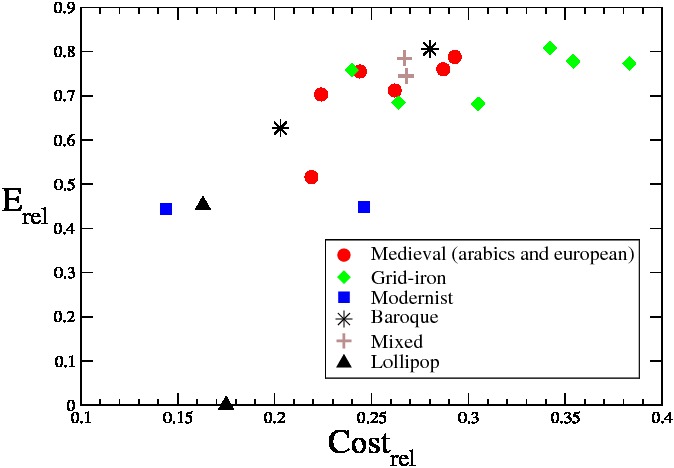,width=0.9\linewidth,clip=}
\end{tabular}
\caption{ Relative efficiency versus relative cost for $20$ different
cities in the world. In this plot the point $(0,0)$ corresponds to the MST and the
point $(1,1)$ to the greedy triangulation. From \cite{Cardillo:2006}.}
\label{fig:cardillo}
\end{figure}

Measures of efficiency are relatively well correlated with the alpha
index but displays broader variations demonstrating that small
variations of the alpha index can lead to large variations in the
shortest path structure. Cardillo et al plotted the relative
efficiency (see section \ref{sec2B2d})
\begin{equation}
E_{rel}=\frac{E-E^{MST}}{E^{GT}-E^{MST}}
\end{equation}
versus the relative cost 
\begin{equation}
C_{rel}=\frac{C-C^{MST}}{C^{GT}-C^{MST}}
\end{equation}
where GT refers to the greedy triangulation (the maximal planar
graph) and MST to the minimal spanning tree. The cost is here
estimated as the total length of segments $C\equiv\ell_T$ (see section \ref{sec2B2d}).
The obtained result is shown in Fig.~\ref{fig:cardillo} which
demonstrates two things.
First, it shows -- as expected -- that efficiency is increasing with the
cost with an efficiency saturating at $\sim 0.8$. In addition, this
increase is slow: typically, doubling the value of $C$ shifts the
efficiency from $\sim 0.6$ to $\sim 0.8$.  Second, it shows that most
of the cities are located in the high cost-high efficiency region. New
York City, Savannah and San Francisco have the largest value of the
efficiency ($\sim 0.8$) with a relative cost value around $\sim
0.35$. It seems however at this stage difficult to clearly identify
different classes of cities and further studies with a larger number
of cities is probably needed in order to confirm the typology proposed
in \cite{Cardillo:2006}.

Finally, we mention the recent study \cite{Strano:2010} on the street
and subway networks of Paris and London. The accessibility in these
cities is studied in terms of self-avoiding random walks displaying
several differences. In particular, Paris seems to have a larger
average accessibility than London, probably due to a large number of
bridges.


\subsubsection{Power grids and water distribution networks}
\label{sec3B2}

Power grids are one of the most important infrastructure in our
societies. In modern countries, they have evolved for a rather long time
(sometimes a century) and are now complex systems with a large variety
of elements and actors playing in their functioning. This complexity leads
to the relatively unexpected result that their robustness is actually
not very well understood and large blackouts such as the huge August
$2003$ blackout in North America call for the need of a better
understanding of these networks.

\begin{figure}[h!]
\centering
\begin{tabular}{c}
\epsfig{file=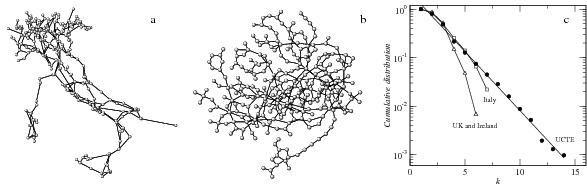,width=1.0\linewidth,clip=}\\
\end{tabular}
\caption{ (a) Map of the Italian power grid. (b) Topology of the
  Italian power grid. (c) Degree distribution for the European network
  (UCTE), Italy and the UK and Ireland. In all cases the degree
  distribution is peaked and can be fitted by exponential. From \cite{Sole:2008}.}
\label{fig:pgitaly}
\end{figure}

\begin{figure}[h!]
\begin{tabular}{c}
\includegraphics[angle=0,scale=.40]{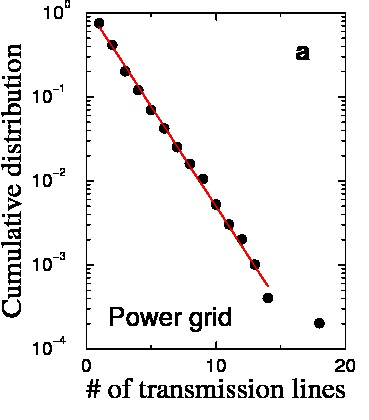}\\
\includegraphics[angle=270,scale=.30]{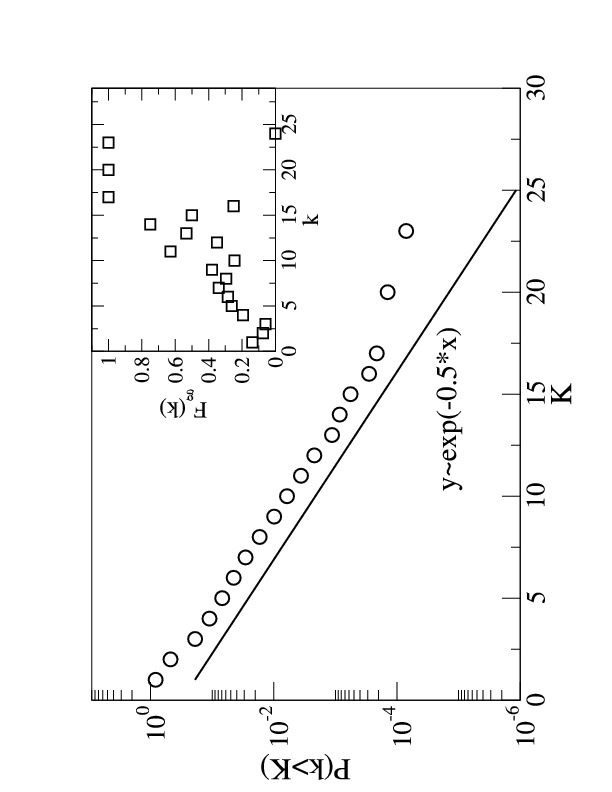}
\end{tabular}
\caption{ Degree distribution of substations in Southern California
  (top panel) and for the North American power grid (bottom panel). In
  both cases, the lines represent an exponential fit. From
  \cite{Amaral:2000} and \cite{Albert:2004}, respectively.}
\label{fig:naka1}
\end{figure}

The topological structure of these networks was studied in different
papers such as \cite{Amaral:2000,Albert:2004,Sole:2008}. In
particular, in \cite{Amaral:2000} and \cite{Albert:2004}, the authors
study the Southern Californian and the North American power grids. In
these networks the nodes represent the power plants, distributing and
transmission substations, and the edges correspond to transmission
lines.
These networks are typically planar (see for example the Italian case,
Fig.~\ref{fig:pgitaly}) and we expect a peaked degree distribution,
decreasing typically as an exponential of the form $P(k)\sim
\exp(-k/\langle k\rangle)$ with $\langle k\rangle$ of order $3$ in
Europe and $2$ in the US. 
The other studies on US power grids confirm that the degree
distribution is exponential (see Fig.~\ref{fig:naka1}).
In \cite{Albert:2004}, Albert, Albert, and Nakarado also studied the
load (a quantity similar to the betweenness centrality) and found a
broad distribution. The degree being peaked we can then expect very
large fluctuations of load for the same value of the degree, as
expected in general for spatial networks (see section
\ref{sec2B1d}). These authors also found a large redundance in this
network with however $15\%$ of cut-edges.

\begin{figure}[h!]
\begin{tabular}{c}
\epsfig{file=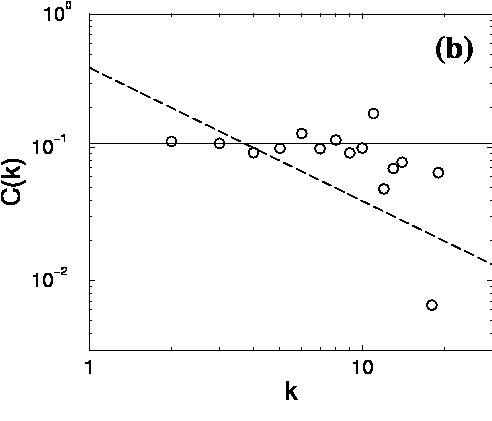,width=0.7\linewidth,clip=}
\end{tabular}
\caption{ Scaling of the clustering $C(k)$ for the Power Grid of the
  Western United States. The dashed line has a slope $-1$ and the solid
line corresponds to the average clustering coefficient. From \cite{Ravasz:2003}.}
\label{fig:power}
\end{figure}

Also, as expected for these networks, the clustering coefficient is
rather large and even independent of $k$ as shown in the case of the
power grid of Western US (see Fig.~\ref{fig:power}).

\begin{figure}[h!]
\begin{tabular}{c}
\epsfig{file=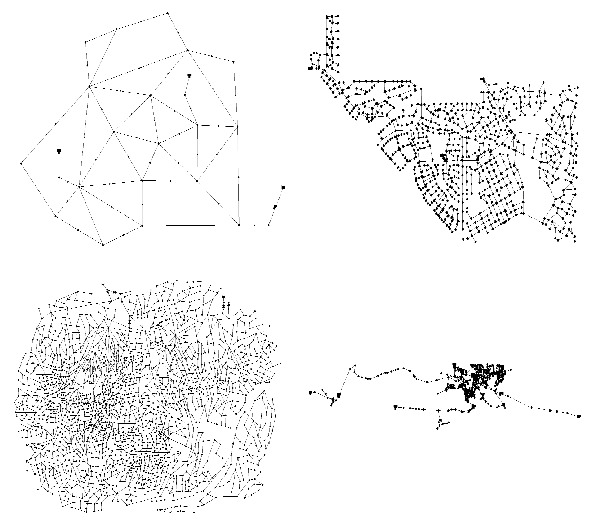,width=1.0\linewidth,clip=}
\end{tabular}
\caption{ Representation of water distribution networks. Left panels
  (from top to bottom):  Synthetic networks (`Anytown' \cite{Walski:1987}, and `EXNET'
  \cite{Farmani:2004}). Top-right panel: Colorado Spring Utilities
  network. Bottom-right panel: Richmond (UK) water distribution
  network. From \cite{Yazdani:2010}.}
\label{fig:water}
\end{figure}

Beside the distribution of electricity, our modern societies also rely
on various other distribution networks. The resilience of these
networks to perturbations is thus an important point in the design and
operating of these systems. In \cite{Yazdani:2010}, Yazdani and
Jeffrey study the topological properties of the Colorado Springs
Utilities and the Richmond (UK) water distribution networks (shown in
Fig.~\ref{fig:water}).
Both these networks (of size $N=1786$ and
$N=872$, respectively) are sparse planar graphs with very peaked
distributions (the maximum degree is $12$).

\subsubsection{The Internet}
\label{sec3B3}

The Internet is the physical layout for the WWW which organizes the
information and makes it accessible. At a small scale, the
Internet is a network of routers connected by wires. At a larger
scale, the routers are grouped according to an administrative
authority, the autonomous system (AS) and at this level, the Internet
can be seen as a set of interconnected AS.

There is obviously a very large literature in computer science (but
also in statistical physics \cite{Pastor:2003}) about the Internet and
its structure. Here we will briefly give some characteristics of this
important network that we believe are related to the spatial aspects
of the Internet.

The nodes of this network (routers) are distributed among different
countries and one expects a non-uniform distribution depending on
socio-economical factors. Indeed, this distribution as been shown in
\cite{Yook:2002} to form a fractal set of dimension $d_f\simeq 1.5$
strongly correlated with the population density.\footnote{The fractal
  dimension is here the usual one- see for example the book
  \cite{Bunde:1991} and is not the fractal dimension that could be
  defined for a network \cite{Song:2005}.} Concerning the links, there
is an obvious cost associated with their length and one expects a
rapidly decreasing length distribution. In most models (in particular
for the Waxman model \cite{Waxman:1988}, see also section
\ref{sec4B4}), the link distribution is assumed to be an exponential
of the form $e^{-d/d_0}$ but the authors of \cite{Yook:2002} seemed to
observe results consistent with a slower decay $P(d)\sim 1/d$ over one
decade.

\begin{figure}[h!]
\begin{tabular}{c}
\epsfig{file=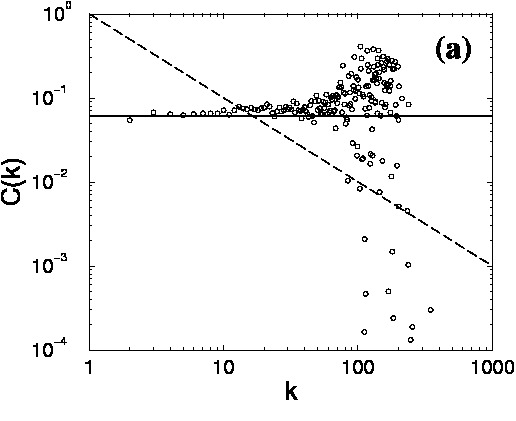,width=0.8\linewidth,clip=}
\end{tabular}
\caption{ Scaling of the clustering $C(k)$ for the Internet at
the router level. The dashed line has a slope $-1$ and the solid
line corresponds to the average clustering coefficient. From \cite{Ravasz:2003}.}
\label{fig:ckinternet}
\end{figure}

The clustering coefficient for the Internet is maybe more
interesting. At the AS level, it varies \cite{Ravasz:2003} as
$C(k)\sim 1/k$, a behavior typical of many scale-free networks
\cite{Dorogovtsev:2002a}. However in sharp contrast with this result,
if we do not aggregate the nodes and stay at the router level, we
observe \cite{Ravasz:2003}, a clustering coefficient approximately
constant (Fig.~\ref{fig:ckinternet}).

All these different results suggest that the location of nodes is 
given exogenously and should be the input of a realistic model. Once
the location of the nodes given, the link length distribution and the
behavior of the clustering coefficient suggest that there is a
competition between preferential attachment and a spatial dependence,
model that we will discuss in the next chapter \ref{sec4B}.

\subsubsection{Geography in social networks}
\label{sec3B4}

With the development of information technologies, we can now have an
idea about some aspects of the structure of social networks. In
particular, as expected,  space is also important in social networks. It
is indeed reasonable to think that in order to minimize the efforts
and to maintain a social tie \cite{Zipf:1949} most individuals will
connect with their spatial neighbors. As a measure of the social tie,
Lambiotte et al. \cite{Lambiotte:2008} used mobile phone data for
$3.3$ millions customers in Belgium. For each customer, a certain
number of attributes is known and in particular the zip code of the
address where the bill is sent. After some filtering procedures, the
resulting network is made of $2.5$ millions nodes (customers) and
$5.4$ millions links, giving an average degree $\langle
k\rangle=4.3$. The degree distribution is peaked (the power law fit
proposed in \cite{Lambiotte:2008} gives an exponent $\gamma\approx
5$) and the probability $P(d)$ that two connected individuals are separated by an
euclidean distance $d$ is found to behave as \cite{Lambiotte:2008}
\begin{equation}
P(d)\sim d^{-2}
\end{equation}
over distances in the range $1-100$ kms. The exponent describing the
decay with distance is thus equal to $2.0$ in this example and
supports the idea of a gravity law (see section \ref{sec3C3}) even in
the context of social ties. This result is also consistent with the
measure obtained for inter-city phone intensity exponent (see
\cite{Krings:2009} and section \ref{sec3C3d}). According to
Kleinberg's navigability theorem \cite{Kleinberg:2000}, this result
also shows that this social network is navigable (see section \ref{sec5C} for
a discussion on navigability).

However, another study \cite{Liben-Nowell:2005} on a social network of
more than one millions bloggers in the USA shows convincingly that the
proportion of pairs of nodes at distance $d$ which are friends decays
as
\begin{equation}
P(d)-\varepsilon\sim d^{-\alpha}
\end{equation}
where $\alpha\approx 1$ and where $\varepsilon\approx 5\times
10^{-6}$. In addition, this exponent $\alpha$ of order one is
confirmed in another studies \cite{Goldenberg:2009,Backstrom:2010} on
Facebook users and on email communications. Interestingly, a recent
paper \cite{Hu:2010} showed that this spatial scaling $P(d)\sim 1/d$
could result from the fact that most individuals tend to maximize the diversity
of their friendships (which can be seen as the maximization of information
entropy). The plateau observed in \cite{Liben-Nowell:2005} at
$\varepsilon>0$ shows that above a certain distance (here of the order
$\approx 1,000$ kms), the probability $P(d)$ flattens to a constant
value independent of distance and which probably results from complex
processes acting in social networks. For this network, on an average
of eight neighbors, we then have on average $5.5$ friends living in
the proximity while only $2.5$ result from non-geographic processes
(for a discussion about navigability on this network, see section
\ref{sec5C5}). This is consistent with the results of
\cite{Goldenberg:2009} where it is shown that about $40\%$ of emails
were sent within the same city (for this particular panel of
individuals).

Although there is no agreement so far (but we expect rapid
developments on this topic) on the nature of the decay and the value
of the exponent, the existence of an important spatial component in
social networks is rather clear and at this point it seems that on
average the majority of our friends are in our spatial
neighborhood. This is confirmed in a recent paper \cite{Scellato:2010}
which measured a variety of parameters on four different online social
networks (BrightKite, FourSquare, LiveJournal, Twitter) and showed in
particular that there is a majority of short-distance links between
users. This dependence with space seems to go even beyond
friendship and applies to other social processes such as
collaborating and writing a paper together \cite{Chandra:2007}. In
fact, the strong association between friendship and spatial location
led Backstrom, Sun, Marlow \cite{Backstrom:2010} to propose an
algorithm -which exceeds IP-based geolocation- to predict the location
of people in Facebook who didn't provide this information.

\begin{figure}[h!]
\begin{tabular}{c}
\includegraphics[angle=-90,scale=.30]{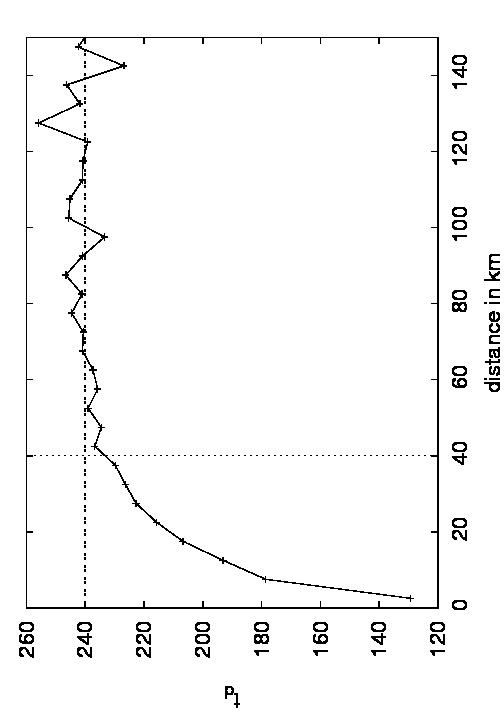}
\end{tabular}
\caption{ Average duration $t_d$ of a phone call as a function of the
  distance $d$. A saturation is observed for distances larger than
  $40$ kms. From \cite{Lambiotte:2008}.}
\label{fig:lambiotte}
\end{figure}

Clustering seems also to be large as shown in
\cite{Backstrom:2010,Lambiotte:2008}. In particular, Lambiotte et
al. \cite{Lambiotte:2008} studied the probability $c(d)$ that a link
of length $d$ belongs to a triangle. This probability decreases with
distance and reaches a plateau at around $0.32$ above $40$ kms.  It is
interesting to note also that the average duration of a phone call
depends also on the distance (see Fig.~\ref{fig:lambiotte}) and
saturates to a value of about $4$ minutes for the same value $d>40$
kms obtained for $c(d)$.
These results led these authors to suggest that two communication
regimes exist: a short distance `face-to-face' regime with short
communication durations and a large clustering coefficient and another
long-distance regime with smaller clustering and longer
communications.

\subsection{Origin-destination matrix and mobility networks}
\label{sec3C}

\subsubsection{Importance of human mobility}
\label{sec3C1}

Knowing the flows of individuals from one point to the other in a city
or a country is a very important piece of information upon which many
models and studies depend. First, it is related to the fundamental
problem in geography and spatial economics of the location of
activities and their spatial distribution. Second, it is a key
ingredient in the epidemiology of infectious diseases (see section
\ref{sec5E}). These diseases transmit between humans in close
proximity and contaminate the population because they travel and
interact. Clues on the statistics of humans movements and interactions
is thus fundamental. At a global level for example, the spread of
infectious diseases such as the flu is -- in addition to seasonal
effects -- largely controlled by air travel patterns (see
\cite{Colizza:2006} and section \ref{sec5E}). Third, statistics about
individual movements are important for more commercial applications
such as geomarketing for example. Finding the best spot to place your
advertisement depends on the number of people going through this
location and thus implies to know the flows, or at least the
dominant flows of individuals in the area under consideration.

A simple way to formalize this problem is to divide the area of
interest in different zones labelled by $i=1,\dots,N$ and to count the
number of individuals going from location $i$ to location $j$. This
number is called the origin-destination (OD) matrix and is denoted in
this review by $T_{ij}$. This quantity is at the core of many
transportation models and its properties have been discussed in a
large number of papers. It defines a network, directed and weighted,
and in the general case is time-dependent. This quantity is of course
very different from a `segment' measurement where we can easily count
the number of individuals going through an individual segment of the
transportation system under consideration (it can be one of the link
of the airline network, a segment of road, etc.). In contrast, the OD
matrix is usually extremely difficult and costly to obtain and
measure.

Despite all the difficulties in measuring the OD matrix, recent
technological advances such as the GPS, the democratization of mobile
phones together with geosocial applications, etc. allow for precise
measurements on large datasets and point to the possibility to
understand quantitatively urban movements. In the next sections we
will then review some recent results in this field.

\subsubsection{Distribution of the trip duration and length}
\label{sec3C2}

The duration and length of trips in urban areas is obviously of great
importance and in particular, daily trips can characterize the
economical efficiency of an urban area.  In particular the duration of
daily journey-to-work trips give a good indication of the economical
health and efficiency of a given region. For many different reasons,
it is then important to understand the factors governing the behavior
of the statistics of the daily trips. In particular in the perspective
of convergence to sustainable cities, we need to understand such
factors in order to reduce energy and environmental problems. We are
now at the stage where always more data is available and it will soon
become possible to characterize the statistics of human movements at
many different scales and in particular at the urban scale.

The important case of the statistics of the daily journey-to-work
travel is an old problem and discussed at length in the transportation
research literature. One can find various empirical observations for
different urban areas (see for example \cite{Mokhtarian:2004} a short
review on this subject) and also some theoretical discussions (see for
example \cite{Mokhtarian:2004}). An important theoretical discussion
was triggered by Zahavi \cite{Zahavi:1977} who suggested that the
average daily travel time at the regional scale varies little over
space and time. This hypothesis is interesting in the sense that it is
the first step towards some `physical laws' ruling human
movement. This hypothesis of a constant time (and money)
budget can be rephrased as the `rational locator
hypothesis' states that individuals maintain (if they can choose)
steady journey-to-work travel times by adjusting their home-work
distance (see for example \cite{Levinson:2005}. 

Koelbl and Helbing \cite{Koelbl:2003} proposed that the travel
time can be understood in terms of a travel energy budget. More
precisely, they found emprically that the average travel time for each daily
transportation mode considered (walk, bicycle, car, bus, train) is
approximately constant for over $27$ years (the data considered was
obtained for the UK in the period $1972-1998$, see the end of this
chapter for a discussion on this point). This result suggests that
there is a travel time budget associated with each transportation mode
and the main idea put forward in \cite{Koelbl:2003} is that the energy
consumed by individuals for travel is constant and mode-independent
\begin{equation}
E\approx \overline{t}_ip_i
\label{eq:expen}
\end{equation}
where $E\approx 615 kJ$ is the estimated average daily energy budget,
$\overline{t}_i$ the average travel time for mode $i$, and $p_i$ the energy
consumption rate for mode $i$ (which varies from $\approx 4kJ/$min for
a train passenger to $\approx 15kJ/$min for walking).

\begin{figure}[!ht]
\begin{tabular}{c}
\includegraphics[angle=-90,scale=.30]{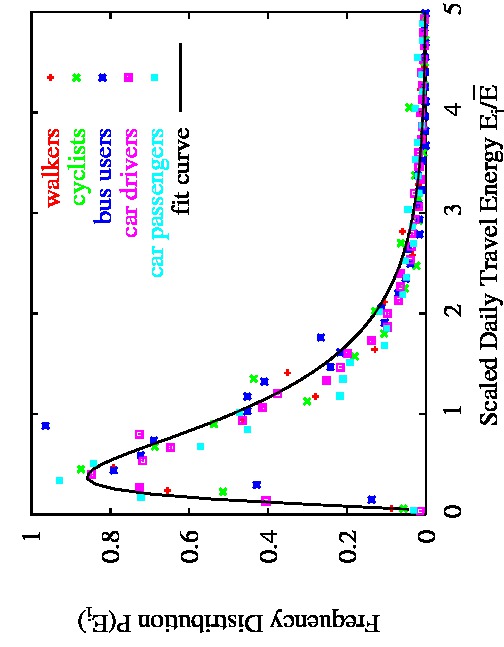}\\
\includegraphics[angle=-90,scale=.30]{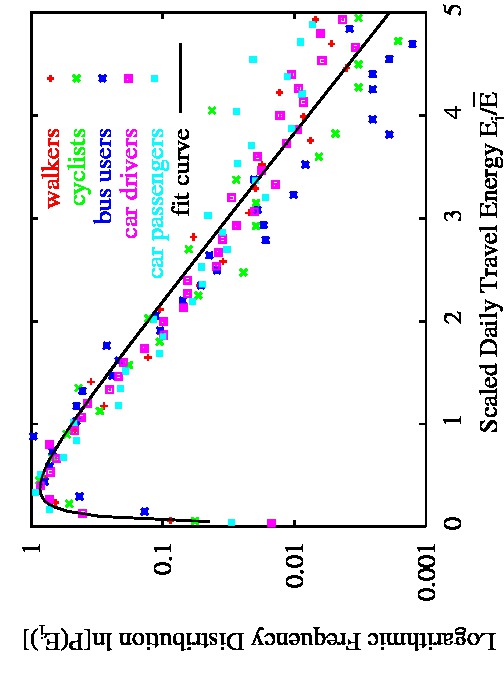}
\end{tabular}
\caption{ Universal travel energy distribution in (a) linear
 representation and in (b) semilogarithmic representation. The solid
line is the fit obtained for $\alpha=0.2$ and $\beta=0.7$. From
 \cite{Koelbl:2003}.}
\label{fig:universal}
\end{figure}

A further observation is that the individual travel time (for a mode
$i$) rescaled by the average time of the corresponding mode
$\tau_i=t/\overline{t}_i$ follows a universal probability distribution
which in terms of energy $E_i=p_it=\overline{E}\tau_i$ reads
\begin{equation}
P(E_i)\sim e^{-\alpha\overline{E}/E_i-\beta E_i/\overline{E}}
\end{equation}
where $\alpha$ and $\beta$ are fitting parameters. This distribution
reflects the variation between individuals but displays a universal
behavior among modes. The collapse is shown in
Figure~\ref{fig:universal}.
The exponential term $e^{-\beta E_i/\overline{E}}$ corresponds to the
canonical energy distribution when the average energy is fixed (and
can be recovered by maximizing the entropy with the normalization and
the fixed average energy constraints). The term $\exp(-\alpha
\overline{E}/E_i)$ suppresses the short trips reflecting what is
called the {\it Simonson effect} (see \cite{Koelbl:2003} and
references therein) and leads to the selection of a less costly
transportation mode. Indeed if the trip is too short for a given
transportation mode, the energy spent for preparation can be larger
than the energy needed for the trip itself and a less costly
transportation mode is then more efficient.

These results of Koelbl and Helbing have however been challenged in
other empirical studies. In \cite{Levinson:2005}, Levinson and Wu
showed that the average daily travel seems to be constant over time
for a given city but varies from one city to another. The average
travel time for a given city then depends on structural parameters
characterizing the city (such as population, density, transportation
networks, congestion, etc). In another study \cite{Hubert:2007} on
three travel surveys (for France, UK, and Belgium), Hubert and Toint
criticize the validity of the hypothesis of constant average daily
human energy expenditure Eq.~\ref{eq:expen}. However, they show that
the distribution of the normalized travel time seems to be the same
for the three different countries of this study pointing to the
possible existence of universal features of human movement.

It is clear at this point that a `unified theory' of human travel
behavior is still needed as well as other empirical studies, but the
analysis presented in \cite{Koelbl:2003} at least shows that tools
from statistical physics can play a role in transportation research
and more generally in uncovering universal patterns in human behavior.

\paragraph{Mobile phone and GPS studies.}
\label{sec3C2a}

As discussed above, obtaining OD matrices is a very difficult and
crucial task. While traditional surveys are costly and incomplete,
mobile phone data give more complete informations. In particular, in
large urban areas, the density of antennas is large enough so that
triangulation gives a relatively accurate indication of the users'
location (mobile devices such as phones are regularly in contact with
emitters and triangulation allows to determine the location of the
device at a resolution scale given by the local density of
emitters). For all these reasons, mobile phone data were recently used
to detail individual trajectories \cite{Ahas:2005}, to identify `anchor
points' where individuals spend most of their time
\cite{Aasa:2008,Eagle:2009}, or statistics of trip patterns
\cite{Gonzalez:2009,Sevtsuk:2010}. Results and studies are so numerous
now, that we will focus on a small subset only and refer the
interested reader to \cite{Sevtsuk:2010} and references therein.

Gonzalez et al \cite{Gonzalez:2009} used a set of
mobile phone data at a national level and found that the
distribution of displacement of all users can be fitted by a Levy law
of the form
\begin{equation}
P(\Delta r)\sim \frac{1}{(\Delta r_0+\Delta r)^\beta}e^{-\Delta
  r/\kappa}
\end{equation}
where $\beta\simeq 1.75$, $\Delta r_0\simeq 1.5$ km. The cutoff value
$\kappa$ depends on the protocol used and varies from $\kappa\simeq
400$ kms (for protocol $D_1$) and $\kappa\simeq 80$ kms for the
protocol $D_2$. The protocol $D_1$ consists in recording trajectories
(of $100,000$ users) when the user initiates or receives a call or
SMS. The protocol $D_2$ captures the location of $206$ mobile phones
every two hours for an entire week. The origin of the data is not
indicated in this paper and in particular could mix both rural and
urban areas which could be the cause of the observed
behavior. However, this type of approach is very promising and will
certainly lead in the future interesting insights about human movement
behavior in various areas.

We note that the measured value for the exponent $\beta $ is not far
from the value $1.6$ observed in \cite{Brockmann:2006}. In this study,
Brockmann et al. used data for the movements of banknotes across the
USA using the `Where is George' database, a system based on voluntary
reporting. They found that the distance distribution is also Levy
distributed. However, the presence of a given banknote at a given
location and at a certain time depends on many processes such as the
diffusion of the bills, but also on the probability to actually use a
banknote, and of course to report its location to the
Whereisgeorges.com website. Such crucial factors were not taken into
account in \cite{Brockmann:2006} and it is therefore difficult to
assess the validity of the obtained Levy law.

We note here that in a recent paper \cite{Song:2010}, Song et
al. proposed a model of human mobility in order to understand the
power laws observed in \cite{Gonzalez:2009,Brockmann:2006}. In
particular, this model includes the two following ingredients in the
description human displacements: First, the tendency to explore
additional locations decreases with time, and second, there is large
probability to return a location visited before. This model gives
relations between different scaling exponents and values which are
consistent with empirical fits \cite{Song:2010}. However, it is still
unclear at this stage at which scale this study applies and how it can
be reconciled with the previous studies showing a peaked distribution
of travel times.

GPS is another interesting tool in order to characterize individual
trajectories. Recent studies \cite{Rambaldi:2007,Bazzani:2010} used
GPS data of private vehicles for the city of Florence (Italy). 
\begin{figure}[!h]
\centering
\begin{tabular}{c}
\includegraphics[angle=0,scale=.35]{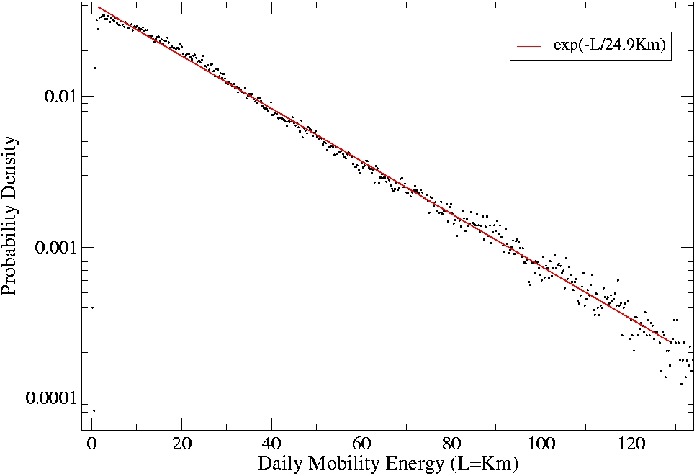}
\end{tabular}
\caption{Distribution of total length of daily trips. The exponential
  fits gives a slope $L_0\simeq 25$kms. From \cite{Bazzani:2010}.}
\label{fig:bazza1}
\end{figure} 
and have shown that the
{\it total} daily length trip is exponentially distributed (see Fig.~\ref{fig:bazza1})
\begin{equation}
P(\ell_T)\sim e^{-\ell_T/L_0}
\end{equation}
and seems to be independent of the structure of road network pointing
to a general mechanism which need to be uncovered. In particular, the
authors suggest a relation with this distribution and Maxwell
distributions. This seems to be actually reminiscent of the general
argument proposed by Koelbl and Helbing (see \cite{Koelbl:2003} and previous chapter \ref{sec3C2}) and
indeed points to the possible existence of more general principles
governing human movements.

\paragraph{RFIDs.}
\label{sec3C2b}

At an intra-urban scale and even at the scale of social networks,
RFIDs might provide interesting insights. RFID stands for Radio
Frequency Identification and is a technology similar to the bar code
and is composed of a tag which can interact with a transceiver that
process all informations contained in the tag. RFIDs are used in many
different instances from tagging goods to the Oyster card system in
London. In fact it can be used anywhere a unique id is needed and the
tag can carry simple or more complex information about the
carrier. RFIDs can also be used for dynamical measures of social
networks such as in the sociopatterns project (see
\url{http://www.sociopatterns.com}).

\begin{figure}[!h]
\centering
\begin{tabular}{c}
\includegraphics[angle=0,scale=.60]{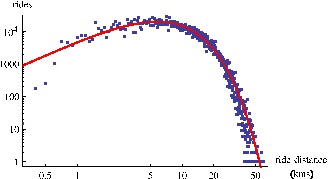}\\
\includegraphics[angle=0,scale=.60]{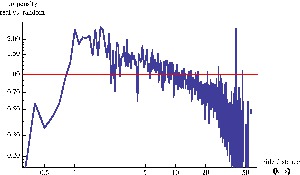}
\end{tabular}
\caption{ Ride distance distribution. (a) Plot of the histogram of
  distances for observed rides. This distribution can be fitted by a
  negative binomial law of parameters $r = 2.61$ and $p = 0.0273$,
  corresponding to a mean $\mu = 9.28$ kms and standard deviation
  $\sigma=5.83$ kms.  This distribution is not a broad law (such as a
  Levy flight for example), in contrast with previous findings using
  indirect measures of movements \cite{Brockmann:2006,Gonzalez:2009}. (b) Ride distance
  propensity.  Propensity of achieving a ride at a given distance with
  respect to a null-model of randomized rides. From \cite{Roth:2010}.}
\label{fig:roth}
\end{figure}

The Oyster card system in London provides information about
instantaneous flows of individuals in the subway system. Since $2003$,
some $10$ million RFID cards have been issued to commuters using the
London transport network. In \cite{Roth:2010}, these individual
trajectories were analyzed, displaying evidence for a polycentric
organization of activity in this urban area. These authors also found
that in agreement with many other transportation networks, the
traffic is broadly distributed (according to a power law with
exponent $\approx 1.3$) but also that the displacement length
distribution is peaked (see Fig.~\ref{fig:roth}).

\subsubsection{The gravity law}
\label{sec3C3}

The origin-destination matrix contains a large amount of information
and allows to test some ideas about the structure of human
movements. In particular, it has been suggested (see for example the
book \cite{Erlander:1990}) more than $50$ years ago that the number of
trips from location $i$ to location $j$ follows the `Gravity' law
\begin{equation}
T_{ij}=K\frac{P_iP_j}{d_{ij}^\sigma}
\end{equation}
where $d_{ij}=d_E(i,j)$ is the euclidean distance between these two
locations, $P_{i(j)}$ is the population at location $i$ ($j$) and
where $\sigma$ is an exponent whose value actually depends on the
system. This idea was generalized to many other situations such as the
important case in economics of international trade
\cite{Anderson:1979,Bergstrand:1985}. In this case, the volume of
trade between two countries is given in terms of their economical
activity and their distance.

More generally the gravity law (see the theoretical discussion in the
section \ref{sec3C3e}) is written under the form
\begin{equation}
T_{ij}\sim P_iP_jf(d(i,j))
\end{equation}
where the deterrence function $f$ describes the effect of space.

In the next sections, we will focus on the most recent measures and
concern highways \cite{Jung:2008}, commuters \cite{Balcan:2009}, and
cargo ship movements \cite{Kaluza:2010}, and phone communications
\cite{Krings:2009}. We then end this chapter with a theoretical
discussion on the gravity law.

\paragraph{Worldwide commuters.}
\label{sec3C3a}

Balcan et al. \cite{Balcan:2009} studied recently flows of commuters
at a global scale. They studied more than $10^4$ commuting flow
worldwide between subpopulations defined by a Voronoi decomposition
and found that the best fit is obtained for a gravity law of the form
\begin{equation}
T_{ij}=CP_i^\alpha P_j^\gamma e^{-d_{ij}/\kappa}
\end{equation}
where $C$ is a proportionality constant, and where the exponents are:
for $d\leq 300$ kms, $(\alpha,\gamma)\simeq (0.46,0.64)$,
$\kappa=82$ kms, and for $d>300$ kms: $(\alpha,\gamma)\simeq
(0.35,0.37)$.  We note an asymmetry in the exponent at small scales
which probably reflects the difference between homes and offices and
which does not hold at large scale where homogenization seems to
prevail.

At this granularity level, there is then a dependence of the traffic
on populations and distances with specific exponents and with
exponentially decreasing deterrence function. At a smaller scale,
different results for US commuters were obtained in \cite{Viboud:2006}
and as suggested in \cite{Balcan:2009} the observed differences might
found their origin in the different granularities used in these
studies (a problem known as the `modifiable areal unit problem' in
geography). Indeed, in \cite{Balcan:2009}, the granularity is defined
by a Voronoi decomposition, while in \cite{Viboud:2006}, counties are
used which are administrative boundaries not necessarily well
spatially consistent with gravity centers of mobility processes.


\paragraph{Korean highways.}
\label{sec3C3b}

\begin{figure}[h!]
\begin{tabular}{c}
\epsfig{file=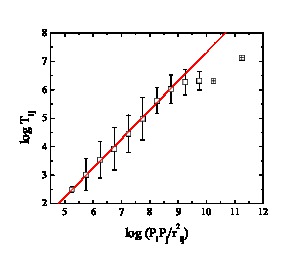,width=0.85\linewidth,clip=}
\end{tabular}
\caption{ Traffic flow between $i$ and $j$ as a function of the
  variable $P_iP_j/d_{ij}^2$. The line has a slope equal to
  $1.02$. From \cite{Jung:2008}.}
\label{fig:jung1}
\end{figure}

In \cite{Jung:2008}, Jung, Wang, and Stanley studied the traffic on the Korean
highway system for the year $2005$. The system consists in $24$ routes
and $238$ exits, and the total length of the system is about $3,000$
kms.  The highway network is described by a symmetrized weight matrix
$T_{ij}$ which represents the traffic flow between $i$ and $j$. The
in- and out-traffic are well correlated with population, as already
seen in the worldwide airline network \cite{Colizza:2006} where
the population $P_i$ of city $i$ scales with the strength $s_i$ as
\begin{equation}
P_i\sim s_i^\alpha
\end{equation}
with $\alpha\approx 0.5$ while it is close to one in
\cite{deMontis:2005} and \cite{Jung:2008}.
For $30$ cities with population larger than $200,000$, Jung et
al. study the traffic flow $T_{ij}$ as a function of the population of
the two cities $P_i$ and $P_j$, and the distance $d_{ij}$ between $i$
and $j$ and used the original formulation of the gravity law
\begin{equation}
w_{ij}\sim \frac{P_iP_j}{d_{ij}^\sigma}
\end{equation}
with $\sigma=2$ (see Fig.~\ref{fig:jung1}).

\paragraph{Global cargo ship.}
\label{sec3C3c}

In \cite{Kaluza:2010}, Kaluza et al. applied the gravity law to the
global cargo ship network. They used the general form
of the gravity law, where the number of trips $T_{ij}$ is given as
(see the discussion in the section \ref{sec3C3e})
\begin{equation}
T_{ij}=A_iB_jO_iI_jf(d_{ij})
\end{equation}
where $O_i$ is the total number of departures from node $i$ and $I_j$
the total number of arrivals at node $j$. The coefficients $A_i$ and
$B_j$ are normalization prefactors such that $\sum_jT_{ij}=O_i$ and
$\sum_iT_{ij}=I_j$.

\begin{figure}[h!]
\begin{tabular}{c}
\epsfig{file=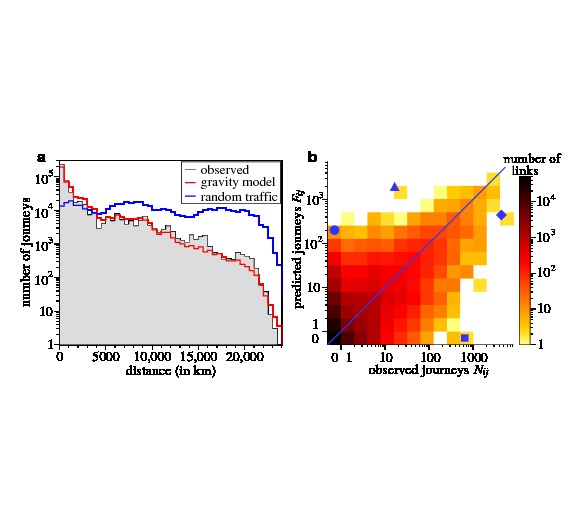,width=1.0\linewidth,clip=}
\end{tabular}
\caption{ (a) Histogram of port to port distances traveled in the global
  cargo ship network. The red line represents the gravity model, the
  blue line a random null model. (b) Count of port pairs with $N_{ij}$
  observed and $F_{ij}$ predicted journeys. From \cite{Kaluza:2010}.}
\label{fig:cargo1}
\end{figure}

The deterrence function used in \cite{Kaluza:2010} is a two parameters
truncated power law $f(d)=d^{-\sigma}\exp (-d/\kappa)$. The best
correlation was obtained for $\kappa=4,900$ kms and $\sigma=0.59$.
As shown in Fig.~\ref{fig:cargo1}, this fit compares extremely
well with the empirical data.
In order to test the dispersion on every pairs of origin-destination,
Kaluza et al. \cite{Kaluza:2010} plot the observed number $N_{ij}$ of trips between
$i$ and $j$ with the prediction $F_{ij}$ obtained with the gravity
model. A perfect agreement would give a diagonal $F_{ij}=N_{ij}$ and we observe
here some dispersion (with a small Kendall's tau of $\tau=0.433$)
showing that even if the gravity law is able to reproduce the
histogram, it is not accurate enough in order to predict the actual
flow between a given pair of origin and destination.

\paragraph{Inter-city phone communications.}
\label{sec3C3d}

Mobile phone data open a new way to analyze geographical structure at
various scales and in particular allows to exhibit spatial features of
social networks. In a recent paper \cite{Krings:2009}, Krings et
al. study the network of mobile phone customers and analyze the
geographical pattern of $2.5$ millions customers. After aggregating by
city, these authors found that the inter-city communication intensity
follows a gravity law
\begin{equation}
L_{ij}\propto\frac{P_iP_j}{d_E(i,j)^2}
\end{equation}
where $L_{ij}$ is the total communication time between cities $i$ and
$j$ (for a period of six months). This result showing a power law
decay with exponent $2$ is in agreement with other studies on the
influence of space on social networks (see \cite{Lambiotte:2008} and
section \ref{sec3B4}). More generally, going beyond the topological
structure of the social network and including other socio-economical
parameters is probably a very interesting research direction which
will be developed in the future.

\paragraph{Theoretical discussion.}
\label{sec3C3e}

The problem of estimating the number of trips from one zone to another
is an old problem which excited the curiosity of many scientists a
long time ago. It is reasonable to think that the number of trips
between two locations $i$ and $j$ is proportional to the number of
possible contacts given by $P_iP_j$ and also decreases with distance,
or equivalently with time and more generally with the cost of
travelling from $i$ to $j$. Readers interested in theoretical
developments of this model can find a full account in the book
\cite{Erlander:1990}. It is interesting to note that in this book
devoted to the gravity model and which is one of the most cited
reference in this field, all the chapters are devoted to theoretical
and mathematical discussions without any attempt to compare the
results with empirical observations.

There are obviously many factors which control the origin-destination
matrix: land use, location of industries and residential areas,
accessibility, etc and it seems difficult to provide a general and
simple enough model. The gravity law in its simplest form retains the
population and the distance as relevant factors (more elaborated forms
with up to dozen of parameters were used in practical
applications). However, given the importance of this law for many
practical applications and its extensive use in simulations (in
epidemiology for example), we believe that an extensive meta-analysis
for different modes, geographical regions and granularities is greatly
needed. Thanks to the increasing availability of data, we can
expect in the future a more systematic investigation of this problem
and we try to provide here a first step in this direction and we show
in Table \ref{Table:gravitylaw} a list of various (essentially recent)
results obtained empirically.
\begin{table*}[ht!]
  \centering
\begin{tabular}{|c|c|c|c|}

\hline

Network [ref] & $N$ & gravity law form & Results\\

\hline
\hline

Railway express \cite{Zipf:1946} & $13$ & $P_iP_j/d_{ij}^\sigma$ & $\sigma=1.0$\\

\hline

Korean highways \cite{Jung:2008} & $238$ & $P_iP_j/d_{ij}^\sigma$ & $\sigma=2.0$\\

\hline

Global cargo ship \cite{Kaluza:2010} & $951$ & $O_iI_jd_{ij}^{-\sigma}\exp(-d_{ij}/\kappa)$ &  $\sigma=0.59$\\

  \hline

Commuters (worldwide) \cite{Balcan:2009} & n/a & $P_i^\alpha P_j^\gamma \exp(-d_{ij}/\kappa)$ & $(\alpha,\gamma)=(0.46,0.64)$ for $d<300$ kms\\
 &  &  & $(\alpha,\gamma)=(0.35,0.37)$ for $d>300$ kms\\

 \hline

US commuters by county \cite{Viboud:2006} & $3109$ & $P_i^\alpha P_j^\gamma /d_{ij}^\sigma$ & $(\alpha,\gamma,\sigma)=(0.30,0.64,3.05)$ for $d<119$ kms\\
 &  &  & $(\alpha,\gamma,\sigma)=(0.24,0.14,0.29)$ for $d>119$ kms\\

  \hline

Telecommunication flow \cite{Krings:2009} & $571$ & $P_iP_jd_{ij}^{-\sigma}$ &  $\sigma=2.0$\\

  \hline

\end{tabular}

\caption{\label{Table:gravitylaw} List of various empirical studies on
  the gravity  law (we essentially focused on recent and illustrative results).}
\end{table*}
As we can observe in this table, there are multiple choices for the
deterrence function and various values of the exponent as
well. Particularly puzzling is the existence of different values for
the exponent $\sigma$ (which can vary from $\approx 0.5$ to $\approx
3$) and the different deterrence function for commuters. As suggested
in \cite{Balcan:2009} the observed differences might found their
origin in the different granularities used in these studies. Indeed,
in \cite{Balcan:2009}, the granularity is defined by a Voronoi
decomposition, while in \cite{Viboud:2006}, counties are used which
are administrative boundaries not necessarily spatially consistent
with mobility processes (a problem known as the {\it modifiable areal unit
problem} in geography).  In addition, the different exponents could
depend on the transportation mode used, of the scale, or other effects
linked to the heterogeneity of users and trips.

In this short discussion, we thought that it could be useful to recall
the classical optimization problem and one of the most important
derivation of the gravity law which uses entropy maximization and also
to give a simple statistical argument which could shed light on the
most important mechanisms in this problem.

\medskip
{\it Optimization}

We first recall the classical approach which is at the basis of many
studies (see for example \cite{Erlander:1990}. We are interested in
this problem in determining the OD matrix $T_{ij}$ given the constraints
\begin{align}
\sum_jT_{ij}&=T_i\\
\sum_iT_{ij}&=T_j
\end{align}
These represent $2N$ constraints for $N^2$ unknowns and as long as
$N>2$ many different choices for $T_{ij}$ are possible. If we assume
that the transport from $i$ to $j$ has a cost $C_{ij}$ we can then
choose $T_{ij}$ such that the total cost
\begin{equation}
C=\sum_{ij}T_{ij}C_{ij}
\end{equation}
is minimum. This is the classical transportation problem and can be
traced back to the $18^{th}$ century and Monge
\cite{Erlander:1990}. Another approach consists in requiring that
$T_{ij}=T_{ij}^0r_is_j$ where $T^0$ is a given set of interzonal
weights and where $s_j$ and $r_i$ are given constants. For an
extensive discussion on this latter approach, see
\cite{Erlander:1990}.

\medskip
{\it Entropy maximization}

Interestingly enough, the gravity model can be shown to result
essentially from the maximization of entropy \cite{Wilson:1967}. 
Wilson, a physicist who got interested in transportation research very
early proposed that the trips $T_{ij}$ are such that the quantity
\begin{equation}
\Omega=\frac{T!}{\Pi_{ij}T_{ij}!}
\end{equation}
is maximal which corresponds to trip arrangements with the largest
number of equivalent configurations (or microstates in the
statistical physics language). In this expression, $T=\sum_{ij}T_{ij}$ is
the number of total trips and
the maximization is subject to the natural constraints on the
origin-destination matrix
\begin{align}
\sum_jT_{ij}&=T_i\\
\sum_iT_{ij}&=T_j
\end{align}
and to a cost constraint
\begin{equation}
\sum_{ij}T_{ij}C_{ij}=C
\end{equation}
where $C_{ij}$ is the cost to travel from $i$ to $j$ and where $C$ is
the total quantity of resources available. This maximization is easy
to perform with the help of Lagrange multipliers and one obtains
\begin{equation}
T_{ij}=A_iB_jT_iT_je^{-\beta C_{ij}}
\end{equation}
where $A_i$, $B_j$ and $\beta$ are such that the constraints are
met. Of course, even if in this expression we do not have explicit
expressions of $A_i$ and $B_j$, the most important problem at this point
is how to express the cost. In order to recover a power law
distribution, one needs a logarithmic dependence on distance:
$C_{ij}=a\ln d_E(i,j)$ which leads to $T_{ij}\propto d_E(i,j)^{-\beta
  a}$. If the cost is proportional to distance, the number of trips
decays exponentially with distance. We thus recover two of the most
important forms used in empirical studies and in model, but the exact
form of the cost dependence with distance remains unsolved.

There is a long discussion about the validity of this approach in
\cite{Erlander:1990} but we note that it assumes in particular that
all individuals act independently from each other. This is obviously
not correct when we introduce congestion which induces correlations
between individuals. In such conditions, it is clear that individual
choices are correlated and that this entropy maximization can give
reasonable results in the limit of small traffic only.


\medskip
{\it A simple statistical argument}

As proposed in \cite{Samaniego:2008}, a statistical approach could be
helpful in the determination of important features of roads and
traffic and more generally can give insights into the structure of
cities. As discussed above, the exact empirical measure of $T_{ij}$ is extremely
difficult but we have access to a more coarse-grained information such
as for example the total number $\ell_d$ of miles driven in $357$ US
cities\footnote{ Source: \url{http://www.census.gov}}. For a given
city of total area $A$ and population $P$, it is natural to test for
simple scalings such as one observed in other systems
\cite{West:1997,Banavar:1999}. In particular, the natural length scale
for a city of area $A$ is given by $\sqrt{A}$ and we thus expect a
scaling for $\ell_d$ of the form
\begin{equation}
\frac{\ell_d}{\sqrt{A}}\sim P^{\beta}
\end{equation}
As suggested in \cite{Samaniego:2008}, we can
explore two extreme cases of mobility in urban areas. If every
individual is going to its next nearest neighbor (located at a typical
distance $1/\sqrt{\rho}$ where $\rho=P/A$ is the average density of
the city), the total distance $\ell_d$ is given by
\begin{equation}
\ell_d=P\times\frac{1}{\sqrt{\rho}}=\sqrt{AP}
\end{equation}
implying that $\ell_d/\sqrt{A}$ scales as $P^\beta$ with $\beta=1/2$. 

On the other hand, if the individuals are going to random points in
the city, the typical distance is given by $\sqrt{A}$ and the total
vehicle miles is given by
\begin{equation}
\ell_d\sim P\times\sqrt{A}
\end{equation}
which implies $\beta=1$.

\begin{figure}[ht!]
\begin{tabular}{c}
\epsfig{file=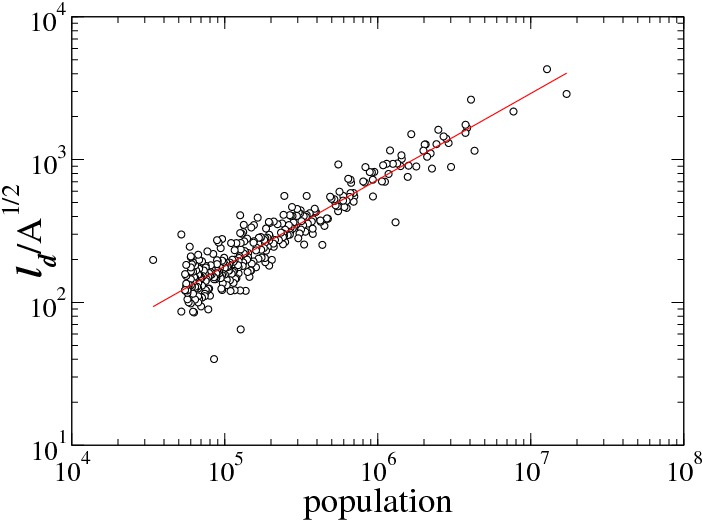,width=0.7\linewidth,clip=}\\
\epsfig{file=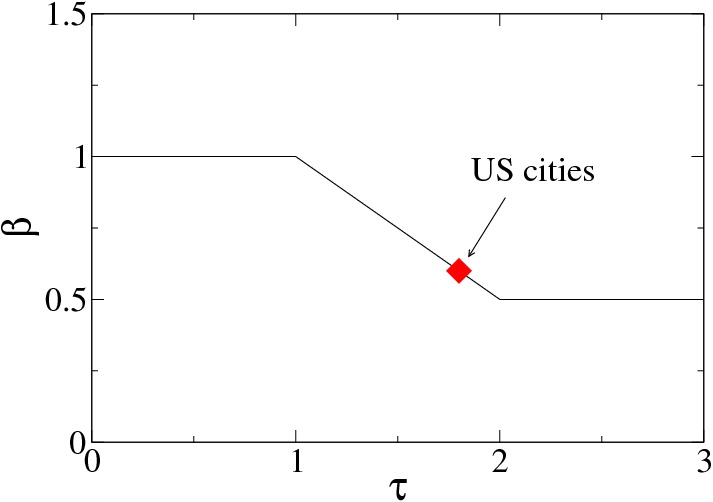,width=0.7\linewidth,clip=}
\end{tabular}
\caption{ (a) Total rescaled vehicule-miles $\ell_d/\sqrt{A}$ versus $N$ for
  $357$ US cities. The line is a power law fit with exponent
  $\beta=0.60$ (the correlation coefficient is $r^2=0.94$).(b)
  Exponent $\beta$ versus $\tau$. The red diamond corresponds to the
  empirical measure $\beta\approx 0.6$ predicting $\tau\approx 1.8$.
}
\label{fig:gravitylaw}
\end{figure}

Following \cite{Samaniego:2008}, we plot in
Fig.~\ref{fig:gravitylaw}a the total number $\ell_d$ of miles driven
by the whole population $P$ of the city and rescaled by $\sqrt{A}$ for all
the $367$ US cities in the database.
We obtain the result $\beta\simeq 0.6$ (in \cite{Samaniego:2008} this
exponent is $0.66$ for a slightly larger dataset) with an excellent correlation
coefficient. We are thus in an intermediate situation between the two
extremes described above showing that the commuting pattern is
probably a mixture of centralized trips (to a central business
district for example) and more local trips, or could also result from
centralized patterns to many different centers and could thus be a
signature of a polycentric organization.  

The total number of miles of built lanes was also studied in
\cite{Samaniego:2008} and essentially scales as $\sqrt{PA}$ which
means that the road network is essentially a lattice of spacing
$1/\sqrt{\rho}$ in agreement with the results shown in section \ref{sec3B1a}.

Using hand-waving arguments, we can explore the consequences of this
empirical result $\beta\simeq 0.6$ on the general properties of the
trip distribution. Since $T_{ij}$ is the number of trips from a zone
$i$ to a zone $j$, the total distance traveled is given by
\begin{equation}
\ell_d=\sum_{i,j}T_{ij}d_{ij}
\label{eq:l}
\end{equation}
and we can assume that total number of trips is proportional 
to the total number of individuals
\begin{equation}
P=\sum_{i,j}T_{ij}
\label{eq:N}
\end{equation}
(in these equations (\ref{eq:l}) and (\ref{eq:N}) some irrelevant
constant factors are  omitted). The simplest assumption is that the trip
distribution depends on the distance $T_{ij}=f(d_{ij})$ and we can
assume that $f(x)=Cx^{-\tau}$. This is in fact a very general case,
as the exponential case is in a statistical sense reproduced by this
form with $\alpha>2$.

We will assume that the distance is uniformly distributed and we
consider the continuous limit which reads
\begin{eqnarray}
P&=&\sum_{ij}f(d_{ij})\simeq C\int_a^Lx^{-\tau}dx\\
\ell_d&=&\sum_{ij}f(d_{ij})d_{ij}\simeq C\int_a^Lx^{1-\tau}dx
\end{eqnarray}
where $a$ is the smallest distance given by $a\sim 1/\sqrt{\rho}\sim L/\sqrt{P}$
and where the upper bound is the maximum distance $L\sim\sqrt{A}$. We have to separate
three cases depending on the value of $\tau$ (in the following we will keep 
the dominant terms, knowing that $a\ll L$).

\medskip
(i) Case $\tau<1$.
\medskip
In this case, the leading terms are
\begin{eqnarray}
\ell_d&\sim& C L^{2-\tau}\\
P&\sim& C L^{1-\tau}
\end{eqnarray}
leading to $\ell_d\sim P\sqrt{A}$.

\medskip
(ii) Case $1<\tau<2$.
\medskip
In this case, the dominant terms are
\begin{eqnarray}
\ell_d\sim C L^{2-\tau}\\
P\sim C a^{1-\tau}
\end{eqnarray}
leading to $\ell_d\sim P^{\frac{3-\tau}{2}}\sqrt{A}$.

\medskip
(iii) Case $\tau>2$.
\medskip
In this case, the dominant terms are
\begin{eqnarray}
\ell_d\sim C a^{2-\tau}\\
P\sim C a^{1-\tau}
\end{eqnarray}
leading to $\ell_d\sim \sqrt{PA}$.

These results can be summarized in the Fig.~\ref{fig:gravitylaw}b. In
particular, we can see on this plot that the empirical result
$\beta=0.6$ corresponds to the value $\tau=1.8$. If we had an
exponential behavior we would have observed a value $\tau>2$ but
instead the simple argument presented here suggests the behavior
\begin{equation}
T_{ij}\sim \frac{1}{d_{ij}^\tau}
\end{equation}
with $\tau=1.8$ slightly less than $2$. This simple scaling argument
leads to a result in agreement with the gravity law and signals a slow
decay of the trip volume with distance and the simultaneous existence
of trips at many different length scales. Aggregated and scaling
arguments could thus help in understanding the commuting patterns, and
reveal interesting features of spatial patterns in the city.

Various factors could be at the origin of the different behaviors
observed in gravity law studies (different deterrent functions,
different exponents, etc.). In particular, we can list the following
factors which could play an important role:
\begin{itemize}
\item{} {\it Importance of transportation modes.} It is at this point
  unclear if the gravity law is the same for all modes. Also, the cost
  of a given transportation mode can vary in time (which was the case
  for air travel when jets appeared and is the case for cars with
  increasing oil prices) and the gravity law could in principle vary
  in time.
\item{} {\it Importance of scale and discretization.} As discussed in
  this chapter, it is important to use a discretization which is
  consistent with mobility processes. Also, a discussion of scale
  is probably important as it can already be seen on the trip length
  distribution which seems to display peaked laws at the urban scale
  and broad distribution at national scales.
\item{} {\it Nature of the trip. Heterogeneity of users.} These
  factors are well-known in transportation research, and it could be
  necessary to distinguish different trips (commuting, travel,
  etc.) in order to get consistent gravity laws. Users are also very
  heterogeneous and it is unclear at this stage if we need to
  distinguish between different populations with different behaviors.
\end{itemize}

\subsection{Neural networks}
\label{sec3D}

\begin{figure}[h!]
\begin{tabular}{c}
\epsfig{file=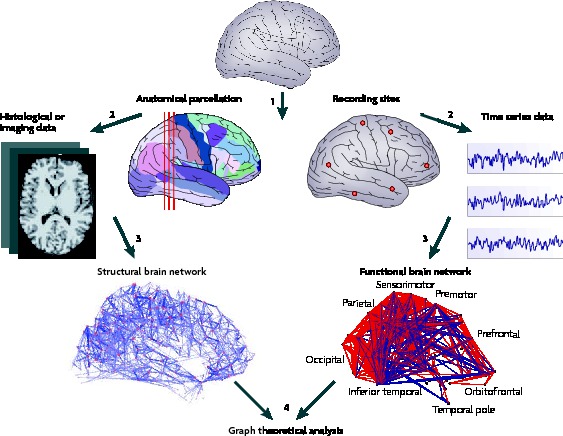,width=1.0\linewidth,clip=}
\end{tabular}
\caption{ Structural and functional brains can be studied with graph
  theory by following different methods shown step-by-step in this figure. From \cite{Bullmore:2009}.}
\label{fig:brain}
\end{figure}

The human brain with about $10^{10}$ neurons and about $10^{14}$
connections is one of the most complex network that we know of. The
structure and functions of the brain are the subjects of numerous
studies and different recent techniques such as
electroencephalography, magnetoencephalography, functional RMI, etc.,
can be used in order to produce networks for the human brain (see
Fig.~\ref{fig:brain} and for a clear and nice introduction see for
example \cite{Bullmore:2009} and \cite{Chavez:2010}).

Brain regions that are spatially close have a larger probability of
being connected than remote regions as longer axons are more costly in
terms of material and energy \cite{Bullmore:2009}. Wiring costs
depending on distance is thus certainly an important aspect of brain
networks and we can expect spatial networks to be relevant in this
rapidly evolving topic. So far, many measures seem to
confirm a large value of the clustering coefficient, and a small-world
behavior with a small average shortest-path length
\cite{Eguiluz:2005,Kaiser:2004b}. It also seems that neural networks
do not optimize the total wiring length but rather the processing
paths thanks to shortcuts \cite{Kaiser:2005}. This small world
structure of neural networks could reflect a balance between local
processing and global integration with rapid synchronization,
information transfer, and resilience to damage \cite{Lago:2000}.

In contrast, the nature of the degree distribution is still under
debate and a recent study on the macaque brain \cite{Modha:2009}
showed that the distribution is better fitted by an exponential rather
than a broad distribution. Besides the degree distribution, most of
the observed features were confirmed in latest studies such as
\cite{Zalesky:2010} where Zalesky et al. propose to construct the
network with MRI techniques where the nodes are distinct grey-matter
regions and links represent the white-matter fiber bundles. The
spatial resolution is of course crucial here and the largest network
obtained here is of size $N\approx 4,000$. These authors find a large
clustering coefficients with a ratio to the corresponding random graph
value of order $10^2$ (for $N\approx 4,000$). Results for the average shortest
path length $\langle\ell\rangle$ are however not so clear due to
relatively low values of $N$. Indeed, for $N$ varying from $1,000$ to
$4,000$, $\langle\ell\rangle$ varies by a factor of order $1.7-1.8$
\cite{Zalesky:2010}. A small-world logarithmic behavior would predict
a ratio
\begin{equation}
r=\frac{\langle\ell\rangle(N=4000)}{\langle\ell\rangle(N=1000)}
\sim \frac{\log(4000)}{\log(1000)}\approx 1.20
\end{equation}
while a $3$-dimensional spatial behavior would give a ratio of order
$r\approx 4^{1/3}\approx 1.6$ which is closer to the observed value. Larger
sets would however be needed in order to be sure about the behavior of
this network concerning the average shortest path and to distinguish a
$\log N$ from a $N^{1/3}$ behavior expected for a three-dimensional
lattice.

Things are however more complex than it seems and even if functional
connectivity correlates well with anatomical connectivity at an
aggregate level, a recent study \cite{Honey:2010} shows that strong
functional connections exist between regions with no direct structural
connections, demonstrating that structural and functional properties
of neural networks are entangled in a complex way and that future
studies are needed in order to understand this extremely complex
system.


\subsection{Summary: Existence of general features}
\label{sec3E}

Through the different examples presented in this chapter, we see that
some general features emerge. Space implies a cost associated to
distance and the formation of long links must be justified for a `good
economical' reason, such as a large degree for example. This simple
observation leads to various effects that we summarize here.

In the table \ref{table2}, we summarize the main empirical features
discussed in this chapter.
\begin{table*}[!ht]
\centering
\begin{tabular}{|c|c|c|c|c|c|c|c|}

\hline

Network [ref.] & $N$ & $\langle k\rangle$ & $\langle C\rangle$ & $\langle\ell\rangle$ & $P(k)$; $P(w)$ & $\beta$ & $\mu$ \\

\hline

Transportation networks  \\

\hline

Chinese Airline network \cite{Li:2003} & $128$ &  $18.2$ &  $0.733$ & $2.1$ & unclear; broad & n/a & n/a\\

Indian Airline network \cite{Bagler:2008} & $79$ &  $5.77$ &  $0.66$ & $2.3$ & PL ($2.2$); broad & $1.4$ & n/a\\

USA Airline network \cite{Dallasta:2006} & $935$ &  $8.4$ &  n/a & $4.0$ & PL ($1.9$); broad & $1.7$ & $1.2^*$\\

Global Airline network \cite{Guimera:2004} & $3883$ & $6.7$ & $0.62$& $4.4$ & PL ($2.0$); n/a & n/a  & n/a$^*$\\

Global Airline network \cite{Barrat:2004b} &  $3880$ &  $9.7$ & n/a & $4.4$  & PL ($1.8$); broad & $1.5$ & n/a$^*$\\

Boston Subway \cite{Latora:2001a}  & $124$    & n/a          & n/a & $15.5$ & n/a; n/a & n/a & n/a\\   


Seoul Subway \cite{Lee:2008}  & $380$    & $2.23$ & n/a &  $20.0$ & n/a ; broad & n/a & n/a\\   

Beijing Subway \cite{Xu:2007c}         & $132$  & $3.2$ & $0.15$ & $12.6$ & peaked; peaked & $1.1$ & n/a\\   
Shanghai Subway \cite{Xu:2007c}     & $151$  & $4.6$ & $0.21$ &  $7.1$ & peaked; peaked & $1.1$ & n/a\\   
Nanjing Subway \cite{Xu:2007c}       & $47$    & $2.9$ & $0.09$ & $12.4$ & peaked; peaked & $1.1$ & n/a\\   

Polish transportation \cite{Sienkiewicz:2005}  & $[152,1530]$ & $[2.48,3.08]$ & $[0.055,0.161]$ & $[6.83,21.52]$ & peaked; n/a & n/a & $\approx 2.0$\\   

World cities Subway \cite{vonFerber:2009}  & $[2024,44629]$ & $[2.18,3.73]$ & $\sim 10^{-2}$ & $[6.4,52]$ & peaked; n/a & n/a & $2-3$\\   


Polish Rail \cite{Kurant:2006b} & $1533$       & $2.4$ & $0.0092$ & $28.1$ & peaked; broad & n/a & n/a\\
Swiss Rail \cite{Kurant:2006b} & $1613$        & $2.1$ & $0.0004$ & $46.6$ & peaked; broad & n/a & n/a\\
European Rail \cite{Kurant:2006b} & $4853$  & $2.4$ & $0.0129$ & $50.9$ & peaked; broad & n/a & n/a\\

Indian Rail \cite{Sen:2003b} & $587$  & $2.4$ & $0.69$ &  $2.2$ & peaked; n/a & n/a & n/a\\

China Cargo ship \cite{Xu:2007} & $162$ & $3.1$ & $0.54$ & $5.9$ & $k<20$; broad & n/a & $1.0^{**}$\\        

Worldwide Cargo ship \cite{Hu:2009b} &  $878$  & $9.0$ & $0.40$ & $3.60$ & PL ($0.95$); PL ($0.9$) & $1.3$ & $1.66^{**}$\\

Worldwide Cargo ship \cite{Kaluza:2010} & $951$  & $76.5$ & $0.49$ & $2.5$ & broad; PL ($1.7$) & $1.46$ & n/a$^{**}$\\

\hline

\hline

Mobility networks  \\

\hline

Location network \cite{Chowell:2003} & $181206$  & $29.9$ & $0.058$ & $3.1$ & PL ($2.4$); broad & $\approx 1.0$ & n/a\\

\hline

Commuters Sardinia \cite{deMontis:2005} & $375$  & $86.6$ & $0.26$ & $2.0$ & peaked; PL ($1.8$) & $1.9$ & n/a\\


\hline

\hline

Infrastructure networks \\

\hline


Road network \cite{Buhl:2006} & $[45,339]$ &  $[2.02,2.87]$ & $M\in [0.009,0.211]$ & $[4.6,13.6]$ & peaked; n/a & n/a & n/a \\

Road network \cite{Cardillo:2006} & $[32,2870]$ &  $[2.09,3.38]$ & $M\in [0.014,0.348]$ & $[27.6,312.1]$ & peaked; n/a & n/a & n/a \\

Power grid \cite{Crucitti:2004} & $341$ & $3.03$ & n/a & n/a & $k<12$; n/a & n/a & n/a$^*$\\

Power grid \cite{Watts:1998} & $4941$ & $2.67$ & $0.08$ & $18.7$ & peaked; n/a & n/a & n/a\\

Water network  \cite{Yazdani:2010} & $1786$ & $2.23$ & $0.0008$ & $25.94$ & peaked; n/a & n/a & n/a\\

Water network  \cite{Yazdani:2010} & $872$ & $2.19$ & $0.0402$ & $51.44$ & peaked; n/a & n/a & n/a\\

Internet \cite{Vazquez:2002b} &$3700-10500$ &  $3.6-4.1$& $0.21-0.29$& $3.7$ & PL ($2.1$); n/a & n/a & n/a\\

  \hline

Biological networks \\

\hline

Human brain \cite{Eguiluz:2005} & $31503$ & $13.4$ & $0.14$ & $11.4$ & PL ($\sim 2.0$); n/a & n/a & n/a\\
Human brain \cite{Eguiluz:2005} & $17174$ & $6.3$   & $0.13$ & $12.9$ & PL ($\sim 2.1$); n/a & n/a & n/a\\
Human brain \cite{Eguiluz:2005} & $4891$    & $4.1$   & $0.15$ & $6.0$ & PL ($\sim 2.2$); n/a & n/a & n/a\\

C. Elegans \cite{Watts:1998} & $282$  & $7.7$  & $0.28$ & $2.65$   & peaked; n/a & n/a & n/a\\

C. Elegans \cite{Kaiser:2005} & $277$  & $15.2$  & $0.167$ & $4.0$ & n/a; n/a & n/a & n/a\\

Macaque VC \cite{Hilgetag:2000} & $32$    & $9.9$  & $0.55$ & $1.8$ & peaked; n/a & n/a & n/a\\

Macaque \cite{Kaiser:2005} & $95$    & $50.6$  & $0.643$ & $1.9$ & n/a; n/a & n/a & n/a\\

Cat Cortex \cite{Hilgetag:2000} & $65$    & $17.5$  & $0.54$ & $1.9$ & peaked; n/a & n/a & n/a\\

  \hline
\end{tabular}

\caption{\label{table2} List of different networks with their main
  characteristics: number of nodes $N$; average degree $\langle k\rangle$; average clustering
  coefficient $\langle C\rangle$; average shortest path $\langle\ell\rangle$; the nature of the degree and weight
  distributions $P(k)$, $P(w)$;  the scaling of the strength with the degree $s(k)\sim
  k^\beta$; and the  relation between the centrality and the degree when it exists
  $g(k)\sim k^\mu$. `PL($x$)' means that a power law fit was measured
  and $x$ is the value of the exponent. $^*$ existence of many
  anomalies and $^{**}$ few anomalies. For road networks
  $M$ denotes the meshedness (denoted by $\gamma$ in section \ref{sec2B2b}). }
\end{table*}

\begin{itemize}

\item{} {\it Network typology}\\
  We can roughly divide spatial networks in two categories. The first
  one consists in planar networks. These networks possess many
  features similar to lattices but in some cases (such as the road
  network) display very distinctive features which call for new
  models. The other category consists in spatial, non-planar networks such as
  the airline network, the cargo ship network, or the Internet where
  nodes are located in spaces, edges have a cost related to their
  length, but where we can have intersecting links.  We note here that
  in some cases such as for the Internet, the point distribution is
  not always uniform and could play an important role.

\item{} {\it Effect of space on $P(k)$}\\
Spatial constraints restrict the appearance of large degrees and
$P(k)$ is usually peaked. Constraints are stronger for planar networks
and for non-planar spatial networks such as airlines the degree
distribution can be broad.

\item{} {\it Effect of space on the link distribution}\\
Here also the spatial constraints limit severely the length of
links. For planar networks such as roads and streets, the distribution is peaked while in
other networks such as the Internet or the airline network, the
distribution can be broader.

\item{} {\it Effect of space embedding on clustering and assortativity}\\
Spatial constraints implies that the tendency to connect to hubs is
limited by the need to use small-range links which explains the almost
flat behavior observed for the assortativity. Connection costs also
favor the formation of cliques between spatially close nodes and thus
increase the clustering coefficient.

\item{} {\it Effect of space on the average shortest path}\\
  The average shortest path for $2d$ planar networks scales as
  $N^{1/2}$ as in a regular lattice. When enough shortcuts (such as in
  the Watts-Strogatz model) are present, this $\sqrt{N}$ behavior is
  modified and becomes logarithmic. In some three-dimensional
  situations such as in the brain for example, it is however difficult
  in some experiments to distinguish between a $\log N$ from a
  $N^{1/3}$ behavior.

\item{} {\it Effect of spatial embedding on topology-traffic correlations}\\
  Spatial constraints induce strong non-linear correlations between
  topology, traffic, and distance. The reason for this behavior is
  that spatial constraints favor the formation of regional hubs and
  reinforces locally the preferential attachment, leading for a given
  degree to a larger strength than the one observed without spatial
  constraints. Essentially long-range links can connect to hubs only,
  which yields a value $\beta_{d}>1$. The existence of constraints
  such as spatial distance selection induces some strong correlations
  between topology (degree) and non-topological quantities such as
  weights or distances.

\item{} {\it Effect of space embedding on centrality}\\
Spatial constraints also induce large betweenness centrality
fluctuations. While hubs are usually very central, when space is
important central nodes tend to get closer to the gravity center of
all points. Correlations between spatial position and centrality
compete with the usual correlations between degree and centrality,
leading to large fluctuations of centrality at fixed
degree.

\item{} {\it Structure of mobility networks}\\
  The different discussions and empirical examples considered in this
  chapter showed that the average trip duration and length distribution actually depend on
  many factors such as the scale considered (urban, inter-urban,
  national, or global), the population, congestion, or on the transportation mode, etc. Also, it
  seems that at the urban scale the trip duration and length
  distributions are peaked law and that at a larger scale the trip length
  is distributed according to a broad law with exponent of order
  $1.6$. The identification of relevant factors and the constitution
  of a clear typology is however still an open problem and we can expect in
  the future a wealth of new results in this area.

\end{itemize}

Although many results and measures are missing due to the lack of data
availability, we see that the main expected effects of spatial
constraints are present for most of these networks. Also, traffic
properties seem to be relatively ubiquitous, with a broad traffic
distribution, and a non-linear strength-degree law with exponent
$\beta$ distributed around $1.5$. In the next chapter, we will discuss
the various models which attempt to reproduce these different effects.

\section{Models of spatial networks}
\label{sec4}

In this chapter we will describe the most important models of spatial
networks. We basically divide these models into five large classes
(which of course could present non empty overlaps):

\begin{itemize}

\item{} The first class describes {\it geometric graphs} which are
  probably the simplest models of spatial networks. They are obtained
  for a set of vertices located in the plane and for a set of edges
  which are constructed according to some geometric condition. There
  is a large body of literature on this rather general subject and we
  will here focus on a small set of these studies that concern more
  precisely graph constructed with an Euclidean proximity condition.

\item{} The second class concerns the Erdos-Renyi model and its spatial
generalization, including the spatial hidden variables models. These
networks are obtained when the probability to connect two nodes
depends on the distance between these nodes. An important example of
this class is the Waxman model~\cite{Waxman:1988} for the Internet structure.

\item{}
The third class comprises spatial variants of the Watts-Strogatz model
\cite{Watts:1998} and could be coined as {\it spatial
  small-worlds}. In these cases, the starting point is a
$d$-dimensional lattice and random links are added according to a
given probability distribution for their length.

\item{}
The fourth class concerns {\it spatial growth models} which can
be considered as spatial extensions of the original growth model
proposed by Barabasi and Albert \cite{Albert:1999}.

\item{} Finally, the last class concerns {\it optimal networks}
  obtained by the minimization of a `cost' function. These networks
  were considered already a long time ago in different fields
  (mathematics, transportation science, computer science) and are now
  back with the explosion of studies on complex networks.

\end{itemize}

\subsection{Geometric graphs}
\label{sec4A}

A random geometric graph is obtained when the points located in the
plane are connected according to a given geometric rule.  The simplest
rule is a proximity rule which states that nodes only within a certain
distance are connected. There is an extensive mathematical literature
on geometric graph and the random case was studied by physicists in the
context of continuum percolation (see for example
\cite{Balberg:1985,Quantanilla:2000}).

\subsubsection{The simplest random geometric graph}
\label{sec4A1}

In the framework of complex networks, one of the first studies on
random geometric graphs is provided in \cite{Dall:2002} where the
nodes are small spheres of radius $r$ and two nodes are connected by
an edge if they are separated by a distance less than $2r$ (which is
the condition for the intersection of spheres of radius $r$). This
network is related to the unit disk graph defined in geometric graph
theory, as the intersection graph of a family of unit circles in the
Euclidean plane: if we draw a vertex at the center of each circle, we
connect two vertices if the corresponding circles intersect. This unit
disk graph has been used in computer science to model the
topology of ad-hoc wireless communication networks where the nodes are
wirelessly connected to each other, without a base station
\cite{Huson:1995}. The nodes are located in a two-dimensional plane
and the area within which a signal can be transmitted to another node
is described as a circle (of the same radius if all nodes have the
same power to transmit). These random geometric graphs have also been
used in continuum percolation models
\cite{Balberg:1985,Quantanilla:2000}.

\begin{figure}
\begin{tabular}{c}
\epsfig{file=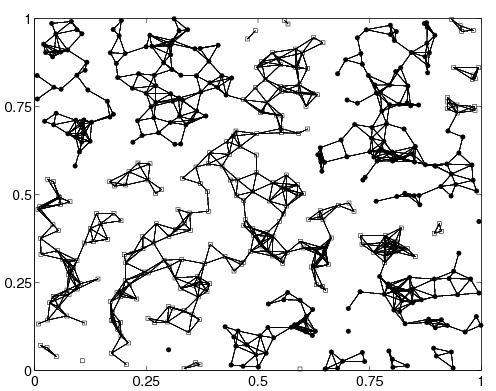,width=0.9\linewidth,clip=}
\end{tabular}
\caption{ Example of a 2d random geometric graph obtained for $N=500$
  spheres and $\langle k\rangle=5$. From \cite{Dall:2002}.}
\label{fig:dall}
\end{figure}

If the volume of the total system is $1$ then the probability $p$ that two
randomly chosen vertices are connected is equal to the volume of the
sphere of radius $R=2r$ which reads in $d$ dimensions
\begin{equation}
p=V(R)=\frac{\pi^{d/2}R^d}{\Gamma(1+d/2)}
\end{equation} 
In the context of continuum percolation this quantity is the excluded
volume $V_e\equiv p$. The average degree is then given by $\langle k\rangle=Np$
and we can then express $R$ as a function of $\langle k\rangle$
\begin{equation}
R=\frac{1}{\sqrt{\pi}}\left[\frac{\langle
    k\rangle}{N}\Gamma\left(\frac{d+2}{2}\right)\right]^{1/d}
\end{equation}
which shows that for a given average degree $\langle k\rangle$ the
nodes (spheres) have to become smaller when more nodes are added.

Similarly to the usual ER random graph, there is a critical average
degree above which there is a non empty giant component. The authors of
\cite{Dall:2002} computed numerically this critical value $\langle
k\rangle_c$ for different dimensions and proposed the scaling $\langle
k\rangle_c=1+bd^{-\gamma}$ with $b=11.78(5)$ and
$\gamma=1.74(2)$. This relation also states that in infinite dimension
the random geometric graph behaves like a ER graph with $\langle
k\rangle_c=1$.

In \cite{Herrmann:2003}, the authors compute analytically the degree
distribution for these random geometric graphs. If we assume that the
points are distributed according to a distribution $p(x)$  and the
condition for connecting to nodes $i$ and $j$ located at positions
$x_i$ and $x_j$, respectively, is $d_E(i,j)\leq R$, we can then
estimate the degree distribution. If we denote by $B_R(x)$ the ball of
radius $R$ and centered at $x$, the probability $q_R(x)$ that a given node is
located in $B_R(x)$ is
\begin{equation}
q_R(x)=\int_{B_R(x)}dx'p(x')
\end{equation}
The degree distribution for a node located at $x$ is thus given by the
binomial distribution
\begin{equation}
P(k;x,R)={N-1\choose k}q_R(x)^k[1-q_R(x)]^{N-1-k}
\end{equation}
In the limit $N\to\infty$ and $R\to 0$, the degree distribution for a
node located at $x$ is Poissonian and reads
\begin{equation}
P(k;x,\alpha)=\frac{1}{k!}\alpha^kp(x)^ke^{-\alpha p(x)}
\end{equation}
where $\alpha=\langle k\rangle/\int dx p^2(x)$ fixes the scale of the
average degree. For example this expression gives for a uniform
density $p(x)=p_0$ a degree distribution of the form
\begin{equation}
P(k)\sim \frac{(\alpha p_0)^k}{k!k^d}
\end{equation}
which decays very rapidly with $k$. In contrast if the density
decays slowly from a point as $p(r)\sim r^{-\beta}$ we then obtain
$P(k)\sim k^{-d/\beta}$ showing that large density fluctuations can
lead to spatial scale-free networks \cite{Herrmann:2003}.

\begin{figure}[!h]
\centering
\begin{tabular}{c}
\includegraphics[angle=0,scale=.40]{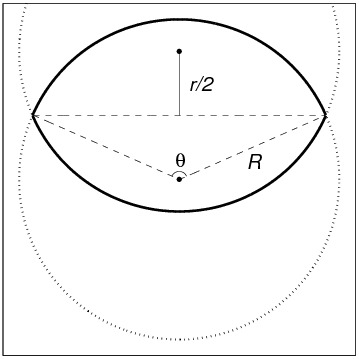}
\end{tabular}
\caption{ The overlap between the two disks (area comprised within the
 bold line) gives the quantity $\rho_2(r)$. From \cite{Dall:2002}.}
\label{fig:dall2}
\end{figure}

The average clustering coefficient can also be calculated analytically
and the argument \cite{Dall:2002} is the following. If two vertices
$i$ and $j$ are connected to a vertex $k$ it means that they are both
in the excluded volume of $k$. Now, these vertices $i$ and $j$ are
connected only if $j$ is in the excluded volume of $i$. Putting all
pieces together, the probability to have two connected neighbors
$(ij)$ of a node $k$ is given by the fraction of the excluded volume
of $i$ which lies within the excluded volume of $k$. By averaging over
all points $i$ in the excluded volume of $k$ we then obtain the
average clustering coefficient. We thus have to compute the volume
overlap $\rho_d$ of two spheres which for spherical symmetry reasons
depends only on the distance between the two spheres. In terms of this
function, the clustering coefficient is given by
\begin{equation}
\langle C_d\rangle =\frac{1}{V_e}\int_{V_e}\rho_d(r)dV
\end{equation}
For $d=1$, we have 
\begin{equation}
\rho_1(r)=(2R-r)/2R=1-r/2R
\end{equation}
and we obtain
\begin{equation}
\langle C_1\rangle=3/4
\end{equation}
For $d=2$, we have to determine the area overlapping in the
Fig.~\ref{fig:dall2} which gives
\begin{equation}
\rho_2(r)=(\theta(r)-\sin(\theta(r)))/\pi
\end{equation}
with $\theta(r)=2\arccos(r/2R)$ and leads to
\begin{equation}
\langle C_2\rangle =1-3\sqrt{3}/4\pi\approx 0.58650
\end{equation}
Similarly an expression can be derived in $d$ dimension \cite{Dall:2002}
which for large $d$ reduces to
\begin{equation}
\langle C_d\rangle\sim 3\sqrt{\frac{2}{\pi
    d}}\left(\frac{3}{4}\right)^{\frac{d+1}{2}}
\end{equation}

The average clustering coefficient thus decreases from the value $3/4$
for $d=1$ to values of order $10^{-1}$ for $d$ of order $10$ and is
independent from the number of nodes which is in sharp contrast with
ER graphs for which $\langle C\rangle \sim 1/N$. Random geometric
graphs are thus much more clustered than random ER graphs. The main
reason -- which is in fact valid for most spatial graphs -- is that
long links are prohibited or rare. This fact implies that if both $i$
and $j$ are connected to $k$ it means that there are in some spatial
neighborhood of $k$ which increases the probability that their
inter-distance is small too, leading to a large $\langle C\rangle$.

\paragraph{Application to ad-hoc networks: Calculation of the giant
  component.}
\label{sec4A1a}

In ad-hoc networks~\cite{Nemeth:2003}, users communicate by means of a
short range radio device. This means that a device can communicate
with another one if their distance is less than a certain distance
which corresponds to the transmission range of the device. The set of
connected devices can be used to propagate information on longer
distance by going from the source to the destination hop by hop
through intermediate nodes. If there is a large density of nodes,
alternate routes are even available which allows to split the
information into separate flows. Usually, the users are mobile and the
network evolves in time and it is important to understand the
condition for the existence of a giant cluster.

In \cite{Nemeth:2003}, Nemeth and Vattay explicitly compute the condition
for the appearance of a giant component using generating functions
such as in \cite{Newman:2001c}. In particular, it is not difficult to
show that the fraction $S$ of nodes which are in the giant component
is given by
\begin{equation}
S=1-G_0(u)
\end{equation}
where the generating function $G_0(x)$ is given by
\begin{equation}
G_0(x)=\sum_kP(k)x^k
\end{equation}
where $P(k)$ is the probability that a node has $k$ neighbors. The
quantity $u$ is the probability of not belonging to the 
giant component and satisfies
\begin{equation}
u=G_1(u)
\end{equation}
where $G_1(x)=G'_0(x)/G'_0(1)$ (see for example \cite{BBVBook:2008}).

In the ad-hoc network case, the typical internode distance $r_0$
depends on the density $\rho=N/A$ as
\begin{equation}
\rho \pi r_0^2\sim 1\;\;\Rightarrow\;\;r_0\sim 1/\sqrt{\pi\rho}
\end{equation}
The range of the radio devices is denoted by $r_T$ and the important
dimensionless parameter is here $\sqrt{r_T/r_0}$. The probability to
have $k$ neighbors is a binomial law of parameter $\rho \pi
r_T^2=\eta$ and in the limit of large $N$ (with $N\eta$ constant) we
thus obtain
\begin{equation}
P(k)=e^{-\eta}\frac{\eta^k}{k!}
\end{equation}
This particular form of $P(k)$ allows to compute $G_0(x)$ and $G_1(x)$
and we eventually obtain an implicit equation for $S$
\begin{equation}
S=1-e^{-\eta S}
\end{equation}
Or, after inversion
\begin{equation}
\eta=\frac{-\log(1-S)}{S}
\end{equation}
This last relation gives the minimum transmission rate in order to get
a giant component of size $NS$.

Also, in this model one can introduce a probability $p(r)$ that the
devices are connected at distance $r$ and different quantities such as
the average clustering coefficient can be computed (see
\cite{Nemeth:2003} for details).

\paragraph{Model of mobile agents for contact networks.}
\label{sec4A1b}

A model for contact networks based on mobile agents was proposed in
\cite{Gonzalez:2004,Gonzalez:2006b,Gonzalez:2006}. In this model,
individuals are described by disks with the same radius and are moving
and colliding in a two-dimensional space. The
contact network is built by keeping track of the collisions (a link
connects two nodes if they have collided) . A collision takes place
whenever two agents are at a certain distance and this model can thus
be seen as a dynamical version of random geometric graphs. In
particular, Gonzalez, Lind, Herrmann \cite{Gonzalez:2006} used this
framework to model the sexual interaction network and interestingly
enough, found in some cases and for a simple collision rule such that
the velocity grows with the number of previous collisions that a
scale-free network with exponent $\gamma=3$ emerges.

\subsubsection{Random geometric graph in hyperbolic space}
\label{sec4A1p}

Motivated by studies on the Internet, a model of random geometric
graph in hyperbolic space was proposed recently (see
\cite{Serrano:2008} or the small review article
\cite{Krioukov:2010}). In their studies, Bogu{\~n}\'a, Krioukov and Serrano
considered the two-dimensional hyperbolic space $\mathbb{H}^2$ of
constant negative curvature equal to $K=-\zeta^2=-1$ and use a polar
representation $(r,\theta)$ for the nodes. They placed $N$ points
distributed uniformly in a disk of radius $R$ and in the Euclidean
disk projection this implies that the nodes have a uniform angle
distribution $U(\theta)=1/2\pi$ for $\theta\in [0,2\pi]$ and that
the radial coordinate is distributed according to
\begin{equation}
\rho(r)=\frac{\sinh r}{\cosh R-1}\approx e^r
\end{equation}
They used then the usual geometric graph rule and connect two nodes if
their hyperbolic distance is less than $R$ which can be written in
terms of the connection probability as
\begin{equation}
p_C(x)=\Theta(R-x)
\end{equation}
where $\Theta$ is the Heaviside function.  The hyperbolic distance $d$
between two nodes $(r,\theta)$ and $(r',\theta')$ is defined by
\begin{equation}
\cosh\zeta d=\cosh\zeta r\cosh\zeta r'-\sinh\zeta r\sinh\zeta
r'\cos(\theta'-\theta)
\end{equation}
The expected degree of a node at point X located at distance $r$
from the origin O is proportional to the area of the intersection of the
two disks of radius $R$ centered at O and X, respectively. The disk
centered at O contains all nodes and the intersection can be
calculated leading to the average degree at distance $r$ (for large $R$)
\begin{equation}
\langle k\rangle (r)\approx \frac{4}{\pi}Ne^{-r/2}
\end{equation}
The average degree of the network is then 
\begin{equation}
\langle k\rangle =\int_0^R\rho(r)\langle k\rangle(r)dr\approx
\frac{8}{\pi}Ne^{-R/2}
\end{equation}
The degree distribution is \cite{Krioukov:2010}
\begin{equation}
P(k)=2
\left(
\frac
{\langle k\rangle}
{2}
\right)^2
\frac
{\Gamma (k-2,\langle k\rangle/2)}
{k!}
\approx k^{-3}
\end{equation}
and behaves as a power law with an exponent equal to three, as in the BA
model. This network heterogeneity however follows directly from the
properties of the hyperbolic geometry and not from a preferential
attachment mechanism.

Krioukov et al. also considered the case where the node density is not
uniform
\begin{equation}
\rho(r)=\alpha\frac{\sinh\alpha r}{\cosh\alpha R-1}\approx e^{\alpha
  r}
\end{equation}
In this case, it can be shown that the degree distribution is also a
power law of the form
\begin{equation}
P(k)\sim k^{-\gamma}
\end{equation}
where
\begin{equation}
\gamma=
\begin{cases}
1+2\alpha/\zeta\;\;&{\rm for}\;\;\frac{\alpha}{\zeta}\geq 1/2\\
2\;\;&{\rm for}\;\;\frac{\alpha}{\zeta}\leq 1/2\\
\end{cases}
\end{equation}
The power law exponent depends then both on the hyperbolicity $\zeta$
and $\alpha$, through the ratio $\alpha/\zeta$, a result expected from
an analogy with trees (see \cite{Krioukov:2010}). We thus obtain
naturally scale-free networks as random geometric graphs on hyperbolic
spaces and conversely, it can also be shown that scale-free networks
can result from a hidden hyperbolic metric \cite{Krioukov:2010}.

In line with random geometric graphs in euclidean space,
there is also a strong clustering for these graphs constructed on
hyperbolic space (unfortunately the clustering coefficient cannot be
computed analytically).

Finally, the authors of \cite{Krioukov:2010} extended their model and
introduce an inverse `temperature' $\beta$ akin to usual statistical
mechanics in the connection probability
\begin{equation}
p_C(x)= \left(1+e^{\beta(\zeta/2)(x-R)}\right)^{-1}
\end{equation}
and showed that this $\mathbb{H}^2$ model reproduces well the Internet
measurements for $P(k)$, the assortativity $k_{nn}(k)$, and the
clustering coefficient $C(k)$ for $\alpha=0.55$, $\zeta=1$
and $\beta=2$.

\subsubsection{A scale-free network on a lattice}
\label{sec4A2}

\begin{figure}
\centering
\begin{tabular}{c}
\epsfig{file=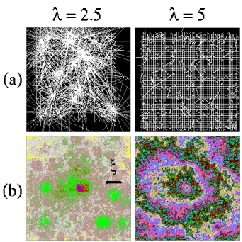,width=0.9\linewidth,clip=}
\end{tabular}
\caption{ Spatial structure of the network obtained by the method
  proposed in \cite{Rozen:2002} for different values of the
  exponent $\lambda$ of degree distribution. In (a) the networks are
  shown and in (b) the corresponding chemical shells of equidistant
  sites form the central node. From \cite{Rozen:2002}.}
\label{fig:rozen}
\end{figure}

As we saw in the section \ref{sec4A1} geometric graphs constructed on
uniformly distributed points naturally lead to networks with degrees
distributed according to a Poisson distribution. It is however
interesting to generate spatial graphs with a broad distribution
(scale-free network) in order to understand the effect of strong
heterogeneity on spatial networks.  In \cite{Rozen:2002}, Rozenfeld et
al. proposed a simple method to construct a scale-free network on
a lattice (and in \cite{Warren:2002} another variant is suggested). This
model is defined on a $d$-dimensional lattice of linear size $R$ and
with periodic boundary conditions. A random degree $k$ is assigned to
each node on this lattice according to the probability distribution
\begin{equation}
P(k)=Ck^{-\lambda}
\end{equation}
The idea is then simple: we connect a randomly chosen node $i$
to all its neighbors until its degree reaches its assigned value
$k_i$. The larger $k$ and the larger the region contained connected
neighbors. The size of this region is then given by
\begin{equation}
r(k)=Ak^{1/d}
\end{equation}
where the prefactor $A$ essentially depends on the density of
nodes. We show in Fig.~\ref{fig:rozen} the obtained networks for two
different values of $\lambda$.
The larger $\lambda$ and the shorter the links (large $k$ and
therefore long links are very rare) and the closer we are to a regular
lattice. We can define a chemical shell $\ell$ as consisting of all
sites at minimal distance $\ell$ from a given site. For large
$\lambda$ these chemical shells are concentric (as in the case of a
regular lattice) while for smaller $\lambda$ the presence of long
links destroys this order. Scaling arguments proposed in
\cite{Rozen:2002} suggest that the minimal length exponent is given
by
\begin{equation}
d_{min}=\frac{\lambda-2}{\lambda-1-1/d}
\end{equation}
This network has then the curious property to have a fractal dimension
which stays identical to the
euclidean dimension, but with a minimal length exponent $d_{min}<1$
for all $\lambda$ and $d>1$.

The authors of \cite{Warren:2002} studied the percolation properties
of such a model and found that for these spatial scale-free networks
with a degree distribution $P(k)\sim k^{-\lambda}$, the percolation
threshold in the limit of infinite networks does not go towards zero
(for $\lambda>2$), in sharp contrast with non-spatial scale-free
network which have $p_c=0$ for $\lambda<3$ (and $N=\infty$).

In a non-spatial scale-free network, there are many short paths
between the different hubs of the system easing the percolation. In
contrast, for spatial (scale-free) networks there is a high local
clustering due to the limited range of links which naturally lengthen
the distance between hubs. This negative assortativity makes it thus
more difficult to achieve percolation in such a system hence the
existence of a non-zero threshold. As noted in \cite{Warren:2002}, the
spread of a disease too would be easier to control than on a
scale-free network (see section \ref{sec5E}) which is expected as
spatial containment is usually easier to set up.

\subsubsection{Apollonian networks}
\label{sec4A3}

\begin{figure}[h!]
\begin{tabular}{c}
\epsfig{file=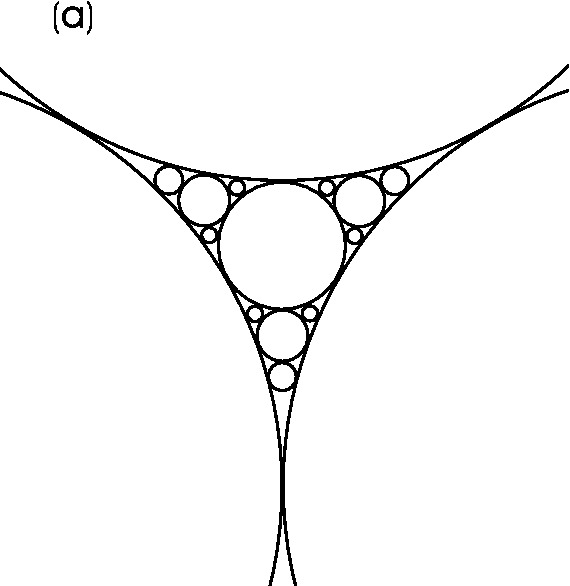,width=0.5\linewidth,clip=} \\
\epsfig{file=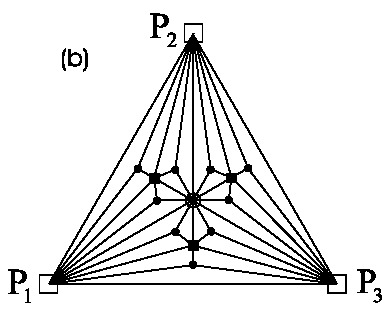,width=0.5\linewidth,clip=}
\end{tabular}
\caption{ Top: Classical Apollonian packing. Bottom: Apollonian
network (showing the first, second and third generation with circles,
squares, dots symbols respectively). From \cite{Andrade:2005}.}
\label{fig:Appollo}
\end{figure}

Other models were proposed in order to obtain a spatial scale-free
network. In particular, Apollonian networks were introduced by
Andrade et al. \cite{Andrade:2005} where they constructed a scale-free
network (Fig.~\ref{fig:Appollo}) from a space-filling packing of
spheres and by connecting the centers of touching spheres by lines.
These networks are simultaneously planar, scale-free (with exponent
$\gamma=1+\ln3/\ln2$, small-world- in fact `ultra-small' with an
average shortest path varying as
\begin{equation}
\langle\ell\rangle \sim (\log N)^{3/4}
\end{equation}
and a clustering coefficient larger than $0.8$ for large $N$. Various
quantities for the Apollonian network and one of its variant are also
computed and discussed in \cite{Zhang:2008,Zhang:2009}. Due to all
these simultaneous properties, Apollonian networks provide an interesting
playground to test theoretical ideas.

\subsection{Spatial generalization of the Erdos-Renyi graph}
\label{sec4B}

\subsubsection{Erdos-Renyi graph}
\label{4B1}

The Erdos-Renyi (ER) graph \cite{Erdos:1959,Erdos:1960,Erdos:1961} is
the paradigm for random graphs and is in many cases used as a null
model. One simple way to generate it is to run through all pairs of
nodes and to connect them with a probability $p$. The average number
of links is then
\begin{equation}
\langle E\rangle = p\frac{N(N-1)}{2}
\end{equation}
giving an average degree equal to $\langle k\rangle=2\langle
E\rangle/N=p(N-1)$. This last expression implies that in order to 
obtain a sparse network, we have to choose a small $p$ scaling
as $p=\langle k\rangle /N$ for large $N$.

The establishment of edges are random, independent events and we
thus obtain a binomial distribution for the degree $k$ of a node
\begin{equation}
P(k)={N-1\choose k}p^k(1-p)^{N-k+1}
\end{equation}
which can be approximated by a Poisson distribution
\begin{equation}
P(k)\approx e^{-\langle k\rangle}\frac{\langle k\rangle ^k}{k!}
\end{equation}
 for large $N$ with $pN=\langle k\rangle$ constant.
Other classical results can be easily derived such as the clustering
coefficient which can be shown to be $C=p$ and the average shortest
path
$\langle \ell\rangle \simeq \log N/\log \langle k\rangle$.
In fact for any generalized uncorrelated random graph characterized by a
probability distribution $P(k)$ it can be shown (see for example
\cite{BBVBook:2008}) that the average clustering coefficient is
\begin{equation}
\langle C\rangle=\frac{1}{N}\frac{(\langle k^2\rangle-\langle
  k\rangle)^2}{\langle k\rangle ^3}
\end{equation}
and that the average shortest path
\begin{equation}
\langle\ell\rangle\approx 
1+
\frac
{
\log N/\langle k\rangle
}
{
\log
\frac{\langle k^2\rangle-\langle k\rangle}{\langle k\rangle}
}
\end{equation}
These last two well-known expressions are useful in the sense that
they provide a reference to which we can compare results obtained
on a specific network in order to understand its features. In
particular, we can expect very different behavior for spatial networks
signalled by different scaling of these quantities with $N$.

\subsubsection{Planar Erdos-Renyi graph}
\label{sec4B2}

\begin{figure}[h!]
\begin{tabular}{c}
\epsfig{file=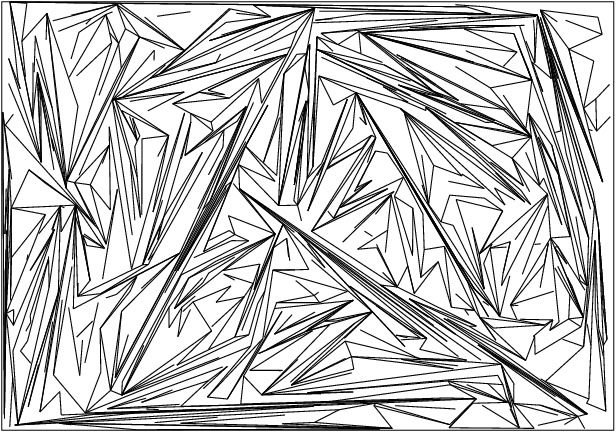,width=0.7\linewidth,clip=}\\
\epsfig{file=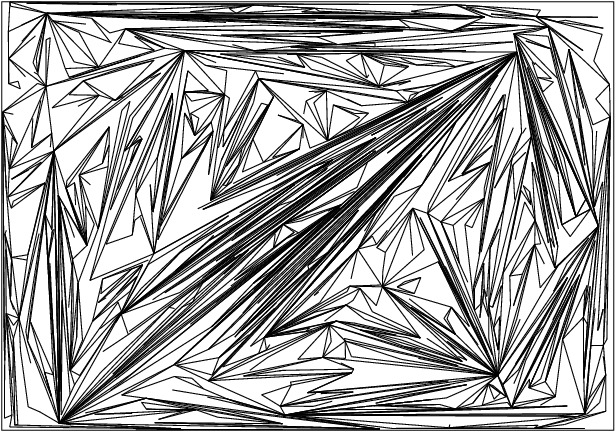,width=0.7\linewidth,clip=}
\end{tabular}
\caption{ Top: Planar Erdos-Renyi network obtained by rejecting links
  if they destroy planarity ($N=1000$). Bottom: Planar Barabasi-Albert network obtained by the
  same rejection method.}
\label{fig:ERplanar}
\end{figure}

The first simple idea to generate a random planar graph would be to
first generate a set of points in the two-dimensional space and then
to construct an Erdos-Renyi graph by connecting randomly the pairs of
nodes. It is clear that in this way we will generate mostly non-planar
graphs but we could decide to keep links which preserve planarity. We
would then obtain something like represented in
Fig.~\ref{fig:ERplanar} (top).
It is easy to imagine even other extensions of this process and we
could construct a planar BA network by adding in the two-dimensional
space nodes (with a random, uniformly distributed location) which will
connect according to the preferential attachment. We would then keep
the node if it preserves the planarity of the system. We note here
that there are some visual similarities with networks obtained by
random sequential adsorption of line segments \cite{Ziff:1990} and it
would be interesting to understand if there are deeper connections
between these problems.

Instead, mathematicians studied closely related networks. If we denote
by $P_N$ the class of all simple labelled planar graphs on $N$
vertices \footnote{A labeled graph refers to a graph where distinct
  labels are assigned to all vertices. The labeling thus adds
  configurations in the counting process of these graphs.}. We can
draw a random planar graph $R_N$ from this class with a uniform
probability \cite{Denise:1996} and ask for some questions
\cite{Gerke:2004,McDiarmid:2005} such as the number of vertices of a
given degree, the number of faces of a given size, etc.  For instance,
the following results have been demonstrated
(\cite{Denise:1996,Gerke:2004,McDiarmid:2005} and references therein)
\begin{itemize}
\item{} The random planar graph $R_N$ is
connected with probability at least $1/e$.
\item{} The number ${\cal N}_{u}$ of unlabelled planar graphs scales as
\begin{equation}
{\cal N}_{u}\sim \gamma_u^{N}
\end{equation}
with $9.48 < \gamma_u < 32.2$ and the number ${\cal N}_{l}$ of labelled
planar graphs as
\begin{equation}
{\cal N}_{l}\sim \gamma_{\ell}(N!)^{1/N}
\end{equation}
where $27.22685 <\gamma_{\ell}<27.22688$.
\item{} The average number of edges of $R_N$ is $\langle E\rangle \geq
  \frac{13}{7}N$.
\item{} The degree distribution decreases at least as $N/\gamma_{\ell}^k(k+2)!$.
\end{itemize}
Other properties can be derived for this class of networks and we
refer the interested reader to
\cite{Denise:1996,Gerke:2004,McDiarmid:2005} and references therein
for more results.

\subsubsection{The hidden variable model for spatial networks}
\label{sec4B3}

In the Erdos-Renyi model, the probability $p$ to connect two nodes is
a constant. In certain situations, we could imagine that a node is
described by a number of attributes (called {\it hidden variables} or
{\it fitnesses}) and the connection between two nodes could depend on
the respective attributes of these nodes
\cite{Caldarelli:2002,Boguna:2003b}. In order to give a concrete
example, we assume that there is only one attribute $\eta$ which is a
real number distributed according to a function $\rho(\eta)$. The
probability of connection for a pair of nodes $(i,j)$ is then given by
$p_{ij}=f(\eta_i,\eta_j)$ where $f$ is a given function. In the case
$f={\text const.}$ we recover the ER random graph. The average degree
of a node with fitness $\eta$ is given by
\begin{equation}
k(\eta)=N\int_0^\infty f(\eta,\eta')\rho(\eta')d\eta'\equiv NF(\eta)
\end{equation}
and the degree distribution is then 
\begin{equation}
\begin{split}
P(k)&=\int\rho(\eta)\delta(k-k(\eta))\\
&=\rho\left[F^{-1}\left(\frac{k}{N}\right)\right]
\frac{d}{dk}F^{-1}\left(\frac{k}{N}\right)
\end{split}
\end{equation}
A surprising result appears if we chose for example an exponential
fitness distribution ($\rho(\eta)\sim e^{-\eta}$) and for the function
$f$ a threshold function of the form
\begin{equation}
f(\eta_i,\eta_j)=\theta [\eta_i+\eta_j-z(N)]
\end{equation}
where $\theta$ is the Heaviside function and $z(N)$ a threshold which
depends in general on $N$. In this case, Caldarelli et
al. \cite{Caldarelli:2002} found a power law of the form $P(k)\sim
k^{-2}$ showing that a scale-free network can emerge even for a peaked
distribution of fitnesses. A spatial variant of this model was
proposed in \cite{Masuda:2005} (and discussed together with other models in
\cite{Hayashi:2006}) where the nodes $i$ and $j$ are connected if the
following condition is met
\begin{equation}
(\eta_i+\eta_j)h[d_E(i,j)]\geq \phi
\end{equation}
where $h[r]$ is a decreasing function and $\phi$ a constant
threshold. For this model, large fitnesses can therefore compensate
for larger distances and we will observe large-fitnesses nodes
connected by long links. If the distribution of fitnesses $f(\eta)$
has a finite support or is strongly peaked around some value, we will
have a typical scale $r_0=h^{-1}(\phi/2\overline{\eta})$ above which
no (or a very few) connections are possible. As a result, the average
shortest path will behave as for a lattice with
$\langle\ell\rangle\sim N^{1/d}$ in a $d$-dimensional space. For an
exponential fitness distribution $f(\eta)=e^{-\lambda\eta}$ and
$h(r)=r^{-\beta}$, the authors of \cite{Masuda:2005} find various
degree distributions according to the value of $\beta$ and goes to a
power-law $p(k)\sim k^{-2}$ for $\beta\to 0$ and to an exponential
distribution for $\beta=d$.  Various other cases were also studied in
\cite{Caldarelli:2002,Masuda:2005} and help to understand within this
model when scale-free distribution could appear.

Finally, we mention here a generalization to other metrics than the
spatial distance \cite{Boguna:2004b,Wong:2006}. In particular, in
\cite{Boguna:2004b}, the probability that two individuals are
connected decreases with a particular distance between these
individuals. This distance is computed in a `social' space and
measures the similarity for different social attributes. This model is
able to reproduce some of the important features measured in social
network such as a large clustering, positive degree correlations and
the existence of dense communities. More recently, Serrano et
al. \cite{Serrano:2008} developed the idea of hidden metric space by
using the one-dimensional circle as an underlying metric space in
which nodes are uniformly distributed. A degree $k$ drawn from a law
$P(k)\sim k^{-\gamma}$ for each node and each pair of nodes is
connected with a probability $r(d;k,k')$ that depends on the distance
$d$ between the nodes and also on their respective degrees $k$ and
$k'$. In particular, they studied the following form
\begin{equation}
r(d;k,k')=\left(1+\frac{d}{d_c(k,k')}\right)^{-\alpha}
\end{equation}
where $\alpha>1$ and where $d_c\sim kk'$ for example. The probability
that a pair of nodes is connected decreases then with distance (as
$d^{-\alpha}$) and increases with the product of their degrees $kk'$. In this case a
long distance can be compensated by large degrees, as it is observed
in various real-world networks. In this model, in agreement with other
models of spatial networks, we observe a large clustering (for $\alpha$ large
enough).

\subsubsection{The Waxman model}
\label{sec4B4}

The Waxman model \cite{Waxman:1988} is a random network topology
generator introduced by Waxman and appears as a spatial variant of the
ER model. In this model, the nodes are uniformly distributed in the
plane and edges are added with probabilities that depend on the
distance between the nodes
\begin{equation}
P(i,j)=\beta e^{-d_E(i,j)/d_0}
\end{equation}
The quantity $d_0$ determines the typical length of the links and
$\beta$ controls the total density of links. In terms of hidden
variables the attribute is here the spatial location of the node and
the pair connection probability depends on the distance between the
nodes. For $d_0\to\infty$, length is irrelevant and we recover the ER
model while for $d_0\sim 1/\sqrt{\rho}$ (where $\rho$ is the
average density of nodes in the plane) long links are prohibited and we
are in the limit of a lattice-like graph.

Even if this model is very simple, it served as a first step towards
the elaboration of more sophisticated models of the Internet
\cite{Zegura:1996}. Also, despite its simplicity, this Waxman model
can be used in order to understand the importance of space in
different processes taking place on this network. We can cite for
example navigation or congestion problems in communication systems
(see section \ref{sec5C}).

A growth model close to the Waxman model was proposed in
\cite{Kaiser:2004} where at each time step a new node $u$ is added
in the plane and is connected to existing nodes $v$ with a probability (if it fails
to connect the node is discarded)
\begin{equation}
P(u,v)=\beta e^{-\alpha d_E(u,v)}
\end{equation}
or even decreasing as power law $P(u,v)\sim
d_E(u,v)^{-\tau}$. Networks generated with this algorithm have a large
clustering coefficient (as expected) and probably a large diameter
(although there is no quantitative prediction in \cite{Kaiser:2004}).

\subsection{Spatial small worlds}
\label{sec4C}

\subsubsection{The Watts-Strogatz model}
\label{sec4C1}

\begin{figure}[h!]
\begin{tabular}{c}
\epsfig{file=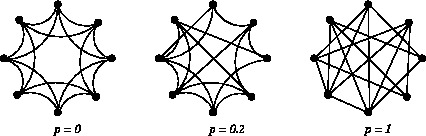,width=0.9\linewidth,clip=}
\end{tabular}
\caption{ Construction of the Watts-Strogatz model for $N=8$ nodes. At
  $p=0$ each node is connected to its four nearest neighbors and by
  increasing $p$ an increasing number of edges is rewired. Adapted
  from Watts and Strogatz \cite{Watts:1998}. }
\label{fig:sw}
\end{figure}

Already in $1977$ spatial aspects of the small-world problem were
considered by geographers in the paper \cite{Stoneham:1977} but we
had to wait until $1998$ when Watts and Strogatz (WS) proposed a
simple and powerful network model \cite{Watts:1998} which incorporates
both a spatial component and long-range links. This model is obtained
by starting from a regular lattice and by rewiring at random the links
with a probability $p$ (Fig.~\ref{fig:sw}).

The degree distribution of this network has essentially the same
features as the ER random graph, but the clustering coefficient and the
average shortest path depend crucially on the amount of randomness
$p$. The average clustering coefficient has been shown to behave as
\cite{Barrat:2000}
\begin{equation}
\langle C(p)\rangle\simeq \frac{3(m-1)}{2(2m-1)}(1-p)^3
\end{equation}
where the average degree is $\langle k\rangle =2m$. 
The average shortest path has been shown to scale as \cite{Barthelemy:1999,Erratum_Barthelemy:1999}
\begin{equation}
\langle \ell\rangle \sim N^*{\cal F}\left(\frac{N}{N^*}\right)
\end{equation}
where the scaling function behaves as 
\begin{equation}
{\cal F}(x)\sim
\begin{cases}
x\;\;\;\;\;&{\rm for}\;\; x\ll 1\\
\ln x\;\;&{\rm for}\;\;x\gg 1
\end{cases}
\end{equation}
The crossover size scales as $N^*\sim 1/p$
\cite{Barthelemy:1999,Erratum_Barthelemy:1999,Barrat:2000} which
basically means that the crossover from a large-world to a small-world
occurs for an average number of shortcuts equal to one
\begin{equation}
N^*p\sim 1
\end{equation}
The network can thus be seen as clusters of typical size $N^*(p)$ connected by
shortcuts.

The interest of these networks is that they can simultaneously present
some features typical of random graphs (with a small-world behavior
$\ell\sim \log N$) and of clustered lattices with a large average
clustering coefficient (while for ER random graph $\langle C\rangle
\sim 1/N$).

\subsubsection{Spatial generalizations}
\label{sec4C2}

One of the first variants of the Watt-Strogatz model was
proposed in \cite{Jespersen:2000,Kleinberg:2000,Sen:2001} and was subsequently generalized to
higher dimensions $d$ \cite{Sen:2002b}. 
\begin{figure}[h!]
\begin{tabular}{c}
\epsfig{file=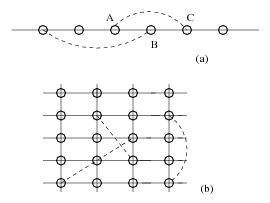,width=0.8\linewidth,clip=}
\end{tabular}
\caption{ Schematic representation of spatial small-world in (a) one
  dimension and (b) two dimensions. The dashed lines represent the
  long-range links occuring with probability $q(\ell)\sim
  \ell^{-\alpha}$. From \cite{Sen:2002b}.}
\label{fig:sswnet}
\end{figure}
In this variant (see Fig.~\ref{fig:sswnet}), the nodes are located on a
regular lattice in $d$-dimensions with periodic boundaries. The
main idea is that if the
shortcuts have to be physically realized there is a cost associated
with their length. A way to model this is to add links with a
probability 
\begin{equation}
q(\ell)\sim \ell^{-\alpha}
\end{equation}
For each node, we add a shortcut with probability $p$ which implies
that on average there will be $pN$ additional shortcuts.

Concerning the average shortest path, it is clear that if $\alpha$ is
large enough, the shortcuts will be small and the behavior of
$\langle\ell\rangle$ will be `spatial' with $\langle\ell\rangle\sim
N^{1/d}$. On the other hand, if $\alpha$ is small enough we can expect
a small-world behavior $\ell\sim \log N$.  In fact, various studies
\cite{Jespersen:2000,Moukarzel:2002,Sen:2002b} discussed the existence
of a threshold $\alpha_c$ separating the two regimes, small- and
large-world. We follow here the discussion of \cite{Petermann:2005}
who studied carefully the behavior of the average shortest path.  The
probability that a shortcut is `long' is given by
\begin{equation}
P_c(L)=\int_{(1-c)L/2}^{L/2}q(\ell)d\ell
\end{equation}
where $c$ is small but non-zero. The critical fraction of shortcuts $p^*N=p^*(L)L^d$
then satisfies
\begin{equation}
P_c(L)p^*(L)L^d\sim 1
\end{equation}
which means that if we have a fraction $p>p^*$ of long shortcuts, the
system will behave as a small-world. We then obtain
\begin{equation}
p^*(L)\sim
\begin{cases}
L^{-d}\;\;\;\;\;&{\rm if}\;\; \alpha<1\\
L^{\alpha-d-1}\;\;&{\rm if}\;\;\alpha>1
\end{cases}
\end{equation}
and a logarithmic behavior $\log L/L^d$ for $\alpha=1$. For a given
value of $p$ we thus have
one length scale 
\begin{equation}
L^*(p)\sim
\begin{cases}
p^{-1/d}\;\;\;\;\;{\rm if}\;\; \alpha<1\\
p^{1/(\alpha-d-1)}\;\;{\rm if}\;\;\alpha>1
\end{cases}
\end{equation}
which in the special case $\alpha=0$ was obtained in \cite{Newman:1999}.
We will then have the following scaling form for
the average shortest path
\begin{equation}
\langle \ell\rangle=L^*{\cal F}_\alpha\left(\frac{L}{L^*}\right)
\end{equation}
where the scaling function varies as
\begin{equation}
{\cal F}_\alpha(x)\sim
\begin{cases}
x\;\;\;\;\;{\rm if}\;\;x\ll 1\\
\ln x\;\;\;\;\;{\rm if}\;\;x\gg 1
\end{cases}
\end{equation}
(or even a function of the form $(\ln x)^{\sigma(\alpha)}$ with
$\sigma(\alpha)>0$ for $x\gg 1$). The characteristic length for
$\alpha>1$ thus scales as
\begin{equation}
L^*(p)\sim p^{1/(\alpha-d-1)}
\end{equation}
and displays a threshold value $\alpha_c=d+1$ a value already obtained
with the average clustering coefficient for $d=1$ in \cite{Sen:2002b}.
For $\alpha>\alpha_c$, the length scale $L^*(p)$ is essentially finite
and less than $1$, which means that for all values of $L$ the system
has a large-world behavior with $\langle\ell\rangle\sim L$. In other
words, the links  in this case cannot be long enough and the graph can
always be coarse-grained to reproduce a regular lattice. In the
opposite case $\alpha<\alpha_c$, the length $L^*(p)$ diverges for
$p\to 0$ and there will always be a regime such that $L^*(p)\gg L$
implying a small-world logarithmic behavior.

Finally, we mention a recent numerical study \cite{Kosmidis:2008} of this model
which seems to show that for $\alpha>d$ there are two regimes. First, for
$d<\alpha<2d$
\begin{equation}
\langle\ell\rangle\sim (\log N)^{\sigma(\alpha)}
\end{equation}
with
\begin{equation}
\sigma=
\begin{cases}
\frac{1/\alpha}{2-\alpha}\;\;{\rm for}\;\;d=1\\
\frac{4/\alpha}{4-\alpha}\;\;{\rm for}\;\;d=2
\end{cases}
\end{equation}
The second regime is obtained for $\alpha>2d$ where the `spatial'
regime $\langle\ell\rangle \sim N^{1/d}$ is recovered. We note that
numerically, the scaling prediction of \cite{Petermann:2005} with two
regimes only and the result of \cite{Kosmidis:2008} are however
difficult to distinguish. For $d=1$ there are no discrepancies
($\sigma=1$ for $\alpha=d=1$) and for $d=2$, the results for
$\alpha=2$ and $\alpha=4$ are consistent with the analysis of
\cite{Petermann:2005}. A problem thus subsists here for $d=2$ and
$\alpha=3$ for which $\sigma=4/3$, a value probably difficult to
distinguish numerically from corrections obtained at
$\alpha=\alpha_c=d+1$.


\subsection{Spatial growth models}
\label{sec4D}

\subsubsection{Generalities}
\label{sec4D1}

In this part we will review models of growing networks which
essentially elaborate on the preferential attachment model proposed by
Albert and Barabasi \cite{Simon:1955,Albert:1999}. In the preferential
attachment, there is a propensity to connect a new node to an already
well-connected one \cite{Simon:1955,Albert:1999} which is probably an
important ingredient in the formation of various real-world networks.

The process to generate such a Barabasi-Albert (BA) network is thus extremely
simple. Starting from a small `seed' network, we introduce a new node
$n$ at each time step. This new node is allowed to make $m$
connections towards nodes $i$ with a probability 
\begin{equation}
\Pi_{n\to i}\propto k_i
\end{equation}
where $k_i$ is the degree of node $i$. We refer the interested reader
to the various books and reviews which describe in detail this model
\cite{Albert:2002,Dorogovtsev:2002a,Pastor:2003,Newman:2003b,Caldarelli:2008,Havlin:2010}
. In particular, the degree distribution behaves as a power-law with exponent
$P(k)\sim k^{-\gamma}$ with $\gamma=3$, the average shortest path
behaves at the dominant order as $\log N$, and the average clustering
coefficient is given by
\begin{equation}
\langle C\rangle =\frac{m}{8N}(\ln N)^2
\end{equation}
while $C(k)\sim 1/k$.

In many networks such as transportation or communication networks,
distance is however a relevant parameter and real-world examples
suggest that when long-range links are existing, they usually connect
to hubs-the well-connected nodes. Many variants of the BA model were
proposed and a few of them were concerned with space. The growth
process is the same as for BA, but in addition one has to specify the
location of the new node. In most models, the location is taken at
random and uniformly distributed in space. The attachment probability is then
written as
\begin{equation}
\Pi_{n\to i}\propto k_iF[d_E(n,i)]
\label{eq:gen}
\end{equation}
where $F$ is a function of the euclidean distance $d_E(n,i)$ from the node $n$
to the node $i$. 

When $F$ is a decreasing function of distance (as in most cases), this
form (Eq.~\ref{eq:gen}) implies that new links preferentially connect
to hubs, unless the hub is too far in which case it could be better to
connect to a less connected node but closer in space. In order to have
long links, the target node must have a large degree in order to
compensate for a small $F(d)$ such that $kF(d)\sim 1$. This is for
instance the case for airlines: Short connections go to small airports
while long connections point preferably to big airports, ie.
well-connected nodes.

\subsubsection{Preferential attachment and distance selection}
\label{sec4D2}

Several models including distance were proposed
\cite{Yook:2002,Rozen:2002,Warren:2002,Dall:2002,Sen:2002b,Jost:2002,Manna:2002,Xulvi:2002,Barthelemy:2003a}
and we review here the main results obtained in these studies. The $N$
nodes of the network are supposed to be in a $d$-dimensional space of
linear size $L$ and we assume that they are distributed randomly in
space with uniform density $\rho$. One could use other distributions:
For instance in cities the density decreases exponentially from the
center \cite{Clark:1951}. The case of randomly distributed points is
interesting since on average it preserves natural symmetries such as
translational and rotational invariance in contrast with lattices.

Essentially two different cases were considered in the literature.

\paragraph{Finite range case.}
\label{sec4D2a}

In this case, the function $F(d)$ decreases sharply with distance,
typically as an exponential \cite{Barthelemy:2003a}
\begin{equation}
F(d)=e^{-d/r_c}
\end{equation}
and thus introduces a new scale in the system, the interaction range $r_c$.
When the interaction range is of the order of the system size (or
larger), the distance is irrelevant and the obtained
network will be scale-free. In contrast, when the interaction range is
small compared to the system size, we expect new properties and a
crossover between these two regimes.

The important dimensionless parameters are here the average number $n$ of points in a
sphere of radius $r_c$
\begin{equation}
n=\rho r_c^d\frac{\pi^{d/2}}{\Gamma(1+\frac{d}{2})}
\end{equation}
and the ratio which controls the importance of spatial effects
\begin{equation}
\eta=\frac{r_c}{L}
\end{equation}
(where $L$ is the system size). It can then be shown \cite{Barthelemy:2003a} that the degree
distribution follows the scaling form valid only for $\eta\ll 1$
\begin{equation}
P(k)\sim k^{-\gamma}f(k/k_c)
\end{equation}
with $\gamma=3$ and where the cut-off $k_c$ behaves as
\begin{equation}
k_c\sim n^{\beta}
\label{kc}
\end{equation}
where $\beta\simeq 0.13$. The distance effect thus limits the choice
of available connections and thereby limits the degree distribution
for large values.

Also, when the distance effect is important we expect a large value of
the average clustering coefficient.  In the limit of small $\eta$, we
can expect the result of random geometric graphs (see section \ref{sec4A1} and
\cite{Dall:2002}) to hold
\begin{equation}
C_0\equiv \langle C\rangle(\eta=0) =1-3\sqrt{3}/4\pi\simeq 0.59
\end{equation}
(for $d=2$). We expect to recover this limit for $\eta\to 0$ and for
an average connectivity $\langle k\rangle=6$ which is a well-known
result in random geometry. If $\eta$ is not too small, the
preferential attachment is important and induces some dependence of
the clustering coefficient on $n$. In addition, we expect that
$\langle C\rangle (\eta)$ will be lower than $\langle C\rangle (0)$
since the longer links will not connect to the nearest
neighbors.
\begin{figure}[h!]
\begin{tabular}{c}
\epsfig{file=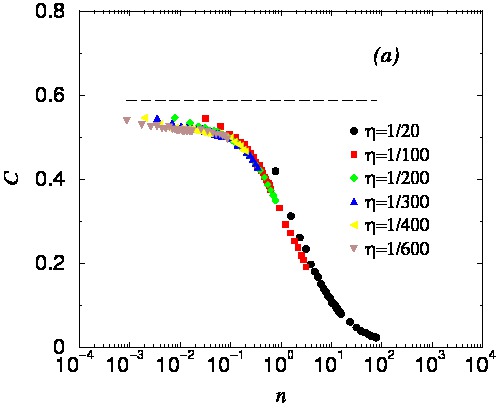,width=0.8\linewidth,clip=}\\
\epsfig{file=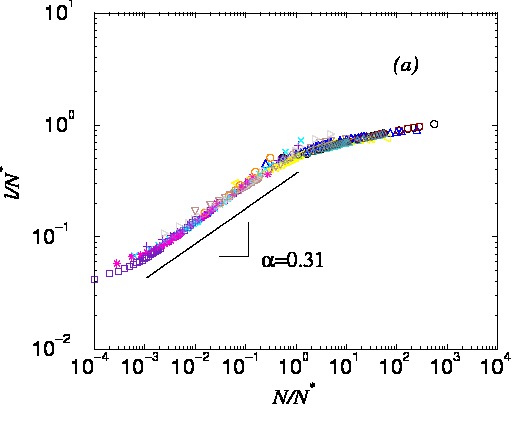,width=0.8\linewidth,clip=}
\end{tabular}
\caption{ (a) Clustering coefficient versus the mean number
  $n=\rho\pi r_c^2$ of points in the disk of radius $r_c$ (plotted in
  Log-Lin). The dashed line corresponds to the theoretical value $C_0$
  computed when a vertex connects to its adjacent neighbors without
  preferential attachment. (b) Data collapse for the average shortest
  path obtained. The first part can be fitted by
  a power-law with exponent $\approx 0.3$ followed by a logarithmic
  regime for $N>N^*$. Both figures from \cite{Barthelemy:2003a}.}
\label{fig:eplmb2}
\end{figure}
Numerical results show that there is a good collapse (see
Fig.~\ref{fig:eplmb2}a) when $\langle C\rangle(\eta)$ is expressed in
terms of $n$ and is a decreasing function since when $n$ increases the
number of neighbors of the node will increase thus decreasing the
probability that two of them will be linked.

The average shortest path is described by a scaling ansatz which governs the crossover
from a spatial to a scale-free network
\begin{equation}
\langle\ell(N,\eta)\rangle=[N^*(\eta)]^\alpha\Phi_{\alpha}\left[\frac{N}{N^*(\eta)}\right]
\label{ansatz}
\end{equation}
with 
\begin{equation}
\Phi_{\alpha}(x)\sim
\begin{cases}
x^\alpha\;\;\;\;\;&{\rm for}\;\; x\ll 1\\
\ln x\;\;&{\rm for}\;\;x\gg 1
\end{cases}
\end{equation}
The typical size $N^*$ depends on $\eta$ and we can find its behavior in
two extreme cases. For $\eta\gg 1$, space is irrelevant and
\begin{equation}
N^*(\eta\gg 1)\sim N_0
\label{eta_big}
\end{equation}
where $N_0$ is a finite constant. When $\eta\ll 1$, the existence of
long-range links will determine the behavior of $\langle\ell\rangle$. If we denote
by $a=1/\rho^{1/d}$ the typical inter-node distance, the transition
from a large to a small-world will be observed for $r_c\sim a(N^*)$ which 
leads to 
\begin{equation}
N^*(\eta\ll 1)\sim \frac{1}{\eta^d}
\label{eta_small}
\end{equation}
In Fig.~\ref{fig:eplmb2}b, the ansatz Eq.~(\ref{ansatz}) together with
the results Eqs.~(\ref{eta_big}), (\ref{eta_small}) is shown. This data
collapse is obtained for $\alpha\simeq 0.3$ and $N_0\simeq 180$ (for
$d=2$).  For $N>N^*$, the network is a small-world: the diameter is
growing with the number of points as $\langle\ell\rangle\sim\log N$. In the opposite
case of the spatial network with a small interaction range, the
network is much larger: To go from a point $A$ to a point $B$, we
essentially have to pass through most of the points in between and the
behavior of this network is much that of a lattice with $\langle\ell\rangle\sim
N^\alpha$, although the diameter is here smaller probably due to the
existence of some rare longer links (in the case of a lattice we
expect $\alpha=1/d$). Probably larger networks and better statistics
are needed here.

This model was extended \cite{Barrat:2005} in the case of weighted
growing networks in a two-dimensional geometrical space. The model
considered consists of growth and the probability that a new site
connects to a node $i$ is given by
\begin{equation}
\Pi_{n \to i}=\frac{s_i^w e^{-d_{ni}/r_c}}{\sum_j s_j^w e^{-d_{nj}/r_c}} ,
\label{sdrive}
\end{equation}
where $r_c$ is a typical scale and $d_{ni}$ is the Euclidean distance
between $n$ and $i$. This rule of {\em strength driven
preferential attachment with spatial selection}, generalizes the
preferential attachment mechanism driven by the strength to spatial
networks. Here, new vertices connect more likely to vertices which
correspond to the best interplay between Euclidean distance and strength.

The weights are also updated according to the following rule already
studied in another paper \cite{Barrat:2004a}
\begin{equation}
w_{ij}\to w_{ij}+\delta\frac{w_{ij}}{s^{w}_i}.
\label{eq:rule}
\end{equation}
for all neighbors $j\in\Gamma(i)$ of $i$.

The model contains thus two relevant parameters: the ratio between the
typical scale and the size of the system $\eta=r_c/L$, and the ability
to redistribute weights, $\delta$.

\begin{figure}
\centering
\begin{tabular}{c}
\includegraphics[angle=0,scale=.30]{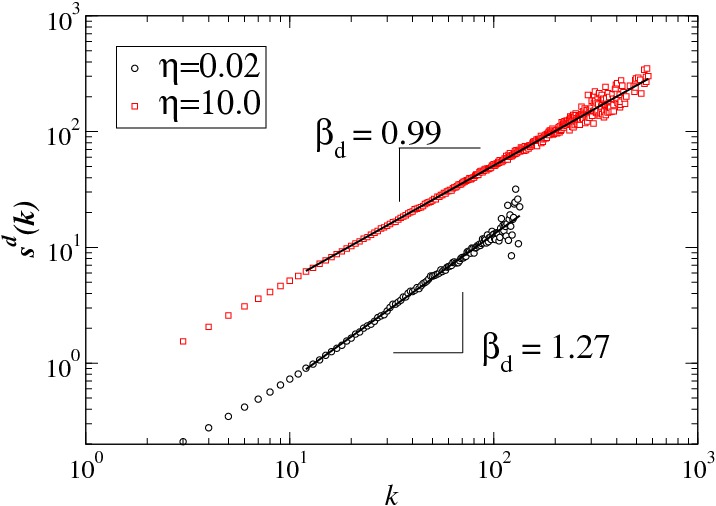}
\end{tabular}
\caption{ Distance strength versus $k$ for two different cases
  ($\eta=0.02$ and $\eta=10.0$).  When $\eta$ is not small, space is
  irrelevant and there are no correlations between degree and space as
  signalled by $\beta_d\approx 1$. When spatial effects are important
  ($\eta=0.02\ll 1$), non-linear correlations appear and
  $\beta_d>1$. We observe a crossover for $k\simeq 10-20$ to a
  power-law behavior and the power-law fit over this range of values
  of $k$ is shown (full lines). From \cite{Barrat:2005}.}
\label{fig:s.vs.k}
\end{figure}

The most important results concerning the traffic are the
following. The correlations appearing between traffic and topology of
the network are largely affected by space as the value of the
exponents $\beta_w$ and $\beta_d$ depend on $\eta$
(for $\beta_d$ see Fig.~\ref{fig:s.vs.k}). Strikingly, the effect of the spatial
constraint is to increase both exponents $\beta_w$ and $\beta_d$ to
values larger than $1$ and although the redistribution of the weights
is linear, non-linear relations $s^w(k)$ and $s^d(k)$ as a function of
$k$ appear. For the weight strength the effect is not very pronounced
with an exponent of order $\beta_w\approx 1.1$ for $\eta=0.01$, while
for the distance strength the non-linearity has an exponent of order
$\beta_d\approx 1.27$ for $\eta=0.02$. 

The nonlinearity induced by the spatial structure can be explained by
the following mechanism affecting the network growth.  The increase of
spatial constraints affects the trend to form global hubs, since long
distance connections are less probable, and drives the topology
towards the existence of `regional' hubs of smaller degree.  The total
traffic however is not changed with respect to the case $\eta=\infty$,
and is in fact directed towards these regional hubs. These
medium-large degree vertices therefore carry a much larger traffic
than they would do if global hubs were available, leading to a faster
increase of the traffic as a function of the degree, eventually
resulting in a super-linear behavior. Moreover, as previously
mentioned, the increase in distance costs implies that long range
connections can be established only towards the hubs of the system:
this effect naturally leads to a super-linear accumulation of $s^d(k)$
at larger degree values.

The spatial constraints act at both local and global level of the
network structure by introducing a distance cost in the establishment
of connections. It is therefore important to look at the effect of
space in global topological quantities such as the betweenness
centrality.  Hubs are natural crossroads for paths and it is natural
to observe a correlation between $g$ and $k$ as expressed in the
general relation $g(k)\sim k^{\mu}$ where the exponent $\mu$ depends
on the characteristics of the network and we expect this relation to
be altered when spatial constraints become important (see section
\ref{sec2B2}). In particular, the betweenness centrality displays
relative fluctuations which increase as $\eta$ decreases and become
quite large.  This can be understood by noticing that the probability to establish
far-reaching short-cuts decreases exponentially in Eq.~(\ref{sdrive})
and only the large traffic of hubs can compensate this decay. Far-away
geographical regions can thus only be linked by edges connected to
large degree vertices, which implies a more central role for these
hubs. The existence of fluctuations means that nodes with small degree may have a
relatively large betweenness centrality (or the opposite), as observed
in the air-transportation network (see Fig.~\ref{fig:bc_na}
and~\cite{Guimera:2004}). 
\begin{figure}
\begin{tabular}{c}
\includegraphics[angle=0,scale=.30]{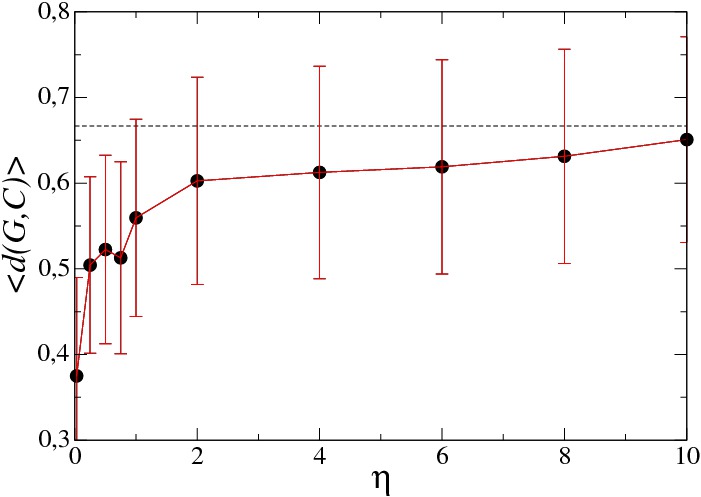}
\end{tabular}
\caption{ Average Euclidean distance between the barycenter $G$ of all
  nodes and the $10$ most central nodes ($C$) versus the parameter
  $\eta$ (Here $\delta=0$, $N=5,000$ and the results are averaged over
  $50$ configurations). When space is important (ie.  small $\eta$),
  the central nodes are closer to the gravity center. For large
  $\eta$, space is irrelevant and the average distance tends to the
  value corresponding to a uniform distribution $\langle r
  \rangle_{unif}=2/3$ (dotted line).  From \cite{Barrat:2005}. }
\label{fig:rvseta}
\end{figure} 
The present model defines an intermediate situation in that we have a
random network with space constraints that introduces a local structure
since short distance connections are favored. Shortcuts and long
distance hops are present along with a spatial local structure that 
clusters spatially neighboring vertices. In Fig.~\ref{fig:rvseta} we
plot the average distance $d(G,C)$ between the barycenter $G$ and the $10$
most central nodes. As expected, as spatial constraints become more important,
the most central nodes get closer to the spatial barycenter of the network.

\paragraph{Power law decay of $F(d)$.}
\label{sec4D2b}

In this case, the function $F$ in Eq.~(\ref{eq:gen}) is varying as
\begin{equation}
F(d)=d^{\alpha}
\end{equation}
This problem was considered in \cite{Manna:2002,Xulvi:2002,Yook:2002}.
The numerical study presented in \cite{Xulvi:2002} shows that in the
two-dimensional case, for all values of $\alpha$ the average shortest
path behaves as $\log N$. The degree distribution is however different
for $\alpha>-1$ where it is broad, while for $\alpha<-1$, it is decreasing
much faster (the numerical results in \cite{Xulvi:2002} suggest according
to a stretched exponential).

\begin{figure}
\begin{tabular}{c}
\includegraphics[angle=0,scale=.60]{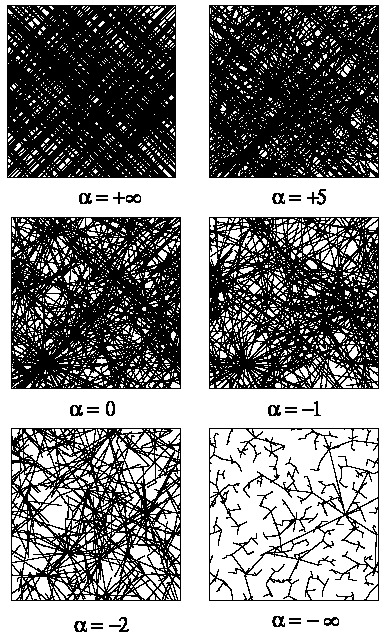}
\end{tabular}
\caption{ Various networks obtained with the rule
  $F(d)=d^{\alpha}$. From \cite{Manna:2002}.}
\label{fig:manna}
\end{figure}

In \cite{Manna:2002}, Manna and Sen study the same model but for various
dimensions and for values of $\alpha$ going from $-\infty$ to
$+\infty$ where the node connects to the closest and the farthest
node, respectively (Fig.~\ref{fig:manna}).
These authors indeed find that if $\alpha>\alpha_c$ the network is
scale-free and in agreement with \cite{Xulvi:2002} that
$\alpha_c(d=2)=-1$ while for large dimensions $\alpha_c$ decreases
with $d$ (the natural guess $\alpha_c=1-d$ is not fully supported by
their simulations). This study was complemented by another one by the same
authors \cite{Sen:2003} in the $d=1$ case and where the probability to connect to a node $i$
is given by (which was already proposed in \cite{Yook:2002})
\begin{equation}
\Pi_{n\to i}\sim k_i^{\beta}d_E(n,i)^{\alpha}
\end{equation}
For $\alpha>\alpha_c=-0.5$ the network is scale-free at $\beta=1$ with
an exponent $\gamma=3$. They also find a scale-free network for
a line in the $\alpha-\beta$ plane and also for $\beta>1$ and
$\alpha<-0.5$.
The degree-dependent clustering coefficient $C(k)$ behaves as
\begin{equation}
C(k)\sim k^{-b}
\end{equation}
where the authors found numerically that $b$ varies from $0$ to $1$
(which is the value obtained in the BA case).

\subsubsection{Growth and local optimization}
\label{sec4D3}

\paragraph{A model for the Internet.}
\label{sec4D3a}

In the Internet, the degree distribution is a power law (see for
example \cite{Pastor:2003} and references therein) and in order to
explain this fact, computer scientists \cite{Fabrikant:2002} proposed
a tree growth model with local optimization. More precisely, at each unit
time, a new node $i$ is connected to an already connected node $j$
such that the quantity 
\begin{equation}
{\cal E}=\lambda d_E(i,j)+h_j
\end{equation}
is minimum. The quantity $h_j$ is a measure of centrality which can be
either the average shortest path length from $j$, or the shortest path
length to a root node, etc. A new node would then typically like to connect to a
very central node but is limited by the cost measured by the
distance. There will thus be an interplay between these two
constraints and we can explore the two extremes: for small $\lambda$,
we obtain a star network. In the opposite case, when $\lambda$ is
large, only distance is important and we obtain some sort of dynamical
version of the Euclidean minimum spanning tree. If $\lambda$ has some
intermediate values ($\lambda$ growing slower than $\sqrt{N}$ and is
larger than a certain constant), we can obtain a network with a power
law degree distribution whose exponent $\gamma$ depends on
$\lambda$. For instance if $\lambda\sim N^{1/3}$ then $\gamma=1/6$ (see
\cite{Fabrikant:2002} for other results on this model).

Even if the resulting network is not obtained by minimizing some
global function, the addition of new nodes minimizes a cost function
and in this sense we can speak of a {\it local} minimization.

\paragraph{A model for distribution networks.}
\label{sec4D3b}

Many networks, including transportation and distribution networks
evolved in time and increased their service area. Clearly, in these
situations the resulting networks are growing and cannot result from a
global optimization but instead, local optimization could be a
reasonable mechanism for explaining the organization of these
structures.

\begin{figure}[h!]
\begin{tabular}{c}
\epsfig{file=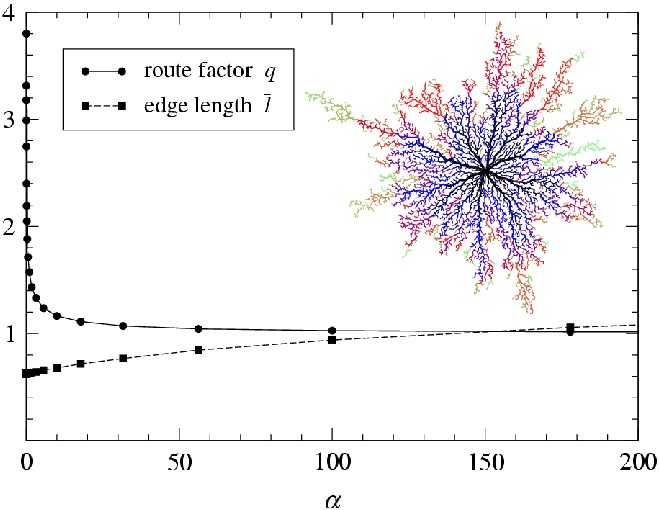,width=1.0\linewidth,clip=}
\end{tabular}
\caption{ Route factor $q$ and average edge length $\overline{l}$
  versus $\alpha$ in Eq.~(\ref{eq:shape}). These results are obtained for
  $N=10^4$ vertices and in the inset the network is obtained for
  $\alpha=12$. From \cite{Gastner:2006b}. }
\label{fig:shape}
\end{figure}

In the example of a transportation network such as the train system,
the nodes represent the train stations and the edges the rail segments
between adjacent stations. In many of these systems, there is also a
{\it root node} which acts as a source of the distribution system or
in the case of the railway can be considered to be the central
station. During the evolution of the network at least two factors
could be considered. First, the total length of the system which
represents the cost of the infrastructure should not be too
large. Space has another important role here: the transportation
system should also allow to connect two nodes in the network through a
shortest path whose length is not too far from the `as crow flies'
distance. This efficiency is for example measured by the route factor
-or detour index - (see section \ref{sec2B2c}) which for two nodes
$i$ and $j$ of the network reads
\begin{equation}
\nonumber
Q(i,j)=\frac{d_R(i,j)}{d_E(i,j)}
\end{equation}
For a system with a root node $0$, one can then compute the route
factor as the average over all nodes except $0$
\begin{equation}
q=\frac{1}{N}\sum_{i\neq 0}\frac{d_R(i,0)}{d_E(i,0)}
\end{equation}
Following these two requirements, Gastner and Newman
\cite{Gastner:2006b} proposed a model of a growing network where
vertices are initially randomly distributed in the two-dimensional
plane and where one vertex is designated as the root node $0$. A
network is then grown from its root by adding an edge between an
unconnected node $i$ to a vertex $j$ which belongs to the network. The
edge is chosen according to a local minimization process such that the
quantity
\begin{equation}
{\cal E}_{ij}=d_E(i,j)+\alpha\frac{d_E(i,j)+d_R(j,0)}{d_E(i,0)}
\label{eq:shape}
\end{equation}
is minimum and where $\alpha>0$ is here a parameter controlling the
importance of the route factor. For $\alpha=0$, the algorithm adds
always a link to the closest vertex and the resulting network is
similar to the MST and has a poor route factor
\cite{Gastner:2006b}. When $\alpha$ increases the route factor
decreases and the average edge length $\overline{l}$ (not be confused
with the average shortest path $\langle\ell\rangle$) increases (see
Fig.~\ref{fig:shape}).
In this figure, we see that the route factor $q$ decreases sharply
when $\alpha$ increases from zero, while the average edge length
-which is a measure of the building cost of the network- increases
slowly. This suggests that it is possible to grow networks with a
small cost but with a good efficiency. 

Gastner and Newman \cite{Gastner:2006b} also studied a simpler version of this
model where the local minimization acts on the quantity
\begin{equation}
{\cal E'}_{ij}=d_E(i,j)+\beta d_R(j,0)
\label{eq:shape2}
\end{equation}
which is similar to the model for the Internet proposed by Fabrikant,
Koutsoupias, Papadimitriou \cite{Fabrikant:2002} (see also section
\ref{sec4D3a}) but where the vertices are added one by one. This model
produces networks similar to the one obtained with
Eq.~(\ref{eq:shape}) and self-organizes to networks with small $q$
which is not imposed here. This model can be applied to the set of
stations of the Boston rail network and produces a network in good
correspondence with the real one. Also, the small value of $q$ is
confirmed in different empirical examples such as sewer systems, gas
pipelines, and the Boston subway \cite{Gastner:2006b} where the ratio
$\overline{l}/\overline{l}_{MST}$ is in the range $[1.12,1.63]$ while
the route factor is less than $1.6$ (and compared to the MST is
improved by a factor in the range $[1.4,1.8]$).

The networks obtained here are trees which is a simplification for
many of the real-world networks which usually contain loops. In
addition, there is also usually an interaction between the density of
points and the network and this co-evolution is not taken into account
here. However, this simple model of local optimization seems to
capture important ingredients and could probably serve as a good
starting point for further improvements.

\paragraph{A model for street and road networks.}
\label{sec4D3c}

In the case of urban street patterns, striking statistical
regularities across different cities have been recently empirically
found, suggesting that a general and detail-independent mechanism may
be in action (see section \ref{sec3B1}). The rationale to invoke a
local optimality principle in this context is as in the case of
distribution networks based on costs: every new road is built to
connect a new location to the existing road network in the most
efficient way~\cite{Bejan:1998}. The locality of the rule is
implemented both in time and space during the evolution and formation
of the street network, in order to reflect evolution histories that
greatly exceed the time-horizon of planners. The self-organized
pattern of streets emerges as a consequence of the interplay of the
geometrical disorder and the local rules of optimality.

\begin{figure}[!t]
   \centering
   \includegraphics[angle=0,scale=.30]{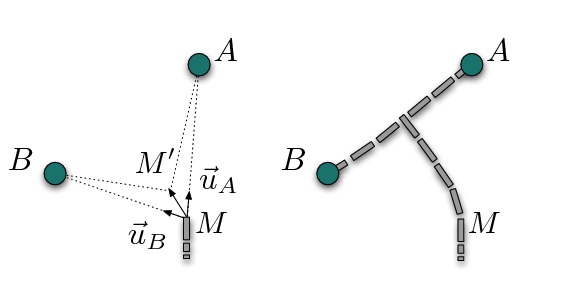}
   \caption{M is the closest network's point to both centers $A$ and $B$. The
     road will grow to point M' in order to maximally reduce the cumulative 
     distance $\Delta$ of A and B from the network. From \cite{Barthelemy:2008}.}
   \label{fig:delta}
\end{figure}

This local optimization process proposed in \cite{Barthelemy:2008} can
be understood with the simple following example
(Fig.~\ref{fig:delta}). We assume that at a given stage of the
evolution, two nodes $A$ and $B$ still need to be connected to the
network.
At any time step, each new node can trigger the construction of a single
new portion of road of fixed (small) length. In order to maximally reduce 
their distance to the network, both $A$ and $B$ would select the closest points 
$M_1$ and $M_2$ in the network as initial points of the new portions of roads to be built. 
If $M_1$ and $M_2$ are distinct, segments of roads are added
along the straight lines $M_1A$ and $M_2B$. If $M_1=M_2=M$, it is not
economically reasonable to build two different segments of roads and
in this case only one single portion $MM'$ of road is allowed. The
main assumption here is that the best choice is to build it in order to
maximize the reduction of the cumulative distance $\Delta$ from $M$ to $A$ and $B$
\begin{equation}
\Delta=[d(M,A)+d(M,B)]-[d(M',A)+d(M',B)]
\end{equation}
The maximization of $\Delta$ is done under the constraint
$|MM'|= \mathrm{const.}\ll 1$ and a simple calculation leads to
\begin{equation}
\overrightarrow{MM'}\propto \vec{u}_A+\vec{u}_B
\label{vec_rule}
\end{equation}
where $\vec{u}_A$ and $\vec{u}_B$ are the unit vectors from M in the
direction of A and B respectively. The rule (\ref{vec_rule}) can
easily be extended to the situation where more than two centers want
to connect to the same point $M$. Already in this simple setting
non-trivial geometrical features appear. In the example of
Fig.~\ref{fig:delta} the road from $M$ will develop a bended shape
until it reaches the line $AB$ and intersects it perpendicularly as it
is commonly observed in most urban settings. At the intersection
point, a singularity occurs with $\vec{u}_A+\vec{u}_B \approx 0$ and
one is then forced to grow two independent roads from the intersection
to A and B. 

Interestingly, we note that although the minimum expenditure principle was not
used, the rule Eq.~(\ref{vec_rule}) was also proposed by Runions {\it et
  al.}~\cite{Rolland:2005} in a study about leaf venation
patterns. 

The growth scheme described so far leads to tree-like
structures which are on one side economical, but which are hardly
efficient. For example the path length along a minimum spanning tree
scales as a power $5/4$ of the Euclidean distance between the
end-points~\cite{Duplantier:1989,Coniglio:1989} and better
accessibility is granted if loops are present. The authors of
\cite{Barthelemy:2008,Barthelemy:2009} followed \cite{Rolland:2005}
and assumed that a new node can trigger the construction of more than
one portion of road per time step, leading to the existence of loops.
\begin{figure}[h!]
\begin{tabular}{c}
\epsfig{file=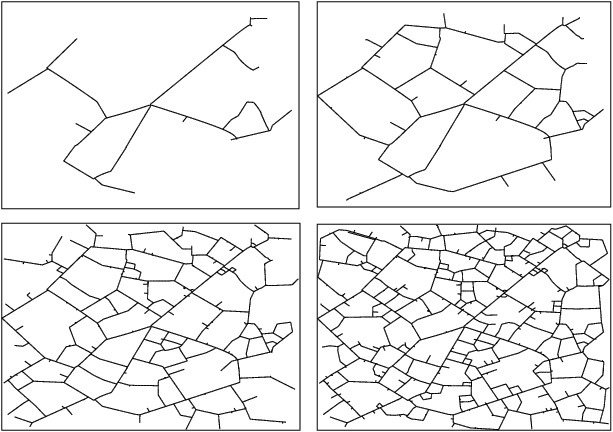,width=0.8\linewidth,clip=}
\end{tabular}
  \caption{Snapshots of the network at different times of its
    evolution: for (a) $t=100$, (b) $t=500$, (c) $t=2000$, (d)
    $t=4000$ (the growth rate is here $\eta=0.1$). At short times, we
    have almost a tree structure and loops appear for larger density
    values obtained at larger times (the number of loops then
    increases linearly with time). From \cite{Barthelemy:2008}.}
\label{fig:time}
\end{figure}
This model produces realistic results (some examples are shown in
Fig.~\ref{fig:time}), which are in good agreement with empirical data
which demonstrates that even in the absence of a well defined
blueprint, non-trivial global properties emerge. This agreement
confirms that the simple local optimization is a good candidate for
the main process driving the evolution of city street patterns but it
also shows that the spatial distribution of nodes $\rho(r)$ is
crucial. Concerning this last aspect, a more general model describing
the co-evolution of the node distribution $\rho(r)$ and the network
was proposed in \cite{Barthelemy:2009}.

\begin{figure}[h!]
\begin{tabular}{c}
\epsfig{file=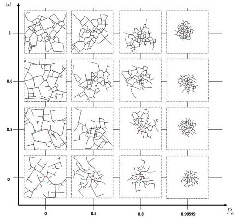,width=1.0\linewidth,clip=}
\end{tabular}
  \caption{Simulations of the model proposed by Courtat, Gloaguen,
    Douady \cite{Courtat:2010}. The different networks are obtained
    for various values of the parameters $P_e$ and $\omega$. At small
    values of $P_e$, the city is organized while for large $P_e$, the
    nodes are added at random. For small $\omega$, a minimum number of
    roads are constructed to connect new nodes while in the opposite
    cases all possible connections are added. From
    \cite{Courtat:2010}.}
\label{fig:courtat}
\end{figure}

Another interesting model was proposed recently by Courtat, Gloaguen and
Douady \cite{Courtat:2010}. In this model, each new `settlement' is
added every time step at a certain location and connects to the
existing infrastructure network in a way similar to
\cite{Barthelemy:2008}. The city generates a spatial field describing
the `attractiveness' of every point. This potential field has a hard
repulsion term at short distances and a large distance behavior
proportional to the total `mass' of the city (measured by the total
length of its roads) and decreasing as a power law with distance. The
sum of all influences of all roads then produce local minima and each
new node has its own policy. Among others, the parameter $\omega$ is
related to the `wealth' of the node and controls the number of roads
constructed to connect a new settlements: if $\omega=1$ all possible
roads are constructed and in the opposite case $\omega=0$, only the
shortest road is built. Another important parameter in this model is
the probability to be at the optimal location controlled by the
potential. For small $P_e$ the city is `unorganized' and nodes are
added randomly, while for large $P_e$ the probability to stick to the
optimal location is large. 
In Fig.~\ref{fig:courtat}, we show the 
We show in Fig.~\ref{fig:courtat} different networks obtained by
varying the two parameters $P_e$ and $\omega$. The variety of networks
obtained by varying two parameters only is remarkable - going from a
`favella-like' organization (for large $P_e$) to a highly organized
area (for small $P_e$)- and suggests that this approach could provide
an interesting first step for the modelling of urban street networks.

Finally, we end this section by noting that there are many other
models for roads and urban structures and that we focused here on a
physicist-like approach with minimal models. In particular, there are
many studies on these problems done by geographers (with usually the help of
cellular automata) and we refer the interested reader to articles in
the handbook \cite{Handbook:2009} for more references.

\subsection{Optimal networks}
\label{sec4E}

Although one of the main pillars in complex systems studies is the
emergence of a collective behavior without any central planning, it
is usually a matter of time scale compared with the typical time
horizon of planners. On a short time scale it is
reasonable to assume that planning plays a role and that the system
under consideration evolves by using an optimization process. On a
larger time scale, most systems result from the addition of successive layers
and even if each of these layers is the result of an optimization
process, it is very likely that the long time result is not an
optimum.

Optimization is however of great importance in many practical
engineering problems and both the problem of optimal networks
\cite{Jungnickel:1999} and of optimal traffic on a network
\cite{Wardrop:1952,Ahuja:1993} have a long tradition in mathematics
and physics. It is well known, for example that the laws that describe
the flow of currents in a resistor network~\cite{Kirchoff:1847} can be
derived by minimizing the energy dissipated by the
network~\cite{snell}. On the other hand, optimal networks have been shown
to be relevant in the study of mammalians circulatory
system~\cite{Mahon:2001}, food webs~\cite{Garlas:2003}, general
transportation networks~\cite{Banavar:1999}, metabolic
rates~\cite{West:1997}, river networks~\cite{Colaiori:1996}, and gas
pipelines or train tracks~\cite{Gastner:2006}. All these studies share
the fact that the nodes of the network are embedded in a
$d$-dimensional euclidean space which implies that the degree is
almost always limited and the connections restricted to `neighbors'
only.

A second broad class of optimal networks where spatial constraints are
absent has been also recently investigated. It has been shown, for
example, that optimization of both the average shortest path and the
total length can lead to small-world networks~\cite{Mathias:2001}, and
more generally, degree correlations~\cite{Berg:2002} or scale-free
features~\cite{Valverde:2002} can emerge from an optimization
process. Cancho and Sole~\cite{Cancho:2001} showed that the
minimization of the average shortest path and the link density leads
to a variety of networks including exponential-like graphs and
scale-free networks. Guimera {\it et al.}~\cite{Guimera:2002b} studied
networks with minimal search cost and found two classes of networks:
star-like and homogeneous networks. Finally, Colizza et
al.~\cite{Colizza:2004} studied networks with the shortest route and
the smallest congestion and showed that this interplay could lead to a
variety of networks when the number of links per node is changed.

\begin{figure}[h!]
\begin{tabular}{c}
\epsfig{file=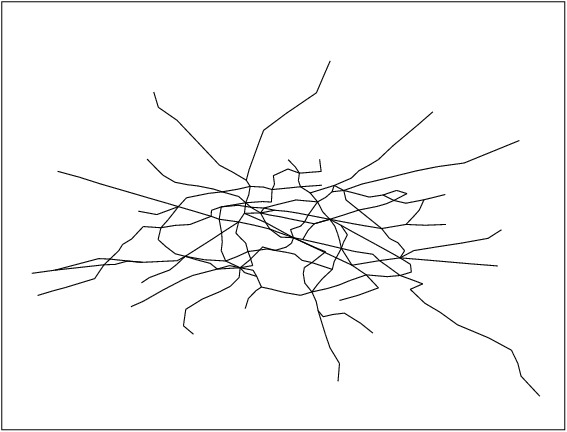,width=0.7\linewidth,clip=}\\
\epsfig{file=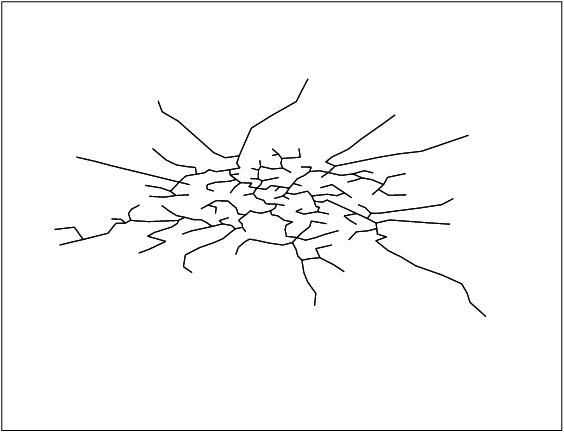,width=0.7\linewidth,clip=}
\end{tabular}
\caption{ Top: Paris subway network (2009). Bottom: Corresponding Minimum spanning tree.}
\label{fig:paris}
\end{figure}

Finally, we note that optimal networks could serve as interesting null
models. For example, the (Euclidean) minimum spanning tree (MST) is
obtained by connecting all the nodes with a minimum total length and
as an illustration we show in Fig.~\ref{fig:paris} the MST obtained
for the set of stations present in the Paris subway in $2009$.
In this subway case the total length is directly connected to the cost
of the network and the MST represents the most economical network. The
MST has however some drawbacks such as a large average shortest path
and a large vulnerability to failure, and the real subway network has
obviously many redundant links. It is however interesting to
understand the interplay between costs and efficiency by comparing the
actual network with the MST.

As we can see optimal networks appear in many different branches such
as mathematics, physics and also in engineering. This subject is in
fact so broad that it would deserve a whole review to explore its
various aspects. In this chapter, we thus made the choice to restrict
ourselves to the most recent and relevant statistical studies involving optimal
networks and space.

\subsubsection{Hub-and-spoke structure}
\label{sec4E1}

\begin{figure}[h!]
\begin{tabular}{c}
\epsfig{file=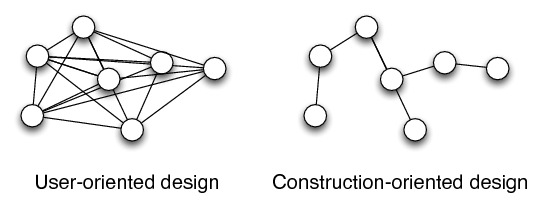,width=0.9\linewidth,clip=}
\end{tabular}
\caption{ Schematic comparison between a transportation network which
  optimizes the user travel time and distance (left) and network which
  minimizes the construction costs (right).}
\label{fig:useroperator}
\end{figure}

An important example of the result of an optimization process is the
hub-and-spoke structure. In this structure, direct connections are
replaced with fewer connections to hubs which form a network at a
larger scale. The hub-and-spoke structure reduces the network costs,
centralizes the handling and sorting, and allows carriers to take
advantage of scale economies through consolidation of flows. Such
networks have widespread application in transportation and became in
particular pregnant in airline transportation and has its cause mainly
in the fact that a carrier has to minimize its costs even if by doing
so the average traveling time for user is not minimized (see the
Fig.~\ref{fig:useroperator} for an illustration of this difference of designs).

\begin{figure}[h!]
\begin{tabular}{c}
\epsfig{file=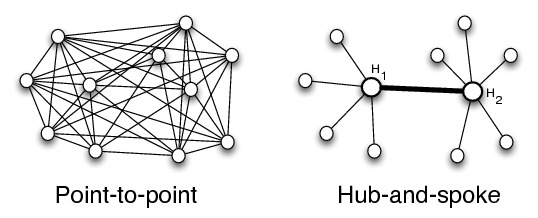,width=0.9\linewidth,clip=}
\end{tabular}
\caption{ Schematic comparison between a point-to-point transit model
  which optimizes the user travel time and distance (left) and a
  hub-and-spoke system (with two hubs $H_1$ and $H_2$) which minimizes the total
  operating costs by bundling together the traffic between the two
  hubs.}
\label{fig:deregu}
\end{figure}

One of the first cases where the hub-and-spoke system was observed is
in the US with Delta Air Lines which had its hub in Atlanta
(GA). FedEx also adopted this system in the $1970$s and after the
airline deregulation in $1978$ most of the airline carriers adopted
it. This system is based on the construction of regional hubs and to
the creation of major routes between these hubs (see Fig.~\ref{fig:deregu}).
One can easily compare the point-to-point (P2P) transit model and the
one hub-and-spoke model for $N$ nodes. In the hub-and-spoke model,
there are of the order $N$ routes to connect all nodes, while in the
P2P the number of routes it is of the order $N^2$. In transportation
systems such as airlines, this will lead to fuller planes, larger
benefits, and expensive operations such as sorting and accounting
packages will be done at the hub only.  Obviously, there are also
drawbacks. The centrality of the hubs is very large and overloading at
a hub can have unexpected consequences throughout the whole
network. In other words, the hub is a bottleneck in the network and in
agreement with general results on networks, hubs also constitute the
vulnerable points in the network, a failure at a hub being able to
trigger large delays which can propagate in the whole network.

An important problem in transportation research is to locate the
hub(s) in a system of $N$ nodes such that a given cost function is
minimum. Analytical research on this {\it hub location problem} began
relatively recently with the study \cite{OKelly:1987} and the
interested reader can find a short review on this subject in
\cite{OKelly:1999}. In the following, we will detail one of the
simplest case, known as the {\it linearized multiple assignement model}
and which can be formulated as follows. We assume that in a system of
$N$ nodes, there is a known origin-destination matrix $T_{ij}$
(ie. the flow between $i$ and $j$, see section \ref{sec3C1}). We
denote by $C_{ij}$ the transportation cost per individual to go from
$i$ to $j$ and the importance of hubs is in the fact that there is a
{\it discount factor} $0\leq \alpha\leq 1$ on the links among
them. For example in the airline case, bundling flows allows to use
for example larger aircrafts and reduce the passenger-mile costs.
This discount factor implies that if $k$ and $m$ are two hubs, then the
interhub cost on $k-m$ is given by $\alpha C_{km}$.  If we denote by
$X_{ij,km}$ the fraction of the flow from $i$ to $j$ which is routed
via the hubs $k$ and $m$, the function to be minimized is then
\begin{equation}
{\cal E}=\sum_{ijkm}T_{ij}[C_{ik}+\alpha C_{km}+C_{mj}]X_{ij,km}
\end{equation}
under the constraints
\begin{align}
&\sum_kZ_k=p\\
&\sum_{km}X_{ij,km}=1\\
&\sum_mX_{ij,km}-Z_k\leq 0\\
&\sum_kX_{ij,km}-Z_m\leq 0
\end{align}
where $Z_k=1$ if $k$ is a hub and zero otherwise. We look for a
solution of this problem with a fixed number $p$ of hubs. The two last
constraints ensure that the flow will not be routed via $k$ and $m$
unless both $k$ and $m$ are hubs. This type of minimization will lead
to solutions schematically shown in Fig.~\ref{fig:deregu} where the
reduction of the number of links is made possible by the establishment
of hubs and the bundling of flows between the hubs. For these optimal
networks, the total network cost is minimum but obviously individual
travel times are larger (as compared with a point-to-point network).

We just gave the flavor here of this type of approaches and there is a
huge amount of studies on this problem that we cannot cite here
because of lack of space. Basically, there is a more practical
research direction which amounts to add more characteristics of
real-world networks \cite{OKelly:1999} such as transfer delays for
example \cite{OKelly:2010}. There is also a more mathematical point of
view (see for example \cite{Aldous:2008a}) where optimal properties
are discussed. At this point however, many problems are still unsolved
and a statistical approach on the large network properties of this
hub-and-spoke minimization problem is probably an interesting
direction for future research.


\subsubsection{Congestion and centralized organization}
\label{sec4E2}

\begin{figure}[h!]
\begin{tabular}{c}
\epsfig{file=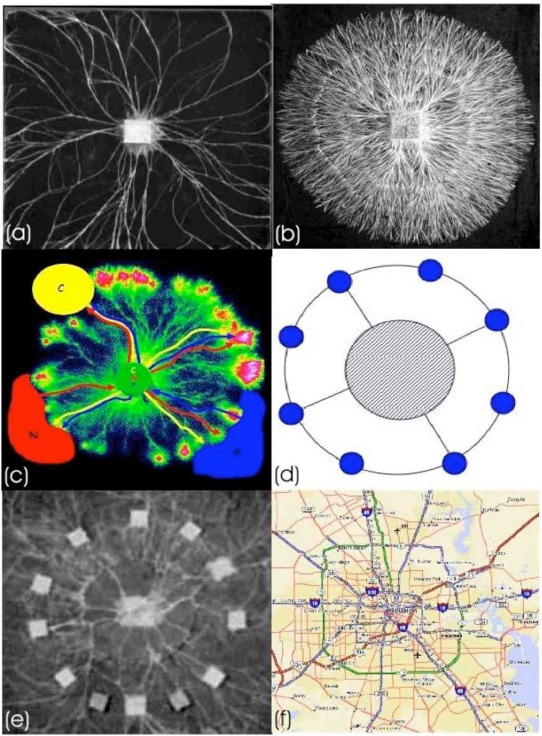,width=0.8\linewidth,clip=}
\end{tabular}
\caption{ Examples of hub-and-spoke structures with rings. (a-c,e):
  Typical fungi networks, in (c) a schematic representation of the
  nutrient flow is shown. (d) The model studied in
  \cite{Dorogovtsev:2000a,Ashton:2004,Jarrett:2006} with spokes radiating
  from a hub. (f) Road network in Houston showing an inner hub with a
  complicated structure. From \cite{Jarrett:2006}. }
\label{fig:ashton}
\end{figure}

\begin{figure}[h!]
\begin{tabular}{c}
\epsfig{file=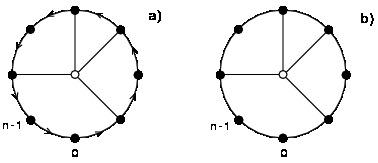,width=0.8\linewidth,clip=}
\end{tabular}
\caption{ Models proposed in \cite{Dorogovtsev:2000a} and studied in
  \cite{Ashton:2004,Jarrett:2006} with congestion. A central site is
  connected to a site on a ring with probability $p$. In (a) all the
  links on the ring are directed and in (b) these links are not
  directed. From \cite{Dorogovtsev:2000a}. }
\label{fig:ashton2}
\end{figure}

In many real-world cases however the pure hub-and-spoke structure is
not present and we observe a ring structure around a complicated core
or an effective hub (see for example Fig.~\ref{fig:ashton}). An interesting discussion on
centralization versus decentralization from the perspectives of the
minimum average shortest path and of the effect of congestion can be
found in \cite{Ashton:2004,Jarrett:2006}.
The idea is then to study the competition between the centralized
organization with paths going through a single central hub and
decentralized paths going along a ring and avoiding the central hub in
the presence of congestion.  A simple model of hub-and-spoke structure
together with a ring was proposed in \cite{Dorogovtsev:2000a}. In this
model, $N$ nodes are on a circle and there is hub located at the
center of the circle (see Fig.~\ref{fig:ashton2}).
A radial link - a spoke - is present with probability $p$. When
computing the shortest path, these spokes are assumed to count for
$1/2$. In other words, the number of hops to go from one site on a
ring to another via the central hub (when it is possible) is then
equal to $1$. In the case of directed links, the shortest path distribution is
easy to compute and one finds \cite{Dorogovtsev:2000a}
\begin{equation}
P(\ell)=\frac{1}{N-1}\left[1+(\ell-1)p+\ell(N-1-\ell
  p^2)p^2\right](1-p)^{\ell-1}
\end{equation}
In particular, we can check that the two limiting cases are correct:
for $p\to 0$, we obtain $P(\ell)\to 1/N-1$ giving
$\langle\ell\rangle=N/2$ and for $p\to 1$, $P(\ell)\to \delta_{\ell,1}$
and $\langle\ell\rangle=1$. The general expression for the average
shortest path is
\begin{equation}
 \langle\ell\rangle=\frac{1}{N-1}\left[\frac{2-p}{p}N-\frac{3}{p^2}+\frac{2}{p}+\frac{(1-p)^N}{p}(N-2+\frac{3}{p})\right]
\end{equation}
In the undirected case, we have more paths going from one site to the
other and the enumeration is a little bit more tedious. The result for
the shortest path distribution now reads \cite{Dorogovtsev:2000a}
\begin{equation}
\begin{split}
P(\ell=1)&=\frac{2}{N-1}\left(1+\frac{N-3}{2}p^2\right)\\
P(\ell\geq
2)&=\frac{1}{N-1}[a_0+a_1p+a_2p^2+a_3p^3\\
&+a_4p^4](1-p)^{2\ell-4}
\end{split}
\end{equation}
where
\begin{equation}
\begin{split}
a_0&=2\\
a_1&=4(\ell-2)\\
a_2&=2(\ell-1)(2N-4\ell-3)\\
a_3&=-2(2\ell-1)(N-2\ell-1)\\
a_4&=\ell (N-2\ell-1)\\
\end{split}
\end{equation}
For both these models a continuous limit can be defined by taking the
limit $N\to\infty$ and $p\to 0$ with $\rho=pN$ and $z\equiv\ell/N$
fixed. The shortest path distribution then converges to (in the
undirected model)
\begin{equation}
NP(\ell)\to 2[1+2\rho z+2\rho^2(1-2z)]e^{-2\rho z}
\end{equation}
The interesting observation made in \cite{Ashton:2004} is that if we
now add a cost $c$ each time a path goes through the central hub, we
could expect some sort of transition between a decentralized regime
where it is less costly to stay on the peripheral ring to a
centralized regime where the cost is not enough to divert the paths
from the central hub. The cost could in general depend on how busy the
center is and could therefore grow with the number of connections to
the hub. In the case of a constant cost $c$, the expression for the
average shortest path is now \cite{Ashton:2004}
\begin{equation}
\begin{split}
P(\ell)=
\begin{cases}
\frac{1}{N-1}\;\;\; &{\rm for}\;\; \ell\leq c\\
\frac{1}{N-1}[1+b_1p+b_2p^2](1-p)^{\ell-c-1}\;\;\;
&{\rm for}\;\; \ell >c
\end{cases}
\end{split}
\end{equation}
where $b_1=\ell-c-1$ and $b_2=(N-1-\ell)(\ell-c)$. For paths of length
$\ell\leq c$, there is no point to go through the central hub. In the
opposite case, when $\ell >c$, we recover a distribution similar to
the $c=0$ case in \cite{Dorogovtsev:2000a}. The average shortest path
is then
\begin{equation}
\begin{split}
\langle\ell\rangle&=\frac{(1-p)^{N-c}[3+(N-2-c)p]}{p^2(N-1)}\\
&+\frac{p[2-2c+2N-(c-1)(c-N)p]-3}{p^2(N-1)}+\frac{c(c-1)}{2(N-1)}
\end{split}
\end{equation}
In the continuous limit ($p\to 0$, $N\to\infty$ and $z\equiv\ell/N$
and $\rho=pN$ fixed), the average shortest path is a function of the
various parameters $\langle\ell\rangle=\langle\ell\rangle(\rho,c,N)$. In the case of
costs increasing linearly with $\rho$, the average shortest path
displays a minimum when $\rho$ is varied ($N$ and $c$ being
fixed). Indeed for $\rho\to 0$, there are no spokes and
$\langle\ell\rangle$ scales as $N$. In the opposite case $\rho$ large
the cost is also large and it is less costly to go along the ring. In
\cite{Ashton:2004}, the authors used a simple approximation and found
that (with $c\equiv k\rho$), the optimal value of $\rho$ is
\begin{equation}
\rho^*\approx \sqrt{\frac{N}{k}}
\end{equation}
a result that is confirmed numerically. This result can actually be
rewritten as
\begin{equation}
pc(\rho)\sim 1
\end{equation}
which means that the optimal situation is obtained when the average
cost of a radial trip through the central hub is of order one: when
$c$ is too large, this trip is too costly and when $p$ is too small,
the existence of this path is too unlikely. The same argument applied
to nonlinear cost $c\sim k\rho^2$ gives the scaling
\begin{equation}
\rho^*\sim(N/k)^{1/3}
\end{equation}
For this optimal value, the minimum average shortest path is then of
the order the cost
\begin{equation}
\langle\ell\rangle_{min}\sim c(\rho^*)
\end{equation}
In the linear case $c=k\rho$, one obtains
\begin{equation}
\langle\ell\rangle_{min}\sim \sqrt{kN}
\end{equation}
and in the nonlinear case $c\sim k\rho^2$, one obtains
\begin{equation}
\langle\ell\rangle_{min}\sim (kN^2)^{1/3}
\end{equation}
These expressions and arguments apply essentially both to the directed and
non-directed model.

This study \cite{Ashton:2004} was generalized in \cite{Jarrett:2006}
to the case of a more complicated cost function such as
$c(\rho)=C\rho+B\rho^2+A\rho^3$ where the authors observe different
behaviors and a phase transition according to the values of the
coefficients $A$, $B$, and $C$.

These studies show how congestion could have an important impact on
the a priori optimal hub-and-spoke structure and favor the
transport along a ring. From a more general perspective it would
indeed be interesting to observe empirically this transition from a purely radial
structure and the appearance of a ring for a congestion large enough,
as it is observed in many urban structures for example.

\subsubsection{From the MST to the SPT}

There is a obviously a huge literature on optimal networks and we are
obliged to restrict ourselves to a small subset of subjects. We will
discuss here the prototypes of optimal networks allowing us to
introduce some of the main objects such as the minimum spanning tree
or the shortest path tree. Concerning the topic of connected networks
over random points which minimize quantities such as the total
route-length, the interested reader can find a recent and modern
account about mathematically tractable models in a series of papers by
Aldous \cite{Aldous:2008a,Aldous:2008,Aldous:2010} and in the
references therein.

\begin{figure}
\centerline{
\epsfysize=0.60\columnwidth{\epsfbox{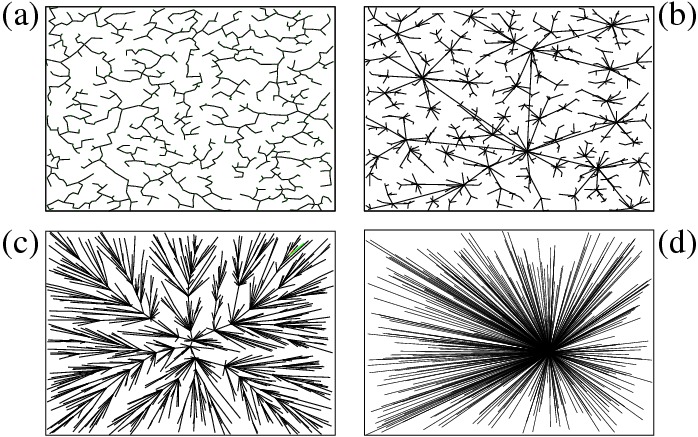}}
}
\caption{ Different spanning trees obtained for different values of
$(\mu,\nu)$ in Eq.~(\ref{eq:munu}) obtained for the same set of $N=1000$
nodes. (a) Minimum spanning tree obtained for $(\mu,\nu)=(0,1)$. In
this case the total distance is minimized. (b) Optimal traffic tree
obtained for $(\mu,\nu)=(1/2,1/2)$. In this case we have an interplay
between centralization and minimum distance resulting in local
hubs. (c) Minimum euclidean distance tree obtained for
$(\mu,\nu)=(1,1)$. In this case centrality dominates over distance and a
`star' structure emerges with a few dominant hubs. (d) Optimal
betweenneess centrality tree obtained for $(\mu,\nu)=(1,0)$. In this
case we obtain the shortest path tree which has one star hub (for the
sake of clarity, we omitted some links in this last figure). From \cite{Barthelemy:2006}.}
\label{fig:examples}
\end{figure}

We will not introduce the different networks one by one but we will
adopt a different point of view by introducing a generalized energy inspired
by studies on optimal traffic networks \cite{Barthelemy:2006}.  The
problem is, given a set of $N$ nodes located in a $2$-dimensional
plane, to find the tree connecting them and which minimizes the
following quantity
\begin{equation}
{\cal E}_{\mu\nu}=\sum_{e\in{\cal T}} b_{e}^{\mu}d_{e}^{\nu}
\label{eq:munu}
\end{equation}
where $d_{e}$ is the Euclidean length of a link (or it could be another measure of
its cost) and $b_{e}$ is the betweenness centrality of the link (see
section \ref{sec2B1d}). The exponents $\mu$ and $\nu$ control the
relative importance of distance against topology as measured by
centrality. Fig.~\ref{fig:examples} shows examples of spanning trees
obtained for different values of $(\mu,\nu)$.

\medskip
{\it Case $(\mu,\nu)=(0,1)$.} 
\medskip
The energy is then
\begin{equation}
{\cal E}_{0,1}=\sum_{e\in{\cal T}} d_{e}
\end{equation}
and represents the total length of the network. In real-world cases,
the cost of the network is connected to this quantity and it makes
sense to try to minimize it. Since the nodes are in space, the optimal
resulting tree is called Euclidean MST. An example of the
two-dimensional MST is shown in the Fig.~\ref{fig:examples}(a) and
many quantities can be calculated in the limit of a large number of
nodes. For example, for the $2d$ case, the length $M(N)$ of the
longest edge scales as \cite{Penrose:1997}
\begin{equation}
M(N)\sim \sqrt{\frac{\ln N}{N}}
\end{equation}
which is larger by a non-trivial factor $\sqrt{\ln N}$ than the typical internode distance which
scales as $1/\sqrt{N}$. Also, the total length $\ell_T(N)$ of the
$d$-dimensional MST scales as
\cite{Steele:1988}
\begin{equation}
\ell_T(N)\sim N^{1-1/d}\int_{\mathbb{R}^d}\rho(x)^{\frac{d-1}{d}}dx
\end{equation}
where $\rho(x)$ denotes the point distribution in $\mathbb{R}^d$. For
a uniform distribution in $d=2$, we thus obtain a scaling of the form
$\ell_T(N)\sim \sqrt{N}$. This scaling can be understood with the
following simple argument. In the case of a uniform density $\rho$,
the typical inter-node distance is given by $1/\rho^{1/d}$ which
scales as $1/N^{1/d}$ and we have $E\propto N$ links which reproduces
the scaling $\ell_T(N)\sim N\times N^{-1/d}$.

We note here that an interesting study \cite{Wu:2006} shows that the
minimum spanning tree for weighted networks can be partitioned into
two distinct components: the `superhighways' with large betweenness
centrality nodes and the `roads' with lower centrality. It would be
interesting to study the consequences of these results for the
euclidean minimum spanning tree.

There are many other studies on the MST and it is impossible to quote
them all here. We just end this paragraph by noting that the MST has
many interesting connections with other problems of statistical
physics such as invasion percolation, directed polymers in random
media, random resistor networks and spin glasses (see for example
\cite{Dobrin:2001,Read:2005,Jackson:2010}).

\begin{figure}
\centerline{ \epsfysize=0.6\columnwidth{\epsfbox{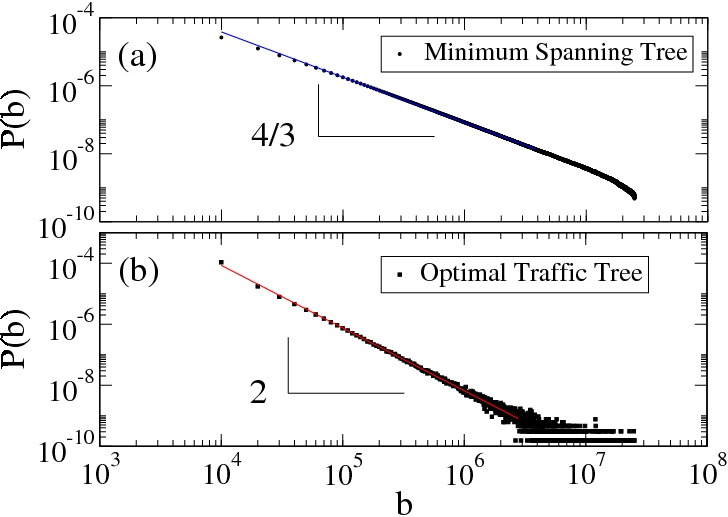}} }
\caption{ Betweenness centrality distribution for the MST and for the
OTT. The lines are power law fits and give for the MST the
theoretical result $\delta=4/3$ and for the OTT the value
$\delta\simeq 2.0$ ($N=10^4$, $100$ configurations). From \cite{Barthelemy:2006}.}
\label{fig:bc}
\end{figure}

\medskip
{\it Case $(\mu,\nu)=(1/2,1/2)$.}
 \medskip
We obtain here the optimal traffic tree (OTT) shown in
Fig.~\ref{fig:examples}(b) and which displays an interesting interplay
between distance and shortest path minimization
\cite{Barthelemy:2006}. It has been shown that trees can be classified
in `universality classes'~\cite{Takayasu:1991,Manna:1992} according to
the size distribution of the two parts in which a tree can be divided
by removing a link (or the sub-basins areas distribution in the
language of river network). We define $A_i$ and $A_j$ as the sizes of
the two parts in which a generic tree is divided by removing the link
$(i,j)$. The betweenness $b_{ij}$ of link $(i,j)$ can be written as
$b_{ij}=\frac{1}{2} [ A_i (N-A_i) + A_j (N-A_j) ]$, and the
distributions of $A's$ and $b's$ can be easily derived one from the
other. It is therefore not surprising that the same exponent $\delta$
characterizes both $P(A)\sim A^{-\delta}$ and $P(b)$. While we obtain
the value $\delta=4/3$ for the MST~\cite{Takayasu:1991}, for the OTT
we obtain (Fig.~\ref{fig:bc}) an exponent $\delta\simeq 2$, a value
also obtained for trees grown with preferential attachment
mechanism~\cite{Albert:2002} (see also~\cite{deLosRios:2001} for a
supporting argument). Interestingly, most real-world networks are also
described by this value $\delta\simeq 2$~\cite{Goh:2002}. The OTT thus
tends to have a more uniform centrality with respect to the
MST~\cite{Wu:2006}, with important consequences on the vulnerability
of the network since there is no clearly designated `Achille's heel'
for the OTT.

\begin{figure}
\centerline{
\includegraphics*[angle=+90,width=0.40\textwidth]{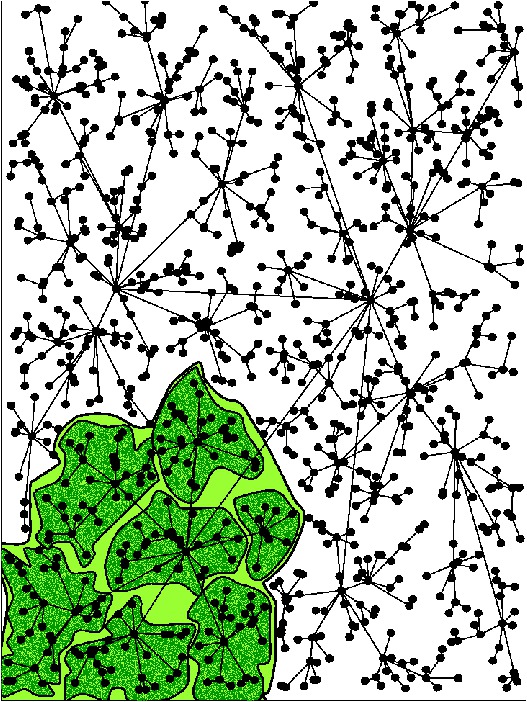}
}
\caption{Hierarchical organization emerging for the optimal traffic
tree $(\mu,\nu)=(1/2,1/2)$ ($N=10^3$ nodes). Longer links lead to
regional hubs which in turn connect to smaller hubs distributing
traffic in smaller regions. From \cite{Barthelemy:2006}.}
\label{fig:regions}
\end{figure}

The traffic properties of the OTT are also interesting as the traffic
scales as $T_{ij}\sim d_{ij}^{\tau}$ with $\tau\approx 1.5$ showing
that large traffic is carried over large distance and is then
dispatched on smaller hubs that distribute it on still smaller
regions. Despite the limited range of degrees, we also observe for the
strength~\cite{Barrat:2004b} $s_i=\sum_j T_{ij}$ a superlinear
behavior with the degree. This result demonstrates that the existence
of degree-traffic correlations as observed for the airport network
\cite{Barrat:2004b} could emerge from a global optimization
process. The spatial properties of the OTT are also remarkable and
displays (Fig.~\ref{fig:regions}) a hierarchical spatial organization
where long links connect regional hubs, that, in turn are connected to
sub-regional hubs, etc.

\medskip
{\it Case $(\mu,\nu)=(1,1)$.}
\medskip
The energy is then given by $\sum_eb_ed_e$ which is proportional to
the average shortest path with weights given by the euclidean
distance. If we think of the betweenness centrality as a proxy for the traffic, the energy
also represents the total length traveled on the system. An example
of such a tree is shown in Fig.~\ref{fig:examples}(c).

\medskip
{\it Case $(\mu,\nu)=(1,0)$.}
\medskip
In this case, the energy Eq.~(\ref{eq:munu}) is proportional to the
average betweenness centrality which is also the average shortest path
$\sum_e b_e\propto \ell$.  The tree $(1,0)$ shown in
Fig.~\ref{fig:examples}d is thus the shortest path tree (SPT) with
an arbitrary `star-like' hub (a small non zero value of $\nu$ would
select as the star the closest node to the gravity center). The total
length here scales as
\begin{equation}
\ell_T(N)\sim N
\end{equation}
with a prefactor proportional to the linear size of the system. Also,
we note that if we were not restricted to trees, the solution would of
course be the complete graph which gives an average shortest minimal
and equal to one.

The minimization of Eq.~(\ref{eq:munu}) thus provides a natural
interpolation between the MST and the SPT, a problem which was
addressed in previous studies~\cite{Khuller:1995}. The degree
distribution for all cases considered above (with the possible
exception of $(\mu,\nu)=(1,1)$ -- a complete inspection of the plane
$(\mu,\nu)$ is still lacking) is not broad, possibly as a consequence
of spatial constraints.

\paragraph{The Steiner problem.}
\label{sec4E3b}

\begin{figure}[h!]
\begin{tabular}{c}
\epsfig{file=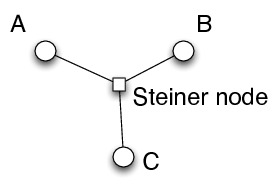,width=0.6\linewidth,clip=}
\end{tabular}
\caption{ Example of the Steiner tree for three points $A$, $B$, and
  $C$. The Steiner point S is located at the Fermat point of the
  triangle $ABC$ and minimizes the total distance
  $d(S,A)+d(S,B)+d(S,C)$. The angles are equal
  $(SB,SA)=(SA,SC)=(SC,SA)=2\pi/3$. }
\label{fig:steiner}
\end{figure}

More generally, finding a subgraph that optimizes a global cost
function is a very important problem. The Steiner tree is one of the
classical (NP-complete) problem which is still studied now
\cite{Bayati:2010} because of its importance in many fields such as
network reconstruction in biology, Internet multicasting, circuit and
distribution network design. It is defined as following: given a graph
with positive weights, it consists in finding a connected subgraph of
minimum weight that contains a selected set of `terminal'
vertices. Such a subgraph may require the addition of some
`non-terminal' nodes which are called Steiner nodes (see
Fig.~\ref{fig:steiner} for an illustration of the three nodes
problem). The difference between the Steiner tree problem and the
minimum spanning tree problem is that in the Steiner tree problem
extra intermediate vertices and edges may be added to the graph in
order to reduce the length of the spanning tree.

In \cite{Bayati:2010}, the authors show that this problem can be
analyzed with tools -- such as the cavity equation -- coming from
statistical physics of disordered, frustrated systems (such as spin
glasses, see for example \cite{Mezard}) and can lead to new optimization
algorithms.

The Steiner problem is obviously important in transportation science
too, where the nodes are considered as cities and the Steiner points
as junctions. Recent results on this topic can be found in
\cite{Aldous:2008} where the author proposes a rigorous study of the
network which minimizes the quantity
\begin{equation}
{\cal E}=\sum_ed_ef(e)^\beta
\end{equation}
where $\beta$ is a positive exponent, $d_e$ is the length of the
link $e$, and where $f(e)$ is a function describing the flow on edge $e$.
In this case, the minimum cost can be shown to scale as
\begin{equation}
\ell_T(N)\sim N^{\alpha(\beta)}
\end{equation}
with
\begin{align}
\alpha(\beta)=
\begin{cases}
1-\beta/2\;\; &{\rm for}\;\;\beta\in ]0,1/2]\\
(1+\beta)/2\;\;&{\rm for}\;\;\beta\in [1/2,1]
\end{cases}
\end{align}
There is a transition at $\beta=1/2$ which corresponds to a change of regime where
short links dominate to another one where long links dominate \cite{Aldous:2008}.

\subsubsection{Adding two antagonistic quantities}
\label{sec4E4}

\begin{figure}[h!]
\begin{tabular}{c}
\epsfig{file=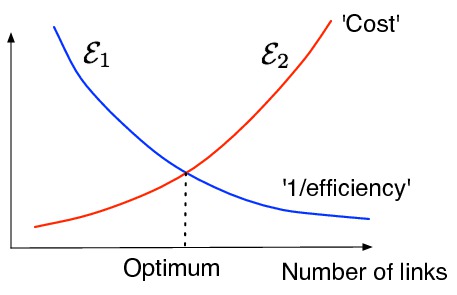,width=0.8\linewidth,clip=}
\end{tabular}
\caption{ Schematic representation of many optimal network models. The
  functional to minimize is the sum of two terms which behave in
  opposite way when adding links. This leads in general to a non-trivial
  optimum.}
\label{fig:sketch}
\end{figure}

After having briefly reviewed the important optimal networks such as
the MST and the SPT, we now discuss some examples which were
recently proposed in the literature. In particular, many models of
optimal networks minimize a functional ${\cal E}$ which is the sum of
two terms which have have antagonistic behavior
\begin{equation}
{\cal E}=\lambda {\cal E}_1+(1-\lambda){\cal E}_2
\end{equation}
where $\lambda\in[0,1]$ and where ${\cal E}_{1}$ is for example a 
decreasing function of the number of links and could represent the
efficiency (such as the average shortest path for example) and where 
${\cal E}_{2}$ is an increasing function of the number of links for
example and could represent the cost as measure by the total length
for example (see Fig.~\ref{fig:sketch}). This interplay will lead in many case to
interesting solutions.

We will now discuss some simple examples that were considered in the literature.
Already in the $1970$s optimal network design was a very active
field (see for example \cite{Billheimer:1973}) and we will focus here
on the more recent results.

\paragraph{Link density and average shortest path.}
\label{sec4E4a}

Ferrer i Cancho and Sol\'e \cite{Cancho:2001} minimized the simplest form which is the
combination of the average (normalized by the maximum) shortest path
\begin{equation}
d=\langle\ell\rangle/\langle\ell\rangle_{max}
\end{equation}
and the average link density $\rho=2E/N(N-1)$ and reads
\begin{equation}
{\cal E}=\lambda d+(1-\lambda)\rho
\label{eq:cancho}
\end{equation}
where $\lambda\in[0,1]$. 
\begin{figure}[h!]
\begin{tabular}{c}
\epsfig{file=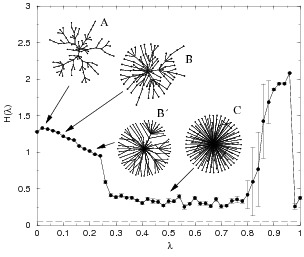,width=1.0\linewidth,clip=}
\end{tabular}
\caption{ Optimal networks obtained with the minimization of
  Eq. (\ref{eq:cancho}) for different values of $\lambda$. The
  $y-$axis represents the degree entropy $H=-\sum_kp_k\log p_k$ and is
  small for a heterogeneous network (and large for a lattice). From \cite{Cancho:2001}. }
\label{fig:cancho}
\end{figure}
For $\lambda=0$, the energy is the density and the optimal network is
a tree. In the opposite case $\lambda=1$, the optimal network
minimizes the average shortest path. For intermediate values of
$\lambda$, Cancho and Sol\'e obtained different networks (see
Fig.~\ref{fig:cancho}) and in particular argue that one of the regimes
is a scale-free network with a broad distribution of degrees. They
obtained for the network denoted by `B' in Fig.~\ref{fig:cancho} a
power law with exponent $-2$, for a network of size $N=100$ (obviously
better statistics on larger networks would be necessary to confirm
this result, but this is not an obvious task). This shows that
optimization might be a candidate for explaining the emergence of
scale-free features.

\paragraph{Total length and average shortest path.}
\label{sec4E4b}

Aldous \cite{Aldous:2008a} studied the network which
minimizes the total length subject to the average number of hops. More
precisely the cost associated to a route $\pi(i,j)=e_1e_2\dots e_m$
going from node $i$ to node $j$ is
\begin{equation}
C(\pi)=m\Delta +a\sum_{e\in \pi}d_e
\label{eq:aldous}
\end{equation}
where $\Delta$ and $a$ are two constants (in \cite{Aldous:2008a}, the
quantity $a$ is chosen to scale as $1/\sqrt{N}$ in order to obtain terms of the
same order of magnitude). The quantity $\Delta$ measures the cost of
connections and we expect that the larger it is and the smaller the number of hubs
(and the longer the spokes). The journey time associated to a
graph $G$ is then
proportional to 
\begin{equation}
T(G)=\sum_{i\neq j}\min_{\pi(i,j)}C[\pi(i,j)]
\end{equation}
(where $\pi(i,j)$ is the set of paths from $i$ to $j$). The total
length of the graph is $\ell_T(G)=\sum_{ij}A_{ij}d_E(i,j)$ and represents
the cost (if we assume that the cost is proportional to the length of
connection between two adjacent nodes). We thus can have an interplay
between the cost represented by the total length and the total journey
time leading to a variety of behaviors. If we constrain the routes to
have $m$ average hops, then the minimal total length scales as \cite{Aldous:2008a}
\begin{equation}
\ell_T(N)\sim N^{\beta(m)}
\end{equation}
where the following bounds are obtained: $\beta(2)=3/2$,
$\beta(3)=13/10$, $\beta(4)=7/6$. The exponent thus decreases when $m$
increases which is expected since a smaller average number of hops
implies a stronger constraint and thus a longer network. Conversely,
if we require the network length $\ell_T$ to scale as $N$, then the average
shortest path scales as $\ln\ln N$ \cite{Aldous:2008a}.

\begin{figure}[h!]
\begin{tabular}{c}
\epsfig{file=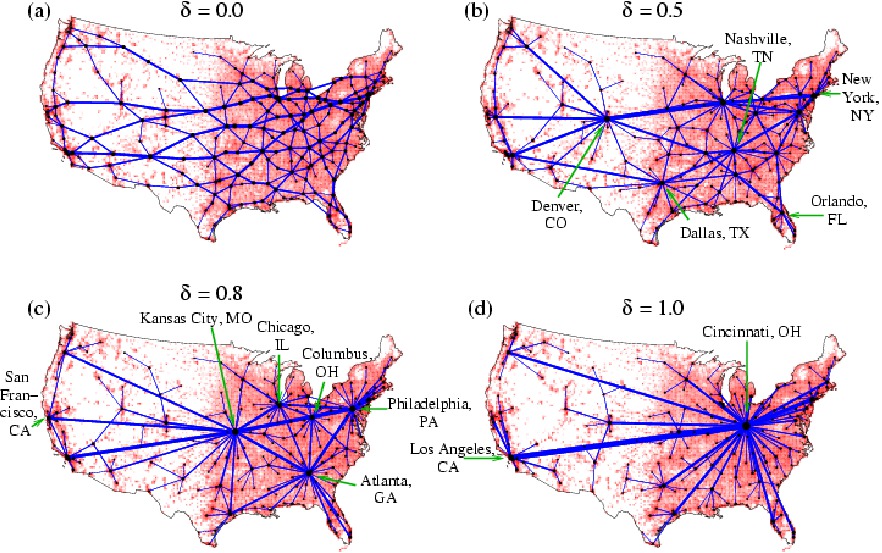,width=0.9\linewidth,clip=}
\end{tabular}
\caption{ Optimal networks obtained with the minimization of
  Eq. (\ref{eq:gastner}) for different values of $\delta$. From \cite{Gastner:2006}. }
\label{fig:gastner}
\end{figure}

In \cite{Gastner:2006} an effective length is assigned to each edge
\begin{equation}
\tilde{\ell}_{ij}=\delta+(1-\delta)d_E(i,j)
\end{equation}
and the effective distance on a route $\pi=e_1e_2\dots e_m$ between
two nodes is given by
\begin{equation}
m\delta +(1-\delta)\sum_{e\in\pi}d_e
\label{eq:gastner}
\end{equation} 
which is Aldous' expression Eq.~(\ref{eq:aldous}) with the
correspondence $\Delta \leftrightarrow\delta$ and $a\leftrightarrow
(1-\delta)$. Gastner and Newman \cite{Gastner:2006} then weight the
distance with a passenger number $w_{ij}$ and minimize a combination
of the form $\ell_T(G)+\gamma T(G)$. We note here that this model was
also used together with a gravity model for computing $w_{ij}$ in
\cite{Nandi:2009} where the authors compared the resulting optimal
network and the real-world Indian airline network and found a good
agreement. Gastner and Newman found then for example optimal networks
shown in Fig.~\ref{fig:gastner}.
In particular, this figure illustrates the expected behavior when the
connection cost increases (similar results were obtained with the same
model in \cite{Gastner:2006c}): for $\delta=0$, the connection cost is
zero, only distance counts and the resulting network is
essentially a minimal length network, such as a planar network of
roads for example. When $\delta$ increases, and thus the cost
associated to connections, we see fewer hubs and longer spokes (the
limit being one hub and the corresponding network is the SPT). As noted
by Aldous \cite{Aldous:2008a}, it would be interesting to obtain more
quantitative results in order to be able to compare with some
real-world data for example. 

\begin{figure}
\begin{tabular}{c}
\epsfig{file=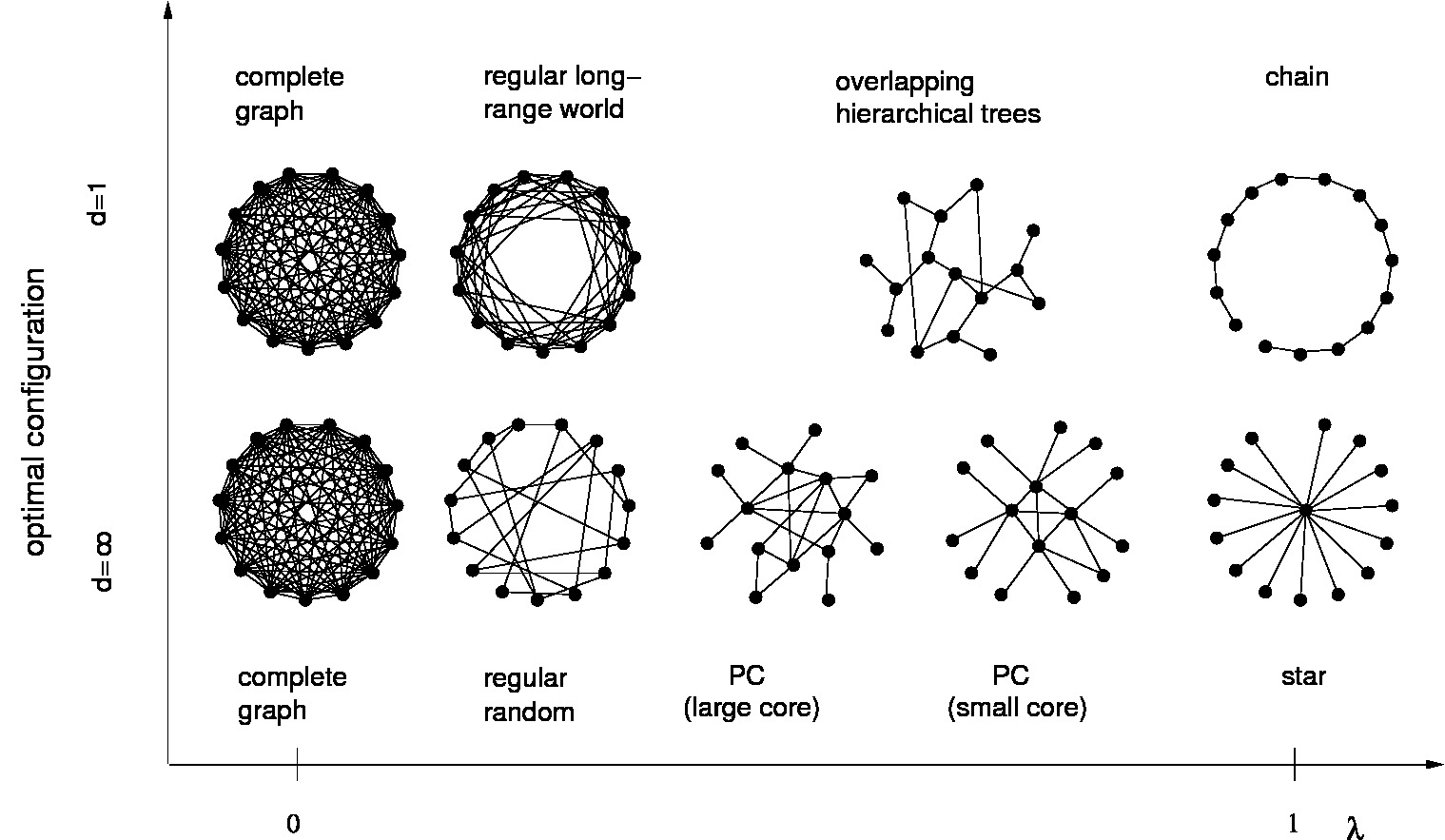,width=0.9\linewidth,clip=}
\end{tabular}
\caption{ Optimal networks obtained with the minimization of
  Eq. (\ref{eq:brede}) for different values of $\lambda$. From \cite{Brede:2010}. }
\label{fig:brede}
\end{figure}

In \cite{Brede:2010}, Brede also revisited recently this problem in
terms of a communication infrastructure with
\begin{equation}
{\cal E}=\lambda\ell_T(G)+(1-\lambda)\langle\ell\rangle
\label{eq:brede}
\end{equation}
In a communication context, the total length represents the cost
associated with the wires and the average shortest path is a measure
of the communication efficiency. The number of links is not limited
and the nodes are located on a regular lattice of dimension $d$ with
periodic boundary conditions. For $\lambda=0$, we obtain the
complete graph which minimizes $\langle \ell\rangle$ and for
$\lambda=1$ we obtain a minimum spanning tree (and depending on the
dimension we can obtain a chain or a star network). For intermediate
values of $\lambda$, we obtain different networks interpolating
between these two extremes (see Fig.~\ref{fig:brede}).

\paragraph{Total length and synchronization.}
\label{sec4E4c}

\begin{figure}[h!]
\begin{tabular}{c}
\epsfig{file=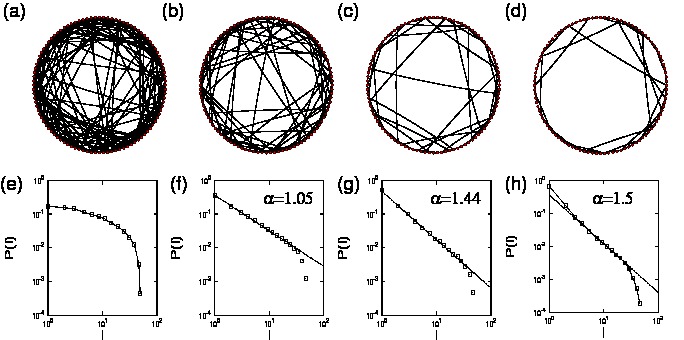,width=1.0\linewidth,clip=}
\end{tabular}
\caption{ (a-d) Examples of networks constructed for different values
  of $\lambda=0.05$, $0.3$, $0.8$, and $0.95$. (e-h) Corresponding
  link length distribution for these different values of
  $\lambda$. Straight lines in (f-h) represent power laws with
  exponent $1.05$, $1.44$, and $1.5$, respectively.  From \cite{Brede:2010b}. }
\label{fig:brede2}
\end{figure}

A wide range of systems can be described by a system of oscillators
coupled through a network of contacts and synchronization problems
have attracted a lot of attention (for a review on synchronization in
complex networks see \cite{Arenas:2008}). If each of the oscillator is
described by its individual dynamics by $\dot{\phi}_i=f(\phi_i)$, the
dynamics of the coupled oscillators is governed by the following
equation
\begin{equation}
\dot{\phi}_i=f(\phi_i)+\sigma\sum_jA_{ij}[h(\phi_j)-h(\phi_i)]
\end{equation}
where $h$ is a given output function. The stability of synchronized
solutions is related to the eigenvalues
$\lambda_0=0\leq\lambda_1\leq\dots\leq\lambda_N$ of the graph
Laplacian. More precisely, Pecora and Carroll \cite{Pecora:1998}
showed that the stability is related to the ratio -sometimes called
`synchronizability' defined as
\begin{equation}
e=\frac{\lambda_N}{\lambda_1}
\end{equation}
and that the smaller $e$, the more stable the synchronized
solutions. Donetti et al. \cite{Donetti:2005} proposed to construct
networks with a fixed number of nodes and average degree and which
optimize synchronizability (ie. minimizing $e$). The optimal networks
are not small-world networks and display peaked distributions of degree,
betweenness, average shortest paths, and of loops. An important
feature of these networks is that there have a very entangled
structure with short $\langle\ell\rangle$, large loops and no
well-defined structure. In a recent article \cite{Brede:2010b}, Brede
extended this problem and proposed to minimize both the total amount
of wire $\ell_T(G)$ to connect the network and the synchronizability
characterized by $e$
\begin{equation}
{\cal E}=\lambda\ell_T(G)+(1-\lambda)e
\end{equation}
Starting with a one-dimensional chain for $\lambda=0$, one obtains
the complete graph which minimizes the synchronizability. For
$\lambda=1$, the $1d$ line is optimal and for intermediate values of
$\lambda$ one obtains the networks shown in Fig.~\ref{fig:brede2}.
Most links are small and the link length $l$ distribution goes from a
peak distribution to a broader law. A power law fit of the form
$P(l)\sim l^{-\alpha}$ gives a value of the exponent in the range
$[1,1.5]$ (for a random graph with the same degree distribution and no
total length constraint, the length distribution is uniform and when
there is a total length constraint, the distribution is
exponential). This result means that small-world networks can display optimal
synchronization but with shortcuts covering many different length
scales.
 
In another paper \cite{Brede:2010c}, the same author studied the
trade-off between enhanced synchronization and total length of wire
but allowing the nodes to rearrange themselves in space with the
constraint that the average Euclidean distance among nodes is
constant. In this model, depending on the cost of wire, one observes
different organization in modules: when the cost of wire $\lambda$
increases one observes an organization in a large number of modules
connected on a ring.


\subsubsection{Beyond trees: noise and loops}
\label{sec4E5}

\begin{figure}[h!]
\begin{tabular}{c}
\epsfig{file=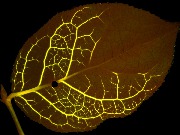,width=0.9\linewidth,clip=}
\end{tabular}
\caption{ Re-routing of the flow around an injury in a lemon
  leaf. From \cite{Katifori:2010}. }
\label{fig:lemon}
\end{figure}

In most examples studied in literature (including the ones presented
above), the optimal networks are trees. However in many natural
networks such as veins in leaves or insect wings, one observe many
loops. Very recently, two studies which appeared simultaneously
\cite{Corson:2010,Katifori:2010} proposed possible reasons for the
existence of a high density of loops in many real optimal networks. In
particular, it seems that the existence of fluctuations is crucial in
the formation of loops. Maybe more related to the botanic evolution of
leaves, the resilience to damage also naturally induces a high density of
loops (see Fig.~\ref{fig:lemon} for an example of flow re-routing after
an injury).

In these studies, the model is defined on a network with
conductances $C_{e}$ on each link and where the total power 
\begin{equation}
P=\frac{1}{2}\sum_k\sum_{j\in\Gamma(k)}C_{kj}(V_k-V_j)^2
\end{equation}
is minimized under the cost condition
\begin{equation}
\frac{1}{2}\sum_k\sum_{j\in\Gamma(k)}C_{jk}^{\gamma}=1
\end{equation}
where in this equation it is assumed that the cost of a conductance
$C_{kj}$ is given by $C_{kj}^{\gamma}$ where $\gamma$ is a real
number.

Following \cite{Katifori:2010}, we can introduce two variants of this
model. The first one which represents the resilience to damage is
defined as follows. We cut a link $e$ and compute for this system the
total dissipated power denoted by $P^{e}$. The resilience to this
damage can then be rephrased as the minimization of the 
functional
\begin{equation}
R=\sum_{e\in E} P^e
\end{equation}
Note that if breaking $e$ disconnects the network, there is therefore
one link with infinite resistance in the system and the dissipated
power $P^e$ is infinite. The finiteness of R implies the existence of
loops in the optimal network. In another model, \cite{Katifori:2010}
Katifori introduces time fluctuating load by introducing a system with
one source at the stem of the leaf and one single moving sink at
position $a$. For this system, one can compute the total dissipated
power $P^a$ and resilience to fluctuations can be rephrased as the
minimization of the functional
\begin{equation}
F=\sum_aP^a
\end{equation}

\begin{figure}[h!]
\begin{tabular}{c}
\epsfig{file=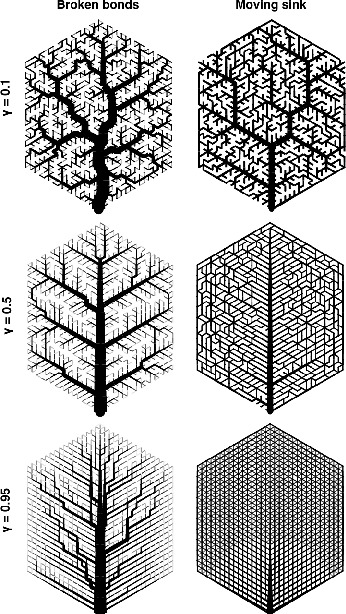,width=0.9\linewidth,clip=}
\end{tabular}
\caption{ Optimal networks for the resilience to damage (left) and to
  a fluctuating load (right panels). From \cite{Katifori:2010}. }
\label{fig:katifori}
\end{figure}

In these models, one observes the formation of loops (see
Fig.~\ref{fig:katifori}), reminiscent of the ones seen in real
leaves. These recent studies shed a new light in the formation and the
evolution of real-world networks and open interesting directions of
research. In particular, it would be very interesting to understand
more quantitatively the condition of appearance of loops for example.

\subsection{Summary: Effect of space on networks}
\label{sec4F}

We presented in this chapter the most important classes of models of
spatial networks formation. When spatial constraints are present, the
network structure is affected and we summarize the main effects below.

\begin{itemize}
\item{} {\it Effect of space embedding on degree distribution,
    assortativity, clustering, and average shortest path}\\

  Spatial constraints imply that the tendency to connect to hubs is
  limited by the need to use small-range links. This explains why the
  degree distribution has a cut-off depending on the node density and
  the almost flat behavior observed for the assortativity. Connection
  costs also favor the formation of cliques between spatially close
  nodes and thus increase the clustering coefficient.  The increasing
  tendency to establish connections in the geographical neighborhood
  leads to a large average shortest path. If the density of shortcuts
  is too low, the average shortest path can behave as $N^{1/d}$ as in
  a $d-$dimensional lattice. In the opposite case, a large density of
  shortcuts leads to a logarithmic behavior typical of small-world
  networks.

\item{} {\it Effect of space embedding on centrality}\\

  Spatial constraints also induce large betweenness centrality
  fluctuations. While hubs are usually very central, when space is
  important central nodes tend to get closer to the gravity center of
  all points. Correlations between spatial position and centrality
  compete with the usual correlations between degree and centrality,
  leading to the observed large fluctuations of centrality at fixed
  degree: many paths go through the neighborhood of the barycenter,
  reinforcing the centrality of less-connected nodes that happen to be
  in the right place; this yields larger fluctuations of the
  betweenness centrality and a larger number of `anomalies'.

\item{} {\it Effect of spatial embedding on topology-traffic correlations}\\

  As we can see with growing network models where preferential
  attachment plays a role, spatial constraints induce strong nonlinear
  correlations between topology and traffic. The reason for this
  behavior is that spatial constraints favor the formation of regional
  hubs and reinforce locally the preferential attachment, leading for
  a given degree to a larger strength than the one observed without
  spatial constraints. Moreover, long-distance links can connect only
  to hubs, which yields a non-linear behavior of the distance-strength
  with values $\beta_{d}>1$: the existence of constraints such as
  spatial distance selection induces some strong correlations between
  topology (degree) and non-topological quantities such as weights or
  distances.

\item{} {\it Space and optimal networks}\\

  Spatial constraints combined with efficiency lead to a large variety
  of networks, with the existence of hub-and-spoke organization. Also,
  the emergence of complex structure in traffic organization could be
  explained by an optimization principle. In particular, strong
  correlations between distance and traffic arise naturally as a
  consequence of optimizing the average weighted shortest path. In the
  optimal network, long-range links carry large traffic and connect
  regional hubs dispatching traffic on a smaller scale ensuring an
  efficient global distribution. These results suggest that the
  organization of the traffic on complex networks and more generally
  the architecture of weighted networks could in part result from an
  evolutionary process.  In the case of optimal networks, fluctuations
  and resilience to attack lead naturally to the formation of loops
  which shed some light on the evolution of various systems.

\end{itemize}


\section{Processes on spatial networks}
\label{sec5}

We will focus in this chapter on various processes which take place on
spatial networks and the guiding idea is to focus on the effects of
space. In this respect, we will not address the general results
concerning complex scale-free networks as there are already many
reviews and books on these (see for example
\cite{BBVBook:2008,Havlin:2010,Newman:2010}) but we will try to
address specifically effects which can be related to the spatial
aspects of networks. In particular, we will essentially focus on
studies using the small-world model with shortcuts located at random
on a $d$-dimensional underlying lattice. This model which interpolates
nicely between a `pure' spatial lattice and a `pure' random graph with
one parameter only, represents an interesting playground to test
theoretical ideas and models.

We will discuss six different processes. We will begin with the study
of Ising model on spatial networks as it represents the archetype of
phase transition and could be used in the modelling of very different
systems. We will then discuss random walks and synchronization on
spatial networks, followed by the important problem of navigating and
searching in a spatial network. We will then discuss the robustness
and resilience of spatial networks, having in mind more practical
examples such as the robustness of power grids for example. Finally,
we will discuss the effect of space on the spread of disease.

\subsection{Ising model}
\label{sec5A}

It is well known in statistical physics that long-range interactions
affect the behavior of most systems (see for example
\cite{Mukamel:2008}) and we expect that long shortcuts in the
small-world lattice will indeed change the thermodynamics of the usual
Ising model on a regular lattice. In this section, we start by
recalling the main results on the Ising model on lattices and we will
then review the results on spatial networks such as the Watts-Strogatz
small-world, and Apollonian networks. We will end this section with a
general discussion on the effect of a spatial embedding on the
critical behavior of phase transitions.

\subsubsection{Generalities on the Ising model on lattices}

The Ising model is a pillar of statistical physics and numerous books
and reviews can be found on this subject. It was developed and studied
in the framework of phase transitions, but the simplicity of the Ising model is the main
reason for its ubiquity and for its various applications to many fields, even far
from traditional physics. Indeed, in a first approach
it is tempting for example to describe the behavior of an agent by a
binary variable. It could be the answer `yes' or `no' to a certain
question or the choice between `A' or `B' for example. The Ising model
then appears as a very useful model for describing collective
phenomena in social systems, such as opinion formation and dynamics
for example (for applications of statistical physics to social
systems, we refer the interested reader to the review
\cite{Castellano:2009} and references therein).

The Ising model is usually defined on a lattice and on each node,
there is an Ising spin which is an object which can take two values
only: $\sigma_i=\pm 1$. The interaction between spins is described by
the Hamiltonian
\begin{equation}
H=-\sum_{i\neq j}J_{ij}\sigma_i\sigma_j
\label{eq:hamil}
\end{equation}
where $J_{ij}$ represents the energy gain if two spins are
aligned. The usual ferromagnetic Ising model on a lattice is obtained
for $J_{ij}=J>0$ for neighbors on a lattice. At zero temperature
$T=0$, the ground state of the Hamiltonian Eq.~(\ref{eq:hamil}) with
ferromagnetic couplings is obtained for all spins aligned and the
magnetization is then $M=\sum_{i}\sigma_i/N=\pm 1$. Depending on the
dimension of the lattice, we obtain different collective behaviors that
we recall here briefly (for the ferromagnetic case).
\begin{itemize}
\item{} For one-dimensional
systems, the ordered ferromagnetic configuration is obtained only for
$T=0$ and there is no phase transition at finite temperature. 
\item{} For $d\geq 2$, the situation is different: at temperatures low
  enough, an ordered state is possible (with $M=\pm 1$) while at high
  temperature this order is destroyed and $M=0$ (for large
  $N$). The competition between the thermal agitation and the
  interaction leads in fact to a non-zero temperature transition $T_c>0$
  (for $N\to\infty$).
\item{} The upper critical dimension of the Ising model is $d_c=4$
  which means that above $4$, the critical exponents are those of the
  mean-field obtained in the infinite dimensional case. 
\item{} In the infinite dimensional case or equivalently in the case of a fully
  connected graph, the behavior of the Ising model is completely
  understood with the so-called mean-field theory. This approach is
  appropriate when the number of neighbors is very large, in which
  case fluctuations are negligible and only the average number of
  spins up and down is important.
\end{itemize}

\subsubsection{Ising model on small-world networks}

The first studies on the Ising model were done on lattices and later
on planar networks. In particular, the critical behavior
on a random planar lattice has been elucidated by
Boulatov and Kazakov in an important paper \cite{Boulatov:1987}. More
recently, Ising models were also studied on complex networks (see for
example \cite{BBVBook:2008}) and we will focus here on the
Watts-Strogatz network where we will discuss the effect of long-range
links on the nature and existence of a transition in the model.

In the complex network field, very quickly after Watts and Strogatz
published their paper, Barrat and Weigt \cite{Barrat:2000} studied the
one-dimensional WS model where at $p=0$, the network is a
one-dimensional ring where each node is connected to its $2m$ nearest
neighbors by ferromagnetic bonds. In this situation the transition
temperature is $T_c=0$ and the question is how the long-range links
will modify this one-dimensional result. In \cite{Barrat:2000}, the authors use
the replica approach to study this problem. They found that at high
temperature, the paramagnetic phase is stable for any value of $p$.
For small values of $p$, an expansion in terms of $p\xi_0$ can be
performed where $\xi_0\sim \exp(Jm(m+1)/T)$ is the correlation length
of the $1d$ system, and implies that the first order approximation
breaks down for $p\xi_0\sim 1$. This result means that for finite $p$,
there is a transition temperature given by \cite{Barrat:2000}
\begin{equation}
T_c\sim \frac{1}{\log p}
\label{eq:Tc}
\end{equation}
We can understand this result using the following argument. In the
one-dimensional Ising model, the correlation length is $\xi_0$ which
is the typical size of domains of correlated spins (ie. having the
same value). The average number of long-range links connected to a
domain is then of order $p\xi_0$ and it is thus clear that when
$p\xi_0\gg 1$ the numerous shortcuts will lead to a mean-field
behavior. 

In \cite{Herrero:2002}, Herrero studied numerically the Ising
model on WS networks constructed with lattice of larger dimensions
($d=2$, $3$). In these networks, the presence of disorder, even small,
leads to a change of universality class going from Ising like at $p=0$
to mean-field type for $p>0$ (as signalled by mean-field values of the
critical exponents).  The $1d$ argument for the transition temperature
is also still correct and $T_c$ is such that the spin correlation
length $\xi_0$ is of order the typical size of small-world clusters
$N^*\sim 1/p$ (see section \ref{sec4C1}) leading to the behavior
\begin{equation}
T_c(p)\sim-1/\log p
\end{equation}
for $d=1$, and for $d>1$:
\begin{equation}
T_c(p)-T_c(0)\sim p^{1/\nu d}
\end{equation}

Chatterjee and Sen \cite{Chatterjee:2006} and Chang, Sun, Cai
\cite{Chang:2007} studied the Ising model on one-dimensional WS
network with shortcuts whose length is distributed according to
$P(\ell)\sim \ell^{-\alpha}$ (see \cite{Kleinberg:2000} and section
\ref{sec4C2}). According to \cite{Chatterjee:2006}, there is a finite
transition temperature for $0<\alpha<2$. Their results seem to
indicate that for $\alpha<1$ the behavior is mean-field and that for
$1<\alpha<2$, the system has a finite-dimensional behavior. This last
result is however still under debate as in \cite{Chang:2007} the
numerical results seem to indicate that for the whole range
$0<\alpha<2$, the behavior is mean-field like.

The Ising model was also studied on Apollonian networks in
\cite{Andrade:2005} where it is shown that for a model with decaying
interaction constants of the form $J_n\propto n^{-\alpha}$ (where $n$
denotes the generation number), the critical temperature depends on
the number of generations $n$ as $T_c(n)\sim n^{\tau(\alpha)}$ where
$\tau$ varies continuously from $\tau(0)=1$ to $\tau(\infty)=1/2$
indicating a new kind of critical behavior.

\subsubsection{Critical fluctuations}

In \cite{Bianconi:2010}, Bradde et al. investigate from a general
point of view the role of the spatial embedding on the critical
behavior of phase transitions. They considered the Ising model as a
prototype for studying the complex behavior induced by space. The
Ising spins are located on the nodes of a $d-$dimensional euclidean
space and two nodes $i$ and $j$ are connected with a probability
$p_{ij}$ which reads
\begin{equation}
p_{ij}\simeq \theta_i\theta_jJ(|r_i-r_j|)
\end{equation}
where $r_{i(j)}$ denotes the position of node $i(j)$. The hidden
variables $\theta_i$ are found by fixing some conditions such as the
average degree for example. 

The important quantity here appears to be the spectrum $\rho(\lambda)$
associated to the matrix $p_{ij}$ (which is similar to the adjacency
matrix). We denote by $\Lambda$ its largest eigenvalue, and by
$\lambda_c$ the spectral edge equal to the average value of the second
largest eigenvalue $\lambda_c=\langle\lambda_2\rangle$ and such that
for $\lambda<\lambda_c$ the spectrum is self-averaging. The spectral
gap is defined as $\Delta_N\equiv\Lambda-\lambda_c$. Under the
assumption that the gap is self-averaging in the thermodynamic limit
and $\Delta_N\to\Delta$, Bradde et al. distinguish two possible
behaviors according to the value $\Delta$. If $\Delta>0$ when $T\to
T_c$, the fluctuations are mean-field. If $\Delta=0$ then the
fluctuations are mean-field or not depending if $\rho(\lambda)$ is
vanishing faster or slower than $\lambda_c-\lambda$ close to the
edge. 

For homogeneous networks (obtained with $\theta_i=\theta$), Bradde et
al. recovered in the exponential case for $J(r)$ the classical
Ginsburg result stating that for $d<4$ critical fluctuations are
relevant. For a power law behavior $J(r)\sim r^{-\alpha}$, the
mean-field approximation is exact for $d<\alpha<3d/2$ and for
$\alpha>3d/2$, non-trivial exponents are expected. For $\alpha<d$, a
non-zero gap appears in the spectrum suggesting a mean-field
behavior. In addition, Bradde et al. studied the case of complex
networks embedded in a $d=2$ space with a power-law degree
distribution with exponent $\gamma$ and with an exponential function
for $J(r)=\exp(-r/d_0)$. For small $d_0$ the effect of space is
important and we expect non-trivial effects of space with a spectrum
close to $\lambda_c$ behaving as $\rho(\lambda)\sim
(\lambda_c-\lambda)^\delta$. For instance, they showed that for
$\gamma=4$ the mean-field behavior holds while for a larger value
$\gamma=6$, the critical fluctuations are not captured by the
mean-field. From a more general perspective, these first results seem
to indicate that this approach opens interesting perspectives in the
understanding of phase transitions in spatial complex networks.

\subsection{Random walks in spatial networks}
\label{sec5B}

Random walks on graphs is an important and old subject in both
statistical physics and in mathematics. Many books can be found on
this subject (see for example \cite{Weiss:1994}) and in this chapter
we will focus on how shortcuts in a Watts-Strogatz network will affect
the usual lattice behavior. For more results and details on random
walks on general complex networks, we refer the interested reader to
the review \cite{Boccaletti:2006} or to the book \cite{BBVBook:2008}. We also note that random
walks are also studied in connection with spectral graph theory and we
refer the interested reader to the book \cite{Chung:1997}.

\subsubsection{Quantifying the effect of shortcuts}
\label{sec5B1}

The random walk can easily be defined on a $d-$dimensional lattice,
for example in $\mathbb{Z}^d$ where the walker jumps at every time
step to one of its $2d$ neighbor with equal probability. We just
recall here the most important results.

If we denote by $p_{ij}$ the probability that a walker at node $i$
will jump to node $j$, the master equation for the probability
$P(i,t|i_0,0)$ that a walker starting at node $i_0$ at time $t=0$ is
visiting the node $i$ at time $t$ is
\begin{equation}
\partial_tP(i,t|i_0,0)=\sum_jA_{ji}p_{ji}P(j,t|i_0,0)-A_{ij}p_{ij}P(i,t|i_0,0)
\label{eq:me}
\end{equation}
where $A_{ij}$ is the adjacency matrix. A usual choice for the
transition rate is given by
\begin{equation}
p_{ij}=\frac{r}{k_i}
\end{equation}
where $k_i$ is the degree of node $i$. This choice corresponds to the
case of uniform jumping probabilities to the neighbors of $i$ and
where $r=\sum_jp_{ij}$ is the total escape rate. With this simple
choice, the stationary solution of Eq.~(\ref{eq:me}) is 
\begin{equation}
P(i,\infty|i_0,0)=\frac{k_i}{N\langle k\rangle}
\end{equation}

\paragraph{Return probability.}
\label{sec5B1a}

Many important quantities can be calculated with the use of the
(modified) Laplacian operator on the network which can defined as
\begin{equation}
L_{ij}=k_i\delta_{ij}-A_{ij}
\end{equation}
where $k_i$ is the degree of node $i$. The link between random walks
on graph and spectral graph theory is known for a long time (see for
example \cite{Chung:1997}). In
particular, the eigenvalue density $\rho(\lambda)$ of this operator
gives the return probability $P_0(t)$ (which is the probability that
starting from the origin the walker returns to it at time $t$)
\begin{equation}
P_0(t)=\int_0^{\infty}e^{-\lambda t}\rho(\lambda)d\lambda
\end{equation}
This relation implies in particular that the long time limit of
$P_0(t)$ is controlled by the behavior of $\rho(\lambda)$ for
$\lambda\to 0$. 

For example, for a $d-$dimensional lattice, the spectral density behaves
as \cite{Economou:2006}
\begin{equation}
\rho_L(\lambda)\sim \lambda^{d/2-1}
\end{equation}
which gives the classical result
\begin{equation}
P_0^{L}(t)\sim \frac{1}{t^{d/2}}
\end{equation}
This quantity allows us to construct a typology of walks for infinite
lattice. For finite lattices, the walk is always recurrent and for
infinite lattice, the nature of the walk can be characterized by the
behavior of
the quantity
\begin{equation}
I=\int_0^TP_0(t)dt
\label{eq:I}
\end{equation}
for $T\to\infty$. We thus find the usual results, namely that for
$d\leq 2$, the walk is recurrent and for $d\geq 3$ it is transient.

In the case of an ER graph, Rodgers and Bray \cite{Rodgers:1988}
showed that the density of states behaves as
\begin{equation}
\rho_{ER}(\lambda)\sim e^{-c/\sqrt{\lambda}}
\end{equation}
leading to the following long time limit
\begin{equation}
P_0^{ER}(t)\sim e^{-at^{1/3}}
\end{equation}
where $a$ and $c$ are non-universal parameters. In the case of the ER
random graph the return probability thus decreases faster than the
lattice: the random links allow the walker to explore other regions
of the network and make its return more difficult. 

If we now think of a $d$-dimensional variant of the Watts-Strogatz
(see section \ref{sec4C1}), we add a fraction $p$ of shortcuts to a
$d-$dimensional lattice and the natural question is how these
shortcuts will affect the diffusion process. This problem was studied
numerically in \cite{Jespersen:2000b} and an analytical answer was
given by Monasson in \cite{Monasson:1999} who showed that the
eigenvalue density for the WS model can be approximately written as
\begin{equation}
\rho_{WS}(\lambda)\sim \rho_L(\lambda)\times\rho_{ER}(\lambda)\sim\lambda^{d/2-1}e^{-p/\sqrt{\lambda}}
\end{equation}
leading to the following behavior for the return probability
\begin{equation}
P_0(t)-P_0(\infty)\sim
\begin{cases}
t^{-d/2}\;\;&t\ll t_1\\
e^{-(p^2t)^{1/3}}\;\; &t\gg t_1
\end{cases}
\end{equation}
where $P_0(\infty)=1/N$, and where $t_1$ is the crossover time. In the first regime, at short times, the walker
didn't encounter shortcuts and we recover the $d$-dimensional
behavior. After a time $t$, the volume explored is of order $V\sim t^{d/2}$
and the number of shortcuts in this volume is given by $pV$. The
crossover time is thus such that $pV\sim 1$ leading to 
\begin{equation}
t_1\sim 1/p^{2/d}
\end{equation}
After this time, the walkers `feel' the small-world nature of the
network and we recover the ER behavior given by a stretched
exponential.

\paragraph{Number of distinct nodes, mean-square displacement and
  first-time return probability.}

In \cite{Almaas:2003}, the authors study the effect of shortcuts on
different quantities such as the number $N_{cov}$ of distinct sites
visited at time $t$ (which was also studied in \cite{Jasch:2001}), the
mean-square displacement $\langle r^2(t)\rangle$ and the first-return
time distribution.

\begin{figure}[!h]
\centering
\begin{tabular}{c}
\includegraphics[angle=0,scale=.40]{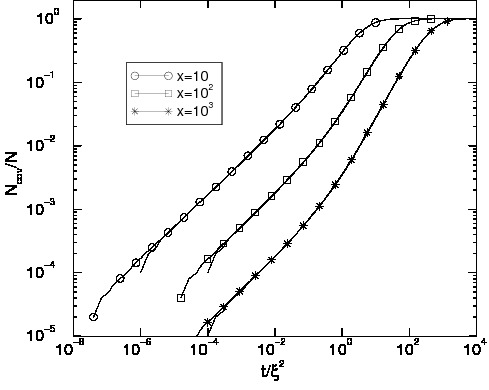}\\
\includegraphics[angle=0,scale=.40]{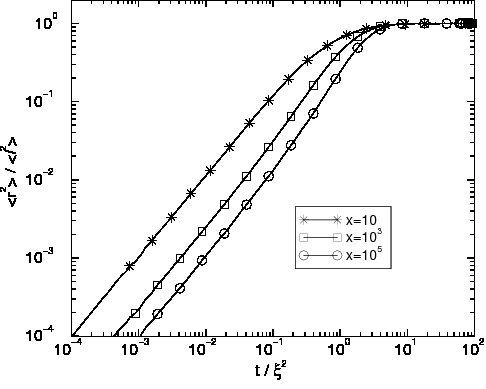}
\end{tabular}
\caption{ (a) Rescaled number of distinct sites visited at time $t$ versus
  the rescaled time. The symbols correspond to different values of $x$
  (circles: $x=10$, squares: $x=10^2$, stars: $x=10^3$) and for each
  value of $x$ two values of $p$ are used ($0.0002$ and $0.01$). The
  collapse is very good making it difficult to identify the two curves
  for each $x$. (b) Rescaled mean-square displacement versus rescaled
  time showing a perfect collapse (for each $x$ there are two curves
  for two different values of $p$). From \cite{Almaas:2003}.}
\label{fig:almaas}
\end{figure}

In the case of $d$-dimensional lattices and with a transition
probability $p_{ij}=1/k_i$, Dvoretzky and Erdos \cite{Dvoretzky:1950}
studied the average number of distinct sites and found the following
behavior
\begin{align}
N_{cov}(t)\sim
\begin{cases}
\sqrt{t}\;\;& d=1\\
t/\ln t\;\; & d=2\\
t\;\; & d\geq 3
\end{cases}
\end{align}
In the case of a one-dimensional WS network with a probability $p$ to
have a shortcut and an average number of shortcuts given by $x=pN$,
these behaviors will be modified and simple scaling arguments can help
us to understand the effect of the shortcuts \cite{Almaas:2003}. In
the WS network, the typical size of a cluster is $\xi\sim 1/p$ (see
section \ref{sec4C1}) and for a time $t\ll \xi^2$, the random walker
doesn't encounter any shortcuts and we expect a lattice-like behavior
with $N_{cov}\sim \sqrt{t}$. For very large times $t\gg N\xi$, the
walker has visited the whole network and we have $N_{cov}\sim N$. For
intermediate times $\xi^2\ll t\ll N\xi$, the walker spends on average
a time $t_1\sim\xi^2$ before encountering a shortcut.  At time $t$ the
number of shortcuts and thus of different segments visited is
$t/t_1\sim t/\xi^2$ and the total number of distinct visited sites is
then $N_{cov}\sim (t/\xi^2)\times \xi^2\sim t$
\cite{Jasch:2001,Almaas:2003}. We can summarize these behaviors in the
following scaling form
\begin{equation}
N_{cov}=NS(\frac{t}{\xi^2},x)
\end{equation}
with 
\begin{align}
S(y,x)\sim
\begin{cases}
\sqrt{y}/x,\;\;&y\ll 1\\
y/x,\;\;& 1\ll y\ll x\\
1,\;\;& y\gg x
\end{cases}
\end{align}
This scaling is confirmed by the excellent collapse observed
numerically (see Fig.~\ref{fig:almaas}).

A similar line of reasoning can be applied to the mean-square
displacement $\langle r^2(t)\rangle$ where the distance is here
computed in terms of hops from the origin (ie. the length of the
shortest path from the origin to the node where the walker is
located).  On a $d$-dimensional lattice the mean-square displacement
behaves as $\langle r^2(t)\rangle\sim t$, a behavior that we expect on
the WS networks for times such that $t\ll\xi^2$. For large times, the
mean-squared displacement saturates to a value equal to the diameter
squared (we recall here that the diameter is defined as the maximum of
all distances between pairs and is well approximated by the average
shortest path $\langle\ell\rangle$) and the saturation thus reads
$\langle r^2(t)\rangle \to \langle\ell\rangle^2$. These behaviors can
be encoded in the following scaling form \cite{Almaas:2003}
\begin{equation}
\langle r^2(t)\rangle=\langle\ell\rangle^2R(\frac{t}{\xi^2},x)
\end{equation}
with 
\begin{align}
R(y,x)\sim
\begin{cases}
y/\langle\ell\rangle^2\;\;& t\ll\xi^2\\
1\;\; & t\gg \xi^2
\end{cases}
\end{align}
which is confirmed numerically (see Fig.~\ref{fig:almaas}).

\subsubsection{Diffusion on power law Watts-Strogatz networks}
\label{sec5B2}


In \cite{Kozma:2005}, Kozma, Hastings and Korniss study diffusion in a $d-$dimensional
Watts-Strogatz networks of linear size $L$ where the shortcuts are
connecting two nodes $i$ an $j$ (located at $r$ and $r'$,
respectively) with a probability decaying as $p|r-r'|^{-\alpha}$. The
exponent $\alpha$ thus characterizes the length of shortcuts and the
$p$ is essentially the probability of a site to have a
shortcut. Analytical calculations are simpler in the case (which is
believed not to modify the universality) where multiple links are
allowed with a Poisson distribution of links with average $p/{\cal
  N}|r-r'|^{\alpha}$ where ${\cal N}$ is a normalization constant.

The equation studied in \cite{Kozma:2005} and which describes the random walk
process is
\begin{equation}
\partial_tP(r,t)=-\sum_{r'}(\Delta_{rr'}+q\Delta_{rr'}^{rand})P(r',t)
\label{eq:memb}
\end{equation}
where $P(r,t)$ is the probability to find the walker at site $r$ at
time $t$, $-\Delta_{rr'}$ is the discretized Laplace (diffusion)
operator. For example for $d=1$, one has
$-\Delta_{ij}M_{j}=M_{i+1}-M_i-(M_i-M_{i-1})$ giving 
\begin{equation}
\Delta_{ij}=2\delta_{i,j}-\delta_{j,i-1}-\delta_{j,i+1}
\end{equation}
The quantity $-\Delta_{rr'}^{rand}$ is the diffusion operator for the
random links with diffusion coefficient $q$. On the $d=1$ lattice, one
has $\Delta_{ij}^{rand}=\delta_{ij}\sum_{l\neq i}J_{il}-J_{ij}$ where
$J_{ij}$ is the number of links connecting the nodes $i$ and $j$.

The Green's function $G(r,r',t)$ for this model is the solution of the
equation (\ref{eq:memb}) with the initial condition $G(r,r',t=0)=\delta(r-r')$.
The Fourier transformed Green's function in the limit $\omega\to 0$
\begin{equation}
G(r,r')=\lim_{\omega\to 0}(i\omega+\Delta+q\Delta^{rand})^{-1}(r,r')
\end{equation}
averaged over disorder gives a translation invariant quantity
depending on $|r-r'|$ only. For $r=r'$, we obtain $\langle
G(0)\rangle$ which is related to the
return probability (and is in fact the quantity
$I$ defined in Eq.~(\ref{eq:I})) and its scaling with the size
$L\to\infty$ determines if the random walk is transient or
recurrent.

\begin{figure}[!h]
\centering
\begin{tabular}{c}
\includegraphics[angle=0,scale=1.0]{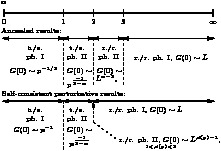}
\end{tabular}
\caption{ Phase diagram for $d=1$. The annealed approximation shows
  some differences with the self-consistent approximation for the
  quenched case. In the quenched case, the walk is transient for
  $\alpha < 2$ and recurrent for $\alpha\geq 2$. From \cite{Kozma:2005}.}
\label{fig:kozma}
\end{figure}

The authors of \cite{Kozma:2005} used a self-consistent perturbation
theory and obtained the phase diagram for all dimensions and values of
$\alpha$ (a detailed version of this work can be found in
\cite{Kozma:2007}). The results in the one-dimensional case can be
summarized in the figure (\ref{fig:kozma}) where a comparison is made with an
annealed approximation where $\Delta^{rand}$ is replaced by its average.
In particular, for $d=1$, these results are in agreement with the
analysis presented in \cite{Jespersen:2000} which predicted $\alpha=2$
as a crossover between a recurrent and transient regime. In other
words for $\alpha\geq 2$ the shortcuts do not affect the lattice
behavior while when $\alpha<2$, there are long shortcuts leading to
entirely different regions of the networks. They modify qualitatively
the lattice behavior and make it look like a large $d$-dimensional lattice, in
agreement with other processes such as spin models where the
mean-field behavior is recovered for $p\neq 0$ (in the case
$\alpha=0$.


\subsection{Synchronization}
\label{sec5Bp}

Many natural phenomena can be described as a set of coupled
oscillators. This is the case for cardiac cells, fireflies, etc. The
synchronization properties depend on the coupling pattern and many
studies were devoted to the effect of the coupling network structure
(see the review \cite{Arenas:2008}). 

Each oscillator $i$ can be described by a scalar degree of freedom
$\phi_i$ and each unit evolves according to an internal dynamics
governed by $\dot{\phi}_i=f(\phi_i)$. If these oscillators are
linearly coupled by a network, the output is a linear superposition of
of the outputs of the neighboring units and the evolution equations
take the form of
\begin{equation}
\frac{d\phi_i}{dt}=f(\phi_i)+\sigma\sum_iC_{ij}h(\phi_j)
\end{equation}
where $C_{ij}$ is the coupling matrix, $\sigma$ is the interaction
strength and $h$ a fixed output function. An important case
corresponds to $C_{ij}=L_{ij}$ where $L$ is the graph Laplacian.
In an important paper, Pecora and Carroll \cite{Pecora:1998}
showed that the stability of synchronized solutions is related to the ratio 
\begin{equation}
e=\frac{\lambda_N}{\lambda_1}
\end{equation}
and that the smaller $e$, the more stable the synchronized
solutions. 

After Watts and Strogatz published their paper \cite{Watts:1998}
synchronization properties were studied on SW networks which showed
that these networks were both able to sustain fast response due to the
long-range links and both coherent oscillations thanks to the large
clustering \cite{Lago:2000}. In particular, the ratio $e$ was
determined for SW networks \cite{Baharona:2002} where additional edges
are randomly wired on a one-dimensional ring. They obtained the
general result that the small-world topology exhibits better
synchronizability properties than ordered rings and that it can be
achieved with a small density of shortcuts \cite{Hong:2002}.

Finally, let us mention the work \cite{Lind:2004} which presents a
study of coherence in various networks, including Apollonian networks
(see \cite{Andrade:2005} and section \ref{sec4A3}). The main result is
that for the Apollonian network, the coupling threshold beyond which
coherence disappears decreases when the heterogeneity is
increased. Also, the transition to coherence in this case is shown to
be of the first order \cite{Lind:2004}.


\subsection{Navigating and searching}
\label{sec5C}

\subsubsection{Searchable networks}
\label{sec5C1}

The original $1967$ experiment of Milgram \cite{Milgram:1967}
showing that the average shortest path in North-America is around $6$
raises a number of questions. The first one is about the structure of
the social network and it is now relatively clear that enough
shortcuts will modify the scaling of $\langle\ell\rangle$ and induce a logarithmic
behavior. Another question raised by Kleinberg \cite{Kleinberg:2000}
is actually how a node can find a target efficiently with only a
local knowledge of the network (the answer being trivial if you know
the whole network). It thus seems that in some way the social network
is search-efficient- or is a {\it searchable} or {\it navigable} network- meaning that
the shortcuts are easy to find, even by having access to local
information only. In these cases, one speaks of navigability or
searchability when the greedy search is efficient.

This problem goes beyond social networks as decentralized searches,
where nodes only possess local information (such as the degree or the
location of their neighbors for example) in complex networks have many
applications ranging from sensor data in wireless sensor networks,
locating data files in peer-to-peer networks, finding information in
distributed databases (see for examples \cite{Thadaka:2007} and
references therein). It is thus important to understand the efficiency
of local search routines and the effect of the network structure on
such decentralized algorithms. 

In the case of social networks, it seems that they are composed by a
spatial part constituted of the neighbors of a node belonging to its
spatial neighborhood (such as in a regular lattice) and a purely
social component, not correlated with space and which can connect
regions which are geographically very far apart. In a search process,
it is thus natural to try these links which open the way to very
different parts of the world (in Milgram's experiment it is indeed
interesting to note that individuals were passing the message only
according geography or proximity in the space of professional
activities \cite{Killworth:1978}, as it was known to them that the
target individual was a lawyer). In order to quantify this, Kleinberg
\cite{Kleinberg:2000} constructed a $d$-dimensional Watts-Strogatz
model where each node $i$ of the lattice
\begin{itemize}
\item{} (i) is connected to all
neighbors such that their lattice distance is less than $p$ (with $p\ge 1$).
\item{} (ii) has $q$ shortcuts to node $j$ with a probability decreasing with the
distance
\begin{equation}
p(i\to j)\sim d_E(i,j)^{-\alpha}
\end{equation}
where $\alpha$ is a tunable parameter.
\end{itemize}

\begin{figure}[!h]
\centering
\begin{tabular}{c}
\includegraphics[angle=0,scale=.40]{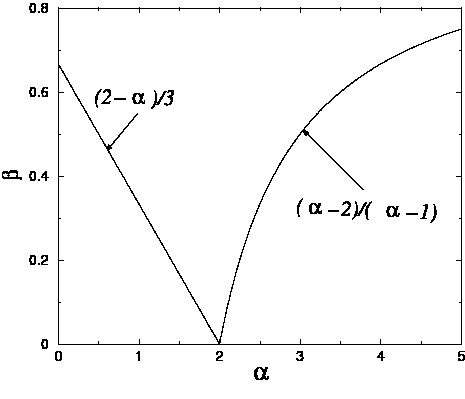}
\end{tabular}
\caption{ Lower bounds of the exponent governing the behavior of the
  average delivery time $\overline{T}$ as a function of the exponent
  $\alpha$ controlling the distribution of shortcuts. After
  \cite{Kleinberg:2000}.}
\label{fig:kleinberg}
\end{figure}

The search process is the following one: a message needs to be sent to
a target node $t$ whose geographical position is known and a node $i$
which receives the message forwards it to one of its neighbor $j$ that
is the closest (geographically) to $t$. This is the simplest
decentralized algorithm that we can construct (and which requires only
geographical information). The most important figure of merit for this
type of algorithm is the delivery time $T$ (or its average
$\overline{T}$ which is easier to estimate analytically) and its
scaling with the number of nodes $N$. Kleinberg found bounds on the
exponent of $\overline{T}$ (see Fig.~\ref{fig:kleinberg}) and
the important result is that the delivery time is optimal for
$\alpha=d$ for which it scales as $\log^2N$ while for $\alpha\neq d$, it
scales faster (as a power of $N$). This behavior can be intuitively understood: for
$\alpha>d$, long links are rare and the network looks essentially as a
lattice (with renormalized spacing). In the opposite case $\alpha<d$,
shortcuts are all long and not necessarily useful. The best case is
obtained when the shortcuts explore all spatial scales, which is
obtained for $\alpha=d$ (This result was extended in
\cite{Roberson:2006} to the case of small-world network constructed by
adding shortcuts to a fractal set of dimension $d_f$).

\subsubsection{Sketch of Kleinberg's proof}
\label{sec5C2}

Inspired from Kleinberg's original rigorous derivation
\cite{Kleinberg:2000b} of the bounds shown in
Fig.~\ref{fig:kleinberg}, we can give the following hand-waving
arguments in order to grasp some intuition on the effects of the link
distribution on the average number of steps to reach a target in a
decentralized algorithm (we will discuss in detail here for the case for $d=2$
but the extension to a generic $d$ is trivial). For the interested
reader, we also note that a detailed study of the $d=1$ case is done
in \cite{Zhu:2004}, that exact asymptotic results were obtained in
\cite{Cartozo:2009}, and that the `greedy' paths connecting a source to
the target were studied in \cite{Sun:2010} by defining a greedy
connectivity.

\begin{figure}[!h]
\centering
\begin{tabular}{c}
\includegraphics[angle=0,scale=.50]{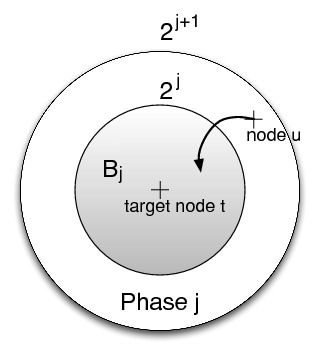}
\end{tabular}
\caption{ Line: jump out of the phase $j$ into the domain $B_j$.}
\label{fig:klein2}
\end{figure}

For $\alpha=2$ (and we assume here that $p=q=1$), the probability to
jump from node $u$ to node $v$ is given by
\begin{equation}
P(u\to v)=\frac{1}{Z}\frac{1}{d_E(u,v)^2}
\end{equation}
where $Z$ is the normalization constant given by
\begin{equation}
\begin{split}
Z=\sum_{v\neq u}\frac{1}{d_E(u,v)^2}&\simeq 2\pi\int_1^{N/2}\frac{rdr}{r^2}\\
                     &\simeq 2\pi\ln N
\end{split}
\end{equation}
implying that $P(u\to v)\sim 1/\ln N d_E(u,v)^2$ (here and in the
following we will use continuous approximation and neglect irrelevant
prefactors-for rigorous bounds we refer to
\cite{Kleinberg:2000b}). Following Kleinberg \cite{Kleinberg:2000b},
we say that the execution of the algorithm is in phase $j$ when the
lattice distance $d$ from the current node (which is holding the message)
is such that $2^j\leq d<2^{j+1}$. The largest phase is then $\ln N$
and the smallest $0$ when the message reaches the target node. The goal at this point is to compute
the average number of steps $\overline{T}$ to reach the target. For this we
decompose the problem in computing the average time duration $\overline{T_j}$ that the message
stays in phase $j$. For this we have to compute the probability that
the message leaves the phase $j$ and jumps in the domain $B_j$ defined
as the set of nodes within a distance $2^j$ to the target node $t$
(see Fig. ~\ref{fig:klein2}).
The size of this set $B_j$ is $|B_j|\sim 2^{2j}$ and the distance
between $u$ and any node of $B_j$ is $d(u,v\in B_j)\leq 2^{j+2}$. The
probability to get out of phase $j$ by using a long-range link is thus
\begin{equation}
P_{out}\sim\frac{|B_j|}{(2^{j+2})^2\ln N}\sim \frac{1}{\ln N}
\end{equation}
(the actual exact bound found by Kleinberg is $P_{out}\ge 1/(136\ln
N)$. We then have $P(T_j=i)=[1-P_{out}]^iP_{out}$ from which we obtain
$\overline{T_j}\sim\ln N$. The average time to reach the target is
then
\begin{equation}
\overline{T}=\sum_{j=0}^{\ln N}\overline{T_j}\sim \ln^2N
\end{equation}
which is the minimum time obtained for a decentralized algorithm for $\alpha=2$ (and $\alpha=d$ for the
general $d$-dimensional case).

\begin{figure}[!h]
\centering
\begin{tabular}{c}
\includegraphics[angle=0,scale=.50]{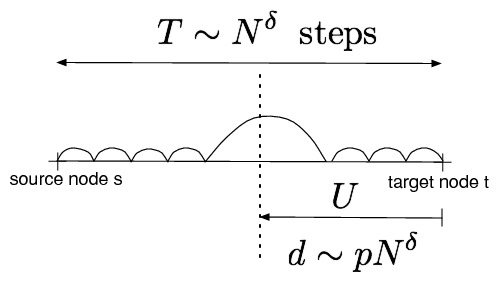}
\end{tabular}
\caption{ One dimensional representation of Kleinberg's theorem in the
  case $\alpha<2$.}
\label{fig:klein3}
\end{figure}

We now consider the minimum scaling of $\overline{T}$ in the case
$\alpha<2$ (and general $p$ and $q$). The normalization constant behaves then as
\begin{equation}
\begin{split}
Z=\sum_{v\neq u}d_E(u,v)^{-\alpha}&\simeq 2\pi\int_1^{N/2}r^{-\alpha}rdr\\
      & \sim \left(\frac{N}{2}\right)^{2-\alpha}
\end{split}
\end{equation}
We assume now that the minimum number of steps to reach the target
scales as $N^\delta$. In this case, there is necessarily a last step
along a long-range link leading to a node which is different from the
target node and which is in the region $U$ centered at $t$ and of size
$\sim pN^\delta$ (Fig. \ref{fig:klein3}).
The probability $P_i$ that this long-range link leading to $U$ at step
$i$ is given by
\begin{equation}
P_i\le \frac{q|U|}{Z}\sim N^{2\delta-2+\alpha}
\end{equation}
and the probability that it happens at any step less than $N^{\delta}$
is 
\begin{equation}
P=\sum_{i\leq N^\delta}P_i\leq N^{3\delta-2+\alpha}
\end{equation}
This probability is non-zero only if $3\delta-2+\alpha\geq 0$ leading
to the minimum possible value for $\delta$ such that $\overline{T}\sim N^\delta$
\begin{equation}
\delta_{min}=\frac{2-\alpha}{3}
\end{equation}
(the $d$-dimensional generalization would give
$\delta_{min}=(d-\alpha)/(d+1)$).

In the last case $\alpha>2$, we will have mostly short links and the
probability to have a link larger than $m$ is given by
\begin{equation}
\begin{split}
P(d_E(u,v)>m)&\sim\int_m^N\frac{rdr}{r^\alpha}\\
&\sim m^{2-\alpha}
\end{split}
\end{equation}
In the following we will use the notation $\epsilon=\alpha-2$. The
probability to have a jump larger than $N^\gamma$ for $T<N^\beta$ is
then given by
\begin{equation}
\begin{split}
P(N^\gamma,T<N^\beta)&\sim qN^\beta (N^\gamma)^{-\epsilon}\\
& \sim qN^{\beta-\gamma\epsilon}
\end{split}
\end{equation}
This probability will be non-zero for
\begin{equation}
\beta-\gamma\epsilon\geq 0
\label{eq:betagamma}
\end{equation}
Also, if at every step during a time $T\sim N^\beta$, we perform a
jump of size $N^\gamma$ the traveled distance must be of order $N$
which implies that $N^\beta N^\gamma\sim N$ leading to the condition
$\beta+\gamma=1$.  This last condition together with
Eq.~(\ref{eq:betagamma}) leads to the minimum value of $\beta$
\begin{equation}
\beta_{min}=\frac{\alpha-2}{\alpha-1}
\end{equation}
which can be easily generalized to $(\alpha-d)/(\alpha-d+1)$ in $d$-dimensions.
We thus recover the bounds $\delta_{min}$ and $\beta_{min}$ shown in
the Fig.~\ref{fig:kleinberg}.

\subsubsection{Searching in spatial scale-free networks}
\label{sec5C3}

Kleinberg thus showed that a simple greedy search passing the message
to the neighbor which is the nearest to the target is then able to
find its target in a logarithmic time for a lattice with
$\alpha=d$. The lattice considered in Kleinberg's paper is a variant
of the WS model and has a low degree heterogeneity. When a large
degree heterogeneity is present it is not clear that the greedy search
used in \cite{Kleinberg:2000} will work well, as it might be best to
jump to a hub even if there is neighbor closer to the target node. In
order to understand the effect of heterogeneity in spatial networks,
Thadakamalla et al. \cite{Thadaka:2007} studied decentralized searches
in a family of spatial scale-free network where the nodes are located
in a $d$-dimensional space:
\begin{itemize}
\item{} With probability $1-p$ a new node $n$ is added and is connected
to an existing node $i$ with a preferential attachment probability weighted by
the distance (see section \ref{sec4D})
\begin{equation}
\Pi_{n\to i}\propto k_iF(d_E(i,j))
\end{equation}
where $F(d)$ is a decreasing function of distance (and can be chosen
as a power law $d^{-\sigma}$ or as an exponential $\exp(-d/d_0)$.
\item{} With probability $p$, a new edge is connecting existing nodes
with probability
\begin{equation}
\Pi_{i\leftrightarrow j}\propto k_ik_jF(d_E(i,j))
\end{equation}
\end{itemize}

The authors of \cite{Thadaka:2007} investigated the following search
algorithms which cover a broad spectrum of possibilities:
\begin{enumerate}
\item{} {\it Random walk:} The message goes from a node to one of its
  randomly chosen neighbor.
\item{} {\it High-degree search:} The node passes the message to the
  neighbor which has the largest degree. This algorithm is already
  very efficient for non-spatial network \cite{Adamic:2001}.
\item{} {\it Greedy search:} This is the algorithm used in Kleinberg's
  study \cite{Kleinberg:2000} and where the node $i$ passes the message to the
  neighbor which is the closest to the target (ie. with the smallest $d_E(i,t)$).
\item{} {\it Algorithms 4-8:} The node passes the message to the
  neighbor which minimizes a function $F[k_i,d_E(i,t)]$ which depends
  both on the degree of the node and its distance to the
  target. The function $F$ considered here are: (i) $F_1[k,d]=d-f(k)$
  where $f(k)$ is the expected maximum length of an edge from a node
  with degree $k$; (ii) $F_2[k,d]\propto d^k$;  (iii) $F_3[k,d]=d/k$;
  (iv) $F_4[k,d]\propto d^{\ln k+1}$; (v) $F_5[k,d]=d/(\ln k+1)$. 
\end{enumerate}

The main result obtained in \cite{Thadaka:2007} is that algorithms
$(4-8)$ perform very well and are able to find a path between the source
and the target whose length is at most one hop more than the average
shortest path. This result is surprising: the calculation of the
shortest average path requires the knowledge of the whole network,
while the algorithms used here have only local information. This
success can be attributed to the fact that the scale-free networks considered
in this study have hubs which allows to find efficiently the target. It should
also be noted that the greedy search performance is not too bad but
with the severe drawback that in some cases it doesn't find the target
and stays stuck in a loop, which never happens with algorithms $(4-8)$.
Similar results were obtained for different values of $p$ and
$\sigma$. These results allow the authors to claim that the class of
spatial networks considered here belong to the class of searchable
networks. The authors checked with these different algorithms that it
is also the case for the US airline network.

Finally, we mention Hajra and Sen \cite{Hajra:2007} who studied the
effect of the transition scale-free/homogeneous network
(\cite{Manna:2002} and section \ref{sec4D2b}) on the navigability for
three different search algorithms. In particular, they showed that the
effect of the transition on navigability is marginal and is the most
pronounced on the highest degree-based search strategy which is less
efficient in the power-law regime.

\subsubsection{Navigability and metric space}
\label{sec5C4}

In \cite{Boguna:2009}, Bogu{\~n}\'a, Krioukov, and Claffy, studied the
navigability on a network constructed on a hidden metric space (see
section \ref{sec4A2}) where the probability to connect two nodes (with
degree $k$ and $k'$ drawn from $P(k)\sim k^{-\gamma}$) lying for
example on a one-dimensional space is given by
\begin{equation}
r(d;k,k')=\left(1+\frac{d}{d_c(k,k')}\right)^{-\alpha}
\end{equation}
where $\alpha>1$ controls the clustering (the larger $\alpha$ and the
larger the clustering coefficient) and where $d_c(k,k')\sim kk'$
(other forms are possible).  Bogu{\~n}\'a, Krioukov and Claffy analyzed the result
of a greedy algorithm on these networks with different values of
$\alpha$ and $\gamma$ and found that the average shortest path
decreases with smaller $\gamma$ and larger $\alpha$. A large
clustering and degree homogeneity thus favors the greedy algorithm, a
somewhat expected result, since hubs allow a fast routing and large
clustering an efficient local search. Strong clustering also favors
the fraction of successful paths (ie. finding the target). If $\gamma$
is too large (larger than $2.6$) the percentage of successful paths
becomes too small which is related to the small number of hubs. These
results can be summarized
\begin{figure}[!h]
\centering
\begin{tabular}{c}
\includegraphics[angle=0,scale=.30]{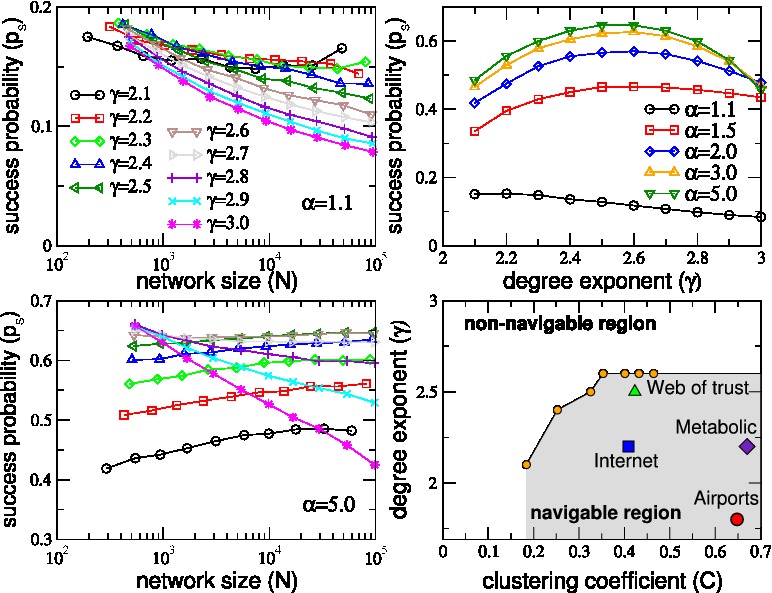}
\end{tabular}
\caption{ (a,b) Success probability $p_S$ versus $N$ for various
  values of $\gamma$ and for (a) weak clustering and (b) strong
  clustering. (c) Success probability $p_S$ versus $\gamma$ for
  various values of $\alpha$ and for fixed system size $N\approx
  10^5$. (d) The solid line separates the navigable from the non-navigable
  region. From \cite{Boguna:2009}.}
\label{fig:serrano}
\end{figure} 
in Fig.~\ref{fig:serrano}(d) where real-world networks are shown in
the clustering-degree distribution exponent plane $(C,\gamma)$. The
solid lines indicate the separation between a navigable region and a
non-navigable one where the efficiency of the greedy algorithm
decreases with the system size. Interestingly enough, we see that all
the real-world examples of communication and transportation, social
and biological networks considered here are in the navigable region.

These various results (summarized in \cite{Krioukov:2010}) show that
many real-world networks are navigable which means that they support
efficient communication without a global knowledge of the
network. From a more theoretical perspective, the authors of
\cite{Krioukov:2010} suggest the interesting idea that real-world
networks have underlying hidden metric space which could be at the
source of their navigability. 

In the case of the Internet it is however important to note that the
navigability is not due to the knowledge of geography alone. Indeed,
the Internet can be mapped to a hyperbolic space \cite{Boguna:2010}
but the distance in this hyperbolic space encodes multiple factors
(such as geography and also political, economical, communities, etc.)
so that navigability results from a combination of all these factors.

\subsubsection{Effect of cost}
\label{sec5C4p}

In \cite{Li:2010}, Li et al. study the navigability on a
Watts-Strogatz network constructed on a two-dimensional square lattice
where the pairs of nodes are connected with a probability $p\sim
1/d^\alpha$. The important difference with the case considered by
Kleinberg \cite{Kleinberg:2000} is that there is cost associated to
the long-range shortcuts and that their number is limited by a total value
$\sum_{ij}d_E(i,j)<\Lambda$ where $\Lambda$ represents the amount of
resources available for the shortcuts. Li et al. study in particular
the average delivery time and found numerically that it has a minimum
value for $\alpha\approx 3$ (and $\alpha=2$ for $d=1$) suggesting that
the optimal value might be $\alpha=d+1$ in contrast with Kleinberg's
result $\alpha=d$. In addition, the optimal delivery time seems to
behave as $N^{1/d}$ in sharp contrast with the behavior $(\log N)^2$
obtained by Kleinberg. These result suggest that cost constraints are
relevant for the navigability condition and it would be interesting to
extend Kleinberg's result in this case and to understand the main
mechanisms and how Kleinberg's derivation is affected by the cost
constraint.

\subsubsection{Routing in social networks}
\label{sec5C5}

In the study \cite{Liben-Nowell:2005}, the authors tested the
navigability on a social network of bloggers and showed that
geographic information is sufficient to perform global routing (as in
Kleinberg's case on a small-world network \cite{Kleinberg:2000}) in a
non-negligible fraction of cases: for $13\%$ of the pairs
source-target, the message reaches the target based on geographic
information only. On the other hand, these authors also showed that
the probability that two nodes separated by a distance $d$ decays as
$P(d)\sim 1/d^{\alpha}$ with $\alpha\approx 1$ (see section
\ref{sec3B4}). We are thus facing an apparent contradiction here: for
a two-dimensional space (which is a good approximation of the US), the
navigability is obtained for $\alpha=2$ according to Kleinberg's
result and not for $\alpha=1$ (see previous section). The authors of
\cite{Liben-Nowell:2005} suggest in fact that the Kleinberg
navigability condition can be generalized for different networks and
is obtained when the probability $P(u,v)$ that two nodes are connected
scales as
\begin{equation}
P(u,v)\sim \frac{1}{R(u,v)}
\label{eq:rank}
\end{equation}
where $R(u,v)$ is the `rank' of $v$ with respect to $u$ and which
counts the number of individuals living between $u$ and $v$. This
result from the argument that distance alone has no meaning and that
one should include the density. Indeed, two individuals separated by a
distance of $500$ meters in rural areas probably know each other,
which is certainly not the case in a dense urban area. If the
two-dimensional density $\rho$ is uniform, the rank then scales with
distance $d$ as
\begin{equation}
R\sim d^2
\end{equation}
which leads to $P(d)\sim 1/d^2$ and the navigability condition
$\alpha=d$ is recovered. The navigability condition of Kleinberg thus
assumes implicitly a uniform density and could be modified for other
distributions. Indeed, the authors of \cite{Liben-Nowell:2005} showed
that networks based on friendship condition of Eq.~(\ref{eq:rank}) are 
indeed navigable and contain discoverable shortcuts. The apparent
paradox is thus solved: for the social network studied in
\cite{Liben-Nowell:2005}, we observe $R\sim d$ which explains
$P(d)\sim 1/d$ and the network is still navigable.

\subsection{Effect of space on robustness and resilience}
\label{sec5D}

Many important infrastructures in our modern societies are structured
in the form of networks and it is natural to question their
robustness. Large-scale collapses such as blackouts or Internet
outages are spectacular illustrations of this problem. Indeed, very
quickly after the first publications on small-worlds and scale-free
networks, the first studies on robustness appeared
\cite{Cohen:2000}. In particular, an important distinction appeared
between robustness in case of random failures and in case of targeted
attacks. In particular, one of the most important results
\cite{Cohen:2000} states that scale-free networks are very resilient
to random failures but very sensitive to targeted attacks on hubs for
example.

We will review in this chapter, essentially two types of studies which
considered the effect of space on the resilience of networks. The
first type concerns percolation studies which analyze the effect of
the underlying lattice on the breaking process. Other studies
considered the effects of additional shortcuts on lattice with the
study of Watts-Strogatz networks. Another set of studies focused on
the vulnerability of power grids and on the failure cascade process.


\subsubsection{Percolation and small-worlds}
\label{sec5D1}

The topological effect of the removal of bonds or nodes can be
understood in the framework of percolation (see for example the books
\cite{Bunde:1991,Stauffer:1992}). In the case of regular lattices and
bond percolation with a probability $p$ that a bond is present, we
observe a percolation transition at a finite, non-universal value
$p_c$ which depends on the lattice. Below the threshold, we observe
finite clusters which size diverges at $p_c$ (which is $p_c=1/2$ for a
$2d$ square lattice and behaves as $p_c\sim 1/2d$ for large dimensions
$d$). At the threshold there is a giant component, or infinite
percolating cluster which has a universal fractal dimension $d_f$
independent from the lattice. In addition, in the vicinity of $p_c$,
the correlation length $\xi$ which measures the linear size of finite
clusters and the probability $P_\infty$ for a node to belong to the
infinite percolating cluster scale as
\begin{align}
\xi &\sim |p_c-p|^{-\nu}\\
P_\infty& \sim (p-p_c)\beta
\end{align}
where the exponent are in the two dimensional case $\beta_{2}=5/36$,
and $\nu_{2}=4/3$ and in the mean-field case (ie. for $d\geq 6$) $\beta_{MF}=1$
and $\nu_{MF}=1/2$.

The natural question to ask at this point is then the effect of
shortcuts on this standard percolation behavior. In
\cite{Callaway:2000,Moore:2000b}, the authors use a generating
function formalism in order to compute various quantities and we
briefly recall this derivation for the one-dimensional WS network
where each site is connected to its $k^{th}$ nearest neighbors and where
additional shortcuts are added between randomly chosen pairs of sites
with probability $\phi$, giving an average of $\phi k N$ shortcuts in
total. The first quantity which is needed is the generating function
\begin{equation}
H(z)=\sum_{n=0}^\infty P(n)z^n
\end{equation}
where $P(n)$ denotes the probability that a randomly chosen node
belongs to a cluster of $n$ sites other than the giant percolating
cluster. In other words, below the transition $H(1)=1$ and above the
percolation threshold we have $H(1)=1-P_\infty$. The quantity $P_0(n)$ which
is the
probability that a randomly chosen node belongs to a cluster of $n$
sites on the underlying lattice is given by (for the one-dimensional
case and for $n>0$)
\begin{equation}
P_0(n)=npq^{n-1}(1-q)^2
\end{equation}
where $q=1-(1-p)^k$. We now define the probability $P(m|n)$ that
there are exactly $m$ shortcuts emerging from a cluster of size $n$
and which is given by 
\begin{equation}
P(m|n)=\binom{2\phi
  kN}{m}\left[\frac{n}{N}\right]^m\left[1-\frac{n}{N}\right]^{2\phi
  kN-m}
\end{equation}
which indeed represents the number of possible ways to choose $m$ end
shortcuts with uniform probability $n/N$ within a total of $2\phi kN$.

\begin{figure}[!h]
\centering
\begin{tabular}{c}
\includegraphics[angle=0,scale=.40]{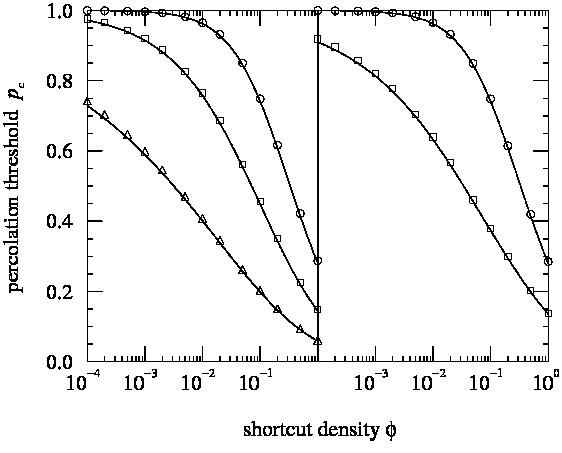}
\end{tabular}
\caption{ Numerical solution for the percolation threshold for
  $N=10^6$ for site and bond percolation (shown left and right,
  respectively). Circles represent the case $k=1$, squares $k=2$, and
  triangles $k=5$. From \cite{Moore:2000b}.}
\label{fig:moore}
\end{figure}

If we assume that there are no loops involving shortcuts, we can now
write a recursive equation on $H(z)$ by noticing that a finite cluster
consists of a local cluster of $n$ sites and with $m$ shortcuts
leading to other clusters
\begin{equation}
H(z)=\sum_{n=0}^\infty P_0(n)z^n\sum_{m=0}^\infty P(m|n)[H(z)]^m
\end{equation}
(this equation can be understood if we note that the prefactor of
$z^l$ is the probability to belong to a finite cluster of size
$l$). For large $N$, we then obtain
\begin{equation}
H(z)=H_0(ze^{2\phi k(H(z)-1)})
\label{eq:moore}
\end{equation}
where 
\begin{equation}
H_0(z)=1-p+pz\frac{(1-q)^2}{(1-qz)^2}
\end{equation}
From this equation (\ref{eq:moore}), we can then estimate various
quantities such as the average cluster size
\begin{equation}
\langle s\rangle=H'(1)=\frac{p(1+q)}{1-q-2k\phi p(1+q)}
\end{equation}
At the percolation threshold $p=p_c$, this size diverges and we
thus obtain an implicit equation for the threshold
\begin{equation}
\phi=\frac{(1-p_c)^k}{2kp_c(2-(1-p_c)^k)}
\end{equation}
The numerical solution for the percolation threshold versus the
density of shortcuts $\phi$ is shown in Fig.~\ref{fig:moore}.
Close to $p_c$ we also obtain $\langle s\rangle\sim
(p_c-p)^{1/\sigma}$ with $\sigma=1$, $P(n)\sim n^{-\tau}e^{-n/n^*}$
with $\tau=3/2$. 

These values $\tau=3/2$ and $\sigma=1$ correspond actually to the
mean-field values of percolation. This result means that the shortcuts
not only modify the percolation threshold (which is expected since
$p_c$ is not universal) but also the universality class: the WS model
resembles more to a random graph in infinite dimension. This result is
actually consistent with what happens for example for the Ising model,
which is mean-field like as long as the shortcut density is non-zero
(see section \ref{sec5A}).

\begin{figure}[!h]
\centering
\begin{tabular}{c}
\includegraphics[angle=0,scale=.40]{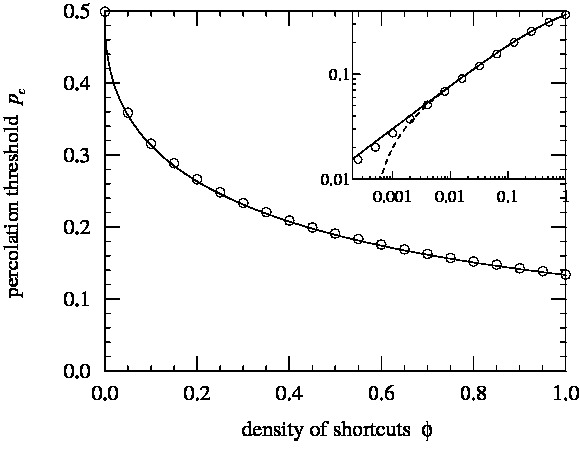} 
\end{tabular}
\caption{ Percolation threshold for a two-dimensional Watts-Strogatz
  network with a fraction $\phi$ of shortcuts (the lines are the
  analytical calculations and the circles represent the numerical
  simulation). The inset is the zoom in loglog on the small $\phi$
  region (the numerical simulation is compared with the analytical
  calculation for the quantity $1/2-p_c$). From \cite{Newman:2002d}.}
\label{fig:ziff}
\end{figure}

This one-dimensional model was extended to the $d=2$ case with the help of
high-order series expansion \cite{Newman:2002d} and Ozana
\cite{Ozana:2001} discussed finite-size scaling for this problem by noting that we
have two length scales in the problem: the length $\xi_{SW}\sim
1/p^{1/d}$ which gives the typical size of clusters connected by the
shortcuts and the length $\xi$ which is the usual cluster size for the
percolation on the underlying lattice. The main result is that
shortcuts indeed lead to a mean-field behavior and the larger their
density, the smaller the percolation threshold (see Fig.~\ref{fig:ziff}).

In \cite{Warren:2002} the authors study the percolation on a model of
lattice-based scale-free network where each node is connected to all
the neighbors in a radius $R$ distributed according to a law $P(R)\sim
R^{-\beta}$ which implies that $P(k)\sim k^{-\gamma}$ with
$\beta=d(\gamma-1)+1$. The main result is that the percolation
threshold is non-zero for $\gamma>2$ in contrast with scale-free
networks with $\gamma<3$ which display the behavior (see for example
\cite{BBVBook:2008}) $p_c(N)\to 0$. In fact, for $\gamma>2$, the
radius distribution is behaving with an exponent $\beta>d+1$ and most
of the links are short. In other words, the behavior should be the one
of an almost regular lattice (with a rescaled lattice spacing of order
$\langle R\rangle<\infty$) leading thus to a finite percolation
threshold. We note here that it would be interesting to extend to the
length for example the study made in \cite{Moreira:2009} where the
probability that an edge depends on the degree of its endpoints,
allowing an interpolation between random failures and targeted
attacks.

In \cite{Auto:2008}, Auto et al. studied percolation on Apollonian
networks (see \cite{Andrade:2005} and section \ref{sec4A3}) using
real-space renormalization. For this two-dimensional spatial,
scale-free, and planar network, the percolation threshold goes to zero
in the thermodynamic limit in agreement with general results for
scale-free networks with $\gamma<3$. The mass of the percolating
cluster however behaves as $M\sim e^{-\lambda/p}$ (where $\lambda$ is
a constant), a result reminiscent of the marginal case $\gamma=3$.

\subsubsection{Cascade of failures in infrastructures}
\label{sec5D2}

A small local failure will lead to a major breakdown when it can
propagate and reinforce itself. A simple illustration of this
phenomenon is displayed in the example of the random fuse network
introduced in $1985$ by de Arcangelis, Redner and Herrmann
\cite{Arcangelis:1985}. The random fuse network is an electrical
metaphor of material breakdown and consists of a lattice where the
bonds are fuses with threshold $i_c$ distributed according to a given
law $P(i_c)$. If the current going through a fuse is larger than its
threshold, the fuse breaks and its conductance goes to zero. The
randomness of the thresholds models the heterogeneity of a material
and the main question here is what happens if a macroscopic current
$I$ is injected in this system. If we assume that a given bond fails,
the current will be redistributed according to Kirchoff's laws on the
neighbors of the failed fuse. For example, for an infinite
two-dimensional square lattice, the current on the nearest neighbors
of the failed bond is multiplied by $\zeta=4/\pi$ (see for example
\cite{Clerc:1990}). If the distribution $P(i_c)$ has a finite support
$[i_m,i_M]$, we can then produce the following argument. In the worst
case, the failed bond had the smallest threshold possible $i_m$. The
current on the neighbors is then $\zeta i_m$ and the failure will
certainly propagate if
\begin{equation}
\zeta i_m> i_M
\end{equation}
This result implies that if the disorder is weak (ie. the difference
$\delta=i_M-i_m$ is small), the failure will very likely propagate across
the system and create a macroscopic avalanche (which in material
science is known as a `brittle' behavior). In the opposite case of
large $\delta$, we will observe the formation of microcracks which will
grow and eventually coalesce when the applied current $I$ is increased
(`ductile' behavior). This simple example shows that the flow
redistribution process after a failure is a crucial element, and
combined with the heterogeneity of the system can lead to a large
variety of behavior. A general model of failure and current
redistribution was for example discussed in \cite{Moreno:2004} but we
will focus in this chapter on cases where space plays a dominant role.

Various studies examined the vulnerability of infrastructures such as
power grids \cite{Albert:2004,Sole:2008,Kinney:2005}, Internet, or
transportation networks \cite{Latora:2005,Dallasta:2006,Wuellner:2009}
and most of these studies are not concerned with spatial aspects but
rather focused on how to measure the damage and the effect of various
attack strategies. In particular, nodes with large betweenness
centrality seem to be the weak points for all these networks, leading
to the idea that global properties of the network are needed in order
to understand the stability of these systems.

\begin{figure}[!h]
\centering
\begin{tabular}{c}
\includegraphics[angle=0,scale=1.0]{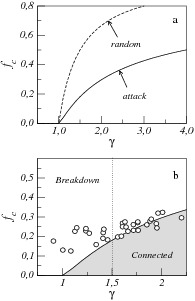}
\end{tabular}
\caption{ (a) Comparison of the critical fraction needed to disrupt an
  exponential network with distribution $P(k)\sim \exp(-k/\gamma)$. (b)
  Estimated values for attacks on $33$ different European power grids
  (circles) and the mean-field prediction of Eq. (\ref{eq:pgmf})
  (continuous line). From \cite{Sole:2008}.}
\label{fig:pgattack2}
\end{figure}

As many spatial networks with strong physical constraints, power grids
have an exponentially distributed degree (see section \ref{sec3B2}) and
in this case we can compute the effect of different removal
strategies. In particular, if we remove nodes at random, the critical
fraction $f_c$ is given by \cite{Cohen:2000}
\begin{equation}
f_c=1-\frac{1}{\kappa-1}
\end{equation}
where $\kappa=\langle k^2\rangle_0/\langle k\rangle_0$ (where the
subscript $0$ indicates that we compute the corresponding quantities
before any node removal).
\begin{figure}[!h]
\centering
\begin{tabular}{c}
\includegraphics[angle=0,scale=.30]{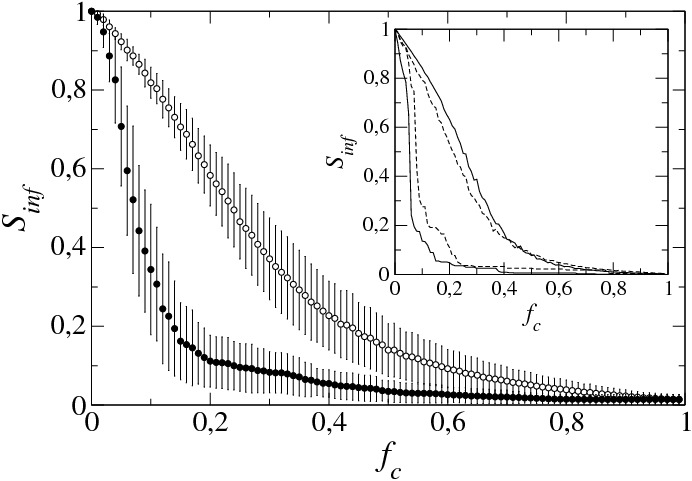}
\end{tabular}
\caption{ Effect of random removal of nodes (white circles) and
  targeted attacks (black circles) on the relative size of the largest
  component of the EU power grid. In the inset, the effect of random
  removal and targeted attacks are shown for Italy (dashed lines) and
  France (continuous line). From \cite{Sole:2008}.}
\label{fig:pgattack}
\end{figure} 
Another strategy consist in attacking the most connected nodes in
which case we can also compute a critical fraction using the argument
developed in \cite{Cohen:2000} (see Fig.~(\ref{fig:pgattack}) where
the two strategies are illustrated in the case of the European power
grid). One obtains for a network with a degree distribution of the
form $P(k)\sim \exp(-k/\gamma)$, the following condition for the
critical fraction $f_c$
\begin{equation}
1+f_c(\ln f_c-1)=\frac{1}{2\gamma-1}
\label{eq:pgmf}
\end{equation}
In the Fig.~\ref{fig:pgattack2}(a), we see that the critical
fraction of node removal is obviously smaller in the targeted attack
strategy. 
In the Fig.~\ref{fig:pgattack2}(b), the authors of \cite{Sole:2008}
compare the theoretical mean-field value of $f_c$ with actual
simulation results. The agreement is relatively good but there are
some deviations for small values of $\gamma$: the simulated values of $f_c$ seem
to be relatively constant (of order $0.2\pm 0.1$) while the mean-field
value goes to zero for $\gamma\to 1$ (which corresponds roughly to the
one-dimensional case).

\subsubsection{Failure of interdependent networks}
\label{sec5D3}

\begin{figure}[!ht]
\centering
\vspace{20 mm}
\begin{tabular}{c}
\includegraphics[angle=0,scale=.30]{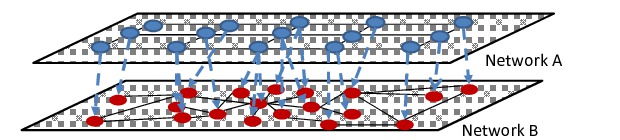}
\end{tabular}
\caption{ Schematic representation of two interdependent networks. Nodes in
  network B (communications network) are dependent on nodes in network
  A (power grid) for power; nodes in network A are dependent on
  network B for control information. From \cite{Buldyrev:2010}.}
\label{fig:inter}
\end{figure}

Space leads naturally to an increased level of interconnection between
critical infrastructure networks. Recent events such as the $2003$
blackout have reinforced the need for understanding the
vulnerabilities of this spatially coupled network. In particular, in
\cite{Buldyrev:2010} the authors modeled the effect of the coupling
between the electrical network and the Internet. They considered that
the electrical blackout in Italy in September $2003$ caused the
shutdown of power stations which led to the failure of Internet nodes,
and which in turn caused the breakdown of other power stations. The
model proposed in \cite{Buldyrev:2010} is made of two interconnected
networks (see Fig.~\ref{fig:inter}).
The two interconnected networks have their own dynamics and can
therefore not be considered as forming one single large network. The
ability of a node in A to function will depend on its B-node (and
vice-versa) which leads to the following process proposed in
\cite{Buldyrev:2010}: all B-nodes connected to A-nodes which do not
belong the giant cluster in A have to be disrupted and
vice-versa. This dynamics leads to surprising effects such the
lowering of critical thresholds, and the appearance of an abrupt
first-order transition with the size of the giant component going
abruptly to zero at the transition point. In particular, for two
connected scale-free networks, the interdependent dynamics leads to an
additional fragility: while the degree heterogeneity is an asset
against failure with many small degree nodes and a few hubs, when two
networks are interdependent, hubs can be dependent on small degree
nodes which leads to an increased vulnerability of the system. We note
here that this problem is not equivalent to one dynamics on two
connected networks, which obviously reduces to the usual case of one
single network.


\subsection{Space and the spread of disease}
\label{sec5E}

The importance of space and mobility networks appears very clearly in
the study of epidemic spread. Infectious diseases indeed spread
because people interact and travel and the modeling of disease spread
thus requires - ideally - the knowledge of the origin-destination
matrix and of the social network.

We first recall here the main results for a prototype of disease
spreading (the SIR model) on lattices and small-world networks. In
particular, we discuss the effect of shortcuts on the spreading
process. We then discuss the metapopulation model which enters the
description of the disease spread among several subpopulations
connected by a network. Finally, we end this chapter by describing
recent studies of a malware propagation among WIFI routers and the
virus spread using Bluetooth and MMS.

\subsubsection{SIR on lattices and small-world networks}
\label{sec5E1}

In theoretical epidemiology, it is customary to divide the population
into compartments, such as infectious ($I$), susceptible ($S$), or
removed individuals ($R$), and where the number of compartments
depends on the specific nature of the disease
\cite{Anderson:1992}. This description assumes that we neglect
fluctuations among individuals in the same state. Two main models have
been studied in the literature which are:
\begin{itemize} 
\item{} The SIS model. In this case, a susceptible individual in
  contact with infected individuals can become himself infected and
  will heal after a certain time and come back to a susceptible state,
  meaning that he can catch again the disease. This could for example
  describe the common cold which does not confer long lasting immunity
  and for which individuals become susceptible again after infection.
\item{} The SIR model. The difference here is that an infected
  individual becomes immunized (or has recovered) after a certain time
  immunized and cannot catch the disease again. This is for example
  the case for infantile disease that we cannot catch a second time
  (in the vast majority of cases).
\end{itemize}

In the homogeneous mixing approximation, the equations governing the
evolution of the number of susceptibles, infected, and recovered are
in the SIR case
\begin{align}
\begin{cases}
\partial_tS&=-\lambda S\frac{I}{N}\\
\partial_tI&=+\lambda S\frac{I}{N}-\mu I\\
\partial_tR&=+\mu I
\end{cases}
\end{align}
along with the condition $S+I+R=N$ and where $\lambda$ is the
probability per unit time to catch the disease and $\mu$ is the
recovery rate (ie. $1/\mu$ is the typical recovery time). The basic
reproductive rate \cite{Anderson:1992} is here given by
$R_0=\lambda/\mu$ and represents the average number of secondary
infections when one infected individual is introduced in a susceptible
population. We thus have an epidemic threshold $\lambda_c=\mu$ below
which the disease is not contagious enough to spread in the population
and above which an extensive fraction of the population gets infected.

In the homogeneous approximation, any pair of individual can interact
which is not realistic in many cases (but which usually allows for an
analytical approach). We have then to introduce the contact network
for which the nodes are the individuals and the presence of a link
between $i$ and $j$ denotes the possibility that the virus can spread
from $i$ to $j$ (or from $j$ to $i$). When we introduce a contact
network, the important question which immediately arises is the one of
the epidemic threshold and the effect of the network structure on its
value. In this section, we will examine two cases. We will first
examine the SIR model in the case of a lattice and we then proceed to
the Watts-Strogatz case where long-range links are added to a lattice.

\begin{figure}[!h]
\centering
\begin{tabular}{c}
\includegraphics[angle=0,scale=.70]{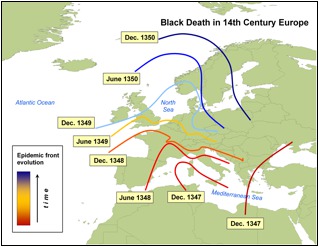}
\end{tabular}
\caption{ Spread of the Black Death during the $14$th century. The
  epidemic front spreads in Europe with a velocity of the order of
$200–400$ miles per year. From \cite{Colizza:2007}.}
\label{fig:blackdeath}
\end{figure}

\paragraph{SIR model in space.}
\label{sec5E1a}
In the pre-industrial times, disease spread was mainly a spatial
diffusion phenomenon. For instance, during the spread of the so-called
Black Death, which occurred in the $14^{th}$ century (see for example
\cite{Noble:1974}), only few travelling means were available and
typical trips were limited to relatively short distances on the time
scale of one day.  Historical studies confirm that the propagation
(see Fig. \ref{fig:blackdeath}) indeed followed a simple scheme, with
a spatio-temporal spread mainly dominated by spatial diffusion.

The simplest way to describe the spatio-temporal evolution of a
disease can be found for example in the book \cite{Murray:1993} who
described a spatial SIR model where $S(x,t)$ denotes the susceptible
individuals and $I(x,t)$ the infected and infectious individuals at
time $t$ and at location $x$. In the non-spatial version the evolution
of the densities $s=S/N$, $i=I/N$ and $r=R/N$ is described by the set of
equations (see for example \cite{Anderson:1992,BBVBook:2008})
\begin{align}
\begin{cases}
\frac{\partial s}{\partial t} &= -\lambda si\\
\frac{\partial i}{\partial t} &= +\lambda si-\mu i
\end{cases}
\end{align}
along with the condition that the sum of the three densities is one
$s+i+r=1$. In these equations, the number of individuals is absorbed
in the definition of the quantity $\lambda$. In the spatial version,
the quantities $s$ and $i$ now depend on space and in the simplest
version it is assumed that simple diffusion is the cause of their
dispersion and that they have the same diffusion constant. The
equations governing their evolution then become
\begin{align}
\begin{cases}
\frac{\partial s}{\partial t} &= -\lambda si+D\nabla^2s\\
\frac{\partial i}{\partial t} &= +\lambda si-\mu i+D\nabla^2i
\end{cases}
\end{align}
When $D\neq 0$ and $\lambda=\mu=0$, the infection front grows as $\sqrt{t}$
while the combination of $D\neq 0$ and $\lambda\neq 0$ leads in fact to a
finite velocity. A simple way to see this is given by the following
argument for the one-dimensional case \cite{Murray:1993}. Using
dimensional arguments, we will use the following rescaling
\begin{equation}
\begin{split}
i\to \frac{I}{S_0},\;\; s\to \frac{S}{S_0},\;\; x\to \sqrt{\frac{\lambda S_0}{D}}x,\;\;t\to \lambda S_0t
\end{split}
\end{equation}
and where $R_0=\lambda S_0/\mu$ is the basic reproductive rate and must
be larger than one in order to observe an outbreak
\cite{Anderson:1992} (the quantity $S_0$ is the initial number of
susceptibles). We then obtain
\begin{equation}
\begin{split}
\begin{cases}
\frac{\partial s}{\partial t} &= -si+\frac{\partial^2s}{\partial x^2} \\
\frac{\partial i}{\partial t} &= +si-\frac{1}{R_0}i+\frac{\partial^2i}{\partial x^2} 
\end{cases}
\end{split}
\label{eq:sir}
\end{equation}
We look now for a traveling wave solution of Eq.~(\ref{eq:sir}) of the form
\begin{align}
\begin{cases}
i(x,t)&=i(x-vt)\\
s(x,t)&=s(x-vt)
\end{cases}
\end{align}
where $v$ is the wavespeed which needs to be determined. Inserting
this form in the system Eq.~(\ref{eq:sir}) we obtain
\begin{equation}
\begin{split}
\begin{cases}
i''+vi'+i(S-\frac{1}{R_0})&=0\\
s''+vs'-is&=0\\
\end{cases}
\end{split}
\end{equation}
where the prime denotes the derivation with respect to the variable
$z=x-vt$. We look for a solution such that $i(-\infty)=i(\infty)=0$
and $0\leq s(-\infty)<s(\infty)=1$. Near the wave front, $s\approx 1$
and $i\approx 0$ and we obtain
\begin{equation}
i''+vi'+i(1-\frac{1}{R_0})\approx 0
\end{equation}
which can be solved and leads to
\begin{equation}
i(z)\propto \exp\left[(-v\pm\sqrt{v^2-4(1-\frac{1}{R_0})})z/2\right]
\end{equation}
This solution cannot oscillate around $i=0$ which implies that the
wavespeed $v$ and $R_0$ must satisfy the conditions $R_0>1$ and 
\begin{equation}
v\geq 2(1-\frac{1}{R_0})^{1/2}
\end{equation}
For most initial conditions, the travelling wave computed in the full
non-linear system will evolve at the minimal velocity and we
eventually obtain the epidemic wave velocity
\begin{equation}
V=2(\lambda S_0D)^{1/2}\sqrt{1-\frac{1}{R_0}}
\end{equation}

We can apply this simple model to the historical case of the Black
Death. Following \cite{Murray:1993}, the number of individuals in
Europe at that time is $\approx 85$ millions which gives a density
$S_0\approx 20/$km$^2$. The quantities $\lambda$ and $\mu$ are more
difficult to determine as the plague transmits with fleas jumping from
rats to humans. In \cite{Noble:1974} these parameters are estimated to
be $\mu\approx 15/$year and $\lambda\approx 1.0$ km$^2/$year. The basic
reproductive rate is then $R_0\approx 1.33$ which is not very
large. The main difficulty here is the estimate of the diffusion
coefficient and if we assume that news spread at a velocity of around
$160$ kms per year, we obtain $D\approx 10^4$ kms$^2/$year. Putting all
the numbers together we then find a velocity of the order $v\approx
700$ kms$/$year which is larger than historical estimates put in the
range $[300,600]$kms$/$year, but given the uncertainties on the
parameter (especially for $D$), the spatial diffusion approximation is
not unreasonable.

\paragraph{SIR model in small-worlds.}
\label{sec5E1b}

\begin{figure}[!h]
\centering
\begin{tabular}{c}
\includegraphics[angle=0,scale=.30]{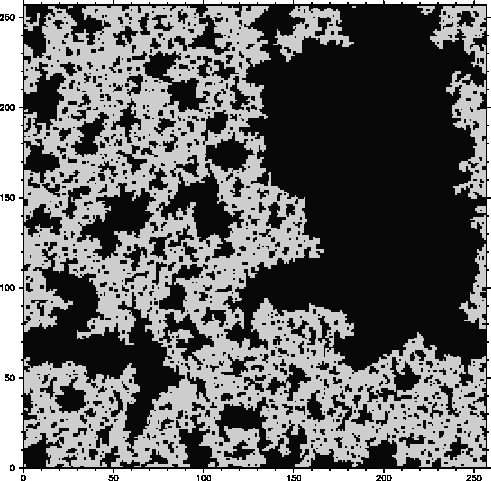}\\
\includegraphics[angle=0,scale=.30]{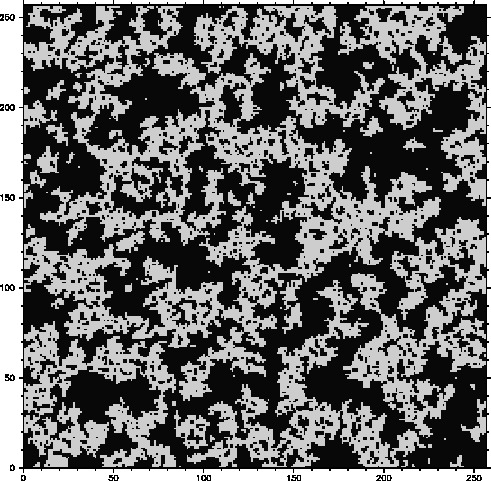}
\end{tabular}
\caption{ Snapshot of the final cluster of recovered nodes (gray
  nodes) after the
  epidemics has died out on (a) a lattice with periodic boundary
  conditions, and (b) on a two-dimensional Watts-Strogatz network with
  a fraction of shortcuts $\phi=0.01$. From \cite{Warren:2002b}.}
\label{fig:warren}
\end{figure}

An important observation which was made by Grassberger
\cite{Grassberger:1983} is that the SIR model on a graph can be mapped
to the corresponding bond percolation problem. This mapping is however
valid only for at least a very peaked distribution of infection time
\cite{Kenah:2007} (the main reason for the failure of the mapping in
%
%
the case of random infection times is the existence of correlations
introduced between neighbors of the same site). As in
\cite{Grassberger:1983}, the authors of
\cite{Warren:2002b,Warren:2002c} consider a SIR model where nodes stay
infectious during a time $\tau$ and that the transmission rate between
two connected nodes is a random variable distributed according to the
distribution $P(\lambda)$. For a given value of $\lambda$, the
probability $1-p$ not to transmit the disease is given by
\begin{equation}
1-p=(1-\lambda \delta t)^{\tau/\delta t}
\end{equation}
which in the continuous time limit and averaged over all values of
$\lambda$ gives
\begin{equation}
p(\tau)=1-\int d\lambda P(\lambda) e^{-\lambda\tau}
\end{equation}
This quantity $p$ gives the fraction of bonds completed in a given
epidemic and we thus obtain a mapping between the SIR model and a bond
percolation problem. If $p$ is larger than the percolation threshold
on the underlying lattice, there is a giant cluster which means that
an extensive number of nodes were infected.  For a peaked distribution
of $\lambda$ the epidemic threshold is given by 
\begin{equation}
\lambda_c=-\tau\ln(1-p_c)
\end{equation}
and indicates the general trend that if $p_c$ is smaller then
$\lambda_c$ is smaller too. On a square lattice, for $p>p_c=1/2$ we
obtain results such as the one shown in Fig.~\ref{fig:warren}(a) where
we see some large regions unaffected by the disease.
The authors of \cite{Warren:2002b,Warren:2002c} studied the SIR model on a
two-dimensional Watts-Strogatz model with a fraction $\phi$ of
shortcuts connecting randomly chosen nodes. The result is shown in the
Fig.~\ref{fig:warren}b and displays evidence that even a small
fraction of long-range links is enough to `homogenize' the system and
to break down the large unaffected regions. In addition, as shown in
\cite{Newman:2002d} the percolation threshold decreases with $\phi$
(see Fig.~\ref{fig:ziff}) which implies that the epidemic threshold
is also smaller due to the shortcuts which facilitate the spread of
the disease.

The same mapping between the SIR model and bond percolation implies
that for the scale-free network with exponent $\gamma$ and constructed
on a lattice (such as in \cite{Warren:2002}) the epidemic threshold is
non-zero for $\gamma>2$, while for a non-spatial scale-free network,
we expect $p_c=0$ (in the limit of infinite networks, see for example
the book \cite{Pastor:2003}). In this spatial lattice, it thus seems
that the addition of shortcuts and the existence of hubs is not enough
to counterbalance the effect of the short links and the underlying
spatial structure.

\subsubsection{From Euclidian space to networks}
\label{sec5E2}

\begin{figure}[!h]
\centering
\begin{tabular}{c}
\includegraphics[angle=0,scale=.40]{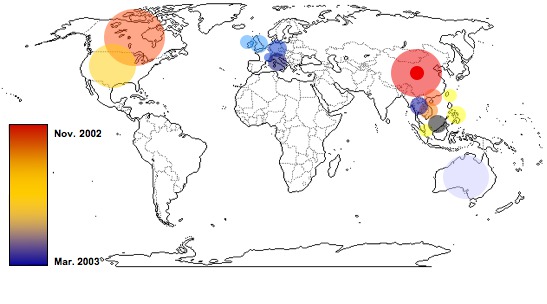}
\end{tabular}
\caption{ Global evolution of the SARS disease which started in China
  in $2002$. The color code corresponds to time and goes from the
  Hong Kong outbreak in November $2002$ to the end of the disease
  spread in
  March $2003$. Courtesy V. Colizza.}
\label{fig:sars}
\end{figure}

In modern times, the simple picture of pure spatial diffusion, with
the possible addition of few shortcuts, does however not hold
anymore. A striking example can be seen in the SARS outbreak in
$2003$. In Fig.~\ref{fig:sars} we show the spatio-temporal evolution
of this disease which started in China.
We can see on this figure that there is a clear spatial component with
the disease spreading in the south-west Asia but also with long-range
jumps to Europe and the US. This clearly shows that pure spatial
diffusion is not a good model anymore and that the global aspect of
transportation networks needs to be included in the modeling. At a
smaller scale, it has also been observed that epizooties such as the
foot-and-mouth disease outbreak in $2001$ in Great Britain's livestock
didn't follow a purely spatial pattern \cite{Keeling:2003}, but
instead, the epidemic showed local stochastic spread and rare
long-distance jumps.

An approach to the problem is provided by metapopulation models which
describe the spread of an infectious disease at a scale
where a transportation network dominates. These models describe
spatially structured populations such as cities or urban areas which
interact through a given transportation network and have been the
subject of many studies (we refer the interested reader to
\cite{BBVBook:2008} and references therein). In the simplest version
of the metapopulation model, the nodes of the network are the cities
(or urban areas) and these cities are connected through a network
defined by a weighted adjacency matrix $p_{ij}$. The element $p_{ij}$
represents the probability per unit time that an individual chosen at
random in city $i$ will travel to city $j$. In the case where we assume that
any individual of a city has the same probability to travel (in other
words if we neglect any social structure), the $p_{ij}$ can then be
estimated as $w_{ij}/N_i$ where $w_{ij}$ is the flow of travelers
between $i$ and $j$ and $N_i$ is the population of city $i$. The
evolution of the number $I_i(t)$ of infected individuals in the city
$i$ at time $t$ is then given by an equation of the form
\begin{equation}
\partial_tI_i(t)=K(S_i,I_i,R_i,...)+\Omega_i(\{I\})
\end{equation}
where $K$ represents the local term and corresponds to the spread of
the disease due to intra-city contamination. The travel term $\Omega$
depends on the mobility network and in the simple case of uniform
travel probability, can be written as (in the limit of large $I_l$)
\begin{equation}
\Omega_i=\sum_j p_{ji}I_j-p_{ij}I_i
\end{equation}
If the probability is constant, $\Omega$ is essentially the Laplacian
on the underlying network and if the mobility graph is a lattice, we recover
the usual Laplacian diffusion term. From a theoretical perspective it
would be interesting to understand precisely the crossovers when we
have different scales and networks interacting with each other.

Theoretical studies on the metapopulation model such as the existence
of a pandemic threshold were done in
\cite{Colizza:2006,Colizza:2007b,Colizza:2007c,Barthelemy:2010}. In
particular, if we assume that the travel probability is constant
$p_{ij}=p$, the pandemic threshold has an expression of the form
\begin{equation}
R_*\propto (R_0-1)p\frac{\langle k^2\rangle}{\langle k\rangle}
\end{equation}
The disease will then spread all over the world if $R_*>1$ which
implies that $R_0>1$ (ie. the disease spreads in each subpopulation)
and that the product $p\langle k^2\rangle$ is large enough. For
scale-free networks (such as for the worldwide airline network), the
quantity $\langle k^2\rangle$ is large which suggests that reducing
travel (ie. decreasing $p$) is not efficient. The fact that reducing
travel is not an efficient strategies was also confirmed in
theoretical and numerical
studies
\cite{Colizza:2007b,Colizza:2007c,Colizza:2007e,Barthelemy:2010}. Other
studies on applications of metapopulation models to the SARS or
Influenza can be found in \cite{Colizza:2007d,Colizza:2007e,Balcan:2009}.

More recently, in \cite{Balcan:2009}, small-scale commuting flows were
added to the large-scale airline network. This is an interesting
problem since we have now the superimposition of networks at
different scales. 
\begin{figure}[!h]
\centering
\begin{tabular}{c}
\includegraphics[angle=0,scale=.20]{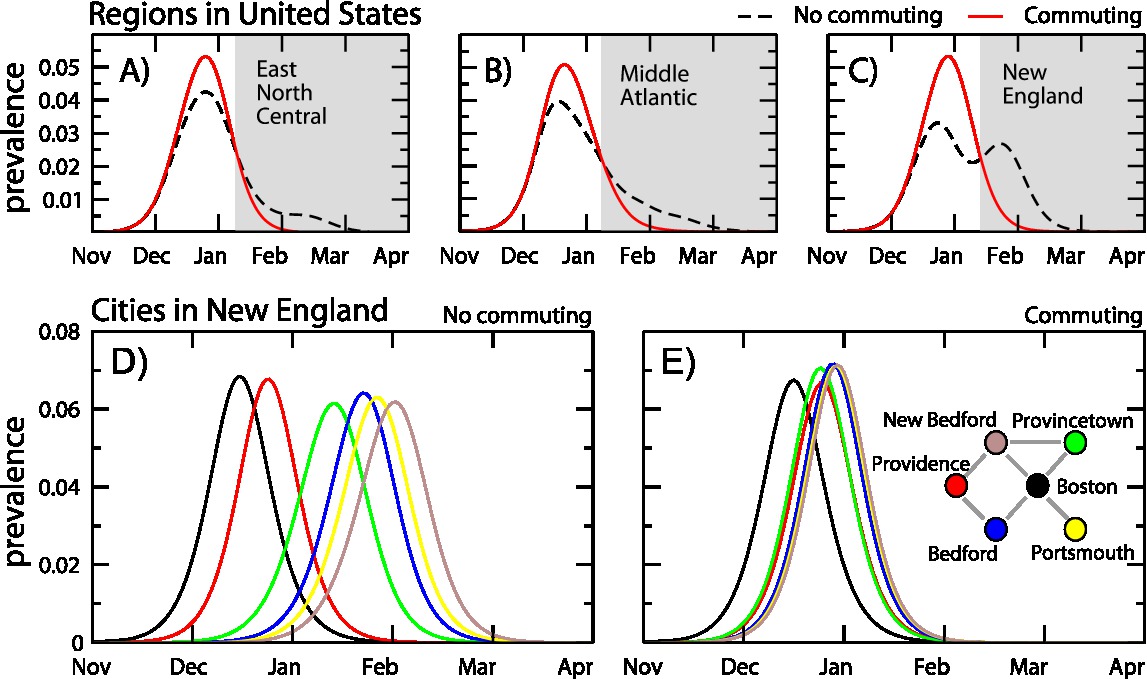}
\end{tabular}
\caption{ Comparison of prevalence profiles without and with commuting
  flows. (a-b-c) Prevalence profiles for three continental US
  regions. (d-e) Prevalence profiles for cities in New England. From \cite{Balcan:2009}.}
\label{fig:cflows}
\end{figure} 
In the Fig.~\ref{fig:cflows}, we show the different prevalence
(defined as the fraction of infected individuals in the population)
profiles obtained with and without commuting flows. At a regional
level, the prevalence is increased when commuting flows are added in
the model but the more spectacular effect is at a local level: the
inclusion of short-range mobility doesn't change the magnitude of
prevalences but largely increases the synchrony between outbreaks in
neighboring cities \cite{Balcan:2009}.


\subsubsection{WIFI and Bluetooth epidemiology}
\label{sec5E3}

\paragraph{WIFI epidemiology.}
\label{sec5E3a}

\begin{figure}[!h]
\centering
\begin{tabular}{c}
\includegraphics[angle=0,scale=.35]{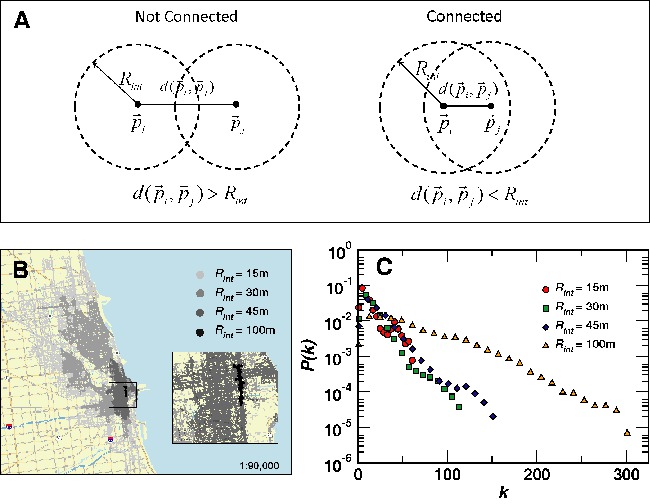}
\end{tabular}
\caption{ (a) Construction of the WIFI network as a random geometric
  graph (see section \ref{sec4A1}). (b) Giant component of the WIFI network in Chicago
  for different values $R_{int}$. (c) Degree distribution for
  different values of $R_{int}$. From \cite{Hu:2009}.}
\label{fig:wifi}
\end{figure}

In densely populated areas, the range of the WIFI routers is such that
the corresponding geometric graph can percolate through the whole
urban area (see Fig.~\ref{fig:wifi}). This fact could be used to spread
a computer virus or a malware that could have large-scale
consequences. An epidemiological model was developed in \cite{Hu:2009}
in order to study various scenarios and to assess the vulnerability of
these networks.
Using a database containing the geographic location of the wireless
routers, Hu et al. \cite{Hu:2009} focused on seven urban areas in the
US. From the set of vertices (which are here the routers), the
geometric network can be constructed for a given value of the router's
range $R_{int}$. In practice, the value of $R_{int}$ fluctuates from
$10$ to $100$ meters depending on the environment and the specific
router type. In this study, this value $R_{int}$ is assumed to be the
same for all routers (but actually the study of geometric random
graphs with random $R_{int}$ could be of interest).  An example of
such geometric network is shown in Fig.~\ref{fig:wifi} for the Chicago
area with $R_{int}=45$ meters.

\begin{figure}[!h]
\centering
\begin{tabular}{c}
\includegraphics[angle=0,scale=.35]{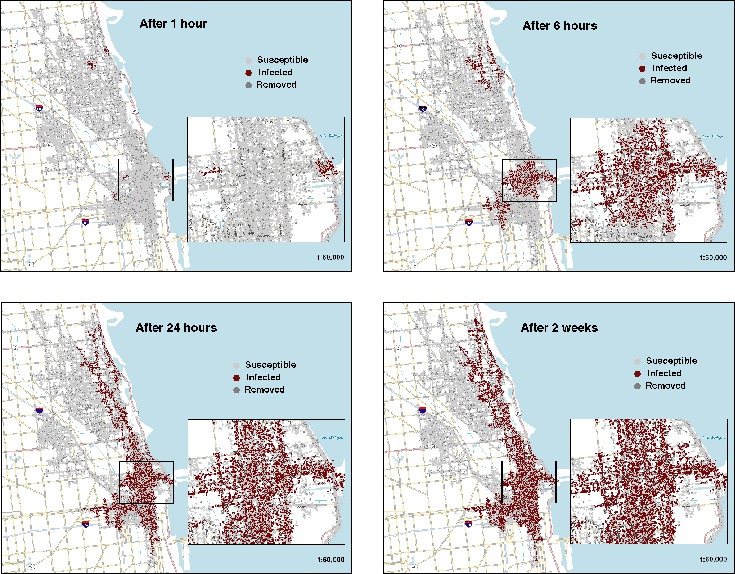}
\end{tabular}
\caption{ Spatial spread of a SIR type disease in the Chicago area
  (results obtained for $R_{int}=45m$). From \cite{Hu:2009}.}
\label{fig:wifi2}
\end{figure}

Hu et al. then use the compartmental approach of standard
epidemiology in order to describe the spread of a computer
virus. The usual scheme must however be enlarged in order to take into
account the heterogeneity observed in the security levels of WIFI
routers. For details of the epidemiological model, we refer the
interested reader to the paper \cite{Hu:2009}. Once this
epidemiological description is set up, the authors studied the spread of
the virus in various settings and on the giant component. An example
of the spread of the virus is shown for the Chicago area in
Fig.~\ref{fig:wifi2}.
This figure displays a clear spatial diffusion of the disease which
takes place on the giant component of the WIFI network. The time scales
are actually very important in this case and Hu et al. estimated that
in the Chicago case, the malware rapidly propagates in a few hours and
infects about $40\%$ two weeks after the beginning of the infection
(and about $10\%$ to $55\%$ for most areas considered in this study).

The fraction of encrypted routers and geometrical constraints of the
urban area may have large consequences on the spread of the
malware. In particular, areas with small bottlenecks may be very
sensitive to the local density of encrypted routers. This study allows for
the future to have an estimate of the immunization threshold in order to stop
the spread of a malware.

\paragraph{Mobile phone disease.}
\label{sec5E3b}

In \cite{Wang:2009}, Wang et al. study the spreading patterns of a
mobile phone virus. A mobile phone virus can spread through two very
different mechanisms.  The first mechanism is through Bluetooth
connections provided that the infected phone and the susceptible one
are separated by a short distance (typically from $10$ to $30$
meters). The second mechanism, a MMS virus can send a copy of itself
to all the numbers found in the infected's phone contact book. We thus
have a short range, spatial infection pattern superimposed with
long-range infections. However, a cell phone virus can infect only
phones with the right operating system. The market share $m$ of an OS
is then an important parameter in the study of mobile phone viruses
spread. The authors of \cite{Wang:2009} used a database of a mobile
phone provider and simulated the spread of disease using a SI type
compartment model and studied the effect of various values of
$m$. They found the existence of a threshold value (of order
$m_c\approx 10\%$) above which there is a giant component of phones
using the same OS, allowing a MMS virus to spread (added to the fact
that even for $m>m_c$, the dynamics still depends on $m$ and the
larger this parameter, the faster the dynamics). At the time of
this study, the share of the largest OS was about $3\%$ and could
explain why we didn't see yet a major mobile virus outbreak.

\section{Summary and outlook}
\label{sec6}

In this article we review the most important effects of space on the
structure of networks. In particular, the existence of a cost
associated to the length of edges leads to many important consequences
such as a large clustering and a flat assortativity. The degree
distribution is usually peaked due to the existence of strong physical
constraints although broad degree distributions can be observed for
spatial, non-planar networks. The betweenness centrality displays
strong fluctuations which can be related to the interplay of space and
degree. These different effects are by now relatively well understood
and as shown in the section \ref{sec4}, there is now a wealth of
various models which incorporate space at different levels. Finally,
in section \ref{sec5} we discussed processes on spatial networks for
various processes such as the random walk, diseases spread etc.

Despite these various advances, there are still many open problems
which could represent interesting research directions both at the
theoretical and the applied levels. We give here a non-exhaustive, subjective list
of such open problems which seems particularly interesting.

\begin{itemize}
\item{} {\it Scale and description}. In problems such as disease
  spread, the description depends strongly on the scale of
  interest. At a small spatial scale (but large enough in order to
  allow for the use of continuous limits), we can use partial
  differential equations. In contrast, at a larger scale, typically of
  the order given by a typical link of the relevant transportation
  network, the system is better described by metapopulation
  models. Given the interest of interdependent scales and networks, an
  interesting question is how we could integrate a `mixed' description
  able to interpolate between a continuous partial differential
  equation to a discrete network, metapopulation-like
  description. Another way to look at this question is too look for a
  continuous, pde-like, description of the spread over a network at a
  large scale.

\item{} {\it Interdependent networks.} Recent studies showed the
  importance of connected networks and in particular that coupled
  dynamics could lead to new and surprising results. In particular, we
  saw that networks are coupled via space and this can lead to cascade
  of failures more important than what could be expected. However, a number
  of questions remained unsolved at this point. For example what are
  the conditions for a coupling to be relevant and to affect the
  single-network behavior ? In particular, in the development of
  failures, it would be interesting to understand the space-time
  properties of blackouts.

\item{} {\it Optimal spatial networks.}

  Optimal networks are very important as we saw in this review (see
  section \ref{sec4E}). In particular, there are many different
  directions such as understanding the hub location problem in
  presence of congestion, and more generally we believe that
  statistical physics could bring interesting insights on these
  problems usually tackled by mathematicians and engineers expert in
  optimization.  Also, as recent studies suggested it, evolution
  through resilience and noise shaped loopy networks and it would be
  interesting to understand quantitatively the formation and the statistics of these
  loops.

\item{} {\it Connection with socio-economical indicators.} Data on
  spatial networks and in particular, road and other infrastructure
  networks are now available and these networks have been the subject
  of many studies. Also, with the emergence of geosocial applications
  on mobile phones for example we can expect interesting studies
  connecting spatial distributions and social behavior. This line of
  research already appeared in recent studies which tried to relate
  topological structures of networks with socio-economical
  indicators. In these studies, an important question concerns the
  correlations between topological quantities and social factors. For
  example, it would be interesting to know if we can understand some
  aspects of the spatial distribution of crime rates in terms of
  topological indicators of the road network (such as the betweenness
  centrality for example).

\item{} {\it Evolution of transportation (and spatial) networks.} How
  transportation networks evolve is an old problem and was already the
  subject of many studies in the $1970$s (see for example
  \cite{Haggett:1969}). However, apart from some exceptions (see for
  example \cite{Levinson:2006} and references therein) this problem is
  still not very well understood. We are now in the position where
  data and tools are available and we can expect some interesting
  developments in this area. In parallel to empirical studies, we also
  need to develop theoretical ideas and models in order to describe
  the evolution of spatial networks.

\item{} {\it Urban studies}. More generally, infrastructure and
  transportation networks are part of urban systems and we believe
  that the current understanding of spatial networks could help in
  understanding the structure and evolution of these systems. In
  particular, our knowledge of spatial networks could help in the
  understanding of important phenomenon such as urban sprawl and in
  the design of sustainable cities.

\end{itemize}

\vskip0.5cm

{\centerline{\bf Acknowledgments}}

I am indebted to A. Banos, A. Barrat, M. Batty, H. Berestycki,
M. Boguna, S. Boettcher, P. Bonnin, P. Bordin, J. Bouttier,
V. Blondel, M. Brede, A. Chessa, V. Colizza, T. Courtat, L. Dall'Asta,
C.-N. Douady, S. Douady, Y. Duan, J.-P. Frey, S. Fortunato,
A. Flammini, M. Gastner, C. Godr\`eche, E. Guitter, D. Helbing,
H.J. Herrmann, S. Havlin, B. Jiang, S.M. Kang, D. Krioukov,
R. Lambiotte, M. Latapy, V. Latora, D. Levinson, J.-M. Luck,
J.-P. Nadal, S. Porta, C. Roth, G. Santoboni, A. Scala, P. Sen,
E. Strano, F. Tarissan, P. Thiran, C. Uzan, A. Vespignani, Z. Zhang,
R.M. Ziff, for discussions and for giving useful suggestions and advices to improve this
manuscript at various stages.

\bibliography{bibfile}		         

\end{document}